\documentclass{elsart1p}

\usepackage{amssymb,longtable,amsmath,graphicx}

\begin{document}

\begin{frontmatter}



\title{Charged-Particle Thermonuclear Reaction Rates:\\ II. Tables and Graphs of Reaction Rates and Probability Density Functions}


\author{C. Iliadis}, \author{R. Longland}, \author{A.~E. Champagne}
\address{Department of Physics and Astronomy, University of North Carolina, Chapel Hill, NC 27599-3255, USA; Triangle Universities Nuclear Laboratory, Durham, NC 27708-0308, USA}
\author{A. Coc}
\address{Centre de Spectrom\'etrie Nucl\'eaire et de Spectrom\'etrie de Masse (CSNSM), UMR 8609, CNRS/IN2P3 and Universit\'e
        Paris Sud 11, B\^atiment 104, 91405 Orsay Campus, France}
\author{R. Fitzgerald}
\address{National Institute of Standards and Technology, 100 Bureau Drive, Stop 8462, Gaithersburg, MD 20899-8462, USA}
 
\begin{abstract}
Numerical values of charged-particle thermonuclear reaction rates for nuclei in the A=14 to 40 region are tabulated. The results are obtained using a method, based on Monte Carlo techniques, that has been described in the preceding paper of this series (Paper I). We present a low rate, median rate and high rate which correspond to the 0.16, 0.50 and 0.84 quantiles, respectively, of the cumulative reaction rate distribution. The meaning of these quantities is in general different from the commonly reported, but statistically meaningless expressions, ``lower limit", ``nominal value" and ``upper limit" of the total reaction rate. In addition, we approximate the Monte Carlo probability density function of the total reaction rate by a lognormal distribution and tabulate the lognormal parameters $\mu$ and $\sigma$ at each temperature. We also provide a quantitative measure (Anderson-Darling test statistic) for the reliability of the lognormal approximation. The user can implement the approximate lognormal reaction rate probability density functions directly in a stellar model code for studies of stellar energy generation and nucleosynthesis. For each reaction, the Monte Carlo reaction rate probability density functions, together with their lognormal approximations, are displayed graphically for selected temperatures in order to provide a visual impression. Our new reaction rates are appropriate for {\it bare nuclei in the laboratory}. The nuclear physics input used to derive our reaction rates is presented in the subsequent paper of this series (Paper III). In the fourth paper of this series (Paper IV) we compare our new reaction rates to previous results.
\end{abstract}


\end{frontmatter}

\section{Introduction}\label{intro}
In the preceding work, referred to as Paper I, we presented a method of evaluating thermonuclear reaction rates that is based on the Monte Carlo technique. The method allows for calculating statistically meaningful values: the recommended reaction rate is derived from the median of the total reaction rate probability density function, while the 0.16 and 0.84 quantiles of the cumulative distribution provide values for the {\it low rate} and the {\it high rate}, respectively. We refer to such rates as ``Monte Carlo reaction rates" in order to distinguish them from results obtained using previous techniques (which we call ``classical reaction rates"). As explained in Paper I, we will strictly avoid using the statistically meaningless expressions ``lower limit" (or ``minimum") and  ``upper limit" (or ``maximum") of the total reaction rate. For detailed information on our method, see Paper I.

In the present work, referred to in the following as Paper II, we present our numerical results of charged-particle thermonuclear reaction rates for A=14 to 40 nuclei on a grid of temperatures ranging from T=0.01 GK to 10 GK. These reaction rates are assumed to involve {\it bare nuclei in the laboratory}. The rates of reactions induced on lighter target nuclei, A$<$14, are not easily analyzed in terms of the present techniques and require different procedures (see, for example, Descouvemont et al. \cite{Des04} for an evaluation of Big Bang nuclear reaction rates using R-matrix theory). The higher target mass cutoff, A=40, was entirely dictated by limitations in resources and time. For use in stellar model calculations, the results presented here must be corrected, if appropriate, for (i) electron screening at elevated densities, and (ii) thermal excitations of the target nucleus at elevated temperatures. Details will be provided below. We emphasize that the present reaction rates are overwhelmingly based on {\it experimental nuclear physics information}. Only in exceptional situations, for example, when a certain nuclear property had not been measured yet, did we resort to nuclear theory. In the subsequent work (Paper III) we will provide the complete nuclear physics data input used to derive our new Monte Carlo reaction rates. In the fourth paper of this series (Paper IV) we compare our new reaction rates to previous results.

Paper II is organized as follows. In Sec. 2 we summarize briefly our Monte Carlo technique. An overview of the literature sources for the nuclear physics input data used to derive our results is provided in Sec. 3. Detailed examples for how to interpret the new Monte Carlo reaction rates are given in Sec. 4. The extrapolation of the laboratory reaction rate to high temperatures is described in Sec. 5, while modifications of the reaction rate that are necessary for use in stellar model calculations are discussed in Sec. 6. The calculation of reverse rates is described in Sec. 7. A summary is given in Sec. 8. Appendix A contains information regarding statistical hypothesis tests. Our Monte Carlo reaction rates are presented in tabular and graphical format in App. B.
\section{Monte Carlo reaction rates}
The expressions used for calculating thermonuclear reaction rates are given in Paper I. Each nuclear input quantity entering in the calculation of the reaction rates is associated with a specific probability density function: a Gaussian distribution for resonance energies; a lognormal distribution for measured resonance strengths, nonresonant S-factors and partial widths; a Porter-Thomas distribution for measured upper limits of partial widths; and so on. Once a probability density function is chosen for each input quantity, the total reaction rate and the associated uncertainty can be estimated using a Monte Carlo calculation. In particular, a random value is drawn for each input quantity according to the corresponding probability density function and the total reaction rate computed from these sample values is recorded. The procedure is repeated many times until enough samples of the reaction rate have been generated to estimate the property of the output reaction rate probability density function with the required statistical precision. Correlations between quantities have been considered carefully: for example, if the strength of a narrow resonance is estimated from a reduced width or a spectroscopic factor, then the uncertainty in the resonance energy enters both in the Boltzmann factor and in the penetration factor. Thus the same random value of the resonance energy, drawn in this case from a Gaussian probability density function, must be used in both expressions.

Our main goal is to calculate the probability density function for the {\it total} reaction rate and to characterize this distribution in terms of certain parameters: for the central (recommended) value of the reaction rate we chose the {\it median} which is equal to the 0.50 quantile of the cumulative distribution, while the low and high values of the reaction rate are chosen to coincide with the 0.16 and 0.84 quantiles, respectively. With this choice the confidence level (or the coverage probability) amounts to 68\%. A reliable estimation of more detailed information, such as the uncertainty contribution of each nuclear input quantity to the total rate, does not seem feasible at this time because of limitations in present-day computing power. See Paper I for details.

A computer code, \texttt{RatesMC}, has been written in order to calculate total reaction rates from resonant and nonresonant input using the Monte Carlo technique. For resonances the code computes reaction rates either from analytical expressions or, if required, by numerical integration. The latter procedure is computationally slow, since one integration has to be performed for each randomly sampled set of input quantities, but gives the most accurate results if the partial widths of a resonance are known. Upper limits of nuclear input quantities and interferences between levels are also taken into account in the random sampling. The user controls the total number of random samples and hence the precision of the Monte Carlo results\footnote{The number of samples used in the present work varied depending on the reaction, but always amounted to more than 5,000.}. For more information, see Paper I. The reaction rate output of the code is discussed in Sec.~\ref{output}. A detailed description of the input file to \texttt{RatesMC} can be found in Paper III.
\section{Overview}
An overview regarding the reaction rates evaluated in the present work is provided in Tab.~\ref{tab:master}. For each of the $62$ reactions listed here we also give the Q-value and some literature sources of the nuclear data input. The list of literature sources provided here is not meant to be comprehensive. For more information on the nuclear physics input, see the captions to the reaction rate tables in App. B. The complete nuclear physics input used in the present work is provided in Paper III.
\begin{center}
\begin{longtable}{@{\extracolsep{\fill}} lll}
\caption{Information on reaction rates evaluated in the present work.}  \label{tab:master} \\
\hline \hline 
	\multicolumn{1}{l}{Reaction$^{a}$} & \multicolumn{1}{c}{Q$^b$ (keV)} & \multicolumn{1}{c}{Some nuclear data sources$^{c}$} \\ \hline 
\endfirsthead
\multicolumn{3}{c}{{\tablename} \thetable{} -- continued} \\
\hline \hline 
	\multicolumn{1}{l}{Reaction$^{a}$} & \multicolumn{1}{c}{Q$^b$ (keV)} & \multicolumn{1}{c}{Some nuclear data sources$^{c}$} \\ \hline 
\endhead
	 \hline \hline
\endfoot
	\hline \hline
\endlastfoot
$^{14}$C(p,$\gamma$)$^{15}$N 		       	& 10207.42$\pm$0.00	 	& G\"orres et al. \cite{GO90}	      \\
$^{14}$C($\alpha$,$\gamma$)$^{18}$O       	& 6226.3$\pm$0.6		 	& Gai et al. \cite{GA87}, Lugaro et al. \cite{LU04}     \\
$^{14}$N($\alpha$,$\gamma$)$^{18}$F       	& 4414.6$\pm$0.5		 	& Tilley et al. \cite{Til95}, G\"orres et al. \cite{Goe00}     \\
$^{15}$N($\alpha$,$\gamma$)$^{19}$F       	& 4013.74$\pm$0.07	 	& de Oliveira et al. \cite{OL96}, Wilmes et al. \cite{WI02} 	\\
$^{15}$O($\alpha$,$\gamma$)$^{19}$Ne       	& 3529.1$\pm$0.6	 		& Mao et al. \cite{Mao95}, Fisker et al. \cite{Fis06}  \\
$^{16}$O(p,$\gamma$)$^{17}$F       		& 600.27$\pm$0.25	 		& Iliadis et al. \cite{Ili07}	\\
$^{16}$O($\alpha$,$\gamma$)$^{20}$Ne       	& 4729.85$\pm$0.00	 	& Tilley et al. \cite{Til98}, Mohr \cite{Moh05}	\\
$^{17}$O(p,$\gamma$)$^{18}$F       		& 5606.5$\pm$0.5	 		& Fox et al. \cite{Fox05}, Chafa et al. \cite{Cha07}	\\
\end{longtable}
\end{center}
$^{a}$ The superscripts $t$, $g$ and $m$ for $^{26}$Al refer to the total, ground state and isomeric state rate, respectively.\\
$^{b}$ Reaction Q-value from Audi, Wapstra and Thibault \cite{Aud03}, unless noted otherwise. The quoted uncertainty represents one standard deviation; an entry ``$\pm$0.00" implies that the experimental uncertainty amounts to a few eV only \cite{Aud03}. Such a small value is entirely negligible in the present astrophysical context. \\
$^{c}$ See Paper III for a complete bibliography. \\ 
$^{d}$ From Eronen et al. \cite{Ero09}.\\ 
$^{e}$ From Mukherjee et al. \cite{Muk04}.\\ 
$^{f}$ Using results from Yazidjian et al. \cite{Yaz07}.\\
$^{g}$ From Yoneda et al. \cite{Yon06}.\\
$^{h}$ Calculated using the new value for the mass of $^{21}$Na from Mukherjee et al. \cite{muk2}.\\
\clearpage
We comment briefly on the Q-values which, except for a few updates, are adopted from Audi, Wapstra and Thibault \cite{Aud03}. It is gratifying to see that for many reactions the uncertainties in $Q$ are less than $1$ keV. In some cases, however, the uncertainties are rather large ($>5$ keV). Notice that resonance energies are frequently calculated by subtracting the Q-value from a measured excitation energy (see Paper I). Thus, even if the excitation energy has been measured precisely, the total uncertainty in the resonance energy may be dominated by a large uncertainty in the Q-value. It is certainly worthwhile to improve such Q-values with large uncertainties in future measurements considering that the resonance energy enters exponentially in the expressions for resonant reaction rates (see Eqs. (1) and (10) of Paper I). 

A number of charged-particle thermonuclear reactions in the A=14 to 40 range have been excluded from the present evaluation. Below we provide a few examples and the reasons for disregarding them so that the reader obtains an impression on the scope of the present work. The $^{15}$N(p,$\gamma$)$^{16}$O and $^{15}$N(p,$\alpha$)$^{12}$C reactions are excluded here since their rates are strongly influenced by interfering resonance tails and nonresonant contributions. Such cases cannot be analyzed easily using the present procedure and require a different approach, such as R-matrix theory. Similar comments apply to the $^{14}$N(p,$\gamma$)$^{15}$O reaction. We disregarded the $^{20}$Na(p,$\gamma$)$^{21}$Mg reaction despite the fact that new information became available recently \cite{Mur06}: (i) the spin-parities of all expected resonances below E$_r^{cm}=1400$ keV are tentative; (ii) the $^{21}$Mg--$^{21}$F mirror state assignments are uncertain; (iii) the spectroscopic factors are unknown; and (iv) there may be missing levels close to the proton threshold in $^{21}$Mg. These sources of error prohibit a reliable estimation of reaction rates based on Monte Carlo techniques. A recent study \cite{Dei08} explored the nuclear structure of $^{27}$Si levels in order to derive a rate for the $^{26}$Al$^m$(p,$\gamma$)$^{27}$Si reaction. Unfortunately, the proton decay of expected low-energy resonances could not be measured (the detection threshold was $\approx$450 keV). In addition, the spin-parities as well as $\gamma$-ray partial widths of most resonances are unknown. We felt that too much information is missing at this time for calculating reliable reaction rates. Finally, we attempted to calculate Monte Carlo rates for the $^{30}$P(p,$\gamma$)$^{31}$S reaction. However, the effort proved futile although several recent studies addressed the nuclear structure of important levels in the $^{31}$S compound nucleus \cite{Jen06,Ma07,Wre07,Wre09a}. At present, the $^{31}$S--$^{31}$P mirror state assignments are uncertain and most of the spectroscopic factors and $\gamma$-ray partial widths are unknown. For these reasons, Monte Carlo rates cannot be calculated reliably.

\section{Results}\label{output}
Numerical values of Monte Carlo thermonuclear reaction rates are given in columns 2-4 of the tables presented in App. \ref{tabgraph} for a grid of temperatures between T=0.01 GK and 10 GK. Each reaction rate table is accompanied by two figures. The first of these displays the reaction rate ratios $N_A\left<\sigma v\right>_{high}/N_A\left<\sigma v\right>_{med}$ and $N_A\left<\sigma v\right>_{low}/N_A\left<\sigma v\right>_{med}$; a visual inspection immediately reveals the reaction rate uncertainty at a given temperature. The second figure shows Monte Carlo reaction rate probability density functions (in red) for six selected temperatures: $T=0.03$, $0.06$, $0.1$, $0.3$, $0.6$ and $1$~GK; they span conditions encountered in red giants, AGB stars, classical novae, massive stars and type I x-ray bursts. The complete nuclear physics input used to compute our reaction rates is presented in Paper III. The new reaction rates are compared with previous results in Paper IV.

The parameters of the lognormal approximations to the reaction rate probability density functions are given in columns 5-6 of the reaction rate tables in App. \ref{tabgraph}. The lognormal distributions are also displayed (in black) in the second figure following a given rate table. Note that the black line does {\it not} represent a fit to the data but its parameters are directly derived from the distribution of randomly sampled reaction rates. That is, the lognormal parameters are computed from the expectation value and the variance for $\ln(x)$ (since $\mu = E[\ln(x)]$ and $\sigma^2 = V[\ln(x)]$; see Sec. 4.2 of Paper I). A measure for the quality of the lognormal approximation, the Anderson-Darling (A-D) test statistic $t_{AD}^*$, is presented in column 6 of the rate tables and is also given in each panel of the following figures (denoted by ``A-D" in both tables and figures). Details regarding the Anderson-Darling test are provided in App. \ref{stattest}. In brief, a value of less than $t_{AD}^*\approx1$ indicates that the Monte Carlo reaction rate probability density function is consistent with a lognormal distribution. For values in the range of $t_{AD}^*\approx1-30$ the lognormal hypothesis is rejected by the Anderson-Darling test. Nevertheless, the lognormal approximation holds reasonably well in this range, as can be seen by inspecting the graphs following the rate tables, and thus seems adequate for use in reaction network calculations. For values in excess of $t_{AD}^*\approx30$ the lognormal approximation is not only rejected by the Anderson-Darling test but, in addition, deviates {\it visually} from the actual Monte Carlo reaction rate probability density function.

Below we give some examples, ordered according to increasing complexity, in order to explain how to interpret and to use our numerical results. The examples are for illustrative purposes only and have been simplified by disregarding minor rate contributions. For the full results, the reader should consult the rate tables.

\subsection{Rate of $^{28}$Si(p,$\gamma$)$^{29}$P at T=0.03 and 0.3 GK}\label{sec:si28pg}
Consider first the reaction rate for the proton capture on $^{28}$Si at a temperature of $T=0.03$ GK. From Tab. \ref{tab:si28pg} we find a recommended rate of $N_A\left<\sigma v\right>_{med}=5.57\times10^{-25}$~cm$^3$mol$^{-1}$s$^{-1}$. This value represents the median Monte Carlo rate and is obtained from the 0.50 quantile of the cumulative reaction rate distribution. The 0.16 and 0.84 quantiles provide the low and high Monte Carlo reaction rates, yielding $N_A\left<\sigma v\right>_{low}=3.79\times10^{-25}$~cm$^3$mol$^{-1}$s$^{-1}$ and $N_A\left<\sigma v\right>_{high}=8.18\times10^{-25}$~cm$^3$mol$^{-1}$s$^{-1}$, respectively. 

\begin{figure}[]
\includegraphics[height=13.5cm,angle=-90]{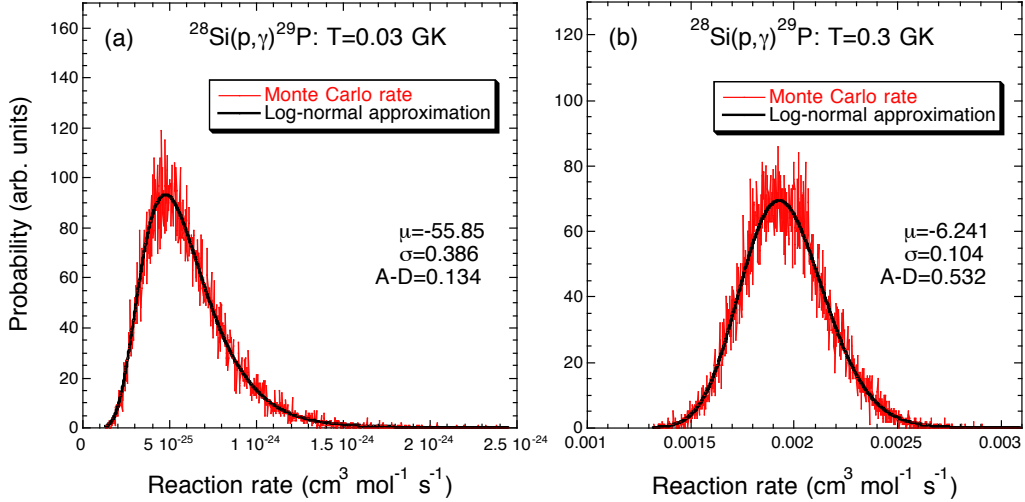}
\caption{\label{fig:pdfsi28pg} 
Total reaction rate probability density functions for $^{28}$Si(p,$\gamma$)$^{29}$P: (a) T=0.03 GK, (b) T=0.3 GK. The red histograms are obtained from the Monte Carlo method, while the black solid curves show lognormal approximations. The lognormal parameters $\mu$ and $\sigma$, as well as the Anderson-Darling test statistic (A-D) are listed for each temperature. The results are obtained with 20,000 samples.}
\end{figure}

The corresponding Monte Carlo reaction rate probability density function is shown in Fig. \ref{fig:pdfsi28pg}a as a red histogram. The black solid line represents a lognormal approximation. The values of the lognormal parameters, $\mu=-55.85$ and $\sigma=0.386$, are given in Tab. \ref{tab:si28pg}. In addition, the Anderson-Darling test statistic, $t_{AD}^*=0.134$, is listed in the table. This small value indicates that in this case the Monte Carlo reaction rate probability density function follows a lognormal distribution, as is also apparent from visual inspection of Fig. \ref{fig:pdfsi28pg}a. Consequently, we can point out a few interesting observations. First, from the relatively large value of $\sigma=0.386$, corresponding to a factor uncertainty of $f.u.=e^{\sigma}=1.47$, it follows immediately that the reaction rate distribution is skewed (see Sec. 4.2 of Paper I). Second, the median Monte Carlo reaction rate is related to the lognormal location parameter by $\mu=\ln(N_A\left<\sigma v\right>_{med})$. Third, the values of the low and high Monte Carlo reaction rates are related to the lognormal spread parameter by $\sigma=\ln([N_A\left<\sigma v\right>_{high}/N_A\left<\sigma v\right>_{low}]^{1/2})$. These last two relationships, which only hold if the total reaction rate distribution is lognormal (see Sec. 5.4 of Paper I), are easily verified from the numerical values provided above. 

The total Monte Carlo reaction rate probability density function follows lognormality  because at this temperature the rate is dominated by the non-resonant direct capture process. Since we assumed a lognormal probability density for the (input) effective S-factor (Sec. 5.1.3 of Paper I), the resulting (output) reaction rate probability density becomes also lognormal. 

Consider now Fig. \ref{fig:pdfsi28pg}b, showing the situation at a temperature of T=0.3 GK. The hypothesis that the Monte Carlo reaction rate probability density follows a lognormal distribution passes again the Anderson-Darling test (since $t_{AD}^*=0.532<t_c$; see App. \ref{stattest}). However, the value of $\sigma=0.104$ indicates that the lognormal distribution is only weakly skewed. In other words, the black solid line is nearly Gaussian (see Sec. 4.2 in Paper I). The reason is that at this temperature the total rate is dominated by a single narrow resonance, located at $E_r^{cm}=357.1$ keV with a strength of $\omega\gamma=0.0020$ eV (see Paper III). The uncertainty in the resonance strength (10\%), although moderate, dominates over the uncertainty in the resonance energy (0.7 keV). Since the resonance strength enters linearly in the narrow resonance reaction rate expression (see Eq. (10) of Paper I) and its uncertainty is relatively small (so that the input probability density function is nearly Gaussian), the total reaction rate distribution as a result becomes also nearly Gaussian.

At this point it is worthwhile to recall the arguments presented in Sec. 5.4 of Paper I of why the lognormal approximation to the total reaction rate probability density function is so useful: (i) the total rate may be dominated by a non-resonant process (as was the case in Fig. \ref{fig:pdfsi28pg}a); (ii) the total rate may be dominated by a single resonance and the uncertainty in the resonance energy dominates over the uncertainty in the resonance strength; (iii) the total rate may be dominated by a single resonance and the uncertainty in the resonance strength dominates over the uncertainty in the resonance energy (as was the case in Fig. \ref{fig:pdfsi28pg}b); (iv) the total rate may be given by the contributions from many resonances. For the last two cases the total reaction rate probability density may tend toward a Gaussian which can be approximated by a lognormal distribution in a straightforward manner (Sec. 4.2 of Paper I). We emphasize that for the overwhelming number of reaction rates computed in the present work the Monte Carlo probability density functions follow a lognormal distribution according to the arguments given above. In the following we will focus on exceptional cases where the resulting probability density functions are {\it not} lognormal. These situations are interesting since they reflect certain properties of the underlying nuclear physics input.

\subsection{Rate of $^{35}$Ar(p,$\gamma$)$^{36}$K at T=0.03 GK}
We will next discuss the reaction rate for the proton capture on $^{35}$Ar at a temperature of T=0.03 GK. At this temperature the total rate is mainly determined by the direct capture process and by a resonance located very close to the proton threshold, at E$_r^{cm}=2$ keV. What makes this case exceptional is the fact that the uncertainty in the resonance energy is very large ($\pm$21 keV; see Paper III). We will explore in the following the consequences of such a large uncertainty. An extra complication is caused by the proximity of the resonance to the particle threshold. A resonance energy of only 2 keV causes major problems for the computation of the Coulomb wave functions according to Eq. (16) of Paper I that are needed in order to integrate Eq. (1) from Paper I numerically. Thus we shifted the resonance to an energy of $-0.1$ keV so that its rate contribution can be obtained in a straightforward manner from the tail of an assumed subthreshold resonance. Test calculations have shown that a moderate energy shift of $\approx2$ keV has no significant effect on the total rate considering a 21 keV uncertainty in the resonance energy. Furthermore, it must be emphasized that the calculation of the rate significantly differs for this example compared to the previous case: first, the rate is obtained using the proton width, which must be calculated from the penetration factor for each sampled energy, instead of the resonance strength; second, the rate is derived from a numerical integration rather than from an analytical expression
(see Sec. 2 of Paper I for details).

\begin{figure}[]
\includegraphics[height=13.5cm,angle=-90]{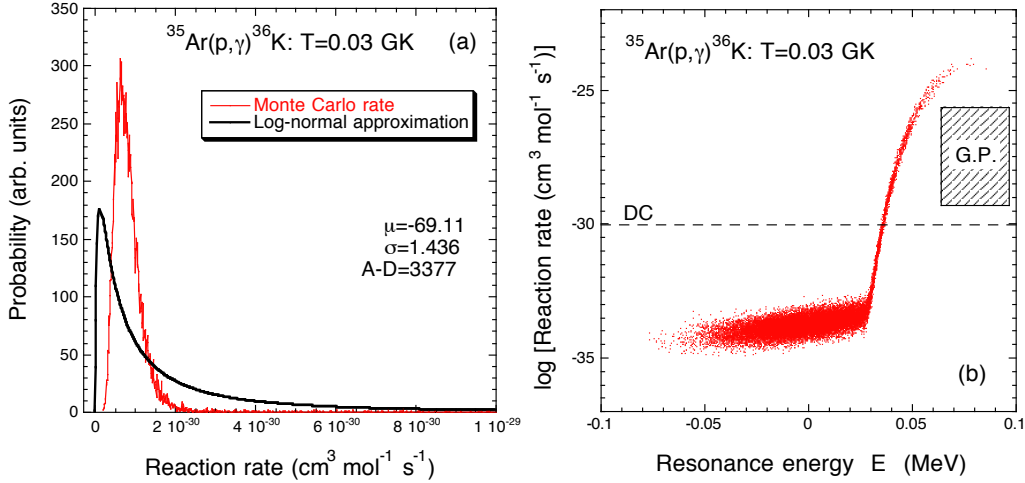}
\caption{\label{fig:ar35pgT003} 
Reaction rates for $^{35}$Ar(p,$\gamma$)$^{36}$K at T=0.03 GK: (a) Probability density function from Monte Carlo sampling (red) and lognormal approximation (black); (b) Reaction rate versus sampled energy of the subthreshold resonance near E$_r^{cm}=-0.1$ keV. In part (b) the horizontal dashed line indicates the mean value of the direct capture rate contribution, while the label ``G.P." refers to the location of the Gamow peak. The results are obtained with 20,000 samples. The resonances at E$_r^{cm}=224-744$ keV (see Paper III) have only a minor impact at this low temperature and have been disregarded for the results shown here. They are, however, included in the full calculation of the rate listed in Tab. \ref{tab:35arpg}.}
\end{figure}

We find a recommended Monte Carlo reaction rate of $N_A\left<\sigma v\right>_{med}=7.65\times10^{-31}$ cm$^3$mol$^{-1}$s$^{-1}$. The low and high rates amount to $N_A\left<\sigma v\right>_{low}=5.16\times10^{-31}$ cm$^3$mol$^{-1}$s$^{-1}$ and $N_A\left<\sigma v\right>_{high}=1.17\times10^{-30}$ cm$^3$mol$^{-1}$s$^{-1}$, respectively. The corresponding Monte Carlo reaction rate probability density function is shown in Fig. \ref{fig:ar35pgT003}a as a red histogram. The peak visible on the left-hand side in the figure is entirely dominated by the direct capture process, for which the mean value amounts to $N_A\left<\sigma v\right>_{mean}^{DC}=8.08\times10^{-31}$~cm$^3$mol$^{-1}$s$^{-1}$. The solid black line in Fig. \ref{fig:ar35pgT003}a represents the lognormal approximation. The lognormal parameters amount to $\mu=-69.11$ and $\sigma=1.436$. Not only is the lognormal distribution highly skewed, but it poorly describes the actual Monte Carlo distribution, shown in red. This observation is supported by the large value of $t_{AD}^*=3377$ for the Anderson-Darling test statistic. Furthermore, since the actual Monte Carlo probability density is clearly not lognormal, the expressions given in Eqs. (39) and (40) of Paper I are not fulfilled anymore, as can be easily verified from the rate values provided above.

The disagreement is caused by the large uncertainty in the resonance energy. It turns out that the red histogram has a very long tail on the right hand side. In fact, the maximum sampled rate amounts to $\approx10^{-24}$ cm$^3$mol$^{-1}$s$^{-1}$. Note that this value exceeds the location of the (red) peak shown in Fig. \ref{fig:ar35pgT003}a by about 6 orders of magnitude. We do not display the entire range of sampled reaction rates in the figure since otherwise the peak would not be resolved on that scale. The lognormal approximation is obtained under the condition of the same expectation value and variance as the actual Monte Carlo probability density function. However, in order to account for the very long tail of the actual rate distribution the lognormal approximation becomes highly skewed. As a result its shape does not resemble that of the Monte Carlo distribution.

More insight can be obtained by considering part (b) of the figure, displaying the resonant Monte Carlo reaction rate contribution versus the sampled energy for the E$_r^{cm}=-0.1$ keV resonance. The horizontal dashed line indicates the mean value of the direct capture rate contribution (see above), while the shaded region marks the location of the Gamow peak for T=0.03 GK (see Eq. (4) of Paper I). Since the resonance is located so close to the threshold, we expect that about half of the sampled energies occur below and half occur above threshold. For sampled energies in the negative region (that is, E$_r$ below threshold), the maximum resonant rate contribution is about $10^{-33}$ cm$^3$mol$^{-1}$s$^{-1}$. This value is 3 orders of magnitude below the direct capture rate contribution and, consequently, it is entirely negligible. The situation is quite different above threshold. Occasionally a relatively large resonance energy is sampled that is located close to the Gamow peak. In such cases the resonant rate contribution will obviously increase dramatically. (A more detailed reason for the kink visible in panel (b) near a sampled resonance energy of $E_r\approx30$ keV will be discussed in Sec. \ref{sec:al23pg}.) It can be seen from Fig. \ref{fig:ar35pgT003}b that a few sampled energies are even located in the Gamow peak, $E_0\pm\Delta/2$ (see Eq. (4) in Paper I), resulting in a resonant rate contribution up to $\approx10^{-24}$ cm$^3$mol$^{-1}$s$^{-1}$. Such values will entirely dominate the total reaction rate, giving rise to the long tail in the reaction rate probability density function (red histogram in Fig. \ref{fig:ar35pgT003}a).   

Our original idea was to cut off the total Monte Carlo reaction rate probability density, say, at a value of $\approx4\times10^{-30}$ cm$^3$mol$^{-1}$s$^{-1}$, and then to calculate new, renormalized, lognormal parameters $\mu$ and $\sigma$ for the truncated distribution. An advantage of this procedure would be that the actual Monte Carlo probability density function (without the long tail) could then be very well approximated by a lognormal distribution. However, a major disadvantage is the fact that disregarding the long tail, which does not represent some artificial statistical ``noise" but rather consists of physically meaningful samples, would introduce a significant bias. Thus we decided not to truncate any of the Monte Carlo probability density functions obtained in the present work. It must be emphasized again that the red histogram and the solid black line shown in Fig. \ref{fig:ar35pgT003}a have the same expectation value and variance. If it turns out that stellar model calculations are sensitive to the {\it shape} of the reaction rate probability density function for $^{35}$Ar(p,$\gamma$)$^{36}$K at T=0.03 GK, then there are two obvious possibilities for proceeding: (i) a new measurement of $^{36}$K excitation energies in order to derive smaller resonance energy uncertainties; or (ii) the use of an analytical function that is more complicated than a lognormal distribution, but approximates the actual reaction rate probability density function more closely.

\subsection{Rate of $^{23}$Al(p,$\gamma$)$^{24}$Si at T=0.09 and 0.2 GK}\label{sec:al23pg}

\begin{figure}[]
\includegraphics[height=12cm]{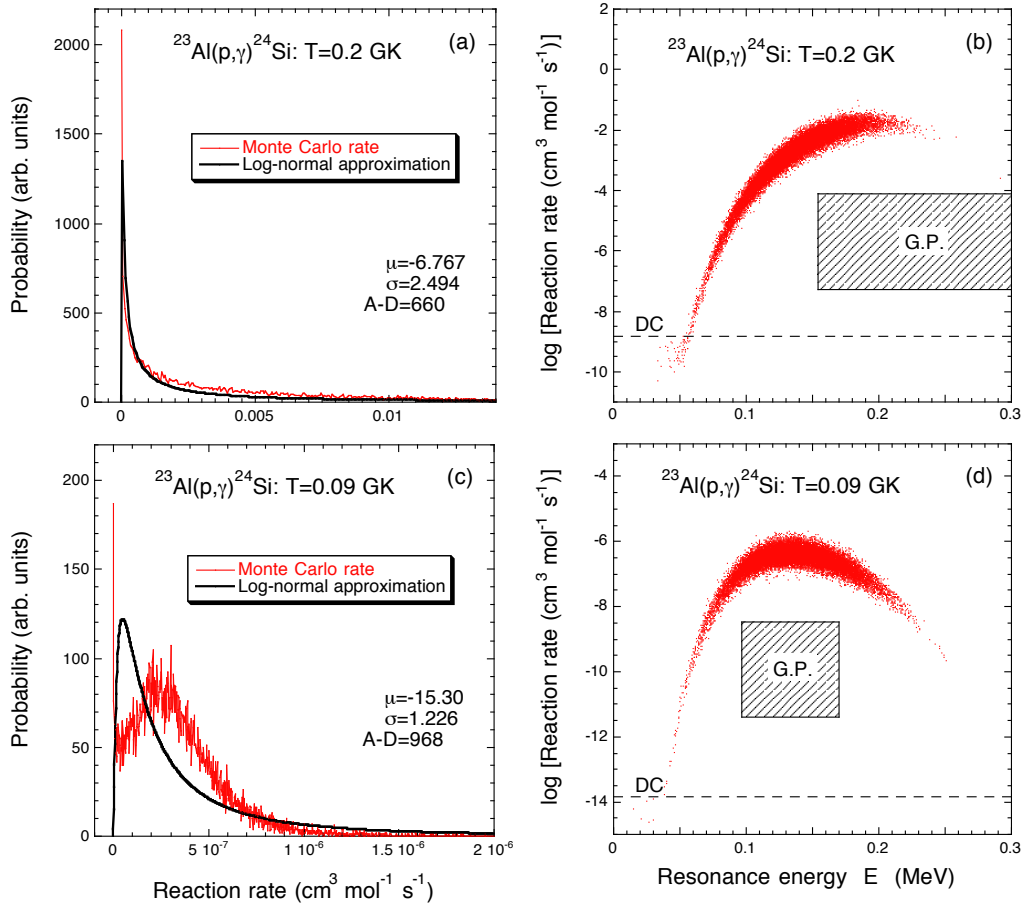}
\caption{\label{fig:al23pgpanel} 
(a), (c) Reaction rate probability density functions for $^{23}$Al(p,$\gamma$)$^{24}$Si at T=0.2 and 0.09 GK from Monte Carlo sampling (red) and lognormal approximation (black); (b), (d) Partial reaction rate versus sampled energy of the E$_r^{cm}=137\pm29$ keV resonance. In parts (b) and (d) the horizontal dashed line indicates the mean value of the direct capture rate contribution, while the label ``G.P." refers to the location of the Gamow peak. The results are obtained with 20,000 samples. The rates shown in the panels are calculated by taking only the E$_r^{cm}=137$ keV resonance into account (see text). Higher-lying resonances (see Paper III) have only a minor impact on the total rate at these temperatures. The latter contributions are, however, included in the full calculation of the total rate listed in Tab. \ref{tab:23alpg}.}
\end{figure}

We will now discuss the influence of a large resonance energy uncertainty on the rate in more detail. Consider, for example, the rates of the $^{23}$Al(p,$\gamma$)$^{24}$Si reaction at T=0.09 and 0.2 GK (Fig. \ref{fig:al23pgpanel}). At these temperatures the total rate is determined by a single, s-wave, resonance, located at E$_r^{cm}=137\pm29$ keV. The resonance strength is determined by the proton partial width (since $\Gamma_p\ll\Gamma_{\gamma}$) for which we adopt a fractional uncertainty of about 50\% (see Paper III for details). Contributions from other resonances or the direct capture process (indicated by the dashed horizontal lines) are negligible for the discussion below. The reaction rate is derived from a numerical integration (see Sec. 2 of Paper I). Note that most resonances considered in the present evaluation do not exhibit such large uncertainties on energy and partial width. But this particular example is interesting since it exemplifies a number of important issues. 

Panel (b) displays the reaction rate versus the sampled energy of the E$_r^{cm}=137$ keV resonance at T=0.2 GK. Here the resonance is located just outside the energy region, $E_0\pm\Delta/2=228\pm72$ keV, of the Gamow peak. A trend is clearly visible: the closer the sampled resonance energy to the Gamow peak maximum, $E_0$, the larger the resulting rate. Most energy samples occur outside the Gamow peak, near the mean resonance energy, where the rate is relatively small. For this reason, the corresponding probability density function in part (a) exhibits a sharp rise at small rates and then declines for increasing rates. This behavior translates into relatively large rate uncertainties. Indeed, we find for the ratio of high to low Monte Carlo rates a factor of $53$ (Tab. \ref{tab:23alpg}). 

Panel (d) shows the rate versus sampled energy at T=0.09 GK where a quite different result is obtained. At this particular temperature the E$_r^{cm}=137$ keV resonance is located in the middle of the Gamow peak, $E_0\pm\Delta/2=134\pm37$ keV. Most of the energy samples occur around $E_r^{cm}$ where the rate is at maximum, while the rate decreases for energies sampled below and above the mean resonance energy. The corresponding probability density function, shown in part (c), shows a broad maximum with similar slopes on the left and right side. For the ratio of high to low Monte Carlo rates we find in this case a factor of $4.7$, which is much smaller than the value obtained at T=0.2 GK. The relative insensitivity of the reaction rate to energy variations if a resonance is located near the middle of the Gamow peak was already pointed out in Refs. \cite{Ili99,Tho99}. Note also that the vertical spread of the distributions in panels (b) and (d) is caused by the uncertainty in the proton partial width (here assumed to be 50\%). 

It may appear at first somewhat surprising that in parts (a) and (c) the lognormal approximation (shown in black) poorly describes the actual Monte Carlo distribution (shown in red), as indicated by the rather large values for the Anderson-Darling test statistic. It is important to realize that none of the arguments given at the end of Sec. \ref{sec:si28pg} hold in this case. For example, we argued under point (ii) that the rate distribution should be lognormal if the rate is dominated by a single resonance for which the uncertainty in the resonance energy dominates over the uncertainty in the resonance strength. This was explained in Sec. 5.4 of Paper I: if the resonance energy, $E_r$, is Gaussian distributed then the rate, proportional to $e^{-E_r/kT}$, is lognormally distributed. However, the point here is that in this particular case the rate is not calculated from a resonance strength, but from a proton partial width that is computed for each sampled energy from the penetration factor. If we approximate the penetration factor for the s-wave resonance at E$_r^{cm}=137$ keV 
by the Gamow factor (Sec. 2.1 of Paper I), then the reaction rate becomes proportional to $e^{-E_r/kT-2\pi\eta}$ (that is, the Gamow peak). Clearly, it is the Sommerfeld parameter $\eta\sim\sqrt{1/E_r}$ (see Eq. (3) of Paper I) in the exponential that causes the reaction rate distribution to deviate strongly from lognormality.

Finally, we will briefly describe a subtle effect that is barely visible in panels (b) and (d). Close inspection reveals that the samples at the lowest energy do not seem to follow the strong energy dependence of the reaction rate. Rather, a kink appears in the distribution near sampled resonance energies of $40-50$ keV. 
Interestingly, the kink disappears if the reaction rate is calculated from an analytical expression (Eqs. (10) and (11) of Paper I) rather than from a numerical integration. The reason is discussed in Sec. 2.3 of Paper I. For a single resonance there are in general two contributions to the reaction rate: first, from the vicinity near the resonance energy $E_r$, and second, from the resonance tail contribution near the Gamow peak $E_0$. For a sampled energy near the Gamow peak the former contribution dominates strongly over the latter. However, the farther the resonance energy is sampled away from the Gamow peak, the smaller the ratio of near-resonance and resonance tail contributions becomes. The kink appears near energies where both contributions are of similar magnitude. For smaller sampled energies the resonance tail contribution, which is much less energy dependent than the near-resonance contribution, dominates the total rate. Note that the far more pronounced kink observed in Fig. \ref{fig:ar35pgT003}b near $E_r\approx30$ keV has precisely the same cause.

\subsection{Rate of $^{26}$Al$^g$(p,$\gamma$)$^{27}$Si at T=0.02 GK}\label{al26gst}
None of the reaction rates discussed so far have any contributions from expected, but yet undetected resonances. In other words, the nuclear input data do not contain any upper limits of resonance strengths. The proton capture on $^{26}$Al$^g$ represents an interesting case since, at low temperatures, the reaction rates are strongly influenced by unobserved, low-energy resonances. At T=0.02 GK the total rates are entirely dominated by a resonance at E$_r^{cm}=69\pm3$ keV \footnote{During the near completion of the present work, we became aware of the recent measurement by Lotay et al. \cite{Lot09}. They locate this particular resonance at E$_r^{cm}=68.3\pm0.7$ keV and find evidence that it is formed via $d$-waves rather than $s$-waves as assumed here. The numerical values of the rates presented in this subsection do not include the latest results of Ref. \cite{Lot09} and thus the present discussion should be regarded as an illustrative example only. However, their latest results are included in the Monte Carlo rates presented in Tab. \ref{tab:al26gpg} and in Papers III and IV.}. Thus we will disregard in the following discussion other resonances and the direct capture process. 

\begin{figure}[]
\includegraphics[height=13.5cm,angle=-90]{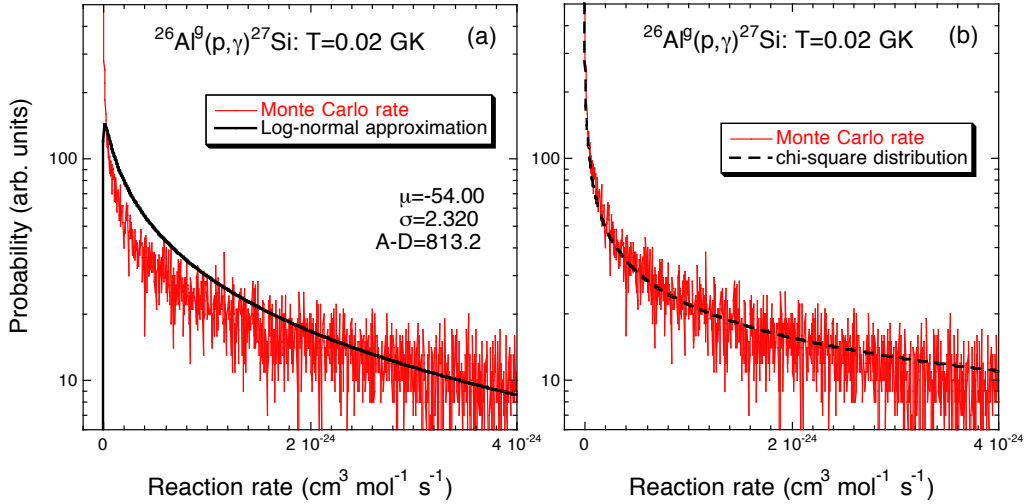}
\caption{\label{fig:al26gpg002} 
Reaction rate probability density function for $^{26}$Al$^g$(p,$\gamma$)$^{27}$Si at T=0.02 GK. (a) Comparison of Monte Carlo rates (red) and lognormal approximation (solid black line); (b) Comparison of Monte Carlo rates (red) and chi-squared distribution with one degree of freedom (dashed black line). The results are obtained with 50,000 samples. At this temperature the reaction rates are dominated by a resonance at E$_r^{cm}=69$ keV for which only an upper limit for the resonance strength can be estimated (see Paper III). All other contributions from resonances and direct capture have been disregarded for the results shown here. They are, however, included in the full calculation of the rate listed in Tab. \ref{tab:al26gpg}.}
\end{figure}

We find a recommended Monte Carlo reaction rate of $N_A\left<\sigma v\right>_{med}=5.42\times10^{-24}$ cm$^3$mol$^{-1}$s$^{-1}$. The low and high rates amount to $N_A\left<\sigma v\right>_{low}=4.58\times10^{-25}$ cm$^3$mol$^{-1}$s$^{-1}$ and $N_A\left<\sigma v\right>_{high}=2.81\times10^{-23}$ cm$^3$mol$^{-1}$s$^{-1}$, respectively. The Monte Carlo reaction rate probability density function is shown in Fig. \ref{fig:al26gpg002}a as a red histogram. The solid black line represents a lognormal approximation, with parameters of $\mu=-54.00$ and $\sigma=2.320$. The Anderson-Darling test statistic amounts to $t_{AD}^*=813.2$, indicating that the lognormal distribution poorly describes the actual Monte Carlo reaction rate probability density function. 

Recall from Paper I how upper limits of resonance strengths have been incorporated into our Monte Carlo sampling: we assume a Porter-Thomas distribution (that is, a chi-squared distribution with one degree of freedom) for the dimensionless reduced width of an unobserved resonance. Since we are considering in this example only a single undetected resonance with a moderate energy uncertainty ($\pm3$ keV), it follows immediately from Eq. (10) of Paper I that the resulting reaction rates must follow the same distribution as the resonance strength (or the corresponding reduced width; see Eq. (13) of Paper I), that is, a chi-squared distribution with one degree of freedom. This distribution is shown as a dashed black line in Fig. \ref{fig:al26gpg002}b. The agreement with the actual Monte Carlo probability density function (red histogram) is apparent. A number of important implications in connection with upper limits of resonance strengths are discussed below.

First, the classical procedure of estimating reaction rates assumes that the value of the ``lower limit" of the rate contribution for an unobserved resonance is exactly zero (see Sec. 3.4 of Paper I). This assumption is presumably based on the idea that the formation probability (or the entrance channel partial width) could in principle be as small as zero, that is, the resonance would not exist in this limit. Thus it may seem surprising at first that in the above example we obtain a non-zero value for $N_A\left<\sigma v\right>_{low}$. One should remember that the classical procedure has no rigorous statistical meaning. Instead, we assume a Porter-Thomas distribution for the dimensionless reduced width, $\theta^2$, of an undetected resonance. In the above example, the distribution of reduced widths determines directly the shape of the reaction rate probability density function. The low Monte Carlo rate is then consistently obtained from the 0.16 quantile of the cumulative reaction rate distribution. The physical meaning of our assumption is the following: the probability density function of the reduced width is given by a chi-squared distribution, implying a higher probability the smaller the value of $\theta^2$. In other words, if a particular level is known to exist in the compound nucleus, then the corresponding partial width, and hence the resonance strength, could indeed become very small, but it is never equal to zero. The coupling of the nucleons will always give rise to some finite value for the reduced width (or the spectroscopic factor).

Second, the above case provides another example for a reaction rate probability density function that is {\it not} lognormal, as is apparent from Fig. \ref{fig:al26gpg002}a. The actual Monte Carlo reaction rate rather follows a chi-squared distribution with one degree of freedom. The lognormal and chi-squared distributions have very different shapes, as can be seen from Fig. 2 of Paper I: for decreasing rates, the former distribution approaches zero probability, while the latter distribution exhibits a pole at the origin; furthermore, the tail of the former distribution falls faster for increasing rates than the tail of the latter distribution. Although the Monte Carlo reaction rate probability density function and the lognormal approximation shown in Fig. \ref{fig:al26gpg002}a have the same expectation value and variance, situations may arise where stellar model calculations could be sensitive to the precise {\it shape} of the reaction rate probability density function for $^{26}$Al$^g$(p,$\gamma$)$^{27}$Si at low temperatures. In such a case there are two possibilities for proceeding: (i) a new measurement of the resonance strength in $^{26}$Al$^g$(p,$\gamma$)$^{27}$Si or of the spectroscopic factor in $^{26}$Al$^g$($^3$He,d)$^{27}$Si, in order to obtain an improved experimental upper limit for the corresponding dimensionless reduced width; or (ii) the use of an analytical function that is more complicated than a lognormal distribution, but approximates the actual reaction rate probability density function more closely.

\begin{figure}[]
\includegraphics[height=13.5cm,angle=-90]{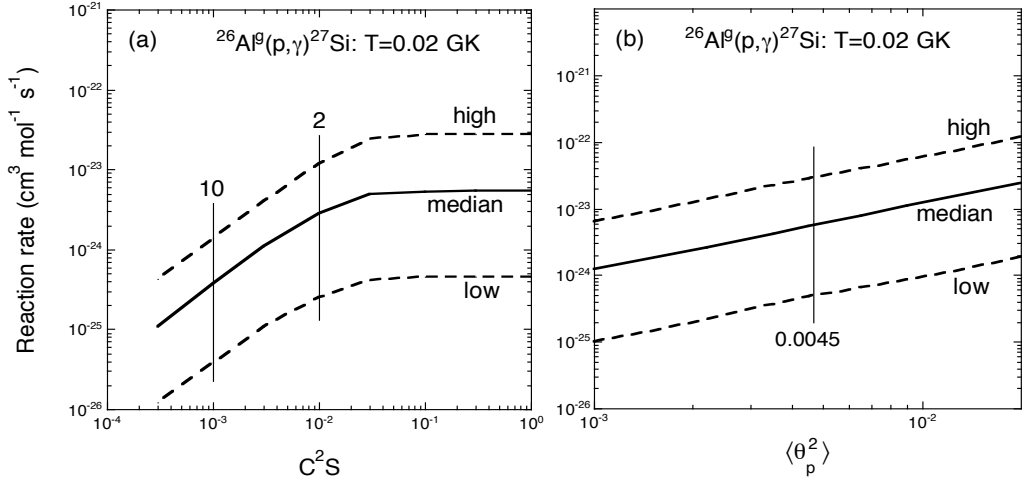}
\caption{\label{fig:al26gpg002c2s} 
Low, median and high Monte Carlo reaction rates for $^{26}$Al$^g$(p,$\gamma$)$^{27}$Si at T=0.02 GK. Only the undetected E$_r^{cm}=69$ keV resonance is taken into account for the results displayed here. (a) Dependence of rates on the value of the experimental upper limit for the spectroscopic factor $C^2S$; the best current estimate for the upper limit is $C^2S=1$. The labels ``2" and ``10" indicate the spectroscopic factor values for which the rates decrease by a factor of 2 and 10, respectively; (b) Dependence of rates on the mean value of the dimensionless reduced proton width $\left<\theta^2_p\right>$; the value adopted throughout the present work amounts to $\left<\theta^2_p\right>=0.0045$ (see Paper I). The results are obtained with 50,000 samples.}
\end{figure}

Third, an obvious question for the experimentalist is how sensitive an experiment needs to be in order to reduce the estimated reaction rate contributions from undetected low-energy resonances. For the above case of the E$_r^{cm}=69$ keV resonance in $^{26}$Al$^g$(p,$\gamma$)$^{27}$Si no experimental upper limit for the spectroscopic factor exists and, consequently, a value of $C^2S\leq1$ has been adopted here. The impact of an improved experimental upper limit of $C^2S$ on the resulting Monte Carlo reaction rates at T=0.02 GK is shown in Fig. \ref{fig:al26gpg002c2s}a. It can be seen that, contrary to the behavior of the classical reaction rates, a reduction of the upper limit by an order of magnitude ($C^2S_{ul}=0.1$) has almost no effect on the resulting Monte Carlo reaction rates. This is explained by the fact that the dimensionless reduced width (or the spectroscopic factor) is randomly sampled according to a Porter-Thomas distribution which predicts increasing probabilities for decreasing values of $\theta^2$. Recall that we adopted a mean value of $\left<\theta^2_p\right>=0.0045$ and thus truncating the distribution, according to Eq. (38) of Paper I, at $C^2S_{ul}=0.1$ is inconsequential\footnote{For an s-wave resonance at E$_r^{cm}=69$ keV in $^{26}$Al$^g$+p the dimensionless reduced proton width and the spectroscopic factor are related by $\theta^2_p\approx0.6~C^2S$ (see Eq. (14) of Paper I). The factor $0.6$ represents the dimensionless single-particle reduced proton width \cite{Ili96}.}. The figure reveals that the experimental upper limit must be reduced significantly before the rates start to change noticeably. For example, if an experimental upper limit of $C^2S_{ul}=0.01$ or $C^2S_{ul}=0.001$ could be achieved in a transfer measurement, the Monte Carlo rates would decrease by a factor of 2 or 10, respectively. Clearly, the counting background must be considered carefully in future transfer experiments in order to achieve the highest possible sensitivities in searches for threshold states.

Fourth, although an improbable conjecture, one may ask how do the present reaction rates for T=0.02 GK change assuming that the E$_r^{cm}=69$ keV resonance {\it has just escaped detection}? In other words, what are the reaction rates if we use $C^2S=1$ instead of regarding this value as an upper limit? We find in this case an increase of the high Monte Carlo rate by a factor of 130, while at the same time the ratio of high over low Monte Carlo rates decreases from a factor of $60$ to $5$. This result underscores the importance of searching for undetected low-energy resonances, especially in cases when the predicted upper limit on $C^2S$ is relatively large.

Fifth, we are now in a position of estimating the sensitivity of the Monte Carlo reaction rates to the mean value of $\left<\theta^2_p\right>=0.0045$ adopted in the present work for protons. The impact of varying this value on the resulting Monte Carlo reaction rates for $^{26}$Al$^g$(p,$\gamma$)$^{27}$Si at T=0.02 GK is shown in Fig. \ref{fig:al26gpg002c2s}b. The nearly straight lines on a log-log scale translate into a linear dependence of the Monte Carlo reaction rates on the value of $\left<\theta^2_p\right>$. For example, if we would have adopted instead a value of $\left<\theta^2_p\right>=0.0090$, the predicted reaction rates would also increase by about a factor of 2. Clearly, changing the mean value of the Porter-Thomas distribution (see Eq. (36) of Paper I) directly influences the quantiles of the cumulative reaction rate distribution. These results emphasize again the importance of future systematic studies of nuclear statistical properties that will provide improved local mean values for dimensionless reduced widths (see Sec. 6 of Paper I).

\subsection{Rate of $^{30}$Si(p,$\gamma$)$^{31}$P at T=0.12 GK}
\begin{figure}[]
\includegraphics[height=8cm,angle=0]{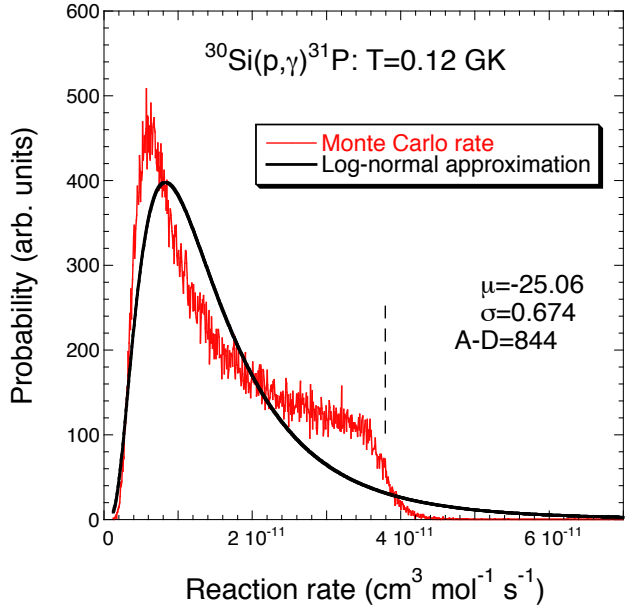}
\caption{\label{fig:si30pg012} 
Reaction rate probability density function for $^{30}$Si(p,$\gamma$)$^{31}$P at T=0.12 GK. The Monte Carlo rates are shown in red, while the lognormal approximation is displayed in black. The results are obtained with 100,000 samples. The step in the distribution, indicated by the vertical dashed line, is caused by the truncation of the Porter-Thomas distribution at an experimental upper limit of $\theta_{ul}^2$, where $\theta_{ul}^2< \left<\theta^2_p\right>$ for the E$^{cm}_r=169$ keV resonance. The location of the step coincides with the ``classical" rate that is computed with the upper limit value of the proton partial width, while the slope of the step is influenced by the energy uncertainty ($\pm2$ keV) of the above resonance.}
\end{figure}

The rate of the proton capture on $^{30}$Si at a temperature of T=0.12 GK has significant contributions from undetected resonances at E$^{cm}_r=52, 169$ and $418$ keV, and from direct capture (see Paper III). We find a recommended Monte Carlo reaction rate of $N_A\left<\sigma v\right>_{med}=1.35\times10^{-11}$ cm$^3$mol$^{-1}$s$^{-1}$. The low and high rates amount to $N_A\left<\sigma v\right>_{low}=6.18\times10^{-12}$ cm$^3$mol$^{-1}$s$^{-1}$ and $N_A\left<\sigma v\right>_{high}=2.82\times10^{-11}$ cm$^3$mol$^{-1}$s$^{-1}$, respectively. The Monte Carlo reaction rate probability density function (red histogram) and the corresponding lognormal approximation (solid black line) are shown in Fig. \ref{fig:si30pg012}. The Monte Carlo distribution displays a maximum, a broad tail for increasing reaction rates, and a sharp decline in probability around a value of $\approx4\times10^{-11}$ cm$^3$mol$^{-1}$s$^{-1}$. The probability density function is clearly not lognormal, a conclusion that is supported by the large value of $t_{AD}^*=844$ for the Anderson-Darling test statistic.

For the three undetected resonances, only upper limit contributions to the total rates can be obtained: we estimate spectroscopic factors (reduced widths) of $C^2S\leq1$ ($\theta_p^2\leq0.67$) for E$^{cm}_r=52$ keV, $C^2S\leq0.003$ ($\theta_p^2\leq0.0009$) for E$^{cm}_r=169$ keV, and $C^2S\leq0.02$ ($\theta_p^2\leq0.011$) for E$^{cm}_r=418$ keV. The E$^{cm}_r=169$ keV resonance is located near the center of the Gamow peak at this temperature and thus has the strongest impact on the total rates. Recall that in these cases we randomly sample the dimensionless reduced proton width according to a Porter-Thomas distribution (with a mean value of $\left<\theta^2_p\right>=0.0045$) that is truncated at the upper limit value, $\theta_{ul}^2$, of the dimensionless reduced width (see Eq. (38) of Paper I). Interestingly, for the E$^{cm}_r=169$ keV resonance we find $\theta^2_{ul} < \left<\theta^2_p\right>$, such that the Porter-Thomas distribution is only sampled in the range of $\theta_p^2\leq0.0009$. As a result, this cutoff value propagates to the reaction rate probability density function, where it gives rise to a step (vertical dashed line in Fig. \ref{fig:si30pg012}). It is also interesting to note that this step occurs at a reaction rate value that is close to the classically calculated ``upper limit" of the rate. This observation is easily explained by the arguments presented in Sec. 3.4 of Paper I. 

We will again address the question of how the present reaction rates for T=0.12 GK change assuming that the E$_r^{cm}=169$ keV resonance has just escaped detection, that is, what are the reaction rates if we use $C^2S=0.003$ instead of regarding this value as an upper limit? We find in this case an increase of the high Monte Carlo rate by a factor of 2.5, while at the same time the ratio of high over low Monte Carlo rates decreases from a factor of $4.5$ to $2.0$. These changes are modest compared to the results we obtained earlier in Sec. \ref{al26gst}. Clearly, the actual detection of a previously unobserved low-energy resonance has less of an impact when the upper limit on $C^2S$ is already relatively small (especially if $\theta_{ul}^2< \left<\theta^2_p\right>$).

\section{Extrapolation of laboratory reaction rates to high temperatures}\label{extrap}
The discussion in the previous sections and in Paper I emphasizes that thermonuclear reaction rates can be estimated reliably if the necessary ingredients, for example, nonresonant and broad-resonance cross sections, resonance energies and strengths, and so on, are known. An extra complication occurs at high temperatures ($T>1$ GK): any experiment exhibits a cutoff at some maximum bombarding energy, $E_{max}^{exp}$. For example, the value of $E_{max}^{exp}$ may be dictated by the highest energy attainable with a particle accelerator. Or the measurement is simply terminated at a bombarding energy where the data analysis becomes intractable, perhaps because resonances start to overlap strongly so that a resonant structure is not discernible anymore. At lower stellar temperatures, where the energy range of effective thermonuclear burning is entirely covered by experiment, the value of $E_{max}^{exp}$ is inconsequential. However, with increasing temperature a point is eventually reached where the reaction rates cannot be calculated anymore using the available experimental information alone since part of the effective energy range has shifted beyond $E_{max}^{exp}$. In order to address this problem a {\it matching temperature}, $T_{match}$, can be determined according to 
\begin{equation}
E^{\prime}(T_{match}) + \Delta E^{\prime}(T_{match}) =  E_{max}^{exp} \label{eq:match}
\end{equation}
where $E^{\prime}$ and $\Delta E^{\prime}$ denote the location and width, respectively, of the effective thermonuclear energy range. For temperatures $T<T_{match}$ the reaction rates are assumed to be based on experimental input alone, while at $T>T_{match}$ insufficient experimental information is available for calculating the total rates. In the latter case, statistical model (Hauser-Feshbach) reaction rates can be renormalized at the matching temperature to the experimental results. The renormalized Hauser-Feshbach reaction rates provide then the extrapolated rates at temperatures beyond $T_{match}$. 

The procedure just described is generally adopted in nuclear astrophysics. (See, for example, Refs. \cite{Ang99,Ili01}, although the details of the matching method are different.) As long as $E^{\prime}$ and $\Delta E^{\prime}$ can be estimated reliably, solution of Eq. (\ref{eq:match}) will yield the value of $T_{match}$. The location and width of the effective thermonuclear energy range have usually been identified with the center and the 1/e width of the Gamow peak (see Eq. (4) of Paper I). However, recent studies emphasized the fundamental shortcomings of the Gamow peak concept for resonances especially at elevated stellar temperatures. We will not repeat the discussion here and the reader is referred to Refs. \cite{New07s,New08} for details. In the following we will summarize the main ideas and results.

Based on the known (directly measured or otherwise estimated) locations and strengths of resonances in a given reaction, the effective thermonuclear energy range can be obtained quantitatively in the following manner: (i) the fractional contribution of each resonance to the total reaction rate is calculated from the known values of resonance energy and strength; (ii) the cumulative distribution function of the fractional resonant rates is computed (which resembles a step function); (iii) the 0.50 quantile of the cumulative distribution, which is equal to the median of the fractional resonant rates, is identified with the energy location of the effective thermonuclear energy range, $E^{\prime}$; (iv) the 0.08 and 0.92 quantiles of the cumulative distribution are calculated. They define an energy range that can be identified with the width of the effective thermonuclear energy range, $\Delta E^{\prime}$. Note that this range covers an integrated rate fraction of 84\%, that is, the same value as the area enclosed between the 1/e points of the (Gaussian approximation of the) Gamow peak (Sec. 2.1 of Paper I).

\begin{figure}[]
\includegraphics[height=10.5cm]{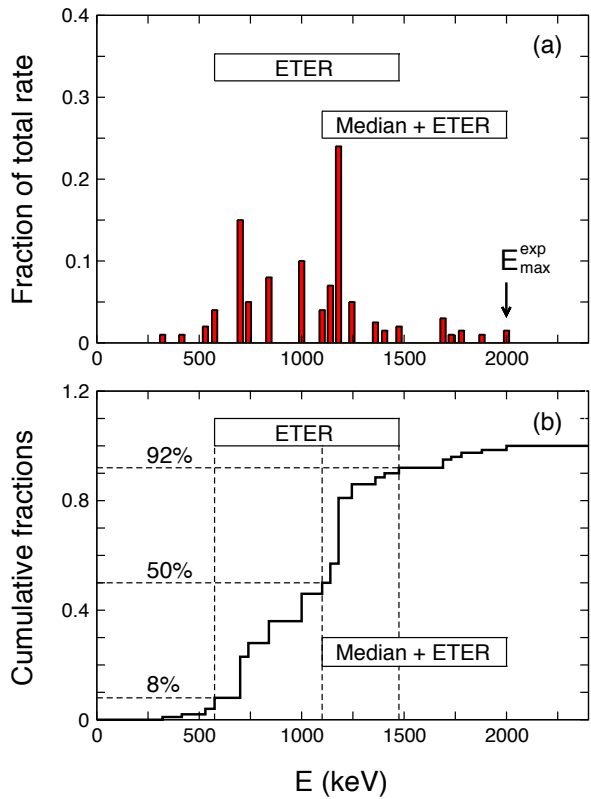}
\caption{\label{fig:eter} 
(Color online) (a) Fractional contributions of narrow resonances to the total reaction rate at a given temperature, $T$. The highest-lying observed resonance corresponds to the experimental cutoff energy, $E_{max}^{exp}$; (b) Cumulative distribution of fractional resonant rates. The 0.08, 0.50 and 0.92 quantiles define the {\it effective thermonuclear energy range}, ETER (upper horizontal bar in panels). For this particular temperature, Eq. (\ref{eq:match}) is fulfilled (see lower horizontal bar in panels), that is, $T=T_{match}$.}
\end{figure}

The situation is schematically displayed in Fig. \ref{fig:eter}. At a given temperature, $T$, the effective thermonuclear energy range (upper horizontal bars) is located well within the experimentally investigated energy region. In this particular case, Eq. (\ref{eq:match}) is precisely fulfilled, that is, $E^{\prime}+\Delta E^{\prime}$ is equal to the experimental cutoff energy, implying $T=T_{match}$ (see lower horizontal bars). For smaller temperatures, the effective thermonuclear energy range shifts to the left and the total reaction rate will be based on experimental information alone. At higher temperatures, the effective thermonuclear energy range shifts to the right such that resonances located beyond the experimentally investigated energy region may significantly contribute to the total rate. In the latter case, the rate beyond the matching temperature, $T_{match}$, may be extrapolated using the Hauser-Feshbach model.

This method yields significantly different results at elevated temperatures compared to those commonly obtained by applying the Gamow peak concept. First, for radiative capture reactions our effective thermonuclear energy range is always located at much {\it smaller} energies. The reason is that, for increasing energy, a point is eventually reached where the entrance channel particle partial width becomes larger than the $\gamma$-ray partial width. Consequently, in such situations a Gamow peak does not exist at all. Second, the value of our matching temperature for radiative capture reactions is always much {\it larger}. This point is important since the matching of experimental and Hauser-Feshbach reaction rates is more reliable at higher temperatures, where the contribution from many resonances is a prerequisite for the applicability of the statistical theory of nuclear reactions. Third, for many reactions our extrapolated rates are significantly different from previous estimates. This implies that for these reactions the experimental and Hauser-Feshbach reaction rates have different temperature dependences. In selected cases, our extrapolated reaction rates differ from previous results by an order of magnitude or more. See Ref. \cite{New08} for details.

At elevated temperatures our tabulated reaction rates are obtained in the following manner. We start by estimating from the known distribution of resonance energies and strengths and the experimental cutoff energy, $E_{max}^{exp}$, a matching temperature, $T_{match}$, according to Eq. (\ref{eq:match}). If $T_{match}$ exceeds $10$ GK, then our tabulated results are based on the Monte Carlo method for all temperatures of interest. Otherwise, at temperatures of $T_{match}\leq T \leq10$ GK the ``median" rate is calculated by normalizing Hauser-Feshbach results \cite{Arn06,Rau00} to the Monte Carlo rate at the matching temperature, $T_{match}$, as described above. At these high temperatures we assume that the reaction rate probability density function is lognormal, as can be easily verified by visual inspection of the plots following the rate tables. (Recall from Paper I that a Gaussian can be described by a lognormal distribution without difficulty). Consequently, the value of the lognormal location parameter can be computed from $\mu=\ln (x_{med})$ (see Eq. (39) of Paper I). The lognormal width parameter is approximated by the value of $\sigma$ at $T_{match}$ and is kept constant for $T>T_{match}$. This assumption can be verified by inspecting the rate tables for the general temperature dependence of $\sigma$ at high values of $T$. The low and high rates are then obtained from $x_{low}=e^{\mu-\sigma}$ and $x_{high}=e^{\mu+\sigma}$ (see Eq. (40) of Paper I). All values of reaction rates and lognormal parameters at $T_{match}\leq T \leq10$ GK that are found from this procedure are placed in parenthesis in the rate tables. No value of the Anderson-Darling statistic can be provided in this case.

A few comments are in order. It is obvious that our {\it extrapolated} reaction rates do not contain the systematic uncertainty caused by adopting Hauser-Feshbach statistical model results. This is especially important for the present rate evaluation involving rather light target nuclei where the statistical model of nuclear reactions may face serious challenges. In fact, for the mass region considered here there even appears to be no reliable estimate available for the model uncertainty of Hauser-Feshbach results. Clearly, the commonly quoted ``factor 2 reliability" of statistical model results \cite{Hof99,Rau00} seems too optimistic as can be seen from Fig. 5 in Iliadis et al. \cite{Ili01}. Furthermore, we find in exceptional cases (for example, $^{21}$Na(p,$\gamma$)$^{22}$Mg) that the extrapolated Hauser-Feshbach rates, normalized in the above manner at $T_{match}$, are {\it smaller} than the (unnormalized) Monte Carlo rates. Obviously this result is unphysical since the extrapolated Hauser-Feshbach rate must account for missing rate contributions. In such cases, we disregarded the Hauser-Feshbach results and simply placed the Monte Carlo rates in parenthesis for temperatures in excess of $T_{match}$.

\section{Reaction rate modifications for stellar model calculations}\label{thermalex}
All of the reaction rates tabulated in this work, $N_A\left<\sigma v\right>$, including the normalized and extrapolated Hauser-Feshbach results, are appropriate for target nuclei in their ground state. However, at elevated temperatures the nuclei will be thermally excited and these excited states may also participate in the nuclear burning. What is of primary astrophysical interest, therefore, is the {\it stellar} rate $N_A\left<\sigma v\right>^*$, which is obtained by taking all the contributions from excited target states into account. Since usually cross sections involving excited target nuclei cannot be measured directly in the laboratory, they have to be estimated using nuclear reaction models. At each temperature the stellar (low, median or high) rate may then be obtained as the product of the laboratory (low, median or high) rate and a stellar enhancement factor, $R_{tt}$. The latter values are frequently computed using the Hauser-Feshbach model (see, for example, Angulo et al. \cite{Ang99} or Rauscher and Thielemann \cite{Rau00}). Numerical examples for deriving stellar from laboratory rates are given in Refs. \cite{Ili01,Ili07} and are not repeated here. 

An inspection of the Hauser-Feshbach stellar enhancement factors tabulated in Ref. \cite{Rau00} reveals that they modify the laboratory rates presented here, even at the highest temperature of T=10 GK, by less than 50\%. However, it must be stressed again that the values of the stellar enhancement factors will have an associated uncertainty, which is difficult to quantify at this time. In any case, the reader should keep in mind that the (laboratory) reaction rate uncertainties and probability density functions presented here may not provide, in general, an impression of the {\it stellar} reaction rate uncertainties and probability density functions. Clearly, more work is needed to assess the reliability of the Hauser-Feshbach model in the region of the light nuclei (A$\leq$40).

In a few selected cases it is possible to estimate stellar enhancement factors without employing the Hauser-Feshbach model. If the total rate of a capture reaction $0+1\rightarrow\gamma +3$ is dominated by a number of narrow resonances labeled by $\rho$, it can be shown that the stellar rate is given by \cite{Ili07}
\begin{equation}
N_A \langle {\sigma v} \rangle^*  =  \frac{1}{G_0^{norm}}\sum\limits_\rho N_A \langle {\sigma v} \rangle_\rho^{\mu_0} \sum\limits_\mu \frac{\Gamma_{\rho \mu}}{\Gamma_{\rho \mu_0}}	\label{}
\end{equation}
where $G_0^{norm}$ is the normalized partition function of target nucleus~$0$ (values are tabulated, for example, in Refs. \cite{Rau00,Ang99}); $N_A \langle {\sigma v} \rangle_\rho^{\mu_0}$ denotes the partial ground state rate for resonance $\rho$; and $\mu$ sums over the levels in the target nucleus including the ground state ($\mu_0$). Thus the total stellar rate is given by the sum over narrow resonance ground state rates, where each resonance term is modified by a factor of $(1 + \Gamma_{\rho \mu_1}/\Gamma_{\rho \mu_0} + \Gamma_{\rho \mu_2}/\Gamma_{\rho \mu_0}  + \ldots)$, with $\Gamma_{\rho \mu}/\Gamma_{\rho \mu_0}$ denoting the ratio of particle partial widths for excited target level $\mu$ and the target ground state $\mu_0$. This branching ratio could either be calculated using the shell model or directly measured by populating the levels in question and measuring the corresponding particle decays to the ground and excited states. For example, it was found in Ref. \cite{Sch05} that the stellar enhancement factor for $^{32}$Cl(p,$\gamma$)$^{33}$Ar based on shell model calculations exceeds the Hauser-Feshbach value by a factor of $5$ near $T=0.2$ GK. We would like to emphasize again the importance of shell model calculations or measurements of particle branching ratios for estimating reliable {\it stellar} rates in the future. An implementation of calculated or measured particle branching ratios and their associated uncertainties into the present Monte Carlo formalism is straightforward.

Finally, it must be pointed out that at elevated densities the {\it bare nuclei} reaction rates presented here have to be corrected for electron screening effects. For more information, see Iliadis \cite{Ili07} and references therein.

\section{Reverse reaction rates} 
Stellar models require as input to their reaction network not only the rate of a forward reaction, but that of the reverse process as well. Forward and reverse reaction rates are intimately connected by the reciprocity theorem. Consider the reactions $0+1\rightarrow 2+3$ and $2+3\rightarrow 0+1$, with a number of excited states in nucleus~$0$ and in nucleus~$3$, which are all in thermal equilibrium with their respective ground states. If we disregard excited states in the light nuclei $1$ and $2$ (usually protons, neutrons and $\alpha$-particles) then it can be shown \cite{Ili07} that for non-identical particles the rates of the forward and reverse reaction are related by
\begin{equation}
\frac{N_A \langle \sigma v \rangle_{23 \to 01}^*}{N_A \langle \sigma v \rangle_{01 \to 23}^*} = 
\frac{g_0 g_1}{g_2 g_3}\frac{G_0^{norm}}{G_3^{norm}}\left(\frac{M_0 M_1}{M_2 M_3}\right)^{3/2} e^{-11.605\,Q/T_9}	\label{reverse1}
\end{equation}
while for reactions involving photons, $0+1\rightarrow \gamma +3$ and $3+\gamma \rightarrow 0+1$, one finds
\begin{equation}
\frac{\lambda_{\gamma}^*(3 \to 01)}{N_A \langle \sigma v \rangle_{01 \to \gamma 3}^*} = 
9.8685 \cdot 10^9\,{T_9^{3/2}} \frac{g_0 g_1}{g_3}\frac{G_0^{norm}}{G_3^{norm}}\left(\frac{M_0 M_1}{M_3}\right)^{3/2} e^{-11.605\,Q/T_9} \label{reverse2}
\end{equation}
where $g_i=(2j_i+1)$ is the ground state statistical weight of nucleus $i$ ($j_i$ denotes the spin); $G_i^{norm}$ is the normalized partition function of nucleus $i$; 
$M_i$ denotes the ground state mass (in u); $Q$ is the ground-state $Q$-value of the forward reaction $0+1 \rightarrow 2+3$ or $0+1 \rightarrow \gamma + 3$~(in MeV); $T_9 \equiv T/10^9$~K; and $\lambda_{\gamma}^*(3 \to 01)$ is the stellar decay constant for photodisintegration of nucleus $3$. Therefore, at any given temperature, the reverse rate can be computed from the forward rate in a straightforward manner once spins, masses, normalized partition functions and the reaction Q-value are known. It must be stressed that although the above relationships are sometimes erroneously applied to {\it laboratory} rates, they are only correct for {\it stellar} reaction rates. Numerical examples for their application are given in Refs. \cite{Ili01,Ili07} and are not repeated here. 

From the present results for rates and probability density functions of a forward reaction it is straightforward to calculate, after proper corrections for thermal target excitations have been taken into account (Sec. \ref{thermalex}), the rates and probability density functions of the corresponding reverse reaction. We may rewrite Eqs. (\ref{reverse1}) and (\ref{reverse2}) as $N_A \langle \sigma v \rangle_R^*=a~(e^{Q})^c~N_A \langle \sigma v \rangle_F^*$, with $a$ equal to the constant factor on the right hand sides excluding the exponential and $c=-11.605/T_9$. The Q-value is expected to follow a Gaussian probability density function, where the value of $Q$ and the uncertainty $\Delta Q$ reported in the literature represent the mean and standard deviation, respectively (Sec. 4.1 in Paper I). Therefore, $e^{Q}$ will be represented by a lognormal distribution. In most cases the forward rate, $N_A \langle \sigma v \rangle_F^*$, is also described by a lognormal distribution, with location and spread parameters of $\mu$ and $\sigma$ that are tabulated here. Consequently, the reverse reaction rate, $N_A \langle \sigma v \rangle_R^*$, will follow a lognormal distribution as well, with location and spread parameters of $\mu_R=\ln (a)+\mu+cQ$ and $\sigma_R^2=\sigma^2+c^2(\Delta Q)^2$, respectively (see Sec. 4.2 in Paper I).

We do not present here numerical values of the lognormal parameters $\mu_R$ and $\sigma_R$ for reverse reactions. Although it would be interesting for some applications to compute the probability density function of a reverse reaction, this information is usually not directly relevant for stellar models. In a reaction network calculation the forward and reverse rates are correlated, as is apparent from Eqs. (\ref{reverse1}) and (\ref{reverse2}). A Monte Carlo reaction network study would involve the following steps: (i) sampling over the rate of the forward reaction, according to a lognormal distribution with parameters of $\mu$ and $\sigma$ that are presented here (after correcting for thermal target excitations); (ii) sampling over the reaction Q-value, which is described by a Gaussian probability density with a mean of $Q$ and a standard deviation of $\Delta Q$ ; (iii) calculation of the resulting reverse rate from Eqs. (\ref{reverse1}) and (\ref{reverse2}) by using the {\it same} value for the rate of the forward reaction that was sampled under step (i). Disregarding the correlation between forward and reverse rate may greatly exaggerate the Monte Carlo uncertainties of selected nuclidic abundances. Finally, note that the procedure just described does not account for the Q-value correlation between forward and reverse rates when the resonance energies for the forward rate are computed from $E_r=E_x-Q$. Such higher-order effects may become important in certain applications.

\section{Summary}
We tabulate charged particle thermonuclear reaction rates and probability density functions for target nuclei in the A=14 to 40 mass range, which are overwhelmingly based on {\it experimental nuclear physics information}. The results are obtained using a Monte Carlo method, which is based on physically motivated probability density functions of all nuclear physics input quantities. For the first time, an evaluation is presented that provides statistically meaningful low, recommended and high reaction rates. They are obtained at each stellar temperature from the 0.16, 0.50 and 0.84 quantiles of the cumulative reaction rate distribution (corresponding to a coverage probability of 68\%). Graphs of reaction rate probability density functions at selected temperatures and of reaction rate uncertainties are provided for visual inspection. 

We approximate the Monte Carlo probability density function of the total reaction rate by a lognormal distribution. At each temperature, the lognormal parameters $\mu$ and $\sigma$ are tabulated. A quantitative measure (Anderson-Darling test statistic) is provided for assessing the reliability of the lognormal approximation. The reasons why the actual Monte Carlo reaction rate probability density function {\it frequently follows} a lognormal distribution are discussed in detail (Sec. \ref{sec:si28pg} and Sec. 5.4 of Paper I). The user can implement the approximate lognormal reaction rate probability density functions directly in a stellar model code for studies of stellar energy generation and nucleosynthesis. 

We provide clear evidence that the reaction rate probability density function does {\it not always follow} a lognormal distribution, as is sometimes erroneously claimed in the literature. This is especially the case when the uncertainty in a resonance energy is large and the rate is calculated from a particle partial width, or if upper limits of partial widths influence the total rate. If a particular reaction rate probability density deviates significantly from lognormality and it turns out that stellar model calculations are sensitive to the {\it shape} of this distribution, then either new measurements need to be performed in order to improve uncertainties of important input quantities, or another analytical function must be used  that is more complicated than a lognormal distribution, but approximates the actual reaction rate probability density function more closely.

The results tabulated here are appropriate for {\it bare nuclei in the laboratory}. For use in stellar model calculations, they must be corrected, if appropriate, at elevated temperatures for thermal target excitations and at elevated densities for electron screening effects. These corrections introduce another source of uncertainty, which is not easily quantified at present. Thus our {\it laboratory} reaction rate uncertainties and probability density functions may not provide, in general, an impression of the {\it stellar} reaction rate uncertainties and probability density functions. More work is called for to assess the reliability of Hauser-Feshbach reaction rates in the region of the light nuclei (A$\leq$40). The probability density function of a reverse rate is easily calculated from the tabulated lognormal parameters of the corresponding forward rate. However, in an actual reaction network study the forward and reverse rates should not be sampled independently since they are correlated. The nuclear physics input used to derive our results is presented in the subsequent paper of this series (Paper III). In the fourth paper of this series (Paper IV) we compare our reaction rates to previous results.

\section{Acknowledgement}
The authors would like to thank Stephane Goriely for providing unpublished Hauser-Feshbach rates for some of the reactions evaluated here. This work was supported in part by the U.S. Department of Energy under Contract No. DE-FG02-97ER41041. 

\clearpage
\appendix
\section{Statistical hypothesis tests}\label{stattest}
The following two questions are of interest in the present work: (i) how can one decide if a given Monte Carlo reaction rate probability density function is consistent with a lognormal distribution? (ii) Can one provide a quantitative criterion for how well the lognormal distribution approximates the actual Monte Carlo probability density function? Both questions are related and are addressed by a branch of statistics that is referred to as statistical test theory. For an introduction to this topic, see Refs. \cite{Cow98,Pre92}. Here we will briefly describe the general ideas for the reader who is not familiar with these concepts. 

\begin{figure}[]
\includegraphics[height=5.5cm]{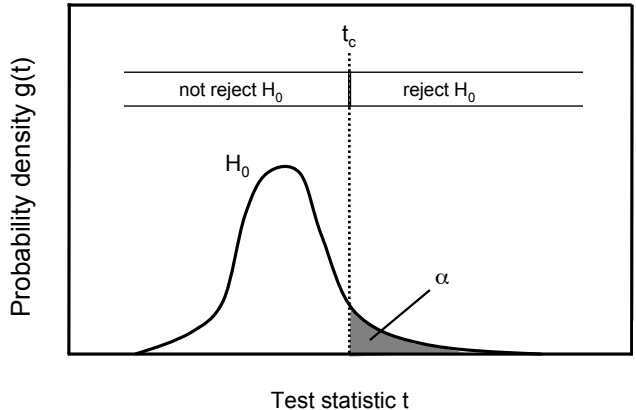}
\caption{\label{fig:hypotest} 
Schematic representation of a statistical test. The curve shows the probability density function of the test statistic $t$ that is obtained under the assumption of the null hypothesis $H_0$ (in our case, that the data follow a specified distribution). The shaded area underneath the curve to the right of the critical value, $t_c$, is equal to the significance level, $\alpha$, and represents the probability that the hypothesis $H_0$ is rejected if it is in fact true. The null hypothesis is rejected if $t$ is observed in the critical region, $t>t_c$. In such a case one concludes that ``$H_0$ is rejected at the $\alpha$\% level".}
\end{figure}

Usually one starts with an assumption, called a null hypothesis, that some data follow a specific distribution. In order to investigate the measure of agreement between the observed (or sampled) data and a given hypothesis, one constructs a function of the measured values, called a test statistic, $t$. This statistic depends on the type of test and has an associated probability density function, $g(t)$. Thus one can define a critical value, $t_c$, by
\begin{equation}
 \alpha = \int_{t_c}^\infty g(t) dt
\end{equation}
such that the null hypothesis is rejected if $t>t_c$. The critical value depends on the probability, $\alpha$, of rejecting the null hypothesis when it is in fact true. This probability is referred to as the significance level of the test. The confidence level of the test, $1-\alpha$, corresponds to the probability that the null hypothesis is not rejected when it is in fact true. For example, a relatively small significance level of $\alpha=0.05$ implies that the null hypothesis is rejected only 5\% of the time when it is in fact true; the confidence level amounts in this case to 95\%. The relationships of the various quantities just introduced are displayed in Fig. \ref{fig:hypotest}.

Frequently, continuous data are binned by grouping events into specific ranges of the continuous variable. The standard statistical test for binned distributions is the chi-squared test. However, binning involves always a loss of information and, furthermore, there is often considerable arbitrariness as to how the bins should be chosen. Thus other, more powerful, tests are preferred that avoid the unnecessary binning of data. One of these is called Kolmogorov-Smirnov (K-S) test which uses as its statistic simply the maximum value of the absolute difference between two cumulative distributions. The K-S test has the limitation that if one estimates any parameters from a data set, then the probability density function of the K-S statistic involving a cumulative distribution that uses the estimated parameters can no longer be obtained in a simple manner. 

An alternative test that is not subject to this limitation is the Anderson-Darling (A-D) test which is adopted in the present work. Recall from Sec. 4.2 in Paper I that a random variable $y=\ln (x)$ is Gaussian distributed if the variable $x$ follows a lognormal distribution. It is computationally simpler to test for a Gaussian instead of a lognormal distribution. Therefore, we transform first our Monte Carlo reaction rate values, $x_i$, by using $y_i = (\ln x_i - \mu)/\sigma$, where $\mu$ and $\sigma$ are the lognormal parameters (equivalent to the mean value and standard deviation for $\ln (x)$; see Paper I). The A-D test statistic is then defined as
\begin{equation}
 t_{AD} = -n-\sum_{i=1}^n \frac{2i-1}{n} \left( \ln F(y_i) + \ln \left[ 1-F(y_{n+1-i}) \right] \right) 
\end{equation}
where $y_i$ are the ordered sample values, $F$ denotes the cumulative distribution for a standard Gaussian (that is, with a mean of zero and a variance of unity), and $n$ refers to the sample size. The critical values, $t_c$, for the A-D test depend on the specific distribution that is being tested. They are tabulated in Ref. \cite{Ste74} for a Gaussian distribution. For example, one obtains in this case $t_c = 0.656, 0.787$ and $1.092$ for significance levels of $\alpha = 0.1, 0.05$ and $0.01$, respectively (corresponding to confidence levels of 90\%, 95\% and 99\%, respectively). For a finite sample size, $n$, the the quantity $t_{AD}$ needs to be replaced by the adjusted A-D test statistic $t_{AD}^*=t_{AD}(1+4/n-25/n^2)$ \cite{Ste74}. The null hypothesis is then rejected if the adjusted A-D statistic, $t_{AD}^*$, exceeds the critical value, $t_c$, that corresponds to a given confidence level.

Numerical values of the A-D test statistic, $t_{AD}^*$, are presented in the reaction rate tables of App. B for each temperature.
In many cases it can be seen that $t_{AD}^*$ exceeds the critical values quoted above. Regarding question (i) posed above we can thus provide proof that in general the Monte Carlo reaction rate probability density function does {\it not} follow a lognormal distribution. Question (ii) posed above can be answered by visually inspecting the graphs of reaction rate probability density functions following the reaction rate tables. We find empirically that for values smaller than $t_{AD}^*\approx30$ the lognormal approximation holds well and seems adequate for use in reaction network calculations. Note that the answers to both questions depend on the sample size: the larger $n$, the better the statistics, and the larger the likelihood that a lognormal approximation will be rejected by the Anderson-Darling test.

\clearpage 
\section{Tables and graphs of Monte Carlo reaction rates}\label{tabgraph}
The Monte Carlo reaction rates listed below have a different meaning from the classical reaction rates, as pointed out in the introduction and as explained in Paper I. The quantities presented in the tables and graphs have the following meaning:\\
\\
\begin{tabbing}
AAAAAAAAAAAAA\= \kill
T				\> Temperature in GK\\  
\\
Low rate			\> 0.16 quantile of the cumulative reaction rate distribution \\
Median rate		\> 0.50 quantile of the cumulative reaction rate distribution\\
High rate			\> 0.84 quantile of the cumulative reaction rate distribution\\
\\
				\> All reaction rates are in units of cm$^3$~s$^{-1}$~mol$^{-1}$\\
\\
$\mu$ 			\> Parameter determining the location of the lognormal reaction \\
				\> rate probability density function\\
$\sigma$   		\> Parameter determining the width of the lognormal reaction \\
				\> rate  probability density function\\
\\
				\> The Monte Carlo reaction rate probability density function is  \\
				\> here approximated by a lognormal distribution (Eq. (25) of  \\
				\> Paper I),
\end{tabbing}
\vspace{6mm}
\begin{equation}
f(x) = \frac{1}{\sigma \sqrt{2\pi}} \frac{1}{x} e^{-(\ln x - \mu)^2/(2\sigma^2)}\notag
\end{equation}
\vspace{3mm}
\begin{tabbing}
AAAAAAAAAAAAA\= \kill
				\> where $x$ stands for reaction rate and ``$\ln$" denotes the natural \\
				\> logarithm. The lognormal parameters are calculated from the \\
				\>  expectation value and variance of $\ln(x)$ by using $\mu = E[\ln(x)]$ \\
				\>  and $\sigma^2 = V[\ln(x)]$ (Sec. 4.2 of Paper I).\\
\\				
				\> When the lognormal approximation is in agreement with the \\
				\> Monte Carlo reaction rate probability density function, which \\
				\> applies in the majority of cases, the parameters $\mu$ and $\sigma$ are \\
				\> related to the low, median and high Monte Carlo reaction rate \\
				\> by $\mu=\ln (x_{med})$ and $\sigma=\ln (x_{high}/x_{low})^{1/2}$ (see Eq. (39) of \\
				\> Paper I). Alternatively, the low, median and high rates can be \\
				\> obtained from the lognormal parameters by using $x_{low}=e^{\mu-\sigma}$, \\
				\>  $x_{med}=e^{\mu}$ and $x_{high}=e^{\mu+\sigma}$ (see Eq. (40) of Paper I). These \\
                                 \> relationships apply to a coverage probability of 68\%. Also, the \\
				\> lognormal parameter $\sigma$ directly indicates the factor uncertainty, \\
				    \> $f.u.=e^{\sigma}$, and the skewness of the reaction rate distribution; a \\
				    \> value of $\sigma<0.1$ corresponds to a nearly symmetric (that is, \\
				    \> Gaussian) distribution, while for larger values the distribution \\
				    \> is noticeably skewed (Sec. 4.2 of Paper I). \\
\\
A-D  				\> Anderson-Darling test statistic, $t_{AD}^*$, indicating how well the  \\
				\> Monte Carlo reaction rates are approximated by a lognormal  \\
				\> distribution. A value in excess of $t_{AD}^*\approx1$ indicates that the \\
				\> reaction rate probability density function is {\it not} lognormal.  \\
				\> For a value in excess of $t_{AD}^*\approx30$ the lognormal approximation  \\
				\> starts to deviate {\it visually} from the reaction rate probability \\
				\> density function, as can be seen by inspecting the graphs \\			
				\> following the reaction rate tables. Note that regardless of the \\
				\>  magnitude of $t_{AD}^*$, the values of $\mu$ and $\sigma$ listed in the tables  \\
				\> define a lognormal distribution of the same expectation value \\
				\> and variance as the actual Monte Carlo probability density  \\
				\> function. If no value of A-D is provided, then the rates either\\
				\> have been found from extrapolation to high temperature (see  \\
				\> below) or are determined by entirely different means (see \\
				\> comments).\\
\\
()				\> Values given in parenthesis are usually {\it not} obtained from the  \\
				\> Monte Carlo method, but are found from extrapolation to  \\
				\> elevated temperatures. In this case, the {\it recommended} rates of  \\
				\> column 3 are calculated by normalizing Hauser-Feshbach results  \\
				\> to the Monte Carlo rate at the matching temperature, $T_{match}$  \\
				\> (see Sec. \ref{extrap}). For $T>T_{match}$ the value of the lognormal location \\
				\> parameter is found from $\mu=\ln (x_{med})$, while the lognormal \\
				\> width parameter is approximated by the value of $\sigma$ at $T_{match}$ \\
				\> and is held constant. The low and high rates are then obtained \\
				\>  from $x_{low}=e^{\mu-\sigma}$ and $x_{high}=e^{\mu+\sigma}$. See Eqs. (39) and (40) of \\
				\> Paper I . No value of A-D is provided in this case. In exceptional \\
				\> cases, the extrapolated Hauser-Feshbach rates become smaller \\ 
				\> than the Monte Carlo rates. Since this result is unphysical, the \\
				\> Hauser-Feshbach results are disregarded and the Monte Carlo \\
				\> rates are placed in parenthesis for $T > T_{match}$. \\
\end{tabbing}

\noindent Each reaction rate table is accompanied by two figures. The first of these displays the reaction rate ratios $N_A\left<\sigma v\right>_{high}/N_A\left<\sigma v\right>_{med}$ and $N_A\left<\sigma v\right>_{low}/N_A\left<\sigma v\right>_{med}$ at temperatures below $T_{match}$; a visual inspection immediately reveals the reaction rate uncertainty at a given temperature. The second figure shows for selected temperatures (T=0.03, 0.06, 0.1, 0.3, 0.6 and 1.0 GK) the Monte Carlo reaction rate probability density functions (in red), together with their lognormal approximations (in black). The latter curves are calculated with the lognormal parameters $\mu$ and $\sigma$ that are listed in columns 5 and 6 of the reaction rate tables. For a given reaction, each panel in the second figure displays the temperature, T, and the Anderson-Darling test statistic, A-D. Note that for each reaction the value of $T_{match}$ can easily be obtained in two ways: it is equal to the lowest temperature for which no value of A-D is listed; it is also equal to the highest temperature shown in the figure displaying the reaction rate uncertainties.\\

\noindent Example: For the $^{23}$Na(p,$\alpha$)$^{20}$Ne reaction at $T=0.3$ GK, we obtain the following numerical results from Tab. \ref{tab:na23pa},\\
\\
$N_A \left< \sigma v \right>_{low}=1.11\times10^{+00}$\\
$N_A \left< \sigma v \right>_{med}=1.26\times10^{+00}$\\
$N_A \left< \sigma v \right>_{high}=1.44\times10^{+00}$\\
$\mu=2.341\times10^{-01}$\\
$\sigma=1.32\times10^{-01}$\\
A-D$=3.54\times10^{+00}$\\
\\
The first three numbers refer to the low, median and high reaction rates. The reaction rate probability density function, assumed to follow a lognormal distribution, is determined by the lognormal parameters $\mu$ and $\sigma$,
\begin{equation}
f(x) = \frac{1}{0.132 \sqrt{2\pi}} \frac{1}{x} e^{-(\ln x - 0.2341)^2/(2 \cdot 0.132^2)}\notag
\end{equation}
This function is plotted as a black line in the second figure following Tab. \ref{tab:na23pa}. The Anderson-Darling statistic amounts to $t_{AD}^*=3.54$ and, since $t_c\approx1$, the A-D test rejects the hypothesis that the reaction rate probability density function is given by a lognormal distribution. The deviation of the actual Monte Carlo reaction rate probability density function, shown as a red histogram, and the lognormal distribution is barely visible in the figure. Although the lognormal approximation is rejected by the A-D test, it is obvious that the black line is in close agreement with the red histogram. Consequently, the lognormal function given above represents a useful approximation for stellar model studies. The reader may also verify that the above numerical results agree with Eqs. (39) and (40) of Paper I. 

%
\clearpage
\setlongtables
\begin{longtable}{@{\extracolsep{\fill}}lccc | ccc}
\caption{Total thermonuclear reaction rates for $^{14}$C(p,$\gamma$)$^{15}$N.}  \label{} \\
\hline \hline 
	\multicolumn{1}{c}{T (GK)} & \multicolumn{1}{c}{Low rate} & \multicolumn{1}{c}{Median rate} & \multicolumn{1}{c}{High rate}  & \multicolumn{1}{c}{lognormal $\mu$} & \multicolumn{1}{c}{lognormal $\sigma$} & \multicolumn{1}{c}{A-D} \\ \hline 
\endfirsthead
\multicolumn{6}{c}{{\tablename} \thetable{} -- continued} \\
\hline \hline 
	\multicolumn{1}{c}{T (GK)} & \multicolumn{1}{c}{Low rate} & \multicolumn{1}{c}{Median rate} & \multicolumn{1}{c}{High rate} & \multicolumn{1}{c}{lognormal $\mu$} & \multicolumn{1}{c}{lognormal $\sigma$} & \multicolumn{1}{c}{A-D} \\ \hline 
\endhead
	 \hline \hline
\endfoot
	\hline \hline
\endlastfoot
0.010 & 2.66$\times$10$^{-19}$ & 3.61$\times$10$^{-19}$ &
      5.06$\times$10$^{-19}$ &  -4.245$\times$10$^{+01}$ &
       3.24$\times$10$^{-01}$  & 1.96$\times$10$^{+00}$ \\ 
0.011 & 1.81$\times$10$^{-18}$ & 2.49$\times$10$^{-18}$ &
      3.46$\times$10$^{-18}$ &  -4.053$\times$10$^{+01}$ &
       3.27$\times$10$^{-01}$  & 1.33$\times$10$^{+00}$ \\ 
0.012 & 9.98$\times$10$^{-18}$ & 1.38$\times$10$^{-17}$ &
      1.91$\times$10$^{-17}$ &  -3.882$\times$10$^{+01}$ &
       3.24$\times$10$^{-01}$  & 1.73$\times$10$^{+00}$ \\ 
0.013 & 4.60$\times$10$^{-17}$ & 6.30$\times$10$^{-17}$ &
      8.84$\times$10$^{-17}$ &  -3.729$\times$10$^{+01}$ &
       3.26$\times$10$^{-01}$  & 1.64$\times$10$^{+00}$ \\ 
0.014 & 1.81$\times$10$^{-16}$ & 2.49$\times$10$^{-16}$ &
      3.57$\times$10$^{-16}$ &  -3.591$\times$10$^{+01}$ &
       3.29$\times$10$^{-01}$  & 3.62$\times$10$^{+00}$ \\ 
0.015 & 6.35$\times$10$^{-16}$ & 8.67$\times$10$^{-16}$ &
      1.22$\times$10$^{-15}$ &  -3.467$\times$10$^{+01}$ &
       3.25$\times$10$^{-01}$  & 2.04$\times$10$^{+00}$ \\ 
0.016 & 1.97$\times$10$^{-15}$ & 2.73$\times$10$^{-15}$ &
      3.83$\times$10$^{-15}$ &  -3.353$\times$10$^{+01}$ &
       3.30$\times$10$^{-01}$  & 1.29$\times$10$^{+00}$ \\ 
0.018 & 1.51$\times$10$^{-14}$ & 2.07$\times$10$^{-14}$ &
      2.92$\times$10$^{-14}$ &  -3.150$\times$10$^{+01}$ &
       3.30$\times$10$^{-01}$  & 1.27$\times$10$^{+00}$ \\
\end{longtable}
Comments: Resonance energies are deduced from excitation energies \cite{Ajz91} and the reaction Q-value (Tab. \ref{tab:master}). For the first five resonances, between E$_r^{cm}=242$ and 597 keV, we use the strengths measured by G\"orres et al. \cite{GO90}, their total widths being unknown or small. The next five resonances are broad and their partial widths are known \cite{BA55,FE59,FR61,KU71,RA72,KU76,NI77}. It must be pointed out that only the $\gamma$-ray transition to the ground state has been measured for the E$_r^{cm}=1085$ and 1230 keV resonances, so that their contribution should be considered as a lower limit. Following G\"orres et al. \cite{GO90}, we calculate the direct capture contribution using experimental spectroscopic factors from Bommer et al. \cite{BO75}. Only five transitions contribute significantly to the total S-factor, which can be parametrized as $S(E) = 5.24 -1.22\times10^{-3}E+2.94\times10^{-7}E^2$~keV~b (with $E$ in keV). (Note that the nonresonant S-factor reported by Ref. \cite{GO90} does not represent the direct capture S-factor, but includes contributions from resonance tails.) 
\begin{figure}[]
\includegraphics[height=8.5cm]{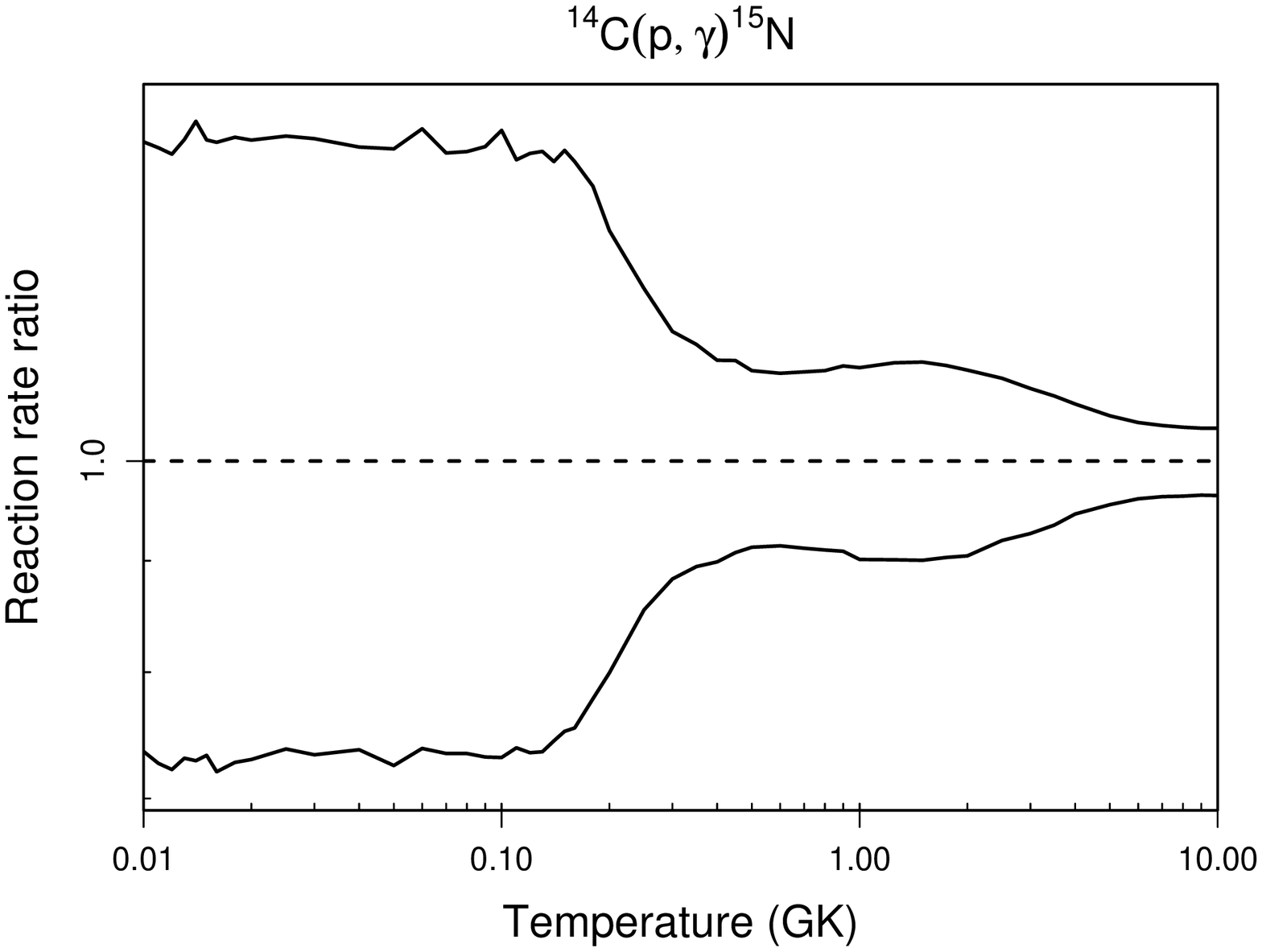}
\end{figure}
\clearpage
\begin{figure}[]
\includegraphics[height=18.5cm]{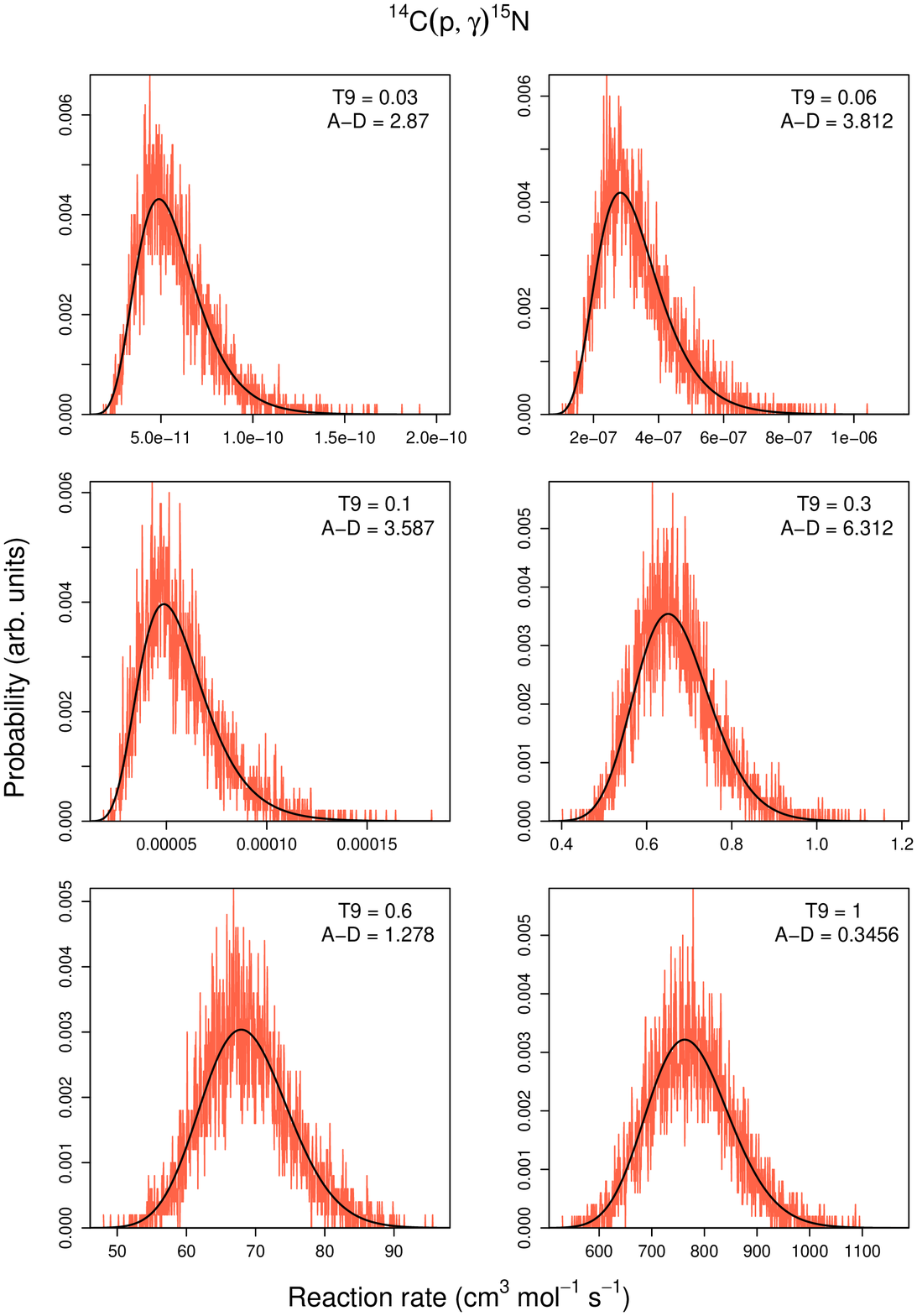}
\end{figure}
\clearpage
\setlongtables
\begin{longtable}{@{\extracolsep{\fill}}lccc | ccc}
\caption{Total thermonuclear reaction rates for $^{14}$C($\alpha$,$\gamma$)$^{18}$O.}  \label{} \\
\hline \hline 
	\multicolumn{1}{c}{T (GK)} & \multicolumn{1}{c}{Low rate} & \multicolumn{1}{c}{Median rate} & \multicolumn{1}{c}{High rate}  & \multicolumn{1}{c}{lognormal $\mu$} & \multicolumn{1}{c}{lognormal $\sigma$} & \multicolumn{1}{c}{A-D} \\ \hline 
\endfirsthead
\multicolumn{6}{c}{{\tablename} \thetable{} -- continued} \\
\hline \hline 
	\multicolumn{1}{c}{T (GK)} & \multicolumn{1}{c}{Low rate} & \multicolumn{1}{c}{Median rate} & \multicolumn{1}{c}{High rate} & \multicolumn{1}{c}{lognormal $\mu$} & \multicolumn{1}{c}{lognormal $\sigma$} & \multicolumn{1}{c}{A-D} \\ \hline 
\endhead
	 \hline \hline
\endfoot
	\hline \hline
\endlastfoot
0.010 & 1.29$\times$10$^{-54}$ & 2.66$\times$10$^{-54}$ &
      6.09$\times$10$^{-54}$ &  -1.233$\times$10$^{+02}$ &
       8.05$\times$10$^{-01}$  & 9.05$\times$10$^{+00}$ \\ 
0.011 & 1.30$\times$10$^{-52}$ & 2.74$\times$10$^{-52}$ &
      6.59$\times$10$^{-52}$ &  -1.187$\times$10$^{+02}$ &
       8.08$\times$10$^{-01}$  & 6.51$\times$10$^{+00}$ \\ 
0.012 & 7.71$\times$10$^{-51}$ & 1.63$\times$10$^{-50}$ &
      3.99$\times$10$^{-50}$ &  -1.146$\times$10$^{+02}$ &
       8.29$\times$10$^{-01}$  & 1.11$\times$10$^{+01}$ \\ 
0.013 & 2.92$\times$10$^{-49}$ & 6.42$\times$10$^{-49}$ &
      1.55$\times$10$^{-48}$ &  -1.109$\times$10$^{+02}$ &
       8.31$\times$10$^{-01}$  & 8.00$\times$10$^{+00}$ \\ 
0.014 & 8.14$\times$10$^{-48}$ & 1.75$\times$10$^{-47}$ &
      4.30$\times$10$^{-47}$ &  -1.076$\times$10$^{+02}$ &
       8.41$\times$10$^{-01}$  & 1.00$\times$10$^{+01}$ \\ 
0.015 & 1.59$\times$10$^{-46}$ & 3.53$\times$10$^{-46}$ &
      8.48$\times$10$^{-46}$ &  -1.046$\times$10$^{+02}$ &
       8.45$\times$10$^{-01}$  & 8.01$\times$10$^{+00}$ \\ 
0.016 & 2.55$\times$10$^{-45}$ & 5.43$\times$10$^{-45}$ &
      1.35$\times$10$^{-44}$ &  -1.019$\times$10$^{+02}$ &
       8.49$\times$10$^{-01}$  & 9.89$\times$10$^{+00}$ \\ 
0.018 & 3.24$\times$10$^{-43}$ & 7.18$\times$10$^{-43}$ &
      1.74$\times$10$^{-42}$ &  -9.699$\times$10$^{+01}$ &
       8.59$\times$10$^{-01}$  & 8.47$\times$10$^{+00}$ \\
\end{longtable}
Comments: Similar to previous evaluations \cite{GA87,GO92,LU04}, we consider the first eight natural-parity levels above, and the first one below, the $\alpha$-particle threshold. Resonance energies are deduced from excitation energies \cite{Til95} and the reaction Q-value (Tab. \ref{tab:master}). Resonance strengths and radiative widths for the seven higher-energy resonances have been measured by Gai et al. \cite{GA87}.  The total widths of the $E_r^{cm}=-28$ and 178 keV resonances, which are equal to the radiative widths, and of the $E_r^{cm}=1987$ and 2056 keV resonances, which are equal to the $\alpha$-particle widths, are known from the literature \cite{Til95}. For the -28 and 178 keV resonances we follow Lugaro et al. \cite{LU04} by adopting for the $\alpha$-particle reduced widths values of $\theta^2_\alpha\approx0.02$ and $\theta^2_\alpha<0.02$, respectively, based on the results of the $\alpha$-particle transfer experiment by Cunsolo et al. \cite{CU81}. We assume a factor of two uncertainty for the adopted $\theta^2_\alpha$ value of the subthreshold resonance. Note that in Lugaro et al. \cite{LU04} there is an apparent confusion between two levels, as it is the 6.2 MeV state that was observed by Ref. \cite{CU81}, not the 6.4 MeV state. The non-resonant S-factor is adopted from the measurement by G\"orres et al. \cite{GO92}. However, it includes contributions from both direct capture and from broad-resonance tails. Since interference effects are expected between the $1^-$ levels \cite{BU07}, we assume a factor of 3 uncertainty in the non-resonant S-factor. Above T = 2.13 GK the total rate is extrapolated using Hauser-Feshbach results.
\begin{figure}[]
\includegraphics[height=8.5cm]{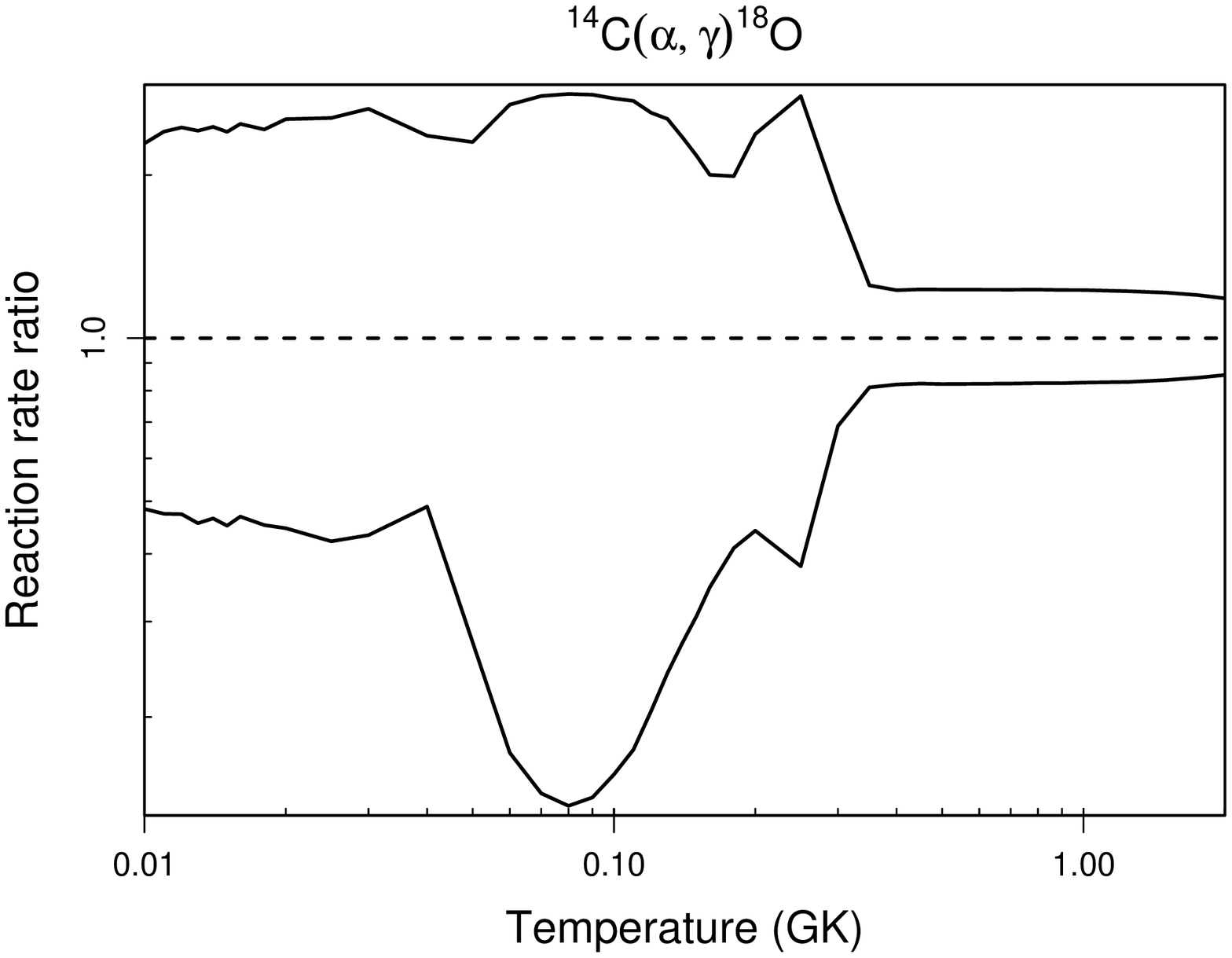}
\end{figure}
\clearpage
\begin{figure}[]
\includegraphics[height=18.5cm]{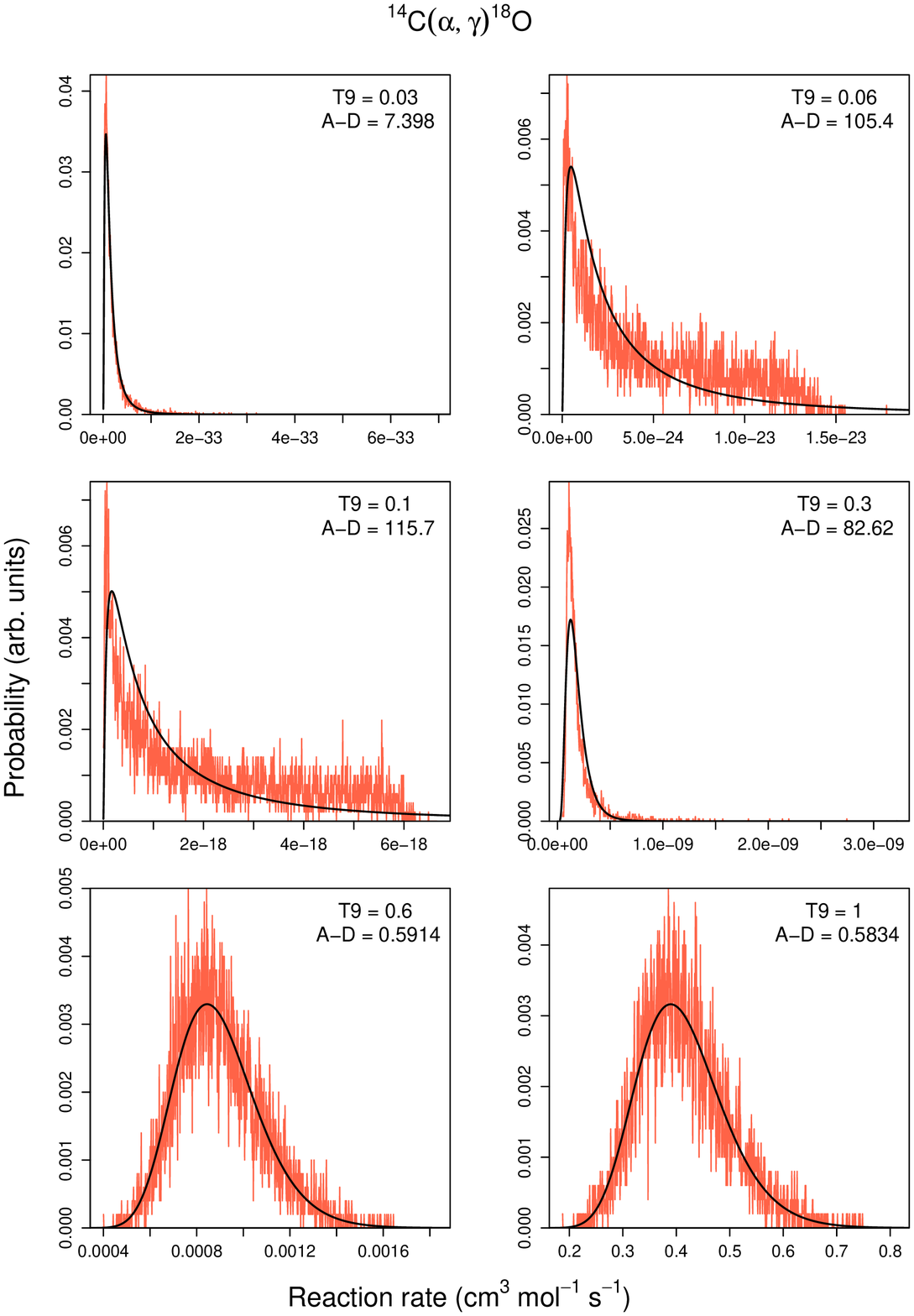}
\end{figure}
\clearpage
\setlongtables
\begin{longtable}{@{\extracolsep{\fill}}lccc | ccc}
\caption{Total thermonuclear reaction rates for $^{14}$N($\alpha$,$\gamma$)$^{18}$F.}  \label{} \\
\hline \hline 
	\multicolumn{1}{c}{T (GK)} & \multicolumn{1}{c}{Low rate} & \multicolumn{1}{c}{Median rate} & \multicolumn{1}{c}{High rate}  & \multicolumn{1}{c}{lognormal $\mu$} & \multicolumn{1}{c}{lognormal $\sigma$} & \multicolumn{1}{c}{A-D} \\ \hline 
\endfirsthead
\multicolumn{6}{c}{{\tablename} \thetable{} -- continued} \\
\hline \hline 
	\multicolumn{1}{c}{T (GK)} & \multicolumn{1}{c}{Low rate} & \multicolumn{1}{c}{Median rate} & \multicolumn{1}{c}{High rate} & \multicolumn{1}{c}{lognormal $\mu$} & \multicolumn{1}{c}{lognormal $\sigma$} & \multicolumn{1}{c}{A-D} \\ \hline 
\endhead
	 \hline \hline
\endfoot
	\hline \hline
\endlastfoot
0.010 & 2.62$\times$10$^{-63}$ & 4.78$\times$10$^{-63}$ &
      9.25$\times$10$^{-63}$ &  -1.435$\times$10$^{+02}$ &
       6.38$\times$10$^{-01}$  & 1.66$\times$10$^{+00}$ \\ 
0.011 & 4.47$\times$10$^{-61}$ & 8.38$\times$10$^{-61}$ &
      1.63$\times$10$^{-60}$ &  -1.383$\times$10$^{+02}$ &
       6.45$\times$10$^{-01}$  & 1.79$\times$10$^{+00}$ \\ 
0.012 & 4.36$\times$10$^{-59}$ & 8.15$\times$10$^{-59}$ &
      1.55$\times$10$^{-58}$ &  -1.337$\times$10$^{+02}$ &
       6.39$\times$10$^{-01}$  & 1.46$\times$10$^{+00}$ \\ 
0.013 & 2.60$\times$10$^{-57}$ & 4.88$\times$10$^{-57}$ &
      9.39$\times$10$^{-57}$ &  -1.296$\times$10$^{+02}$ &
       6.48$\times$10$^{-01}$  & 1.67$\times$10$^{+00}$ \\ 
0.014 & 1.02$\times$10$^{-55}$ & 1.95$\times$10$^{-55}$ &
      3.80$\times$10$^{-55}$ &  -1.260$\times$10$^{+02}$ &
       6.48$\times$10$^{-01}$  & 2.75$\times$10$^{+00}$ \\ 
0.015 & 2.97$\times$10$^{-54}$ & 5.63$\times$10$^{-54}$ &
      1.08$\times$10$^{-53}$ &  -1.226$\times$10$^{+02}$ &
       6.39$\times$10$^{-01}$  & 1.24$\times$10$^{+00}$ \\ 
0.016 & 6.44$\times$10$^{-53}$ & 1.18$\times$10$^{-52}$ &
      2.31$\times$10$^{-52}$ &  -1.195$\times$10$^{+02}$ &
       6.37$\times$10$^{-01}$  & 2.91$\times$10$^{+00}$ \\ 
0.018 & 1.44$\times$10$^{-50}$ & 2.71$\times$10$^{-50}$ &
      5.22$\times$10$^{-50}$ &  -1.141$\times$10$^{+02}$ &
       6.37$\times$10$^{-01}$  & 2.53$\times$10$^{+00}$ \\
\end{longtable}
Comments: In total, 17 resonances are taken into account for calculating the reaction rates. Resonance energies are derived from level energies \cite{Til95} and the reaction Q-value (Tab. \ref{tab:master}), except for the $^{18}$F levels at 5672 and 5790 keV (E$_r^{cm}=1257$ and 1375 keV), for which more accurate values are adopted from Ref. \cite{Cha07}. Resonance strengths are adopted from Rolfs, Charlesworth and Azuma \cite{Rol73b}; Kieser et al. \cite{Kie79a}; Becker et al. \cite{BE82}; and G\"orres et al. \cite{Goe00} (where the latter resonance strength uncertainties listed in their Tab. I have been modified to include an additional 7\% uncertainty from the stopping power). Note that the strengths from these data sets are in mutual agreement. A number of levels have been disregarded: (i) E$_x=4964$ keV (E$_r^{cm}=549$ keV with $J^{\pi};T=2^+;1$) since the formation via $^{14}N+\alpha$ is isospin-forbidden (also, the experimental upper limit for the strength \cite{Cou71} is less than the observed strength of a lower-lying resonance); (ii) levels with $J^{\pi}=0^+$ since their population as resonances in $^{14}N+\alpha$ is forbidden according to angular momentum selection rules; and (iii) E$_x=4848$ keV ($J^{\pi}=5^-$; E$_r^{cm}=434$ keV) since this (unobserved) $\ell=5$ resonance is presumably negligible compared to the observed E$_r^{cm}=445$ keV ($\ell=1$) resonance. For the doublet at E$_x=5603+5605$ keV we use the summed strength adopted by Ref. \cite{Rol73b}. For the lowest-lying observed resonance, E$_r^{cm}=445$ keV, both the resonance strength \cite{Goe00} and the lifetime \cite{Rol73a} have been measured; from these results we deduce values for $\Gamma_\gamma$ and $\Gamma_\alpha$ in order to integrate the rate contribution of this resonance numerically. The direct capture S-factor is adopted from G\"orres et al. \cite{Goe00}. Note that their result is not based on experimental spectroscopic factors, but was estimated using an arbitrary value of $S_\alpha=0.1$ for the $\alpha$-particle spectroscopic factors  of all final states; we assign a factor of 2 uncertainty to the direct capture S-factor (yielding a fractional uncertainty of 0.79; see the numerical example at the end of Sec. 5.1.2 of Paper I).

The case of the undetected, low-energy resonance at E$_r^{cm}=237$ keV (E$_x=4652$ keV with $J^{\pi};T=4^+;1$) requires further mention. The level is weakly populated in the $^{14}$N($^{7}$Li,t)$^{18}$F study of Middleton et al. \cite{Mid68} and the differential cross sections for many levels are listed in their Tab. I. The observed intensity can only be related to an experimental {\it upper limit} of the $\alpha$-particle spectroscopic factor (since the state may have been populated via non-direct transfer). Experimental spectroscopic factors for levels that are strongly populated in $^{14}$N($^{7}$Li,t)$^{18}$F are presented by Tab. 2 of Cooper \cite{Coo86}. From these results we derive spectroscopic factors of S$_\alpha\leq0.0022$ and S$_\alpha=0.018$ for the E$_x=4652$ and 4860 keV levels, respectively. Since the $\alpha$-particle partial width for the latter state is experimentally known ($\Gamma_\alpha=4.5\times10^{-5}$ eV) we find an experimental upper limit of $\Gamma_\alpha^{4652}=(S_\alpha^{4652}/S_\alpha^{4860})(\Gamma_{sp}^{4652}/\Gamma_{sp}^{4860})\Gamma_\alpha^{4860}\leq4.1\times10^{-15}$ eV, where the {\it ratio} of single-particle $\alpha$-widths is found from the ratio of penetration factors. The population of this T=1 state in $^{14}N+\alpha$ is suppressed by isospin selection rules. Consequently, we may not use in this case $\langle \theta^2_{\alpha} \rangle=0.010$ for the mean value of the dimensionless reduced $\alpha$-particle width of the Porter-Thomas distribution (see discussion in Paper I). Since mean values of $\langle \theta^2_{\alpha} \rangle$ for isospin-forbidden population via $\alpha$-particle capture from a T=0 target to a T=1 state do not exist at present, we arbitrarily assume an isospin suppression factor of 0.001 (see also Ref. \cite{Goe00}); thus this level is randomly sampled using $\langle \theta^2_{\alpha} \rangle=0.010\times0.001=1.0\times10^{-5}$. Raising $\langle \theta^2_{\alpha} \rangle$ by an order of magnitude would increase the total rates only at low temperatures, near T=80 MK, by about a factor of 8.
\begin{figure}[]
\includegraphics[height=8.5cm]{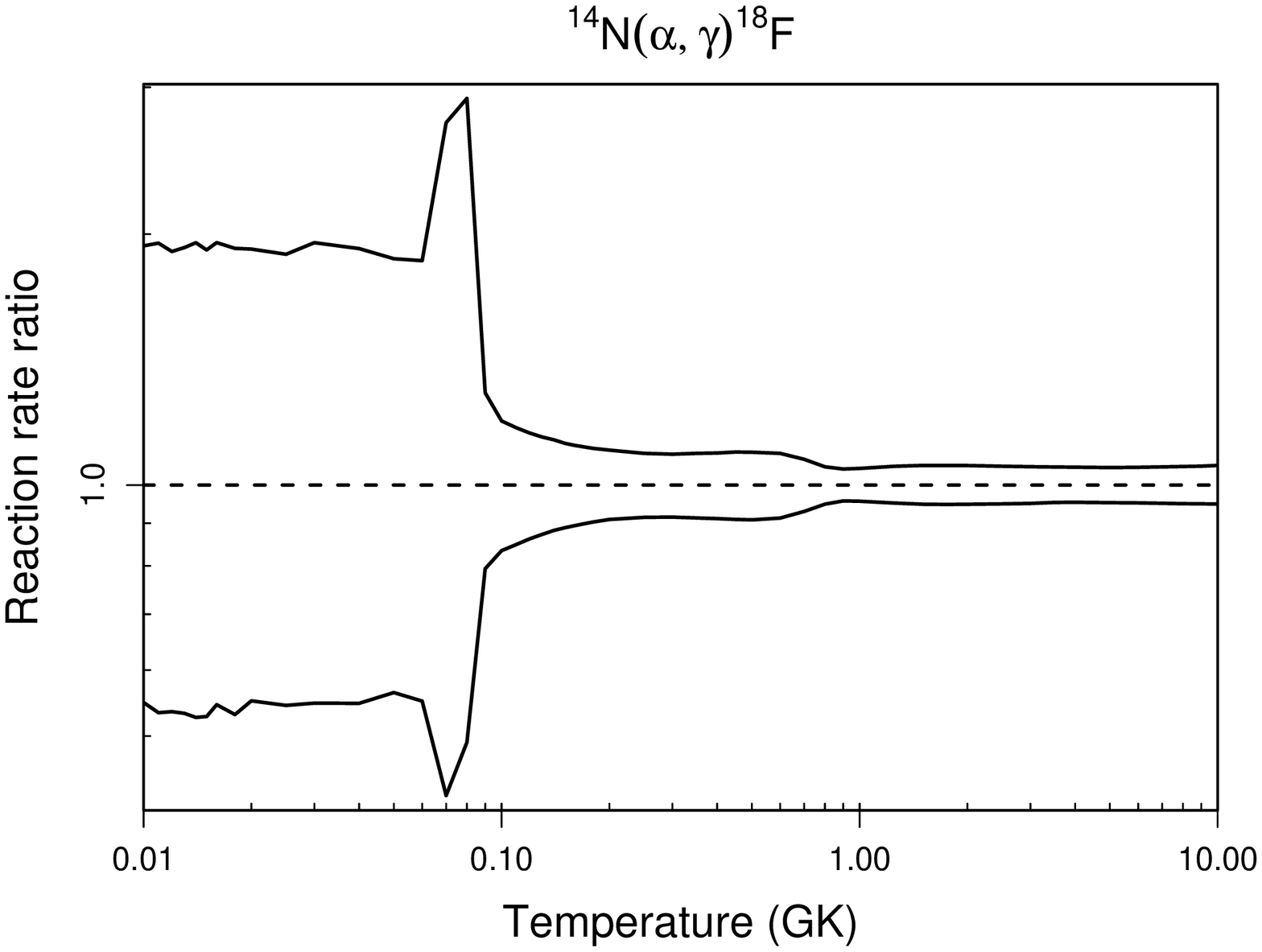}
\end{figure}
\clearpage
\begin{figure}[]
\includegraphics[height=18.5cm]{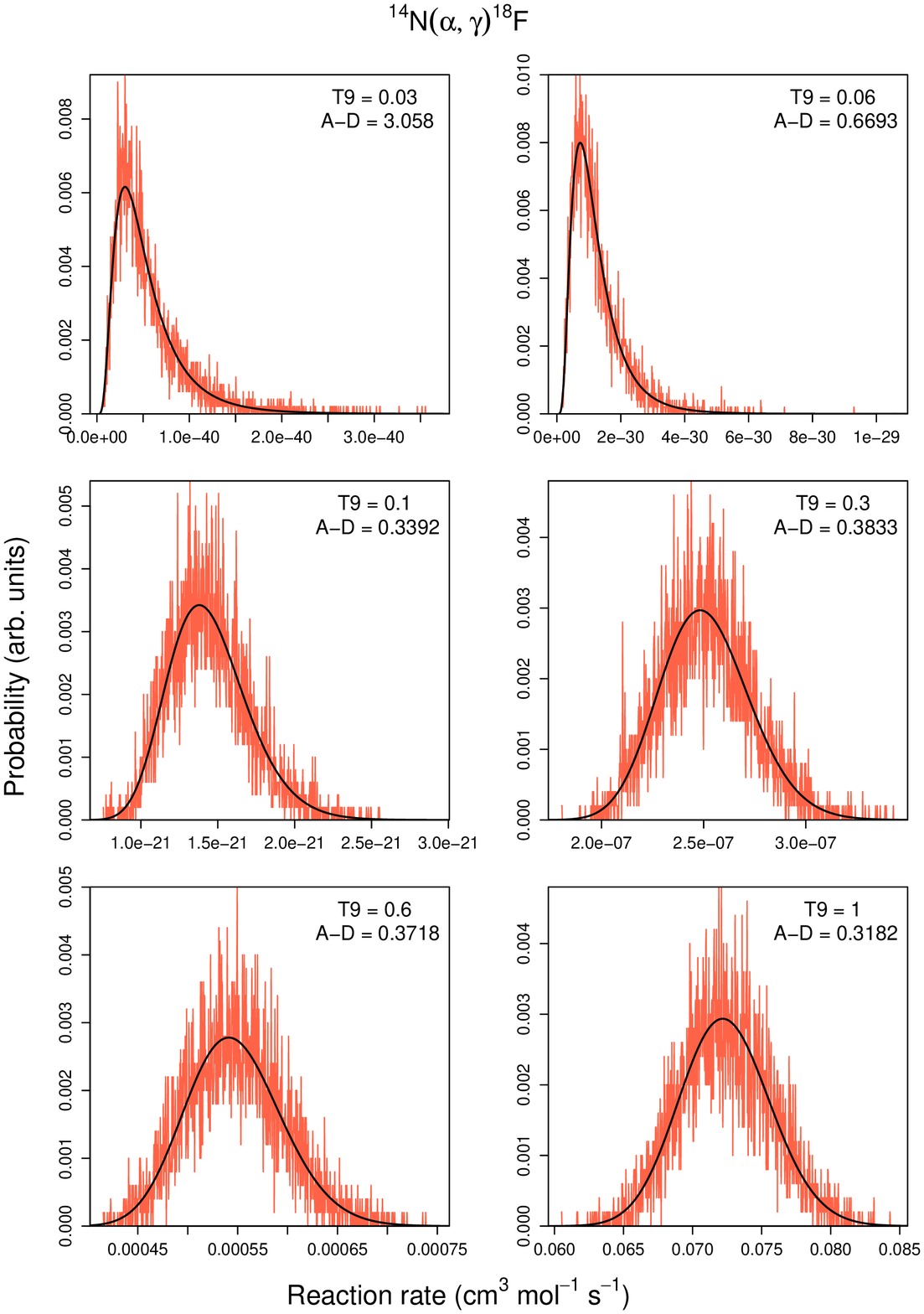}
\end{figure}
\clearpage
\setlongtables
\begin{longtable}{@{\extracolsep{\fill}}lccc | ccc}
\caption{Total thermonuclear reaction rates for $^{15}$N($\alpha$,$\gamma$)$^{19}$F.}  \label{} \\
\hline \hline 
	\multicolumn{1}{c}{T (GK)} & \multicolumn{1}{c}{Low rate} & \multicolumn{1}{c}{Median rate} & \multicolumn{1}{c}{High rate}  & \multicolumn{1}{c}{lognormal $\mu$} & \multicolumn{1}{c}{lognormal $\sigma$} & \multicolumn{1}{c}{A-D} \\ \hline 
\endfirsthead
\multicolumn{6}{c}{{\tablename} \thetable{} -- continued} \\
\hline \hline 
	\multicolumn{1}{c}{T (GK)} & \multicolumn{1}{c}{Low rate} & \multicolumn{1}{c}{Median rate} & \multicolumn{1}{c}{High rate} & \multicolumn{1}{c}{lognormal $\mu$} & \multicolumn{1}{c}{lognormal $\sigma$} & \multicolumn{1}{c}{A-D} \\ \hline 
\endhead
	 \hline \hline
\endfoot
	\hline \hline
\endlastfoot
0.010 & 1.60$\times$10$^{-61}$ & 2.13$\times$10$^{-61}$ &
      2.88$\times$10$^{-61}$ &  -1.397$\times$10$^{+02}$ &
       2.99$\times$10$^{-01}$  & 2.12$\times$10$^{+00}$ \\ 
0.011 & 2.84$\times$10$^{-59}$ & 3.82$\times$10$^{-59}$ &
      5.16$\times$10$^{-59}$ &  -1.345$\times$10$^{+02}$ &
       3.02$\times$10$^{-01}$  & 1.91$\times$10$^{+00}$ \\ 
0.012 & 2.83$\times$10$^{-57}$ & 3.82$\times$10$^{-57}$ &
      5.17$\times$10$^{-57}$ &  -1.299$\times$10$^{+02}$ &
       3.04$\times$10$^{-01}$  & 1.34$\times$10$^{+00}$ \\ 
0.013 & 1.72$\times$10$^{-55}$ & 2.30$\times$10$^{-55}$ &
      3.14$\times$10$^{-55}$ &  -1.258$\times$10$^{+02}$ &
       3.01$\times$10$^{-01}$  & 3.03$\times$10$^{+00}$ \\ 
0.014 & 7.06$\times$10$^{-54}$ & 9.40$\times$10$^{-54}$ &
      1.29$\times$10$^{-53}$ &  -1.221$\times$10$^{+02}$ &
       3.01$\times$10$^{-01}$  & 2.53$\times$10$^{+00}$ \\ 
0.015 & 2.01$\times$10$^{-52}$ & 2.71$\times$10$^{-52}$ &
      3.71$\times$10$^{-52}$ &  -1.187$\times$10$^{+02}$ &
       3.06$\times$10$^{-01}$  & 2.86$\times$10$^{+00}$ \\ 
0.016 & 4.47$\times$10$^{-51}$ & 6.00$\times$10$^{-51}$ &
      8.24$\times$10$^{-51}$ &  -1.156$\times$10$^{+02}$ &
       3.06$\times$10$^{-01}$  & 2.80$\times$10$^{+00}$ \\ 
0.018 & 1.05$\times$10$^{-48}$ & 1.40$\times$10$^{-48}$ &
      1.91$\times$10$^{-48}$ &  -1.102$\times$10$^{+02}$ &
       2.98$\times$10$^{-01}$  & 3.14$\times$10$^{+00}$ \\
\end{longtable}
Comments: Strengths have been measured directly for 61 resonances from $E_r^{cm}=0.536$ MeV up to $E_r^{cm}=6.397$ MeV
\cite{AI70,DI71a,DI71b,DI77,MA87,RO72a,RO72b,RO76,SY78,UN74,WI02}. The experimental data cover the entire temperature range of interest. There is a high level of coherence in these data sets as they all use the  $E_r^{cm}=3.526$ MeV (5/2$^{+}$) or $E_r^{cm}=1.323$ MeV (1/2$^{(+)}$) resonances as references. We use the input data extracted from the same references as in Angulo et al. \cite{Ang99} (NACRE), except for the resonances between $E_r^{cm}=0.536$ and 2.086 MeV, where we adopt the new and more precise measurement of Wilmes et al. \cite{WI02}. Below 0.5 MeV we use (as in NACRE) results from the ($^7$Li,t) transfer reaction experiment of de Oliveira et al. \cite{OL96}, in particular for the $E_r^{cm}=364$ keV resonance whose contribution dominates the reaction rate in the range of 0.1 to 0.2 GK. The published resonance strength, $\omega\gamma=(6^{+6}_{-3})\times10^{-9}$ eV \cite{OL96}, assumes a factor of 2 uncertainty for the DWBA analysis. Resonance energies are obtained from the excitation energies listed in Table 19.9 of Tilley et al. \cite{Til95} and $Q=4013.74\pm0.07$ keV \cite{Aud03}. The largest contribution from the three near-threshold levels is associated with the 3.91 MeV state, but their total absolute contribution is too small to be of any astrophysical importance. The direct capture S-factor was calculated by de Oliveira et al. \cite{OL96} and amounts to $S(0)=6.148$ MeV~b. 
\begin{figure}[]
\includegraphics[height=8.5cm]{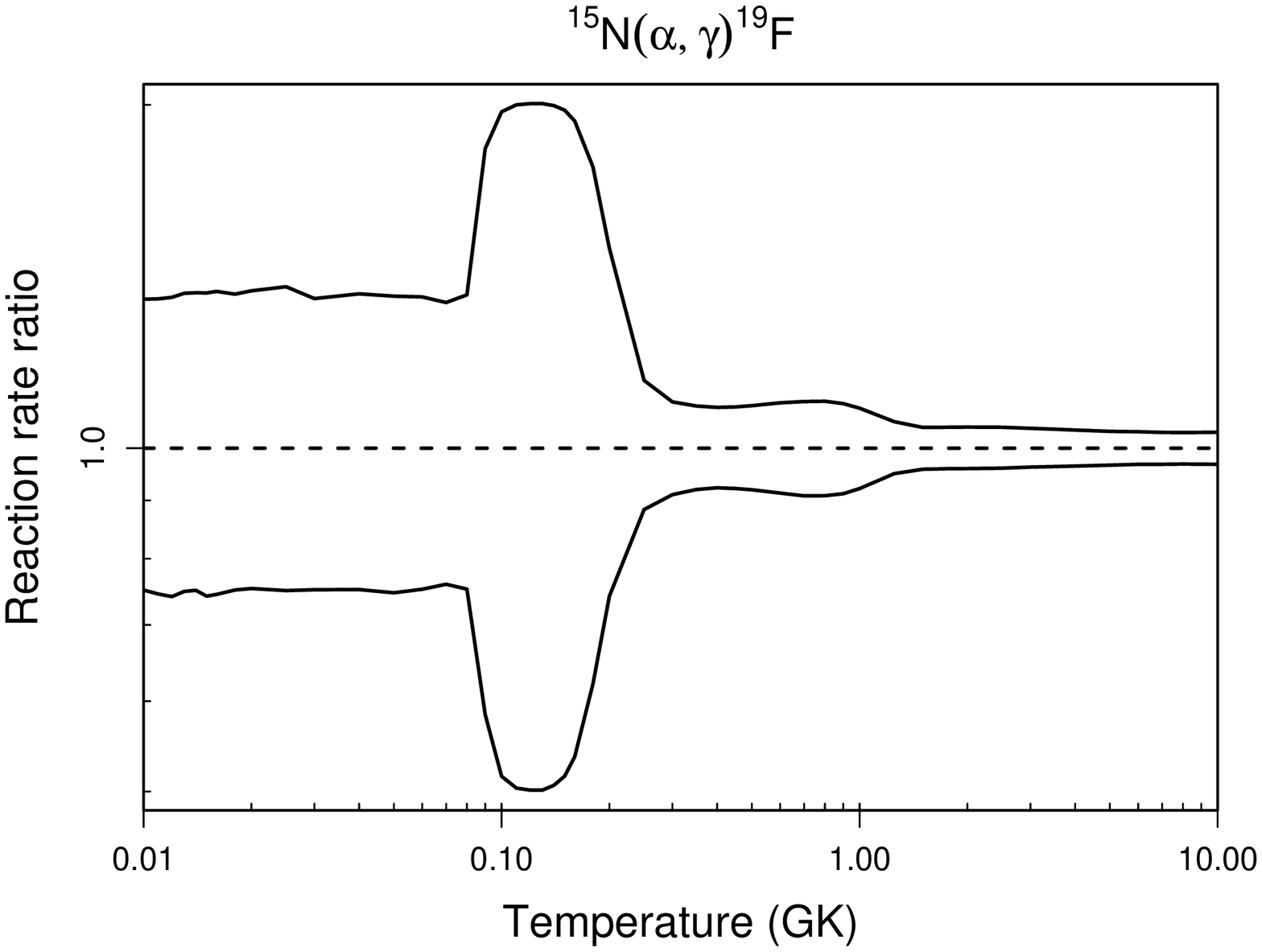}
\end{figure}
\clearpage
\begin{figure}[]
\includegraphics[height=18.5cm]{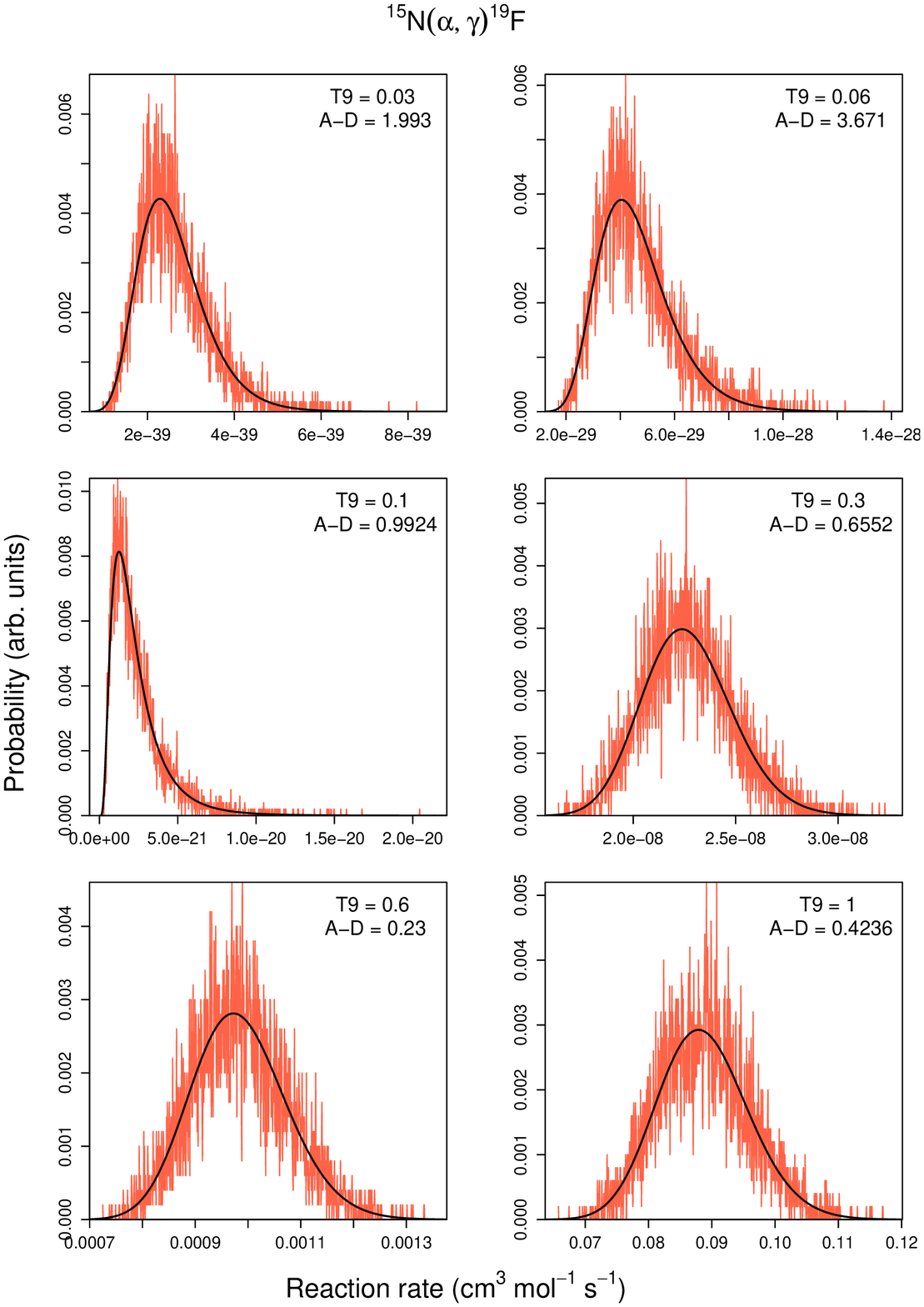}
\end{figure}
\clearpage
\setlongtables
\begin{longtable}{@{\extracolsep{\fill}}lccc | ccc}
\caption{Total thermonuclear reaction rates for $^{15}$O($\alpha$,$\gamma$)$^{19}$Ne.}  \label{} \\
\hline \hline 
	\multicolumn{1}{c}{T (GK)} & \multicolumn{1}{c}{Low rate} & \multicolumn{1}{c}{Median rate} & \multicolumn{1}{c}{High rate}  & \multicolumn{1}{c}{lognormal $\mu$} & \multicolumn{1}{c}{lognormal $\sigma$} & \multicolumn{1}{c}{A-D} \\ \hline 
\endfirsthead
\multicolumn{6}{c}{{\tablename} \thetable{} -- continued} \\
\hline \hline 
	\multicolumn{1}{c}{T (GK)} & \multicolumn{1}{c}{Low rate} & \multicolumn{1}{c}{Median rate} & \multicolumn{1}{c}{High rate} & \multicolumn{1}{c}{lognormal $\mu$} & \multicolumn{1}{c}{lognormal $\sigma$} & \multicolumn{1}{c}{A-D} \\ \hline 
\endhead
	 \hline \hline
\endfoot
	\hline \hline
\endlastfoot
0.010 &  6.29$\times$10$^{-68}$  &  9.20$\times$10$^{-68}$  &
       1.34$\times$10$^{-67}$  &  -1.544$\times$10$^{+02}$  &
       3.82$\times$10$^{-01}$  &  1.80$\times$10$^{-01}$  \\
0.011 &  1.83$\times$10$^{-65}$  &  2.67$\times$10$^{-65}$  &
       3.88$\times$10$^{-65}$  &  -1.487$\times$10$^{+02}$  &
       3.83$\times$10$^{-01}$  &  2.22$\times$10$^{-01}$  \\
0.012 &  2.79$\times$10$^{-63}$  &  4.09$\times$10$^{-63}$  &
       6.01$\times$10$^{-63}$  &  -1.437$\times$10$^{+02}$  &
       3.84$\times$10$^{-01}$  &  3.38$\times$10$^{-01}$  \\
0.013 &  2.52$\times$10$^{-61}$  &  3.68$\times$10$^{-61}$  &
       5.35$\times$10$^{-61}$  &  -1.392$\times$10$^{+02}$  &
       3.82$\times$10$^{-01}$  &  5.08$\times$10$^{-01}$  \\
0.014 &  1.45$\times$10$^{-59}$  &  2.11$\times$10$^{-59}$  &
       3.09$\times$10$^{-59}$  &  -1.351$\times$10$^{+02}$  &
       3.81$\times$10$^{-01}$  &  4.18$\times$10$^{-01}$  \\
0.015 &  5.87$\times$10$^{-58}$  &  8.50$\times$10$^{-58}$  &
       1.24$\times$10$^{-57}$  &  -1.314$\times$10$^{+02}$  &
       3.78$\times$10$^{-01}$  &  3.07$\times$10$^{-01}$  \\
0.016 &  1.72$\times$10$^{-56}$  &  2.48$\times$10$^{-56}$  &
       3.61$\times$10$^{-56}$  &  -1.280$\times$10$^{+02}$  &
       3.72$\times$10$^{-01}$  &  2.79$\times$10$^{-01}$  \\
0.018 &  6.56$\times$10$^{-54}$  &  9.61$\times$10$^{-54}$  &
       1.41$\times$10$^{-53}$  &  -1.221$\times$10$^{+02}$  &
       3.85$\times$10$^{-01}$  &  1.84$\times$10$^{-01}$  \\
\end{longtable}
Comments: Resonance energies are calculated from excitation energies \cite{Til95,Tan05} and the Q-value (Tab. \ref{tab:master}). Total widths are deduced from the averaged DSAM lifetime measured values \cite{RO72a,Tan05,Kan06,Myt08}, except for the first two resonances at 505 and 849 keV, for which the results of Davids et al. \cite{Dav03b} are used. Alpha-particle partial widths are calculated using the $\alpha$-particle branching ratios adopted by Ref. \cite{Dav03b}, except for the two first resonances. For those we use results from $\alpha$-particle transfer experiments to $^{19}$F mirror states \cite{Mao95,OL96} and assume, as in Fisker et al. \cite{Fis06}, the equality of reduced $\alpha$-particle widths between mirror levels. This hypothesis has been questioned \cite{OL97} because of an apparent disagreement \cite{WI95} between the reduced $\alpha$-particle widths of the 849 keV resonance (as derived from the total width and the $\alpha$-particle branching ratio \cite{MA90}) and of the corresponding $^{19}$F mirror state (as obtained from a $^{15}$N($^7$Li,t)$^{19}$F transfer experiment \cite{OL96}). However, a new measurement \cite{Dav03b} of the $\alpha$-particle branching ratio has not confirmed this disagreement: for the first two resonances the measured $\alpha$-particle branching ratios \cite{Dav03b,Tan07} are compatible with the $\Gamma_\alpha$ values deduced from transfer reactions. We adopt a factor of $3$ uncertainty for the resulting $\alpha$-particle partial widths. For the mirror level of the 1563 keV resonance, the available spectroscopic data ($\Gamma_\gamma/\Gamma=0.97\pm0.03$ \cite{PR89} and $\omega\gamma_{\alpha\gamma}$ \cite{WI02}) are not judged sufficient to extract a reliable $\alpha$-particle partial width and thus we disregard this resonance. For the direct capture component we adopt a constant value of 20 MeV~b \cite{Lan86,Duf00,Fis06}, with an assumed uncertainty of 40\%. Above 0.6 GK the total reaction rate is extrapolated using results computed with the TALYS statistical model code \cite{Gor08}.
\begin{figure}[]
\includegraphics[height=8.5cm]{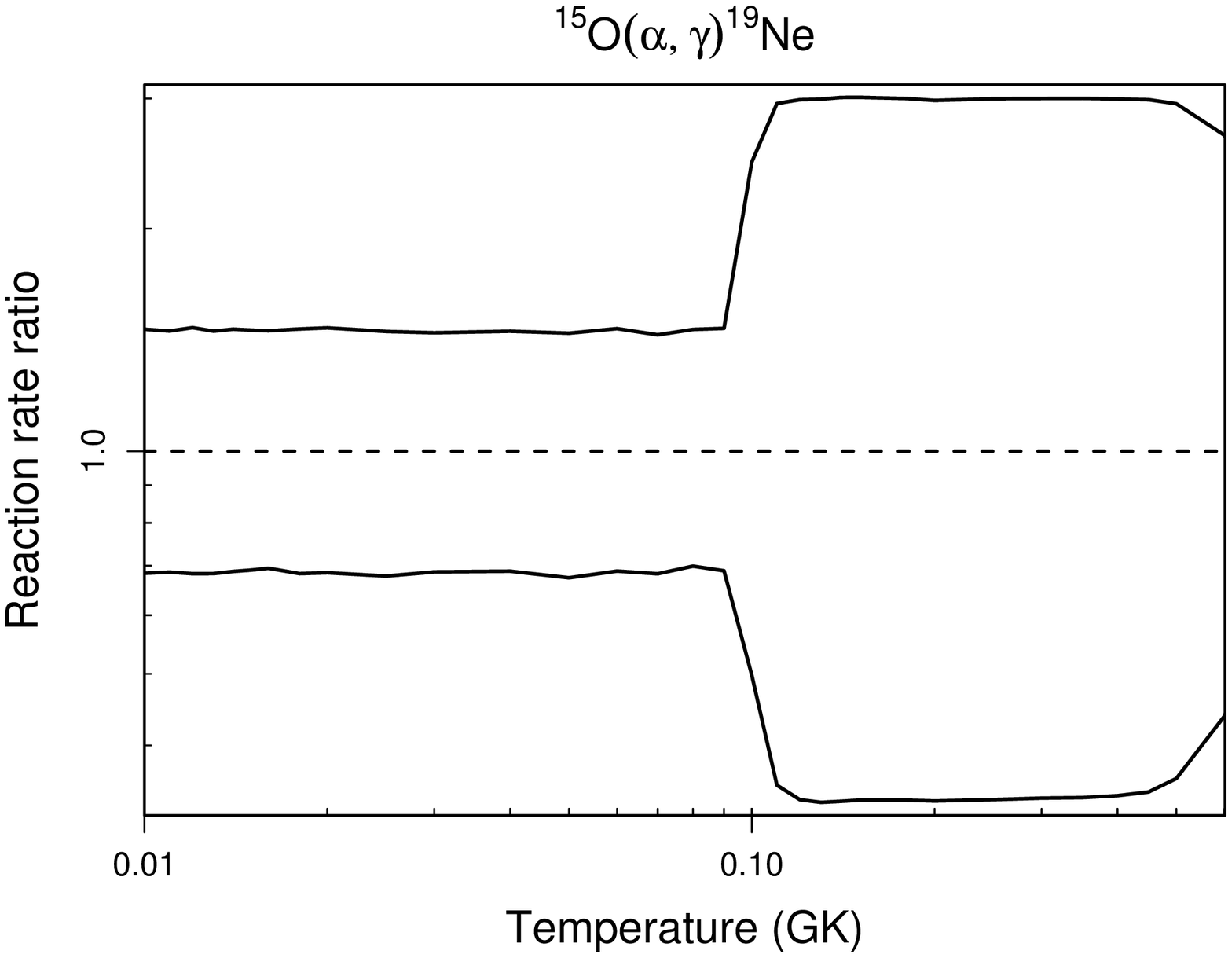}
\end{figure}
\clearpage
\begin{figure}[]
\includegraphics[height=18.5cm]{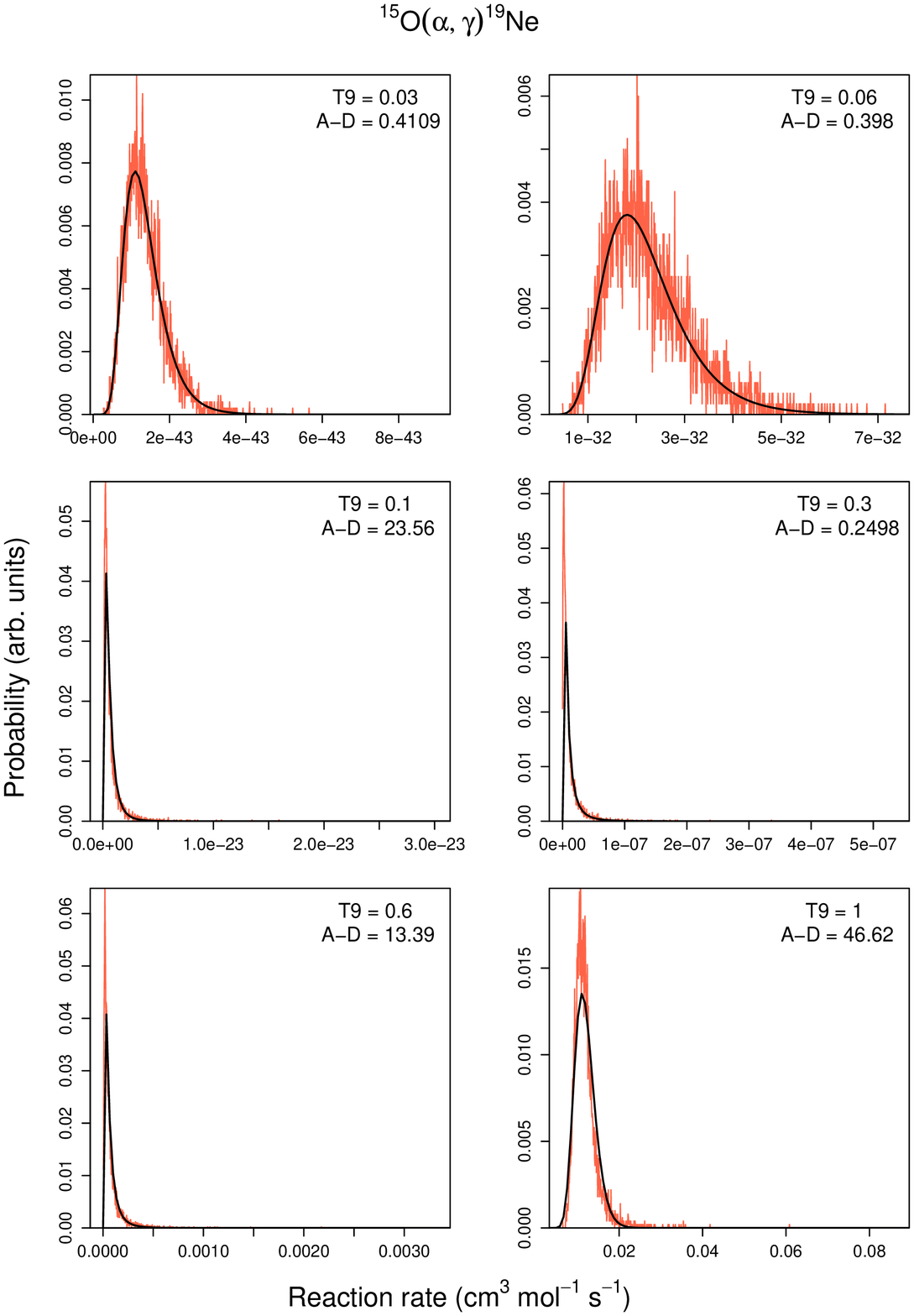}
\end{figure}
\clearpage
\setlongtables
\begin{longtable}{@{\extracolsep{\fill}}lccc | ccc}
\caption{Total thermonuclear reaction rates for $^{16}$O(p,$\gamma$)$^{17}$F.}  \label{} \\
\hline \hline 
	\multicolumn{1}{c}{T (GK)} & \multicolumn{1}{c}{Low rate} & \multicolumn{1}{c}{Median rate} & \multicolumn{1}{c}{High rate}  & \multicolumn{1}{c}{lognormal $\mu$} & \multicolumn{1}{c}{lognormal $\sigma$} & \multicolumn{1}{c}{A-D} \\ \hline 
\endfirsthead
\multicolumn{6}{c}{{\tablename} \thetable{} -- continued} \\
\hline \hline 
	\multicolumn{1}{c}{T (GK)} & \multicolumn{1}{c}{Low rate} & \multicolumn{1}{c}{Median rate} & \multicolumn{1}{c}{High rate} & \multicolumn{1}{c}{lognormal $\mu$} & \multicolumn{1}{c}{lognormal $\sigma$} & \multicolumn{1}{c}{A-D} \\ \hline 
\endhead
	 \hline \hline
\endfoot
	\hline \hline
\endlastfoot
0.01 		& 6.67$\times$10$^{-25}$  	& 7.20$\times$10$^{-25}$  	&  7.73$\times$10$^{-25}$ & -5.559$\times$10$^{+01}$   &  7.36$\times$10$^{-02}$  &\\
0.011 	& 7.03$\times$10$^{-24}$  	& 7.58$\times$10$^{-24}$  	&  8.14$\times$10$^{-24}$ & -5.324$\times$10$^{+01}$   &  7.35$\times$10$^{-02}$  &\\
0.012 	& 5.64$\times$10$^{-23}$  	& 6.08$\times$10$^{-23}$  	&  6.53$\times$10$^{-23}$ & -5.115$\times$10$^{+01}$   &  7.33$\times$10$^{-02}$  &\\
0.013 	& 3.63$\times$10$^{-22}$  	& 3.91$\times$10$^{-22}$  	&  4.20$\times$10$^{-22}$ & -4.929$\times$10$^{+01}$   &  7.31$\times$10$^{-02}$  &\\
0.014 	& 1.94$\times$10$^{-21}$  	& 2.09$\times$10$^{-21}$  	&  2.25$\times$10$^{-21}$ & -4.762$\times$10$^{+01}$   &  7.30$\times$10$^{-02}$  &\\
0.015 	& 8.91$\times$10$^{-21}$  	& 9.60$\times$10$^{-21}$ 	&  1.03$\times$10$^{-20}$ & -4.609$\times$10$^{+01}$   &  7.30$\times$10$^{-02}$  &\\
0.016 	& 3.59$\times$10$^{-20}$  	& 3.87$\times$10$^{-20}$  	&  4.15$\times$10$^{-20}$ & -4.470$\times$10$^{+01}$   &  7.27$\times$10$^{-02}$  &\\
0.018 	& 4.21$\times$10$^{-19}$  	& 4.54$\times$10$^{-19}$ 	&  4.87$\times$10$^{-19}$ & -4.224$\times$10$^{+01}$   &  7.23$\times$10$^{-02}$  &\\
\end{longtable}
\begin{figure}[]
\includegraphics[height=8.5cm]{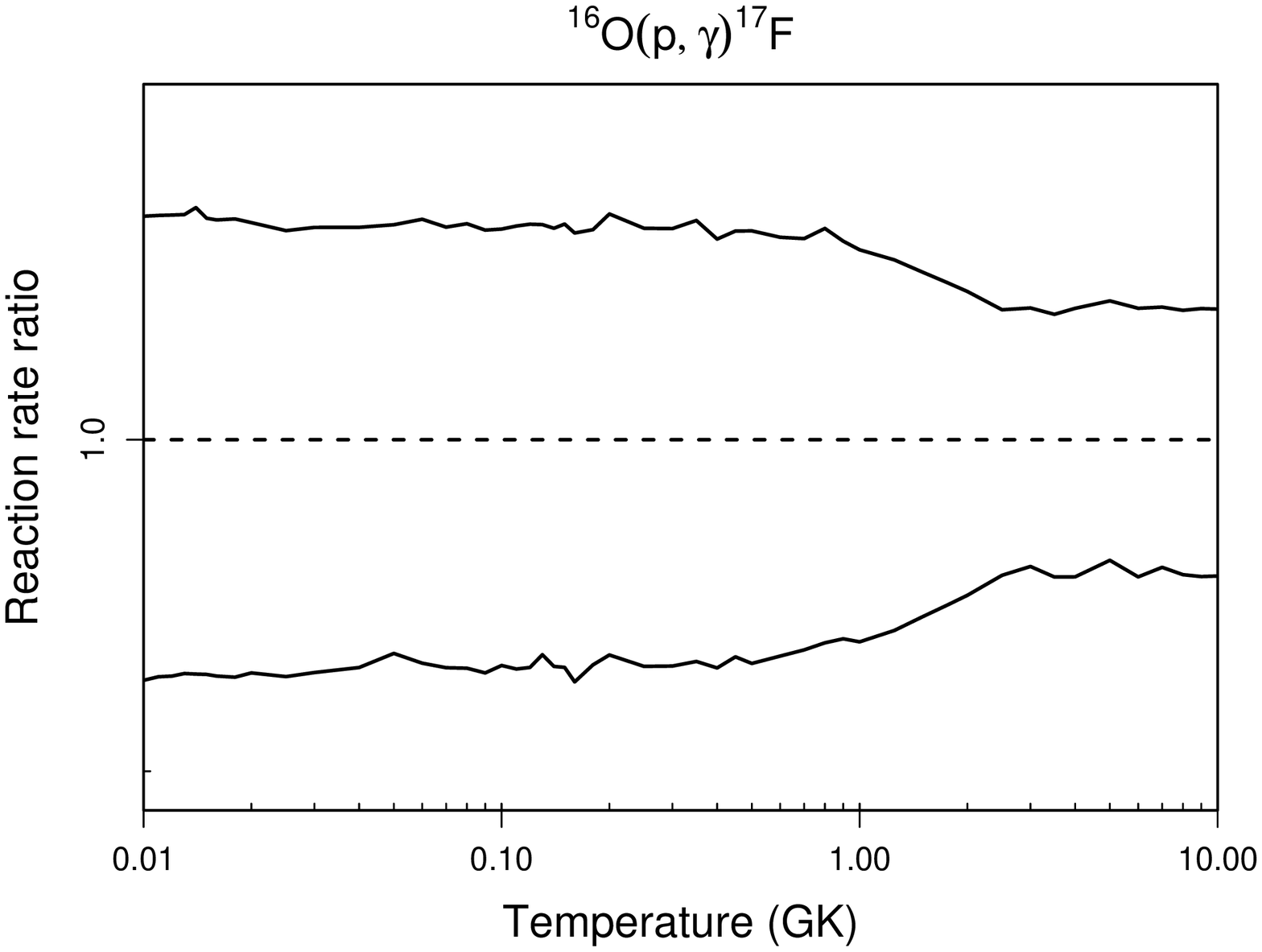}
\end{figure}
Comments: {\it This is the only reaction analyzed here for which the rate uncertainties are not derived from the Monte Carlo method}. The reaction rates, including uncertainties, are adopted from Iliadis et al. \cite{Ili08}. For temperatures of T$\geq$3 GK, the rates are calculated assuming a constant S--factor of $S(0)=4\times 10^{-3}$ MeV~b (see Fig. 2 in Iliadis et al. \cite{Ili08}). The two resonances at $E^{cm}_r=2.50$ and 3.26 MeV (Tilley et al. \cite{Til93}) are negligible for the total rate. The lognormal parameters $\mu$ and $\sigma$ are computed from columns 2-4 by using Eq. (39) from Paper I.
\clearpage
\setlongtables
\begin{longtable}{@{\extracolsep{\fill}}lccc | ccc}
\caption{Total thermonuclear reaction rates for $^{16}$O($\alpha$,$\gamma$)$^{20}$Ne.}  \label{} \\
\hline \hline 
	\multicolumn{1}{c}{T (GK)} & \multicolumn{1}{c}{Low rate} & \multicolumn{1}{c}{Median rate} & \multicolumn{1}{c}{High rate}  & \multicolumn{1}{c}{lognormal $\mu$} & \multicolumn{1}{c}{lognormal $\sigma$} & \multicolumn{1}{c}{A-D} \\ \hline 
\endfirsthead
\multicolumn{6}{c}{{\tablename} \thetable{} -- continued} \\
\hline \hline 
	\multicolumn{1}{c}{T (GK)} & \multicolumn{1}{c}{Low rate} & \multicolumn{1}{c}{Median rate} & \multicolumn{1}{c}{High rate} & \multicolumn{1}{c}{lognormal $\mu$} & \multicolumn{1}{c}{lognormal $\sigma$} & \multicolumn{1}{c}{A-D} \\ \hline 
\endhead
	 \hline \hline
\endfoot
	\hline \hline
\endlastfoot
0.010 &  5.10$\times$10$^{-69}$  &  7.34$\times$10$^{-69}$  &
       1.07$\times$10$^{-68}$  &  -1.569$\times$10$^{+02}$  &
       3.74$\times$10$^{-01}$  &  2.78$\times$10$^{-01}$  \\
0.011 &  1.53$\times$10$^{-66}$  &  2.22$\times$10$^{-66}$  &
       3.24$\times$10$^{-66}$  &  -1.512$\times$10$^{+02}$  &
       3.73$\times$10$^{-01}$  &  3.20$\times$10$^{-01}$  \\
0.012 &  2.43$\times$10$^{-64}$  &  3.51$\times$10$^{-64}$  &
       5.07$\times$10$^{-64}$  &  -1.461$\times$10$^{+02}$  &
       3.70$\times$10$^{-01}$  &  2.46$\times$10$^{-01}$  \\
0.013 &  2.18$\times$10$^{-62}$  &  3.17$\times$10$^{-62}$  &
       4.62$\times$10$^{-62}$  &  -1.416$\times$10$^{+02}$  &
       3.77$\times$10$^{-01}$  &  2.23$\times$10$^{-01}$  \\
0.014 &  1.29$\times$10$^{-60}$  &  1.87$\times$10$^{-60}$  &
       2.71$\times$10$^{-60}$  &  -1.375$\times$10$^{+02}$  &
       3.73$\times$10$^{-01}$  &  3.80$\times$10$^{-01}$  \\
0.015 &  5.19$\times$10$^{-59}$  &  7.54$\times$10$^{-59}$  &
       1.10$\times$10$^{-58}$  &  -1.338$\times$10$^{+02}$  &
       3.81$\times$10$^{-01}$  &  7.61$\times$10$^{-01}$  \\
0.016 &  1.55$\times$10$^{-57}$  &  2.27$\times$10$^{-57}$  &
       3.26$\times$10$^{-57}$  &  -1.304$\times$10$^{+02}$  &
       3.72$\times$10$^{-01}$  &  3.48$\times$10$^{-01}$  \\
0.018 &  6.28$\times$10$^{-55}$  &  9.00$\times$10$^{-55}$  &
       1.31$\times$10$^{-54}$  &  -1.244$\times$10$^{+02}$  &
       3.70$\times$10$^{-01}$  &  5.71$\times$10$^{-01}$  \\
\end{longtable}
Comments: In total, 22 resonances at energies of E$_r^{cm}=892$-7671 keV are considered. Resonance energies are calculated using the excitation energies listed in Tab. 20.17 of Tilley et al. \cite{Til98}. Resonance strengths are adopted from Refs. \cite{Til98,May01,Ang99}. For the E$_r^{cm}=1058$ keV resonance, the $\alpha$-particle and $\gamma$-ray partial widths are obtained from the measured resonance strength together with a value of $\Gamma_{\alpha}=28\pm3$ eV from ($\alpha$,$\alpha$) scattering \cite{Mac80}. For the E$_r^{cm}=892$ keV resonance, the partial widths are deduced from measured values of the resonance strength \cite{Til98} and the mean lifetime \cite{Ajz72}. The $\alpha$-particle width is identified with the larger solution of the quadratic equation, in agreement with the $\alpha$-particle width deduced from the ($^6$Li,d) study of Ref. \cite{Mao96}. The direct capture S-factor is adopted from Fig. 1 of Mohr \cite{Moh05}, where we disregard the transitions arising from the E$_r^{cm}=1058$ keV resonance tail (they are explicitly taken into account in the numerical integration of the resonant reaction rate). 
\begin{figure}[]
\includegraphics[height=8.5cm]{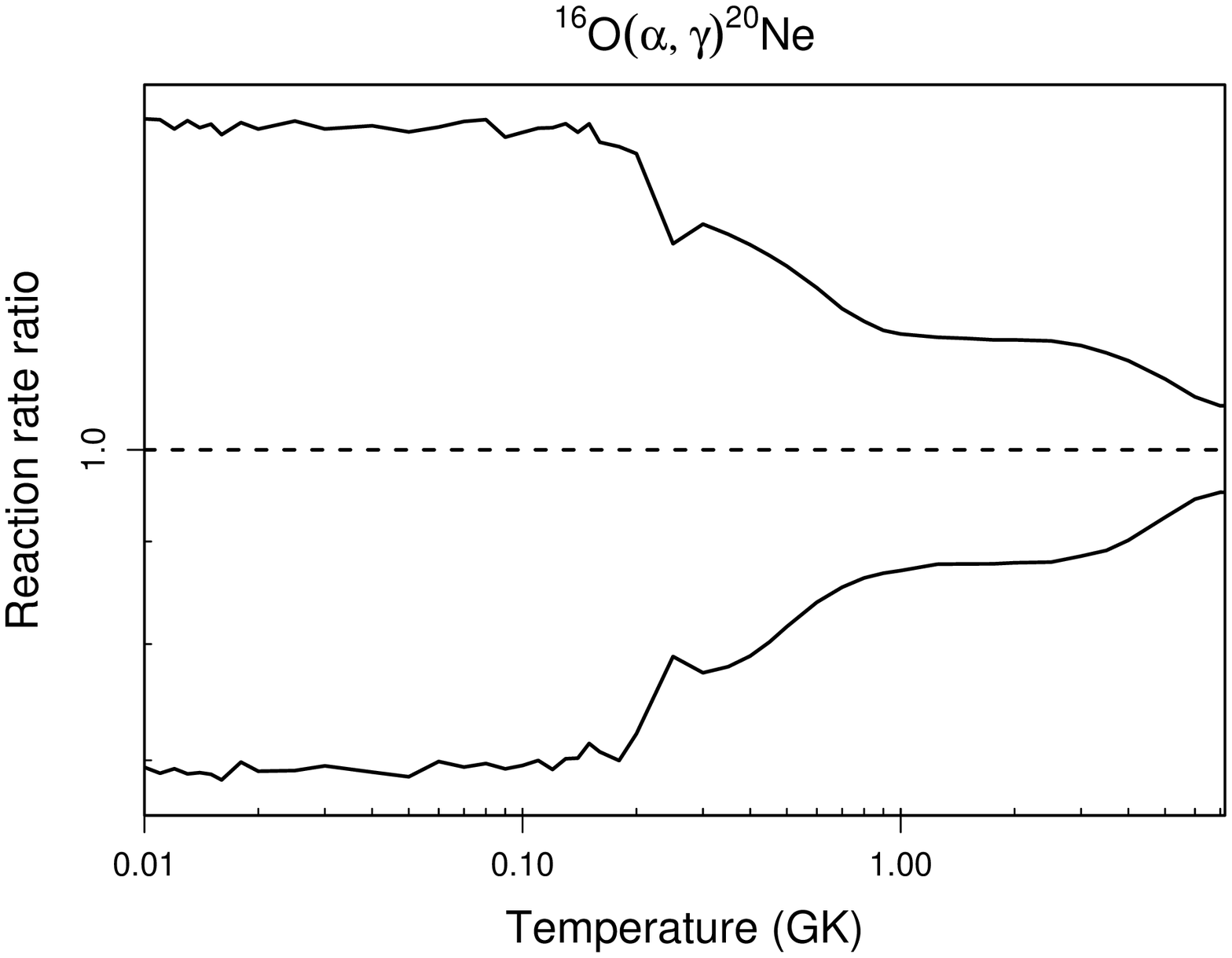}
\end{figure}
\clearpage
\begin{figure}[]
\includegraphics[height=18.5cm]{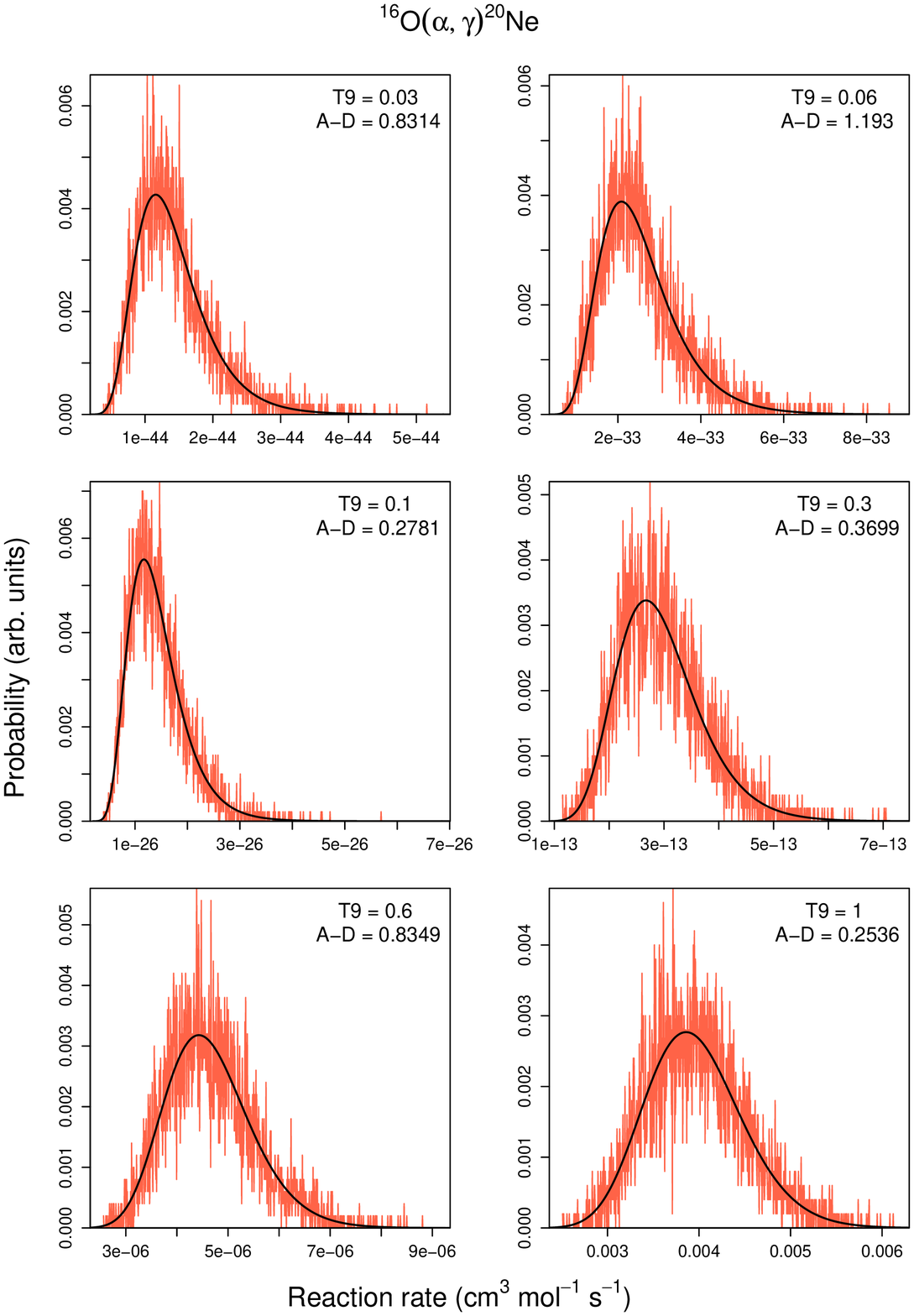}
\end{figure}
\clearpage
\setlongtables
\begin{longtable}{@{\extracolsep{\fill}}lccc | ccc}
\caption{Total thermonuclear reaction rates for $^{17}$O(p,$\gamma$)$^{18}$F.}  \label{} \\
\hline \hline 
	\multicolumn{1}{c}{T (GK)} & \multicolumn{1}{c}{Low rate} & \multicolumn{1}{c}{Median rate} & \multicolumn{1}{c}{High rate}  & \multicolumn{1}{c}{lognormal $\mu$} & \multicolumn{1}{c}{lognormal $\sigma$} & \multicolumn{1}{c}{A-D} \\ \hline 
\endfirsthead
\multicolumn{6}{c}{{\tablename} \thetable{} -- continued} \\
\hline \hline 
	\multicolumn{1}{c}{T (GK)} & \multicolumn{1}{c}{Low rate} & \multicolumn{1}{c}{Median rate} & \multicolumn{1}{c}{High rate} & \multicolumn{1}{c}{lognormal $\mu$} & \multicolumn{1}{c}{lognormal $\sigma$} & \multicolumn{1}{c}{A-D} \\ \hline 
\endhead
	 \hline \hline
\endfoot
	\hline \hline
\endlastfoot
0.010 & 2.95$\times$10$^{-25}$ & 3.58$\times$10$^{-25}$ &
      4.38$\times$10$^{-25}$ &  -5.629$\times$10$^{+01}$ &
       2.02$\times$10$^{-01}$  & 7.75$\times$10$^{-01}$ \\ 
0.011 & 3.10$\times$10$^{-24}$ & 3.78$\times$10$^{-24}$ &
      4.64$\times$10$^{-24}$ &  -5.393$\times$10$^{+01}$ &
       2.01$\times$10$^{-01}$  & 1.77$\times$10$^{+00}$ \\ 
0.012 & 2.51$\times$10$^{-23}$ & 3.09$\times$10$^{-23}$ &
      3.81$\times$10$^{-23}$ &  -5.183$\times$10$^{+01}$ &
       2.08$\times$10$^{-01}$  & 2.58$\times$10$^{-01}$ \\ 
0.013 & 1.64$\times$10$^{-22}$ & 1.99$\times$10$^{-22}$ &
      2.45$\times$10$^{-22}$ &  -4.996$\times$10$^{+01}$ &
       2.03$\times$10$^{-01}$  & 7.69$\times$10$^{-01}$ \\ 
0.014 & 8.83$\times$10$^{-22}$ & 1.08$\times$10$^{-21}$ &
      1.33$\times$10$^{-21}$ &  -4.827$\times$10$^{+01}$ &
       2.05$\times$10$^{-01}$  & 7.02$\times$10$^{-01}$ \\ 
0.015 & 4.08$\times$10$^{-21}$ & 4.98$\times$10$^{-21}$ &
      6.14$\times$10$^{-21}$ &  -4.675$\times$10$^{+01}$ &
       2.07$\times$10$^{-01}$  & 3.50$\times$10$^{-01}$ \\ 
0.016 & 1.65$\times$10$^{-20}$ & 2.02$\times$10$^{-20}$ &
      2.50$\times$10$^{-20}$ &  -4.534$\times$10$^{+01}$ &
       2.05$\times$10$^{-01}$  & 5.34$\times$10$^{-01}$ \\ 
0.018 & 1.98$\times$10$^{-19}$ & 2.42$\times$10$^{-19}$ &
      2.96$\times$10$^{-19}$ &  -4.287$\times$10$^{+01}$ &
       2.03$\times$10$^{-01}$  & 2.96$\times$10$^{-01}$ \\
\end{longtable}
Comments: In total, 16 resonances in the range of E$_r^{cm}\leq1271$ keV are taken into account, including two subthreshold resonances at $-3$ keV ($1^+$) and $-2$ keV ($1^-$). The energy of the resonance at E$_r^{cm}=65.1\pm0.5$ keV is obtained from the corrected excitation energy E$_x=5671.6\pm0.2$ keV reported in Chafa et al. \cite{Cha07}. For the E$_r^{cm}=183$ and 490 keV resonances, the strengths are adopted from Fox et al. \cite{Fox05}. For E$_r^{cm}\ge490$ keV, strengths are reported in Ref. \cite{Rol73}. These results have to be renormalized (see comments in Ref. \cite{Fox05}) by using the correct resonance strength for E$_{r}^{cm}$=609 keV in $^{27}$Al(p,$\gamma$)$^{28}$Si. For the E$_r^{cm}=490$ keV resonance, the strength value from Ref. \cite{Fox05} and the renormalized value from Ref. \cite{Rol73} are in good agreement. Strengths are also presented in Ref. \cite{Sen73}, but have been disregarded since they seem to deviate from the results of Refs. \cite{Fox05,Rol73} by a factor of 2.
Two-level interferences between the $1^-$ resonances at  $-2$ keV and $65$ keV, between the $2^-$ resonances at $183$ keV and $1037$ keV, and between the $2^+$ resonances at $677$ keV and $779$ keV, are explicitly taken into account. Since the signs of the interferences are unknown, the signs are sampled randomly using a binary probability density function (see Sec 4.4 in Paper I). 
The direct capture S-factor is adopted from the recent measurement of Newton et al. \cite{New09}.
\begin{figure}[]
\includegraphics[height=8.5cm]{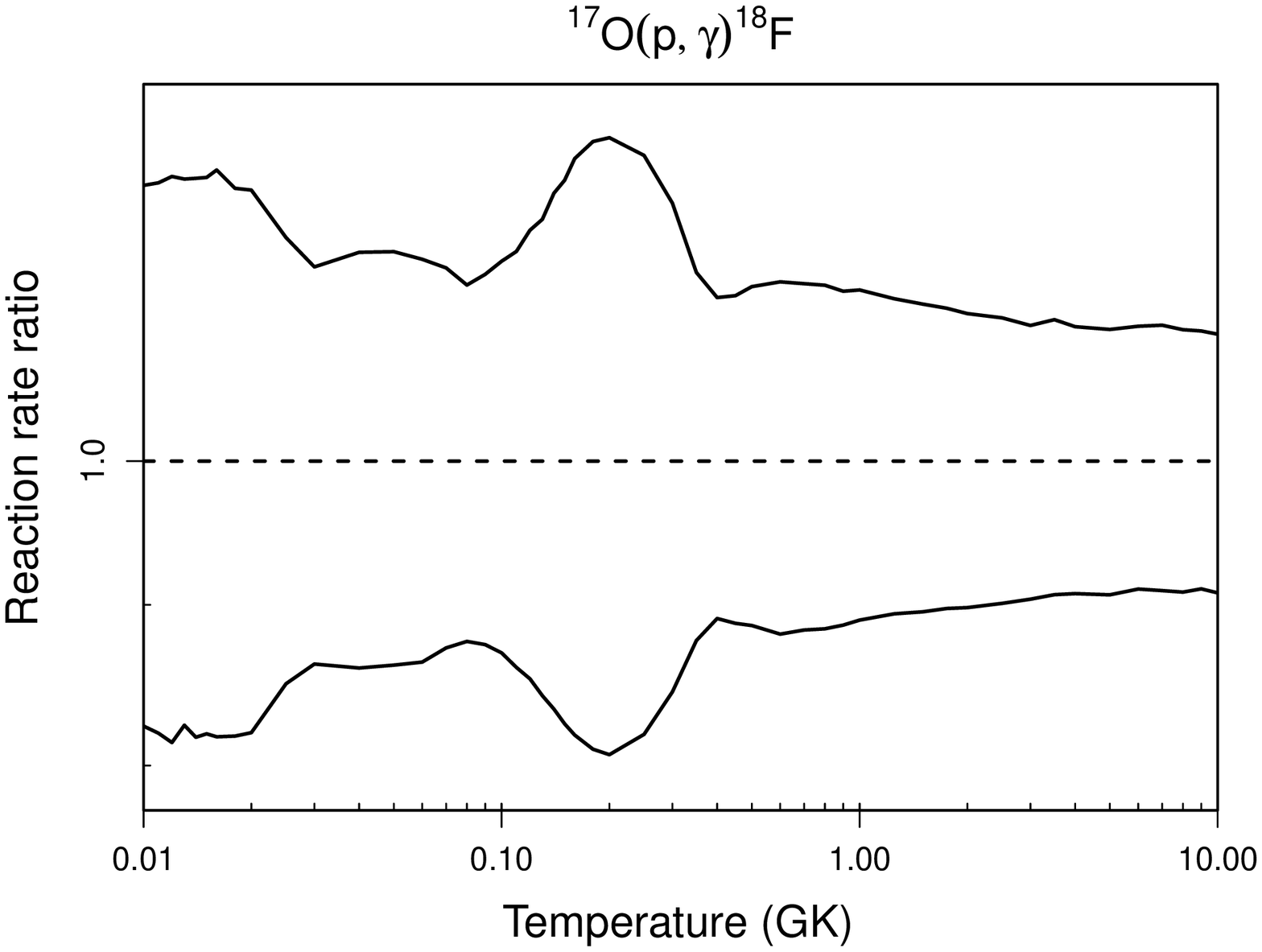}
\end{figure}
\clearpage
\begin{figure}[]
\includegraphics[height=18.5cm]{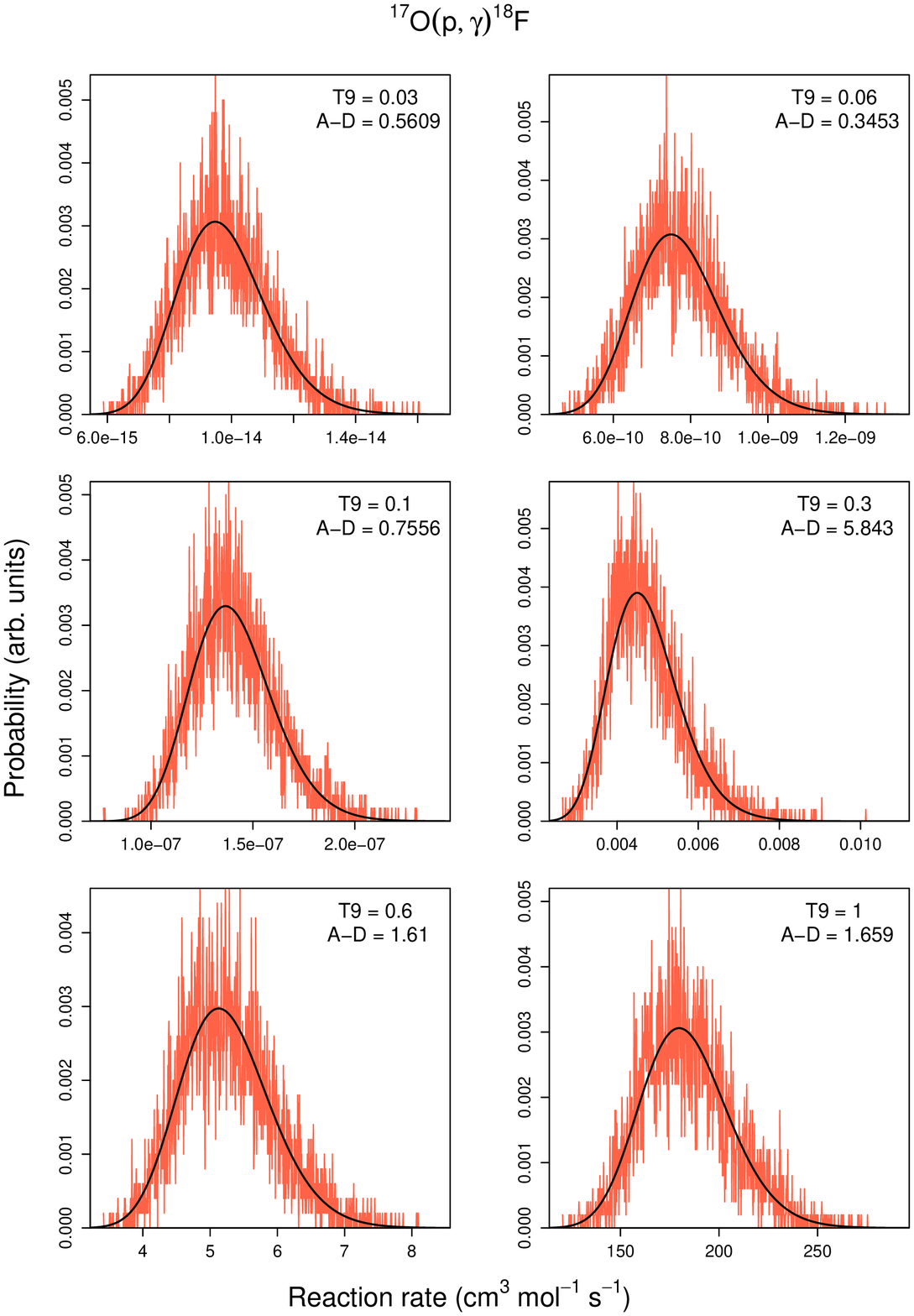}
\end{figure}
\clearpage
\setlongtables
\begin{longtable}{@{\extracolsep{\fill}}lccc | ccc}
\caption{Total thermonuclear reaction rates for $^{17}$O(p,$\alpha$)$^{14}$N.}  \label{} \\
\hline \hline 
	\multicolumn{1}{c}{T (GK)} & \multicolumn{1}{c}{Low rate} & \multicolumn{1}{c}{Median rate} & \multicolumn{1}{c}{High rate}  & \multicolumn{1}{c}{lognormal $\mu$} & \multicolumn{1}{c}{lognormal $\sigma$} & \multicolumn{1}{c}{A-D} \\ \hline 
\endfirsthead
\multicolumn{6}{c}{{\tablename} \thetable{} -- continued} \\
\hline \hline 
	\multicolumn{1}{c}{T (GK)} & \multicolumn{1}{c}{Low rate} & \multicolumn{1}{c}{Median rate} & \multicolumn{1}{c}{High rate} & \multicolumn{1}{c}{lognormal $\mu$} & \multicolumn{1}{c}{lognormal $\sigma$} & \multicolumn{1}{c}{A-D} \\ \hline 
\endhead
	 \hline \hline
\endfoot
	\hline \hline
\endlastfoot
0.010 &  5.05$\times$10$^{-25}$  &  5.89$\times$10$^{-25}$  &
       6.99$\times$10$^{-25}$  &  -5.578$\times$10$^{+01}$  &
       1.65$\times$10$^{-01}$  &  5.08$\times$10$^{+00}$  \\
0.011 &  5.35$\times$10$^{-24}$  &  6.21$\times$10$^{-24}$  &
       7.28$\times$10$^{-24}$  &  -5.343$\times$10$^{+01}$  &
       1.56$\times$10$^{-01}$  &  4.30$\times$10$^{+00}$  \\
0.012 &  4.37$\times$10$^{-23}$  &  5.04$\times$10$^{-23}$  &
       5.85$\times$10$^{-23}$  &  -5.134$\times$10$^{+01}$  &
       1.49$\times$10$^{-01}$  &  2.34$\times$10$^{+00}$  \\
0.013 &  2.89$\times$10$^{-22}$  &  3.30$\times$10$^{-22}$  &
       3.82$\times$10$^{-22}$  &  -4.946$\times$10$^{+01}$  &
       1.42$\times$10$^{-01}$  &  4.30$\times$10$^{+00}$  \\
0.014 &  1.60$\times$10$^{-21}$  &  1.82$\times$10$^{-21}$  &
       2.09$\times$10$^{-21}$  &  -4.775$\times$10$^{+01}$  &
       1.35$\times$10$^{-01}$  &  2.65$\times$10$^{+00}$  \\
0.015 &  7.69$\times$10$^{-21}$  &  8.75$\times$10$^{-21}$  &
       10.00$\times$10$^{-21}$  &  -4.618$\times$10$^{+01}$  &
       1.32$\times$10$^{-01}$  &  5.43$\times$10$^{-01}$  \\
0.016 &  3.32$\times$10$^{-20}$  &  3.77$\times$10$^{-20}$  &
       4.30$\times$10$^{-20}$  &  -4.472$\times$10$^{+01}$  &
       1.29$\times$10$^{-01}$  &  6.47$\times$10$^{-01}$  \\
0.018 &  5.87$\times$10$^{-19}$  &  6.81$\times$10$^{-19}$  &
       7.92$\times$10$^{-19}$  &  -4.183$\times$10$^{+01}$  &
       1.49$\times$10$^{-01}$  &  6.69$\times$10$^{-01}$  \\
\end{longtable}
Comments: In total, 24 resonances in the range of E$_r^{cm}\leq1685$ keV are taken into account, including two subthreshold resonances at $-3$ keV ($1^+$) and $-2$ keV ($1^-$). The energy of the resonance at E$_r^{cm}=65.1\pm0.5$ keV is obtained from the corrected excitation energy E$_x=5671.6\pm0.2$ keV reported in Chafa et al. \cite{Cha07}. For the E$_r^{cm}=183$ keV resonance, the weighted average values of the energies and strengths measured by Chafa et al. \cite{Cha07}, Newton et al. \cite{New07o} and Moazen et al. \cite{Moa07} are adopted. For all resonances above E$_r^{cm}=500$ keV, the partial widths are adopted from the R-matrix analysis (see Tab. 3 of Kieser et al. \cite{Kie79}). Two-level interferences between the $1^-$ resonances at $-2$ keV and $65$ keV, and between the $2^-$ resonances at  $183$ keV and $1203$ keV, are explicitly taken into account. Since the signs of the interferences are unknown, the signs are sampled randomly using a binary probability density function (see Sec 4.4 in Paper I). 
\begin{figure}[]
\includegraphics[height=8.5cm]{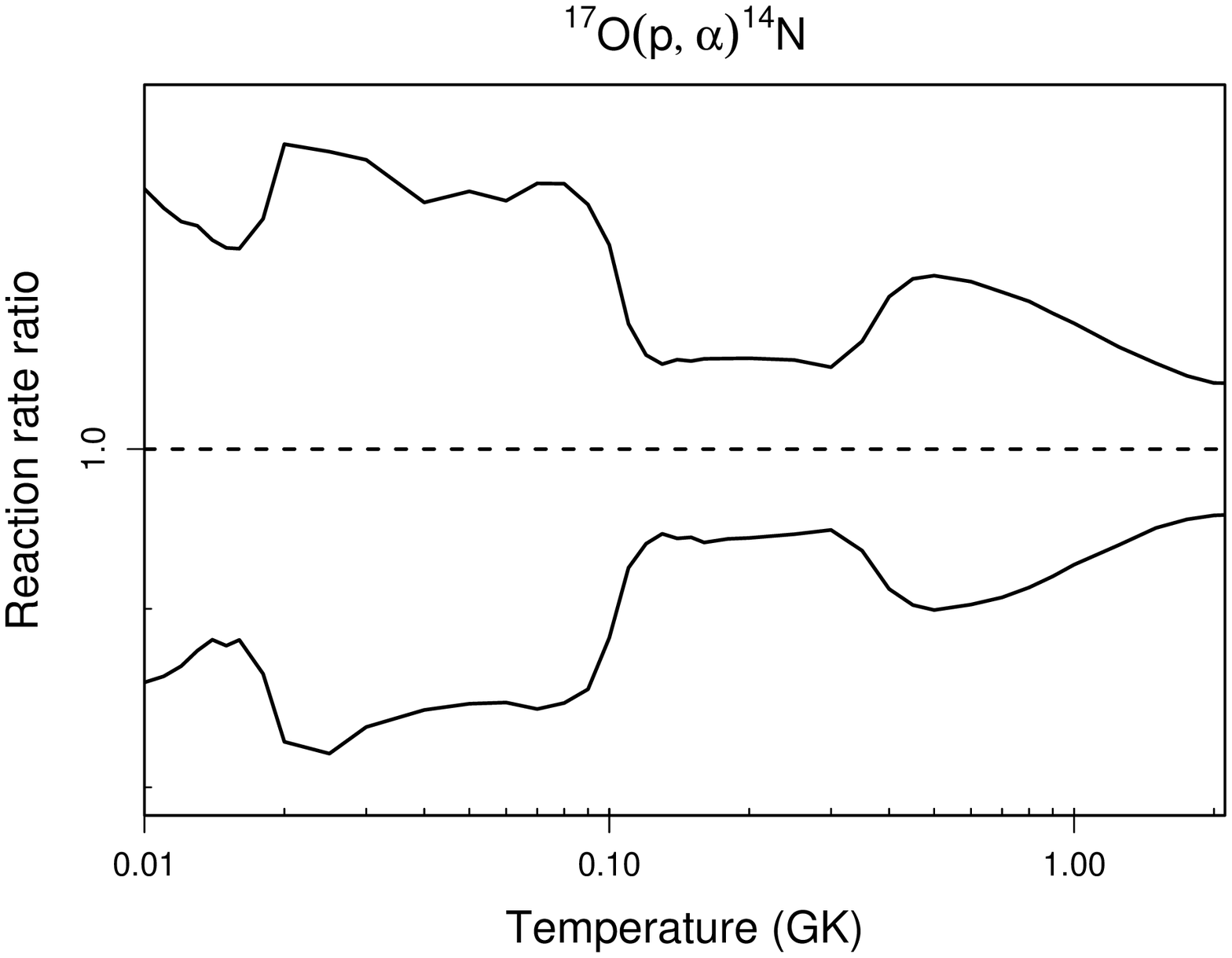}
\end{figure}
\clearpage
\begin{figure}[]
\includegraphics[height=18.5cm]{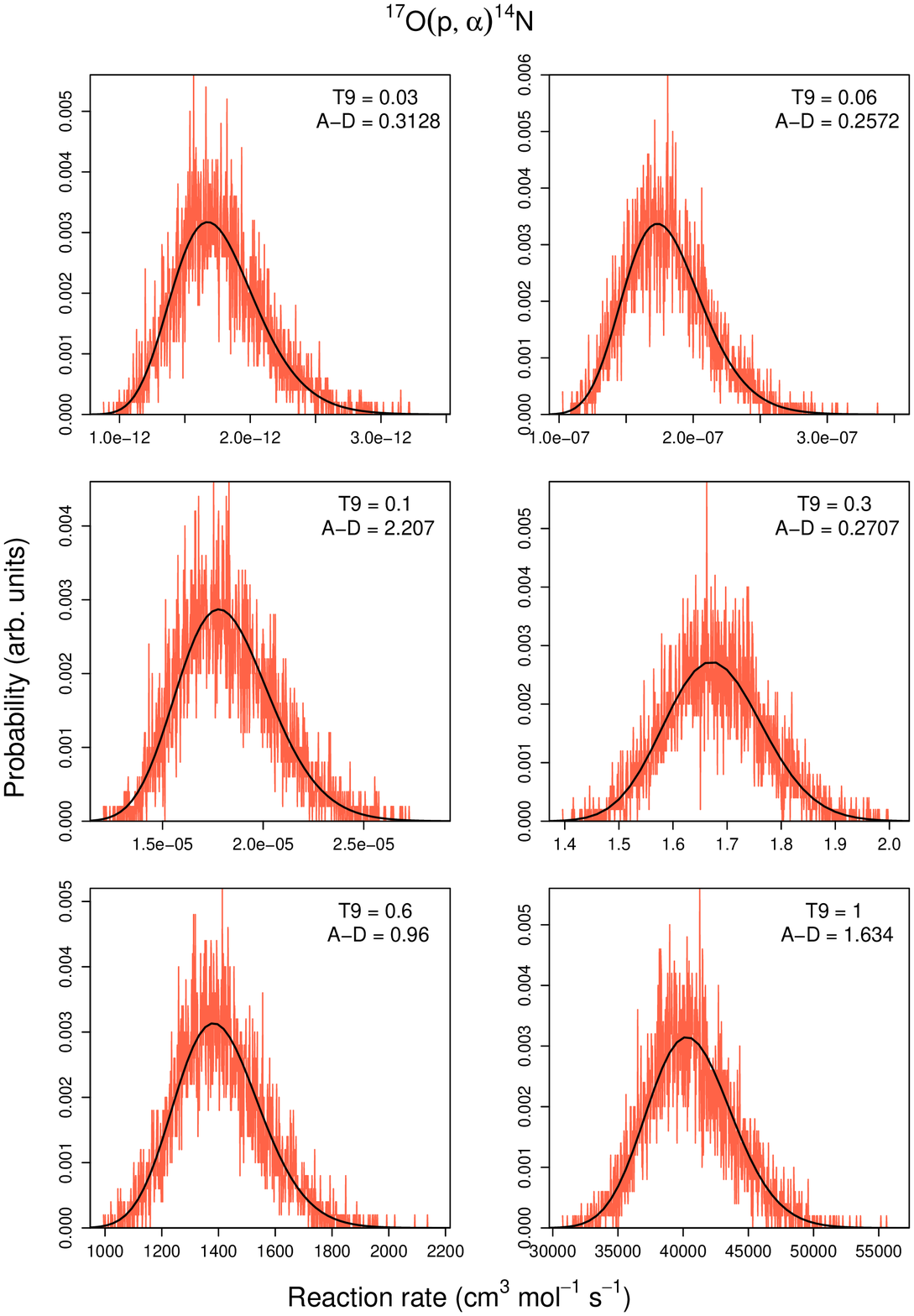}
\end{figure}
\clearpage
\setlongtables
\begin{longtable}{@{\extracolsep{\fill}}cccc | ccc}
\caption{Total thermonuclear reaction rates for $^{18}$O(p,$\gamma$)$^{19}$F.}  \label{tab:o18pgresults} \\
\hline \hline 
	\multicolumn{1}{c}{T (GK)} & \multicolumn{1}{c}{Low rate} & \multicolumn{1}{c}{Median rate} & \multicolumn{1}{c}{High rate}  & \multicolumn{1}{c}{lognormal $\mu$} & \multicolumn{1}{c}{lognormal $\sigma$} & \multicolumn{1}{c}{A-D} \\ \hline 
\endfirsthead
\multicolumn{6}{c}{{\tablename} \thetable{} -- continued} \\
\hline \hline 
	\multicolumn{1}{c}{T (GK)} & \multicolumn{1}{c}{Low rate} & \multicolumn{1}{c}{Median rate} & \multicolumn{1}{c}{High rate} & \multicolumn{1}{c}{lognormal $\mu$} & \multicolumn{1}{c}{lognormal $\sigma$} & \multicolumn{1}{c}{A-D} \\ \hline 
\endhead
	 \hline \hline
\endfoot
	\hline \hline
\endlastfoot
0.010 &  9.63$\times$10$^{-24}$  &  1.75$\times$10$^{-23}$  &
       3.16$\times$10$^{-23}$  &  -5.240$\times$10$^{+01}$  &
       5.88$\times$10$^{-01}$  &  4.90$\times$10$^{-01}$  \\
0.011 &  7.72$\times$10$^{-23}$  &  1.35$\times$10$^{-22}$  &
       2.38$\times$10$^{-22}$  &  -5.035$\times$10$^{+01}$  &
       5.57$\times$10$^{-01}$  &  9.17$\times$10$^{-01}$  \\
0.012 &  4.68$\times$10$^{-22}$  &  7.89$\times$10$^{-22}$  &
       1.34$\times$10$^{-21}$  &  -4.858$\times$10$^{+01}$  &
       5.21$\times$10$^{-01}$  &  2.31$\times$10$^{+00}$  \\
0.013 &  2.37$\times$10$^{-21}$  &  3.71$\times$10$^{-21}$  &
       6.11$\times$10$^{-21}$  &  -4.702$\times$10$^{+01}$  &
       4.74$\times$10$^{-01}$  &  5.17$\times$10$^{+00}$  \\
0.014 &  1.01$\times$10$^{-20}$  &  1.51$\times$10$^{-20}$  &
       2.38$\times$10$^{-20}$  &  -4.562$\times$10$^{+01}$  &
       4.28$\times$10$^{-01}$  &  6.99$\times$10$^{+00}$  \\
0.015 &  3.87$\times$10$^{-20}$  &  5.49$\times$10$^{-20}$  &
       8.19$\times$10$^{-20}$  &  -4.432$\times$10$^{+01}$  &
       3.81$\times$10$^{-01}$  &  7.68$\times$10$^{+00}$  \\
0.016 &  1.34$\times$10$^{-19}$  &  1.83$\times$10$^{-19}$  &
       2.62$\times$10$^{-19}$  &  -4.312$\times$10$^{+01}$  &
       3.39$\times$10$^{-01}$  &  8.60$\times$10$^{+00}$  \\
0.018 &  1.27$\times$10$^{-18}$  &  1.67$\times$10$^{-18}$  &
       2.24$\times$10$^{-18}$  &  -4.092$\times$10$^{+01}$  &
       2.86$\times$10$^{-01}$  &  3.69$\times$10$^{+00}$  \\
\end{longtable}
Comments: 
Resonance energies are calculated from excitation energies \cite{Til95} and the Q-value \cite{Aud03}, except when a more precise value of the resonance energy was measured directly. Resonance strengths have been measured directly for 19 resonances at energies of E$_r^{cm}=89-1892$~keV by Wiescher et al. \cite{WI80} and Vogelaar et al. \cite{VO90}. At variance with Angulo et al. \cite{Ang99}, we use published and derived values of partial widths \cite{YA62,SE69,LO79} in order to integrate broad resonance contributions numerically. Additional data were provided by $^{15}$N+$\alpha$ studies and by other experiments \cite{SY78,PR89}. For E$_r^{cm}=19$ keV, the proton partial width is deduced from indirect measurements; the ($^3$He,d) transfer experiment by Champagne and Pitt \cite{CH86} finds a value of $\Gamma_p$ = 2.0$\times10^{-19}$~eV, while a recent study using the Trojan Horse Method \cite{La08} obtains $\Gamma_p$ = (3.1$\pm1.1)\times10^{-19}$~eV. For the radiative and total width we adopt the values quoted in Wiescher et al. \cite{WI80} as a private communication from K. Allen ($\Gamma_\gamma=2.3$ eV, $\Gamma=2.5$ keV); a 40\% uncertainty is assumed for these values. For E$_r^{cm}=89$ keV, the proton partial width is calculated from the measured (p,$\alpha$) resonance strength \cite{LO79} and the derived value agrees with the result of the recent Trojan Horse study \cite{La08}. The radiative width, $\Gamma_\gamma=0.6$ eV, is again adopted from Ref. \cite{WI80} (private communication from K. Allen), while an upper limit of $\Gamma \leq 3$ keV is given in Ref. \cite{Til95}. For E$_r^{cm}=143$ keV, the resonance energy \cite{BE95}, strength \cite{WI80,BE82,VO90}, total width \cite{BE95} and proton partial width (deduced from $\omega\gamma_{p\alpha}$ \cite{BE95}) have all been measured precisely. The direct capture S-factor is adopted from Wiescher et al. \cite{WI80}. Above $T=5$~GK the rate is extrapolated using Hauser-Feshbach results. 
\begin{figure}[]
\includegraphics[height=8.5cm]{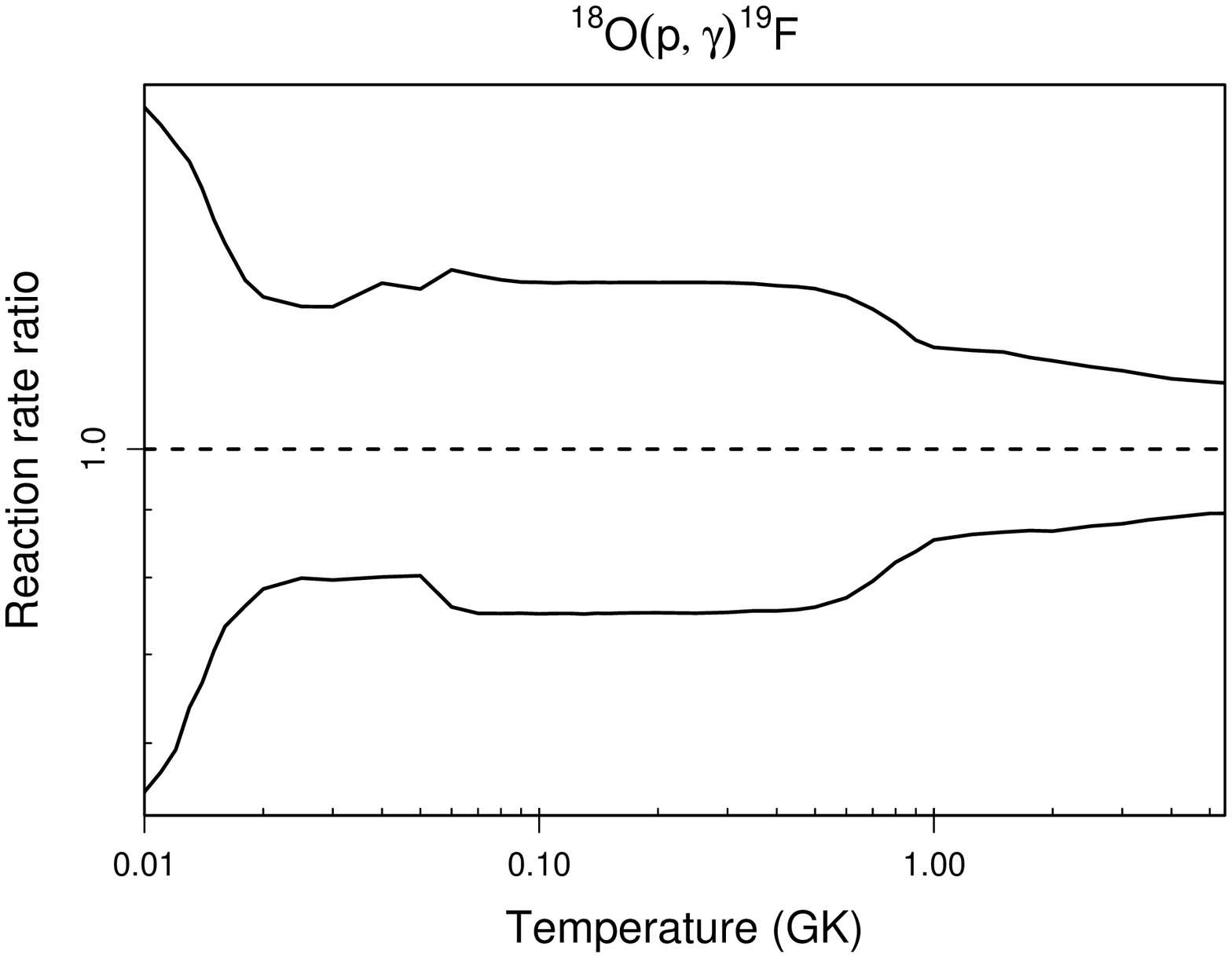}
\end{figure}
\clearpage
\begin{figure}[]
\includegraphics[height=18.5cm]{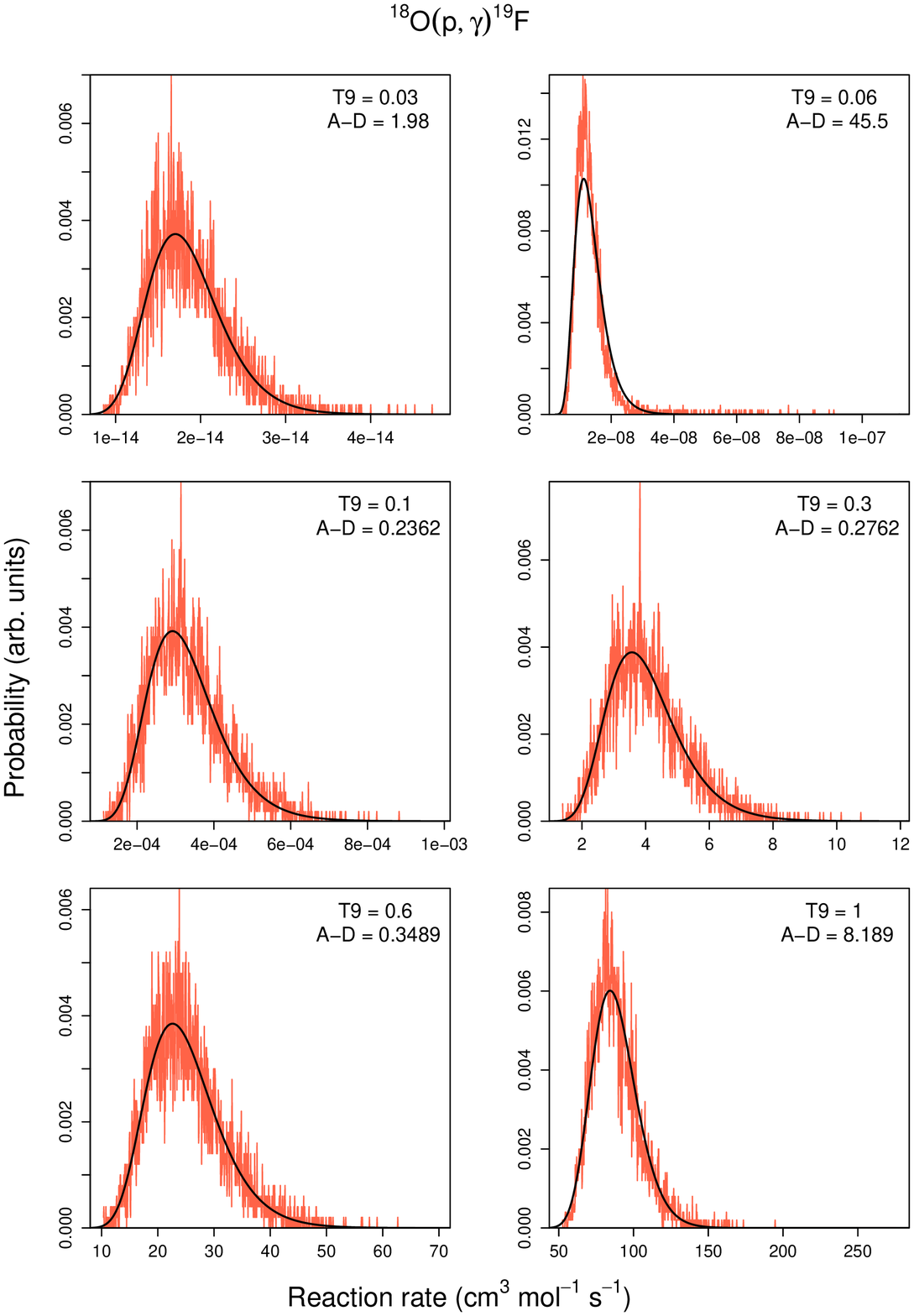}
\end{figure}
\clearpage
\setlongtables
\begin{longtable}{@{\extracolsep{\fill}}cccc | ccc}
\caption{Total thermonuclear reaction rates for $^{18}$O(p,$\alpha$)$^{15}$N.}  \label{} \\
\hline \hline 
	\multicolumn{1}{c}{T (GK)} & \multicolumn{1}{c}{Low rate} & \multicolumn{1}{c}{Median rate} & \multicolumn{1}{c}{High rate}  & \multicolumn{1}{c}{lognormal $\mu$} & \multicolumn{1}{c}{lognormal $\sigma$} & \multicolumn{1}{c}{A-D} \\ \hline 
\endfirsthead
\multicolumn{6}{c}{{\tablename} \thetable{} -- continued} \\
\hline \hline 
	\multicolumn{1}{c}{T (GK)} & \multicolumn{1}{c}{Low rate} & \multicolumn{1}{c}{Median rate} & \multicolumn{1}{c}{High rate} & \multicolumn{1}{c}{lognormal $\mu$} & \multicolumn{1}{c}{lognormal $\sigma$} & \multicolumn{1}{c}{A-D} \\ \hline 
\endhead
	 \hline \hline
\endfoot
	\hline \hline
\endlastfoot
0.010 & 1.49$\times$10$^{-20}$ & 1.93$\times$10$^{-20}$ &
      2.49$\times$10$^{-20}$ &  -4.540$\times$10$^{+01}$ &
       2.57$\times$10$^{-01}$  & 3.13$\times$10$^{-01}$ \\ 
0.011 & 1.14$\times$10$^{-19}$ & 1.52$\times$10$^{-19}$ &
      1.99$\times$10$^{-19}$ &  -4.334$\times$10$^{+01}$ &
       2.78$\times$10$^{-01}$  & 1.34$\times$10$^{+00}$ \\ 
0.012 & 6.64$\times$10$^{-19}$ & 9.00$\times$10$^{-19}$ &
      1.19$\times$10$^{-18}$ &  -4.156$\times$10$^{+01}$ &
       3.00$\times$10$^{-01}$  & 2.63$\times$10$^{+00}$ \\ 
0.013 & 3.11$\times$10$^{-18}$ & 4.35$\times$10$^{-18}$ &
      5.94$\times$10$^{-18}$ &  -3.999$\times$10$^{+01}$ &
       3.22$\times$10$^{-01}$  & 4.75$\times$10$^{+00}$ \\ 
0.014 & 1.21$\times$10$^{-17}$ & 1.80$\times$10$^{-17}$ &
      2.50$\times$10$^{-17}$ &  -3.858$\times$10$^{+01}$ &
       3.59$\times$10$^{-01}$  & 8.34$\times$10$^{+00}$ \\ 
0.015 & 4.21$\times$10$^{-17}$ & 6.60$\times$10$^{-17}$ &
      9.72$\times$10$^{-17}$ &  -3.728$\times$10$^{+01}$ &
       4.05$\times$10$^{-01}$  & 9.10$\times$10$^{+00}$ \\ 
0.016 & 1.33$\times$10$^{-16}$ & 2.24$\times$10$^{-16}$ &
      3.42$\times$10$^{-16}$ &  -3.607$\times$10$^{+01}$ &
       4.55$\times$10$^{-01}$  & 1.29$\times$10$^{+01}$ \\ 
0.018 & 1.03$\times$10$^{-15}$ & 2.01$\times$10$^{-15}$ &
      3.39$\times$10$^{-15}$ &  -3.389$\times$10$^{+01}$ &
       5.52$\times$10$^{-01}$  & 2.84$\times$10$^{+01}$ \\
\end{longtable}
Comments: 
Up to a resonance energy of $E_r^{cm}=2$ MeV, the energies and partial widths are adopted from the same sources \cite{YA62,SE69,LO79,WI80,CH86,BE95,La08} as for the $^{18}$O(p,$\gamma)^{19}$F reaction (see comments of Tab. \ref{tab:o18pgresults}). At higher energies, partial widths have been measured up to $E_r^{cm}=13$ MeV \cite{SE69,OR73,AL75,MU79}. A broad resonance near $E_r^{cm}\approx660$ keV, which has not been detected in the (p,$\gamma$) channel, was observed by Yagi \cite{YA62}, Mak et al. \cite{MA78} and Lorentz-Wirzba et al. \cite{LO79}. These measurements are in agreement regarding the proton width \cite{YA62,LO79}, but not for the $\alpha$-particle width ($\Gamma_\alpha$ = 317 keV \cite{LO79}, 150 keV \cite{MA78} or 90 keV \cite{YA62}). Thus we adopt a value of $\Gamma_\alpha=200\pm110$ keV. The effects of the interference between this resonance and the broad 1/2$^+$ resonance at E$_r^{cm}=798$ keV are included in our results. We do not introduce any nonresonant rate contribution as in Lorentz-Wirzba et al. \cite{LO79} or Mak et al. \cite{MA78}. Presumably this contribution originates from the tails of higher-lying broad resonances that are already taken into account since we numerically integrate the partial rates of all 70 resonances.
\begin{figure}[]
\includegraphics[height=8.5cm]{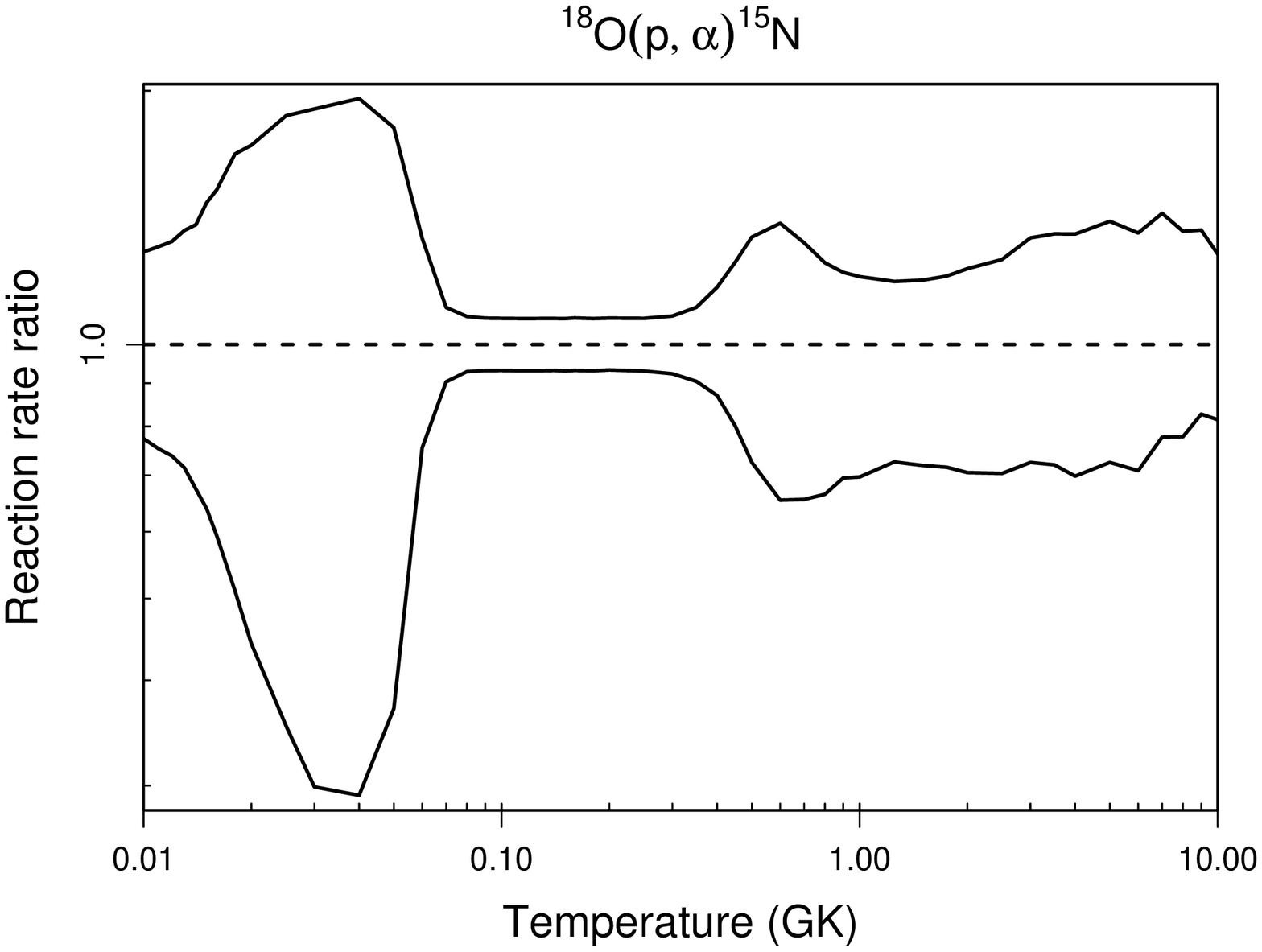}
\end{figure}
\clearpage
\begin{figure}[]
\includegraphics[height=18.5cm]{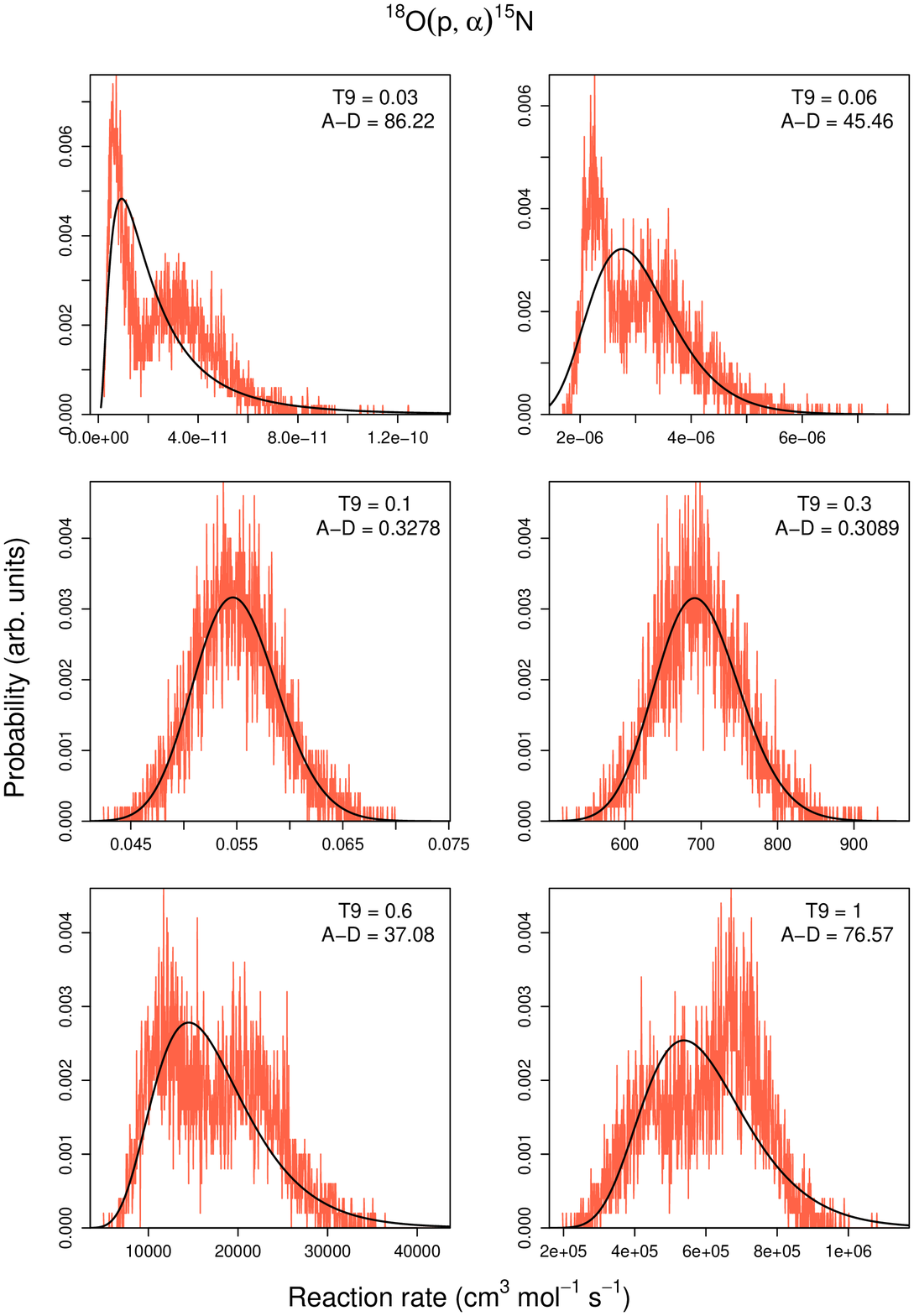}
\end{figure}
\clearpage
\setlongtables
\begin{longtable}{@{\extracolsep{\fill}}cccc | ccc}
\caption{Total thermonuclear reaction rates for $^{18}$O($\alpha$,$\gamma$)$^{22}$Ne.}  \label{} \\
\hline \hline 
	\multicolumn{1}{c}{T (GK)} & \multicolumn{1}{c}{Low rate} & \multicolumn{1}{c}{Median rate} & \multicolumn{1}{c}{High rate}  & \multicolumn{1}{c}{lognormal $\mu$} & \multicolumn{1}{c}{lognormal $\sigma$} & \multicolumn{1}{c}{A-D} \\ \hline 
\endfirsthead
\multicolumn{6}{c}{{\tablename} \thetable{} -- continued} \\
\hline \hline 
	\multicolumn{1}{c}{T (GK)} & \multicolumn{1}{c}{Low rate} & \multicolumn{1}{c}{Median rate} & \multicolumn{1}{c}{High rate} & \multicolumn{1}{c}{lognormal $\mu$} & \multicolumn{1}{c}{lognormal $\sigma$} & \multicolumn{1}{c}{A-D} \\ \hline 
\endhead
	 \hline \hline
\endfoot
	\hline \hline
\endlastfoot
0.010 &  5.80$\times$10$^{-63}$  &  2.42$\times$10$^{-62}$  &
       5.99$\times$10$^{-62}$  &  -1.421$\times$10$^{+02}$  &
       1.40$\times$10$^{+00}$  &  1.06$\times$10$^{+02}$  \\
0.011 &  2.59$\times$10$^{-60}$  &  8.08$\times$10$^{-60}$  &
       1.88$\times$10$^{-59}$  &  -1.363$\times$10$^{+02}$  &
       1.27$\times$10$^{+00}$  &  1.07$\times$10$^{+02}$  \\
0.012 &  3.26$\times$10$^{-58}$  &  1.17$\times$10$^{-57}$  &
       2.77$\times$10$^{-57}$  &  -1.314$\times$10$^{+02}$  &
       1.46$\times$10$^{+00}$  &  1.65$\times$10$^{+02}$  \\
0.013 &  1.54$\times$10$^{-56}$  &  8.43$\times$10$^{-56}$  &
       2.16$\times$10$^{-55}$  &  -1.272$\times$10$^{+02}$  &
       1.77$\times$10$^{+00}$  &  1.91$\times$10$^{+02}$  \\
0.014 &  3.56$\times$10$^{-55}$  &  3.29$\times$10$^{-54}$  &
       1.00$\times$10$^{-53}$  &  -1.237$\times$10$^{+02}$  &
       2.08$\times$10$^{+00}$  &  1.76$\times$10$^{+02}$  \\
0.015 &  5.24$\times$10$^{-54}$  &  7.53$\times$10$^{-53}$  &
       3.03$\times$10$^{-52}$  &  -1.206$\times$10$^{+02}$  &
       2.37$\times$10$^{+00}$  &  1.46$\times$10$^{+02}$  \\
0.016 &  5.44$\times$10$^{-53}$  &  1.12$\times$10$^{-51}$  &
       6.05$\times$10$^{-51}$  &  -1.180$\times$10$^{+02}$  &
       2.61$\times$10$^{+00}$  &  1.17$\times$10$^{+02}$  \\
0.018 &  2.68$\times$10$^{-51}$  &  9.39$\times$10$^{-50}$  &
       8.97$\times$10$^{-49}$  &  -1.135$\times$10$^{+02}$  &
       2.94$\times$10$^{+00}$  &  7.18$\times$10$^{+01}$  \\
\end{longtable}
Comments: 
A total of 21 resonances are taken into account for calculating the reaction rate.  For resonances in the range of E$_r^{cm}=398-628$ keV, resonance energies are adopted from either Endt \cite{End98} or Vogelaar et al. \cite{VO90}, while resonance strengths are adopted from Ref. \cite{VO90} and Dababneh et al. \cite{Dab03}. Note that for the very weakly observed E$_r^{cm}=398$ keV resonance we adopt a strength uncertainty (50\%) that is larger than the originally published value \cite{Dab03}. For the E$_r^{cm}=542$ keV resonance, the $\alpha$-particle partial width is calculated from the measured resonance strength, while the $\gamma$-ray partial width is equal to the total width (see Berg and Wiehard \cite{Ber79}). For resonances in the E$_r^{cm}=947-1800$ keV region, the energies and strengths are adopted from Trautvetter et al. 1978 \cite{Tra78}. For the E$_r^{cm}=1255$ keV resonance, the total width is known ($\Gamma\approx\Gamma_n=25$ keV), but not enough information is available to derive values for $\Gamma_{\alpha}$ and $\Gamma_{\gamma}$. Thus we do not take the tail of this broad resonance into account. For the two low-energy resonances at E$_r^{cm}=57$ and 174 keV, we use the results of the $^{18}$O($^{6}$Li,d)$^{22}$Ne work of Giesen et al. \cite{Gie94}. For E$_r^{cm}=57$ keV we find a 2$^{+}$ assignment more convincing than the originally proposed 3$^{-}$ assignment (see the angular distribution shown in Fig. 6 of Ref. \cite{Gie94}); we adopt a factor of 2 for the uncertainty of the measured $\alpha$-particle spectroscopic factor. The level corresponding to E$_r^{cm}=174$ is only weakly populated in Ref. \cite{Gie94} and, in our opinion, even a $\ell_{\alpha}=0$ transfer cannot be excluded (as can be seen by comparing the angular distributions for nearby levels). Thus we treat the $\alpha$-particle partial width of this resonance as an upper limit, where an s-wave spectroscopic factor of $S_{\alpha}\le0.086$ can be estimated from the data of Ref. \cite{Gie94}). Note that our assumptions differ from those in the original work of Ref. \cite{Gie94}, where a $2^+$ assignment was adopted (in fact, our procedure is more consistent with Giesen's thesis \cite{Gie94b}). The direct capture S-factor is adopted from Trautvetter et al. \cite{Tra78}. It is interesting to note that vastly different results are obtained by Buchmann, D'Auria and McCorquodale \cite{Buc88} and Descouvemont \cite{Des88}. For example, at $T=0.1$ GK, the calculated direct capture rate in Refs. \cite{Des88,Buc88,Tra78} amounts to $N_A\left<\sigma v\right>_{DC}=2.7\times10^{-34}$, $1.5\times10^{-22}$, $1.3\times10^{-25}$ cm$^3$mol$^{-1}$s$^{-1}$, respectively. The direct capture component and the low-energy tail of the broad E$_r^{cm}=1255$ keV resonance are only expected to influence the total rate at very low temperatures of $T=0.018-0.03$ GK. For resonances above E$_r^{cm}=2$ MeV, see the comment section in Paper III.
\begin{figure}[]
\includegraphics[height=8.5cm]{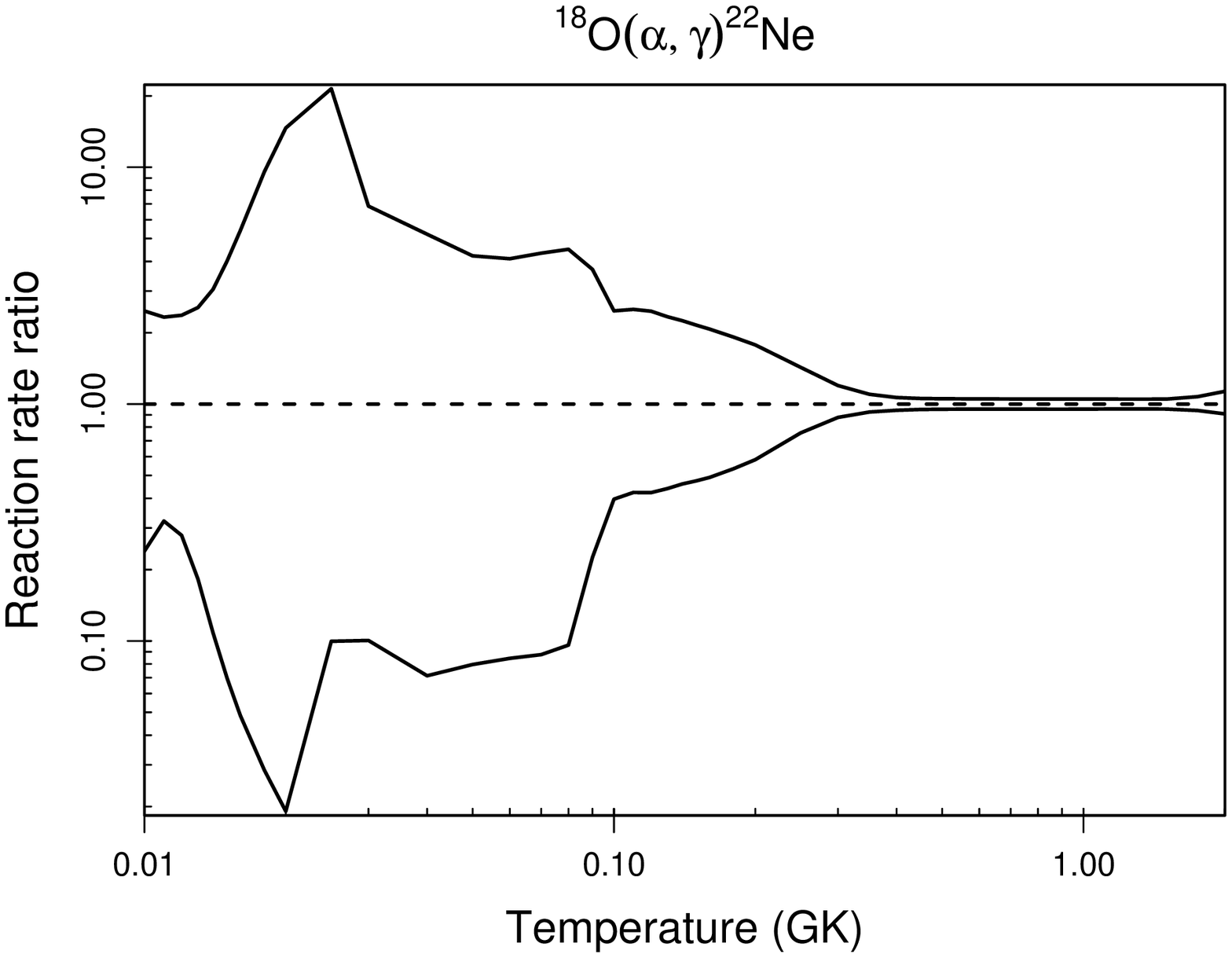}
\end{figure}
\clearpage
\begin{figure}[]
\includegraphics[height=18.5cm]{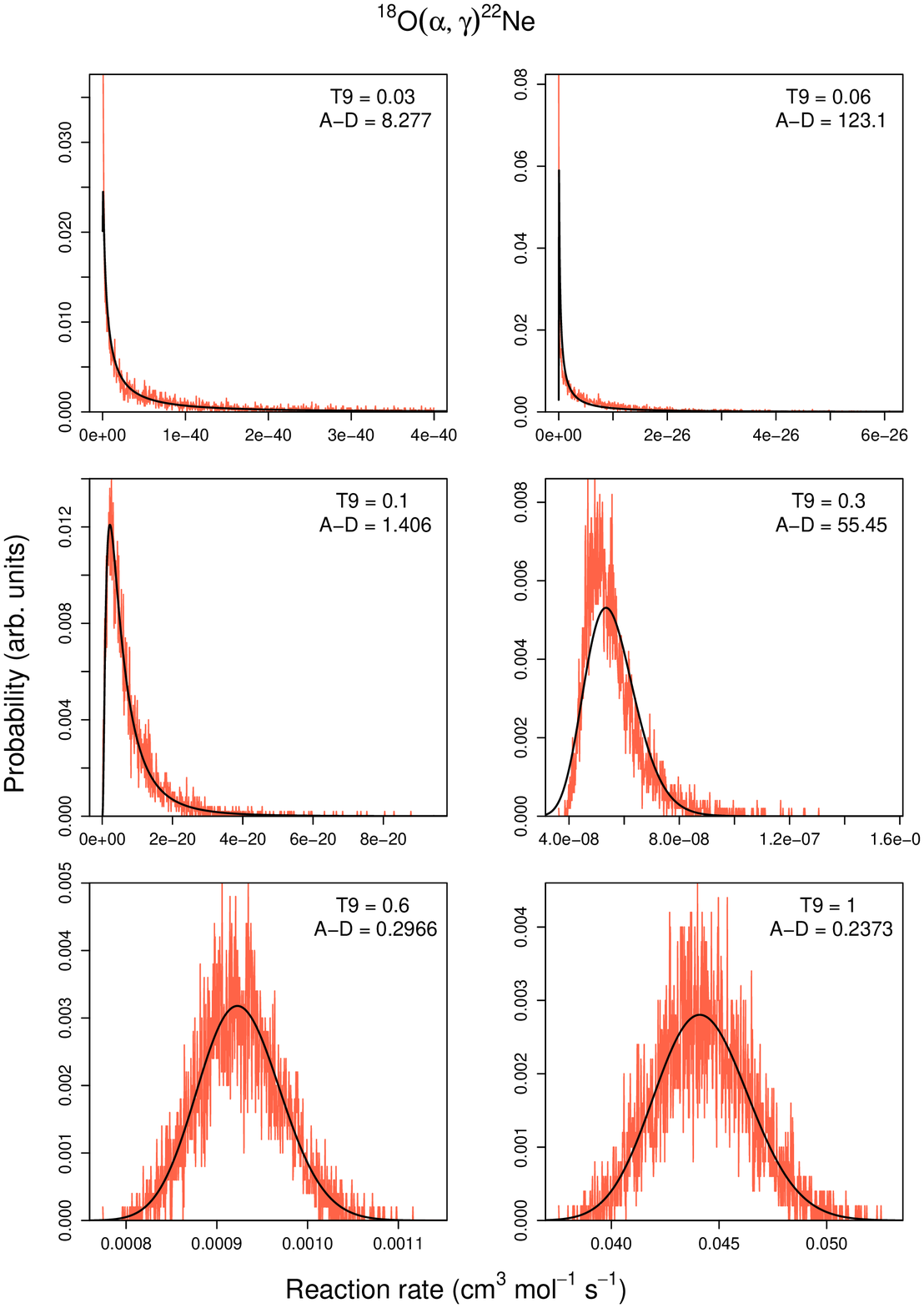}
\end{figure}
\clearpage
\setlongtables
\begin{longtable}{@{\extracolsep{\fill}}cccc | ccc}
\caption{Total thermonuclear reaction rates for $^{17}$F(p,$\gamma$)$^{18}$Ne.}  \label{} \\
\hline \hline 
	\multicolumn{1}{c}{T (GK)} & \multicolumn{1}{c}{Low rate} & \multicolumn{1}{c}{Median rate} & \multicolumn{1}{c}{High rate}  & \multicolumn{1}{c}{lognormal $\mu$} & \multicolumn{1}{c}{lognormal $\sigma$} & \multicolumn{1}{c}{A-D} \\ \hline 
\endfirsthead
\multicolumn{6}{c}{{\tablename} \thetable{} -- continued} \\
\hline \hline 
	\multicolumn{1}{c}{T (GK)} & \multicolumn{1}{c}{Low rate} & \multicolumn{1}{c}{Median rate} & \multicolumn{1}{c}{High rate} & \multicolumn{1}{c}{lognormal $\mu$} & \multicolumn{1}{c}{lognormal $\sigma$} & \multicolumn{1}{c}{A-D} \\ \hline 
\endhead
	 \hline \hline
\endfoot
	\hline \hline
\endlastfoot
0.010 &  2.02$\times$10$^{-28}$  &  2.97$\times$10$^{-28}$  &
       4.29$\times$10$^{-28}$  &  -6.339$\times$10$^{+01}$  &
       3.77$\times$10$^{-01}$  &  3.91$\times$10$^{-01}$  \\
0.011 &  2.60$\times$10$^{-27}$  &  3.85$\times$10$^{-27}$  &
       5.63$\times$10$^{-27}$  &  -6.082$\times$10$^{+01}$  &
       3.83$\times$10$^{-01}$  &  3.85$\times$10$^{-01}$  \\
0.012 &  2.57$\times$10$^{-26}$  &  3.69$\times$10$^{-26}$  &
       5.36$\times$10$^{-26}$  &  -5.856$\times$10$^{+01}$  &
       3.72$\times$10$^{-01}$  &  3.51$\times$10$^{-01}$  \\
0.013 &  1.92$\times$10$^{-25}$  &  2.81$\times$10$^{-25}$  &
       4.10$\times$10$^{-25}$  &  -5.653$\times$10$^{+01}$  &
       3.84$\times$10$^{-01}$  &  2.23$\times$10$^{-01}$  \\
0.014 &  1.19$\times$10$^{-24}$  &  1.75$\times$10$^{-24}$  &
       2.53$\times$10$^{-24}$  &  -5.471$\times$10$^{+01}$  &
       3.73$\times$10$^{-01}$  &  3.97$\times$10$^{-01}$  \\
0.015 &  6.32$\times$10$^{-24}$  &  9.15$\times$10$^{-24}$  &
       1.35$\times$10$^{-23}$  &  -5.304$\times$10$^{+01}$  &
       3.81$\times$10$^{-01}$  &  4.01$\times$10$^{-01}$  \\
0.016 &  2.89$\times$10$^{-23}$  &  4.17$\times$10$^{-23}$  &
       6.13$\times$10$^{-23}$  &  -5.152$\times$10$^{+01}$  &
       3.82$\times$10$^{-01}$  &  3.35$\times$10$^{-01}$  \\
0.018 &  4.19$\times$10$^{-22}$  &  6.07$\times$10$^{-22}$  &
       8.87$\times$10$^{-22}$  &  -4.885$\times$10$^{+01}$  &
       3.79$\times$10$^{-01}$  &  3.65$\times$10$^{-01}$  \\
\end{longtable}
Comments: In total, 7 resonances in the range of E$_r^{cm}$=596-2226 keV are taken into account. Resonance energies are calculated from the $^{18}$Ne excitation energies measured by Hahn et al. \cite{Hah96} and Park et al. \cite{Par99}, except for the expected s-wave resonance at E$_r^{cm}$=600 keV for which the resonance energy has been measured in elastic scattering studies (Bardayan et al. \cite{Bar00}). Proton partial widths are either adopted from experiment (Refs. \cite{Hah96,Par99,Bar00}) or are calculated by using the measured spectroscopic factors of the $^{18}$O mirror states (Li et al. \cite{Li76}). Gamma-ray partial widths are adopted either from the measured mean lifetimes of the $^{18}$O mirror states (Tilley et al. \cite{Til95}), and corrected for the E$^{2L+1}_\gamma$ energy dependence, or are calculated from the shell model. For all resonances considered, the proton partial width exceeds the $\gamma$-ray partial width. For the broad resonances at E$_r^{cm}$=596, 600, 665 and 1182 keV the reaction rate contributions are found from a numerical integration. The direct capture into 5 bound states is computed by using measured spectroscopic factors of $^{18}$O mirror states \cite{Li76}. The resulting total direct capture S-factor below E=2.5 MeV is approximately constant and amounts to S(E)=2.4$\times$10$^{-3}$ MeV~b. Below T=0.5 GK the direct capture dominates the total rates. Our uncertainties in the total rate are significantly lager than the unrealistic values ($\leq$17\% below T=0.5 GK) reported by Ref. \cite{Bar00}. Our reaction rate is in much better agreement with the theoretical calculations reported in Dufour and Descouvemont \cite{Duf04} compared to those described in Chatterjee, Okolowicz and Ploszajczak \cite{Cha06}.
\begin{figure}[]
\includegraphics[height=8.5cm]{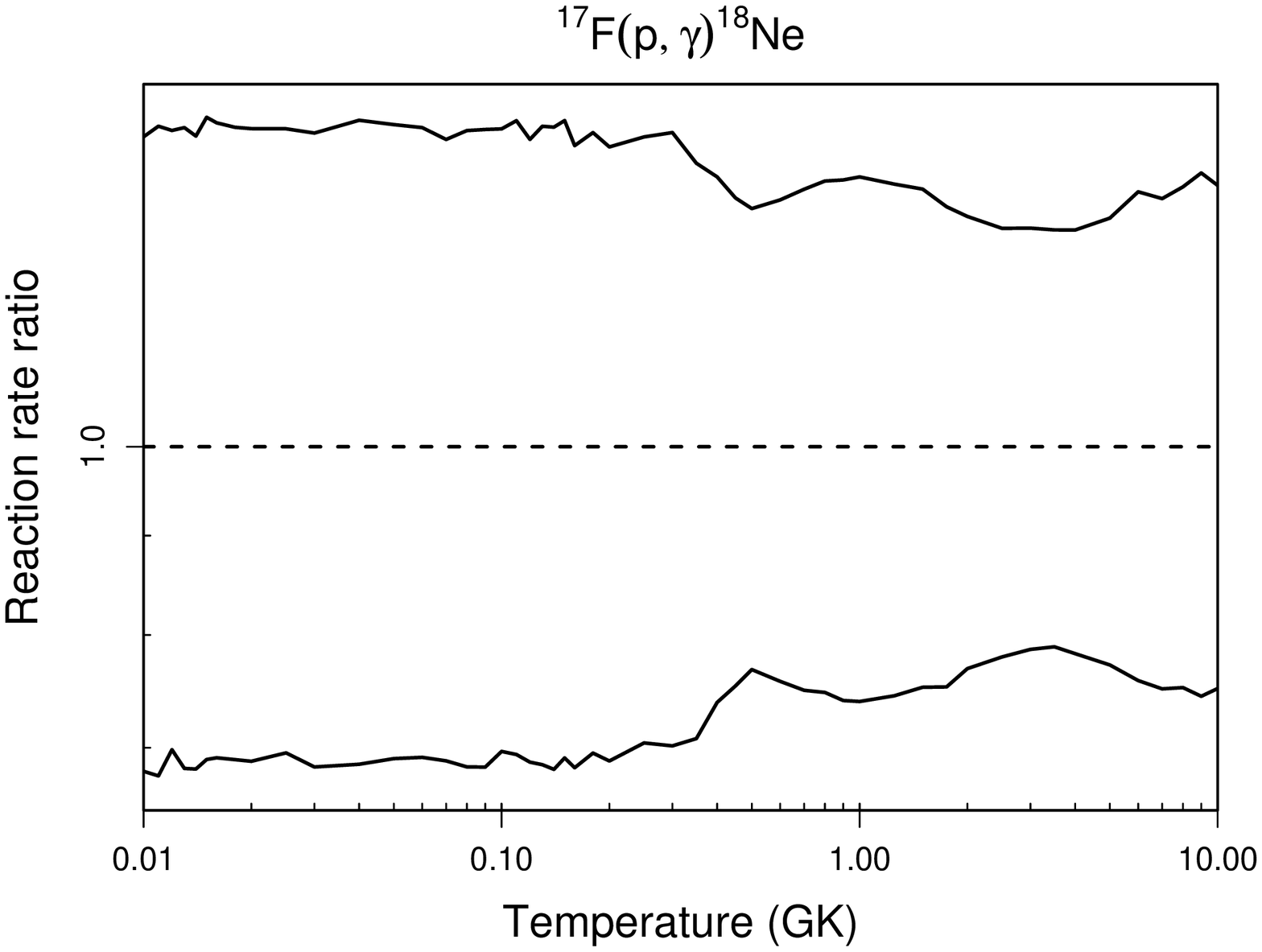}
\end{figure}
\clearpage
\begin{figure}[]
\includegraphics[height=18.5cm]{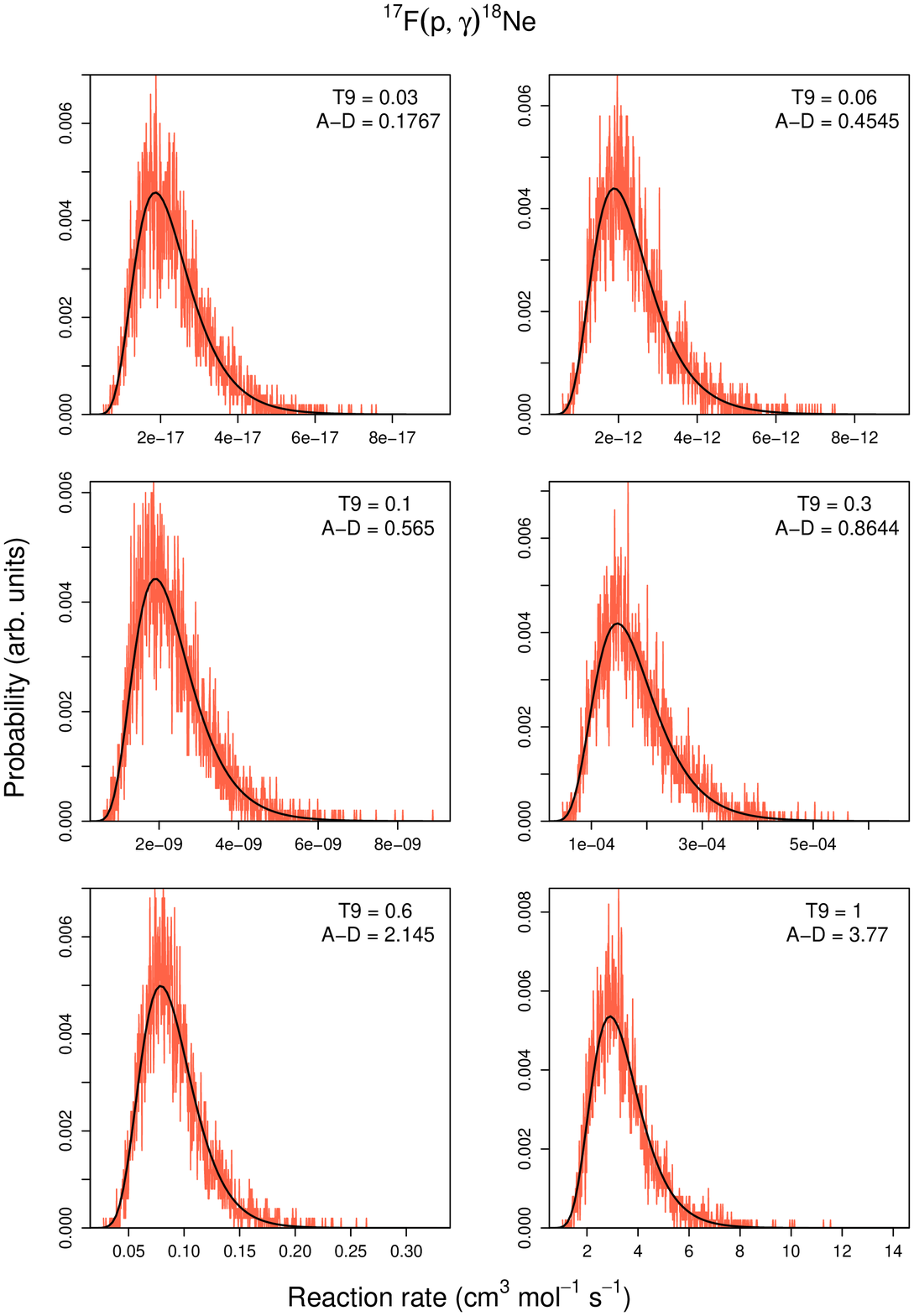}
\end{figure}
\clearpage
\setlongtables
\begin{longtable}{@{\extracolsep{\fill}}cccc | ccc}
\caption{Total thermonuclear reaction rates for $^{18}$F(p,$\gamma$)$^{19}$Ne.}  \label{} \\
\hline \hline 
	\multicolumn{1}{c}{T (GK)} & \multicolumn{1}{c}{Low rate} & \multicolumn{1}{c}{Median rate} & \multicolumn{1}{c}{High rate}  & \multicolumn{1}{c}{lognormal $\mu$} & \multicolumn{1}{c}{lognormal $\sigma$} & \multicolumn{1}{c}{A-D} \\ \hline 
\endfirsthead
\multicolumn{6}{c}{{\tablename} \thetable{} -- continued} \\
\hline \hline 
	\multicolumn{1}{c}{T (GK)} & \multicolumn{1}{c}{Low rate} & \multicolumn{1}{c}{Median rate} & \multicolumn{1}{c}{High rate} & \multicolumn{1}{c}{lognormal $\mu$} & \multicolumn{1}{c}{lognormal $\sigma$} & \multicolumn{1}{c}{A-D} \\ \hline 
\endhead
	 \hline \hline
\endfoot
	\hline \hline
\endlastfoot
0.010 &  4.80$\times$10$^{-26}$  &  2.18$\times$10$^{-25}$  &
       2.07$\times$10$^{-24}$  &  -5.653$\times$10$^{+01}$  &
       1.73$\times$10$^{+00}$  &  4.43$\times$10$^{+01}$  \\
0.011 &  8.74$\times$10$^{-25}$  &  5.25$\times$10$^{-24}$  &
       4.09$\times$10$^{-23}$  &  -5.352$\times$10$^{+01}$  &
       1.73$\times$10$^{+00}$  &  3.05$\times$10$^{+01}$  \\
0.012 &  1.22$\times$10$^{-23}$  &  9.69$\times$10$^{-23}$  &
       5.18$\times$10$^{-22}$  &  -5.081$\times$10$^{+01}$  &
       1.69$\times$10$^{+00}$  &  3.07$\times$10$^{+01}$  \\
0.013 &  1.69$\times$10$^{-22}$  &  1.21$\times$10$^{-21}$  &
       4.63$\times$10$^{-21}$  &  -4.840$\times$10$^{+01}$  &
       1.56$\times$10$^{+00}$  &  3.98$\times$10$^{+01}$  \\
0.014 &  1.69$\times$10$^{-21}$  &  1.01$\times$10$^{-20}$  &
       3.18$\times$10$^{-20}$  &  -4.629$\times$10$^{+01}$  &
       1.42$\times$10$^{+00}$  &  4.68$\times$10$^{+01}$  \\
0.015 &  1.38$\times$10$^{-20}$  &  6.52$\times$10$^{-20}$  &
       1.85$\times$10$^{-19}$  &  -4.440$\times$10$^{+01}$  &
       1.33$\times$10$^{+00}$  &  5.62$\times$10$^{+01}$  \\
0.016 &  9.29$\times$10$^{-20}$  &  3.42$\times$10$^{-19}$  &
       8.23$\times$10$^{-19}$  &  -4.270$\times$10$^{+01}$  &
       1.15$\times$10$^{+00}$  &  4.91$\times$10$^{+01}$  \\
0.018 &  1.98$\times$10$^{-18}$  &  5.59$\times$10$^{-18}$  &
       1.20$\times$10$^{-17}$  &  -3.986$\times$10$^{+01}$  &
       1.01$\times$10$^{+00}$  &  4.77$\times$10$^{+01}$  \\
\end{longtable}
Comments: The same data as for the $^{18}$F(p,$\alpha$)$^{15}$O reaction are used as input. The radiative widths are deduced from analog levels \cite{NE07} when known; otherwise, we calculate $\Gamma_\gamma$ by using $\mu = -0.44$ and $\sigma = 1.2$, which are derived from the statistics of 25 radiative widths in $^{19}$F \cite{Til95} assuming a lognormal distribution. The direct capture contribution, based on measured $^{18}$O+p spectroscopic factors, is adopted from Utku et al. \cite{UT98}, where we assume a factor uncertainty of an order of magnitude.
\begin{figure}[]
\includegraphics[height=8.5cm]{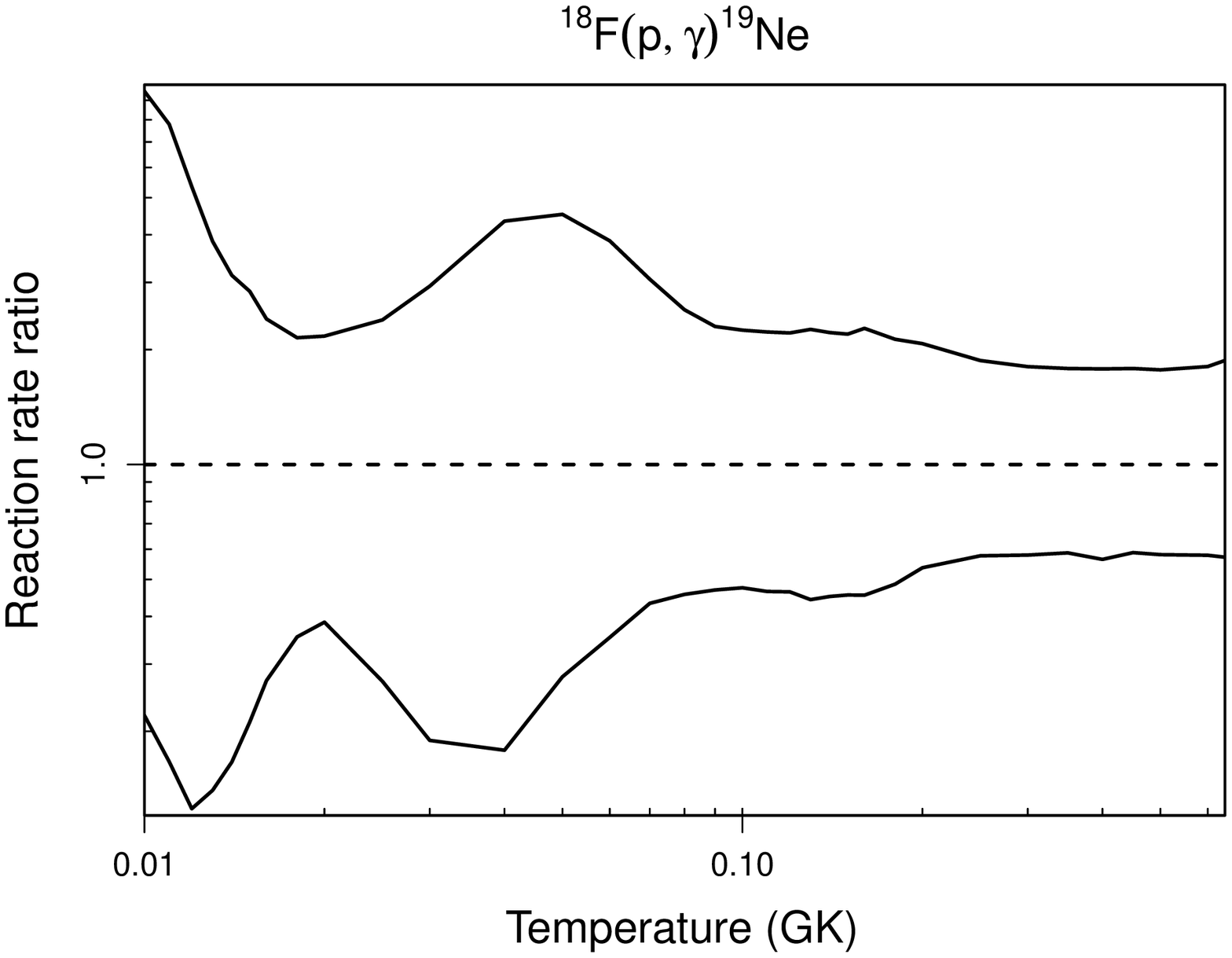}
\end{figure}
\clearpage
\begin{figure}[]
\includegraphics[height=18.5cm]{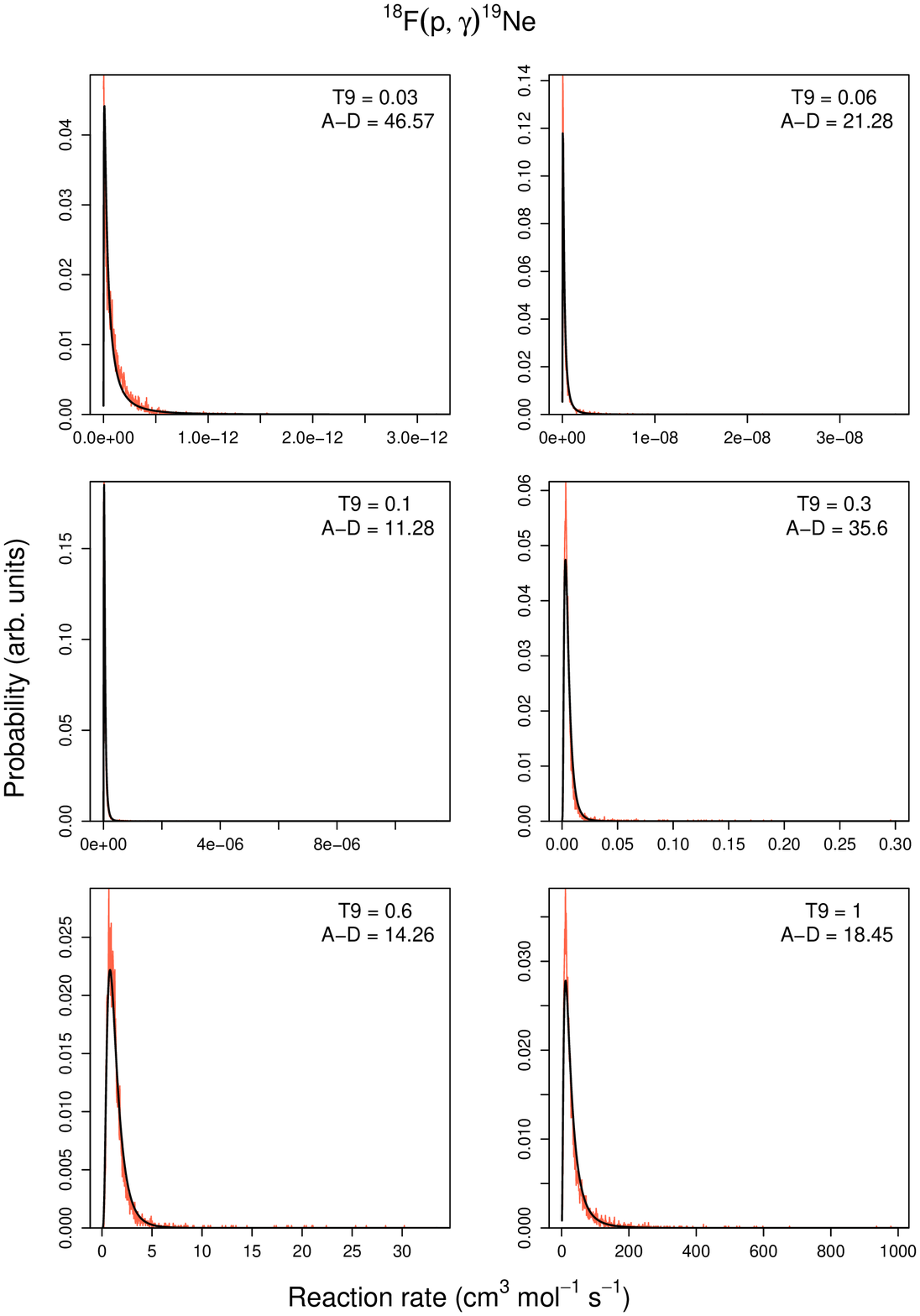}
\end{figure}
\clearpage
\setlongtables
\begin{longtable}{@{\extracolsep{\fill}}cccc | ccc}
\caption{Total thermonuclear reaction rates for $^{18}$F(p,$\alpha$)$^{15}$O.}  \label{} \\
\hline \hline 
	\multicolumn{1}{c}{T (GK)} & \multicolumn{1}{c}{Low rate} & \multicolumn{1}{c}{Median rate} & \multicolumn{1}{c}{High rate}  & \multicolumn{1}{c}{lognormal $\mu$} & \multicolumn{1}{c}{lognormal $\sigma$} & \multicolumn{1}{c}{A-D} \\ \hline 
\endfirsthead
\multicolumn{6}{c}{{\tablename} \thetable{} -- continued} \\
\hline \hline 
	\multicolumn{1}{c}{T (GK)} & \multicolumn{1}{c}{Low rate} & \multicolumn{1}{c}{Median rate} & \multicolumn{1}{c}{High rate} & \multicolumn{1}{c}{lognormal $\mu$} & \multicolumn{1}{c}{lognormal $\sigma$} & \multicolumn{1}{c}{A-D} \\ \hline 
\endhead
	 \hline \hline
\endfoot
	\hline \hline
\endlastfoot
0.010 &  7.75$\times$10$^{-23}$  &  2.77$\times$10$^{-22}$  &
       2.55$\times$10$^{-21}$  &  -4.933$\times$10$^{+01}$  &
       1.57$\times$10$^{+00}$  &  7.45$\times$10$^{+01}$  \\
0.011 &  1.29$\times$10$^{-21}$  &  6.79$\times$10$^{-21}$  &
       5.03$\times$10$^{-20}$  &  -4.635$\times$10$^{+01}$  &
       1.59$\times$10$^{+00}$  &  4.67$\times$10$^{+01}$  \\
0.012 &  1.79$\times$10$^{-20}$  &  1.28$\times$10$^{-19}$  &
       6.21$\times$10$^{-19}$  &  -4.365$\times$10$^{+01}$  &
       1.55$\times$10$^{+00}$  &  5.83$\times$10$^{+01}$  \\
0.013 &  2.51$\times$10$^{-19}$  &  1.59$\times$10$^{-18}$  &
       5.20$\times$10$^{-18}$  &  -4.123$\times$10$^{+01}$  &
       1.41$\times$10$^{+00}$  &  7.38$\times$10$^{+01}$  \\
0.014 &  2.63$\times$10$^{-18}$  &  1.47$\times$10$^{-17}$  &
       3.48$\times$10$^{-17}$  &  -3.909$\times$10$^{+01}$  &
       1.26$\times$10$^{+00}$  &  1.15$\times$10$^{+02}$  \\
0.015 &  2.05$\times$10$^{-17}$  &  9.42$\times$10$^{-17}$  &
       1.88$\times$10$^{-16}$  &  -3.723$\times$10$^{+01}$  &
       1.14$\times$10$^{+00}$  &  1.40$\times$10$^{+02}$  \\
0.016 &  1.50$\times$10$^{-16}$  &  4.99$\times$10$^{-16}$  &
       8.82$\times$10$^{-16}$  &  -3.551$\times$10$^{+01}$  &
       9.76$\times$10$^{-01}$  &  1.57$\times$10$^{+02}$  \\
0.018 &  3.19$\times$10$^{-15}$  &  7.80$\times$10$^{-15}$  &
       1.22$\times$10$^{-14}$  &  -3.269$\times$10$^{+01}$  &
       8.07$\times$10$^{-01}$  &  1.67$\times$10$^{+02}$  \\
\end{longtable}
Comments: Compilations of data or summary tables for $^{18}$F+p can be found in Refs. \cite{UT98,BA04,CH06,NE07}. However, whenever possible, we prefer to consult the original papers. Most resonance energies are calculated from excitation energies as measured by Utku et al. \cite{UT98}, using a proton separation energy of $S_p$ = 6411.2$\pm$0.6 keV \cite{Aud03}. For the lowest-energy resonances, when the analog assignments are known, we use the reanalyzed \cite{BA05} results of Smotrich et al. \cite{SM61} from $^{15}$N($\alpha,\alpha)^{15}$N to estimate the $^{19}$Ne $\alpha$-particle partial widths. Following Ref. \cite{UT98}, the analogs of the 6.419 and 6.449 MeV levels in $^{19}$Ne (corresponding to the 8 and 38 keV resonances) were assumed to be the $^{19}$F states at 6.497 and 6.528 MeV ($J^{\pi}=3/2^+$). They were previously thought \cite{CO00} to possibly dominate the reaction rate at low temperatures; neutron spectroscopic factors were previously extracted from analog transfer reactions \cite{SE05,KO05} and used to calculate proton partial widths. This procedure now seems obsolete with the recent neutron and proton transfer experiment of Adekola et al. \cite{AD09}, who assign $\ell = 1$ to the 8 keV resonance (corresponding to $J^\pi$ = 1/2$^{-}$ or 3/2$^{-}$) and observe a stong population of the subthreshold 6.290 MeV level (E$_{r}^{cm}=-121$ keV, $J^\pi$ = (1/2,3/2)$^+$). With these new experimental results, the analog assignments for the levels corresponding to the E$_{r}^{cm}=-121$, 8 and 38 keV resonances need to be reassessed. Here, we make the following assumptions: (i) the subthreshold 6.290 MeV level has $J^{\pi}=1/2^+$ and is the analog of the 6.255 MeV level in $^{19}$F \cite{AD09}; (ii) the E$_{r}^{cm}=8$ keV resonance has $J^{\pi}=3/2^-$ \cite{AD09}, corresponding to the 6.088 MeV level in $^{19}$F; and (iii) the E$_{r}^{cm}=38$ keV resonance has $J^{\pi}=3/2^+$, as previously assumed \cite{UT98}, and may thus interfere with the E$_r^{cm}=665$ keV resonance \cite{SE08}; the proton widths are adopted from Adekola et al. \cite{AD09}. Needless to say that these assumptions are subject to caution since the spins and parities for these levels are not known unambiguously. Also, there are many missing (unobserved) levels in $^{19}$Ne compared to  $^{19}$F \cite{NE07}, which we prefer to ignore in the absence experimental evidence. For the E$_{r}^{cm}=26$ keV resonance, the total width was measured by Utku et al. \cite{UT98} while an upper limit for the proton width is adopted from Kozub et al. \cite{KO05} (as for the E$_{r}^{cm}=287$ keV resonance). A lower limit of $\Gamma_p/\Gamma>0.007$ was obtained by Visser et al. \cite{VI04} for the E$_{r}^{cm}=450$ keV (7/2$^-$) resonance, yielding $\Gamma_p>8$  eV. Since this value exceeds the Wigner limit, we use the latter result instead. Two important resonances have been directly observed: the 3/2$^+$ resonance at 665 keV, for which we use the precise energy and partial widths measured by Bardayan et al. \cite{BA01}, and the  3/2$^-$ resonance at 330 keV \cite{BA02}. Partial widths (or upper limits) for the next two resonances are extracted from Refs. \cite{BA04,UT98}. Because of conflicting experimental data, the reported resonances above 900~keV are not included in our calculation of the reaction rate. For instance, the spin, parity and proton partial width given by Bardayan et al. \cite{BA04} for the assumed 1009 keV resonance were shown to be incompatible \cite{FO06}, leading to an unrealistic spectroscopic factor. In addition, this resonance was not observed by Murphy et al. \cite{MU90}. The broad 1/2$^+$ resonance, predicted by Dufour and Descouvemont\cite{DU07} at 1.49 MeV above the proton threshold and apparently observed recently \cite{DA09}, was also not detected by Murphy et al. \cite{MU90}. On the other hand, several resonances reported by Ref. \cite{MU90} were not seen in previous work. Accordingly, we match the reaction rate at 0.4 GK with the Hauser-Feshbach result \cite{Gor08} obtained with the TALYS code. The contribution of the E$_{r}^{cm}=8$ keV resonance (with unknown spin) is negligible. On the contrary, the E$_{r}^{cm}=-121$ keV subthreshold resonance could increase the low rate by a factor of 3 near 0.1 GK (if the E$_{r}^{cm}=38$ and 665 keV resonances interfere destructively). Note 
the bimodal rate probability density function near T=0.1 GK, which is caused by the unknown interference sign for these two resonances. 
\begin{figure}[]
\includegraphics[height=8.5cm]{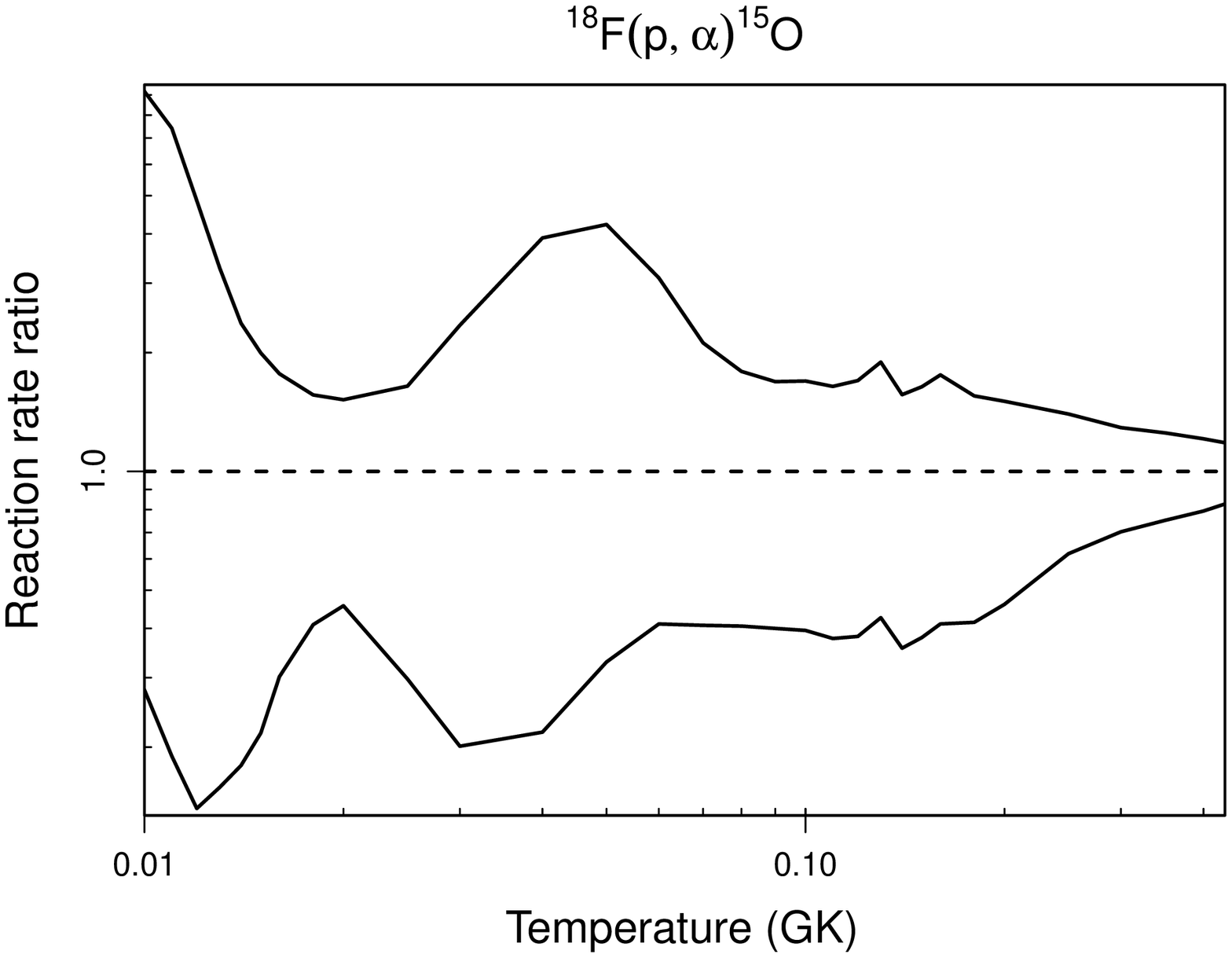}
\end{figure}
\clearpage
\begin{figure}[]
\includegraphics[height=18.5cm]{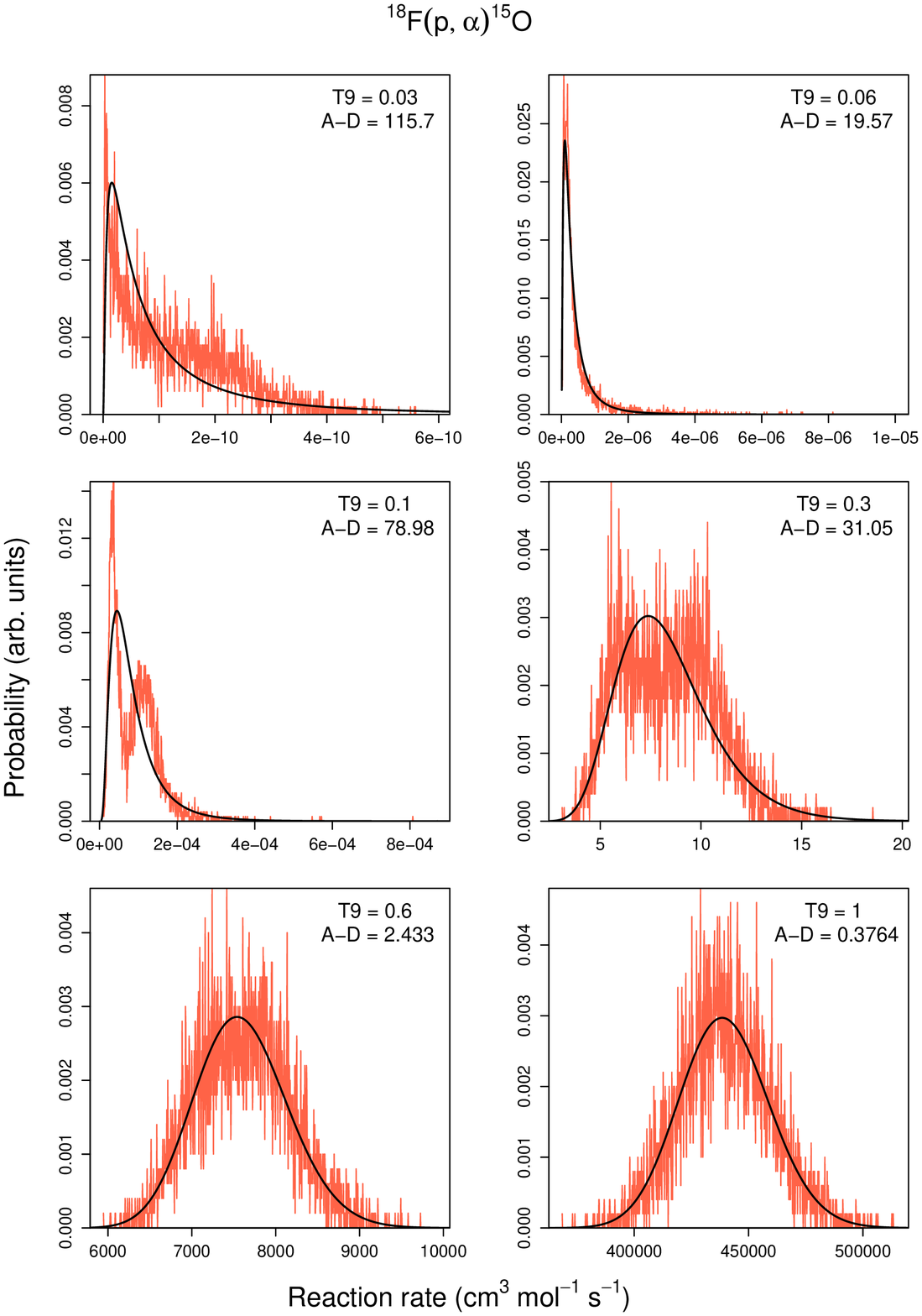}
\end{figure}
\clearpage
\setlongtables
\begin{longtable}{@{\extracolsep{\fill}}cccc | ccc}
\caption{Total thermonuclear reaction rates for $^{19}$Ne(p,$\gamma$)$^{20}$Na.}  \label{} \\
\hline \hline 
	\multicolumn{1}{c}{T (GK)} & \multicolumn{1}{c}{Low rate} & \multicolumn{1}{c}{Median rate} & \multicolumn{1}{c}{High rate}  & \multicolumn{1}{c}{lognormal $\mu$} & \multicolumn{1}{c}{lognormal $\sigma$} & \multicolumn{1}{c}{A-D} \\ \hline 
\endfirsthead
\multicolumn{6}{c}{{\tablename} \thetable{} -- continued} \\
\hline \hline 
	\multicolumn{1}{c}{T (GK)} & \multicolumn{1}{c}{Low rate} & \multicolumn{1}{c}{Median rate} & \multicolumn{1}{c}{High rate} & \multicolumn{1}{c}{lognormal $\mu$} & \multicolumn{1}{c}{lognormal $\sigma$} & \multicolumn{1}{c}{A-D} \\ \hline 
\endhead
	 \hline \hline
\endfoot
	\hline \hline
\endlastfoot
0.010 & 1.71$\times$10$^{-31}$ & 2.46$\times$10$^{-31}$ &
      3.57$\times$10$^{-31}$ &  -7.048$\times$10$^{+01}$ &
       3.65$\times$10$^{-01}$  & 3.03$\times$10$^{-01}$ \\ 
0.011 & 2.69$\times$10$^{-30}$ & 3.86$\times$10$^{-30}$ &
      5.61$\times$10$^{-30}$ &  -6.772$\times$10$^{+01}$ &
       3.67$\times$10$^{-01}$  & 6.62$\times$10$^{-01}$ \\ 
0.012 & 3.11$\times$10$^{-29}$ & 4.49$\times$10$^{-29}$ &
      6.51$\times$10$^{-29}$ &  -6.527$\times$10$^{+01}$ &
       3.67$\times$10$^{-01}$  & 3.19$\times$10$^{-01}$ \\ 
0.013 & 2.79$\times$10$^{-28}$ & 4.00$\times$10$^{-28}$ &
      5.72$\times$10$^{-28}$ &  -6.309$\times$10$^{+01}$ &
       3.69$\times$10$^{-01}$  & 4.09$\times$10$^{-01}$ \\ 
0.014 & 1.95$\times$10$^{-27}$ & 2.83$\times$10$^{-27}$ &
      4.10$\times$10$^{-27}$ &  -6.112$\times$10$^{+01}$ &
       3.68$\times$10$^{-01}$  & 1.32$\times$10$^{+00}$ \\ 
0.015 & 1.19$\times$10$^{-26}$ & 1.70$\times$10$^{-26}$ &
      2.48$\times$10$^{-26}$ &  -5.933$\times$10$^{+01}$ &
       3.67$\times$10$^{-01}$  & 4.21$\times$10$^{-01}$ \\ 
0.016 & 6.03$\times$10$^{-26}$ & 8.71$\times$10$^{-26}$ &
      1.25$\times$10$^{-25}$ &  -5.770$\times$10$^{+01}$ &
       3.70$\times$10$^{-01}$  & 2.13$\times$10$^{-01}$ \\ 
0.018 & 1.09$\times$10$^{-24}$ & 1.56$\times$10$^{-24}$ &
      2.24$\times$10$^{-24}$ &  -5.482$\times$10$^{+01}$ &
       3.65$\times$10$^{-01}$  & 2.45$\times$10$^{-01}$ \\
\end{longtable}
Comments: Because of the limited available spectroscopic information on $^{20}$Na, only four levels are considered for the calculation of the reaction rate. They are located at $E_x$=2645$\pm$6, 2849$\pm$6, 3001$\pm$2 and 3086$\pm$2 keV \cite{Til98}. The higher lying states have well determined spins and parities, have assigned analogs in $^{20}$Ne and $^{20}$F, and their total widths ($\Gamma$=19.8$\pm$2 and 35.9$\pm$2 keV) have been measured \cite{CO94}. The 2645 keV level is assigned a spin and parity of either 1$^+$ or 3$^+$, while 3$^+$ or 3$^-$ are assigned to the 2849 keV level. With the revised $^{20}$Na mass, the proton emission threshold is now at 2193$\pm$7 keV \cite{Aud03} and, consequently, the resonances are located at $E_r^{cm}$ = 452$\pm$9, 656$\pm$9, 808$\pm$7 and 893$\pm$7 keV. Only upper limits have been obtained for the strength of these resonances ($<$15 meV \cite{CO04}, $<$81, $<$102, $<$152 meV, respectively \cite{VA98}). Hence, to calculate the rate, we use shell model results for $\Gamma_\gamma$ \cite{VA98} with an $\pm$50\% estimated uncertainty. The calculated resonance strengths are well within the experimental upper limits except for the first and most important level at 2645 keV. If its spin is 1$^+$, then the calculated strength (6 meV \cite{VA98}) is below the experimental upper limit ($<$15~meV) of Couder et al. \cite{CO04}. However, Fortune et al. \cite{FO00} argue that it is more likely a 3$^+$ state and calculate a lower limit of $\omega\gamma>$16 meV, marginally consistent with the experimental upper limit. Clearly, new experiments are needed to settle this issue and to account for this uncertainty. We adopt a value of $\omega\gamma$=9$\pm$6 meV to cover the interval between the shell model result of 6$\pm$3 meV and experiment. Because of the uncertain parity assignment, we use $\omega\gamma$=8$\pm$7 meV for the second level, based on shell model calculations. The direct capture S-factor is adopted from Vancraeynest et al. \cite{VA98}. (Note that the energy in their Eq. (10) must be in units of MeV although their S-factor is given in units of keVb.) Above $T\approx$1 GK the reaction rate should be extrapolated using the statistical model. Below this temperature, our results are in agreement with Refs. \cite{VA98,CO04}, except that the rate uncertainty below 0.2 GK is higher because of the 40\% uncertainty assigned to the direct capture S-factor, and is lower above 0.2 GK since we adopt shell model results instead of experimental upper limits.
\begin{figure}[]
\includegraphics[height=8.5cm]{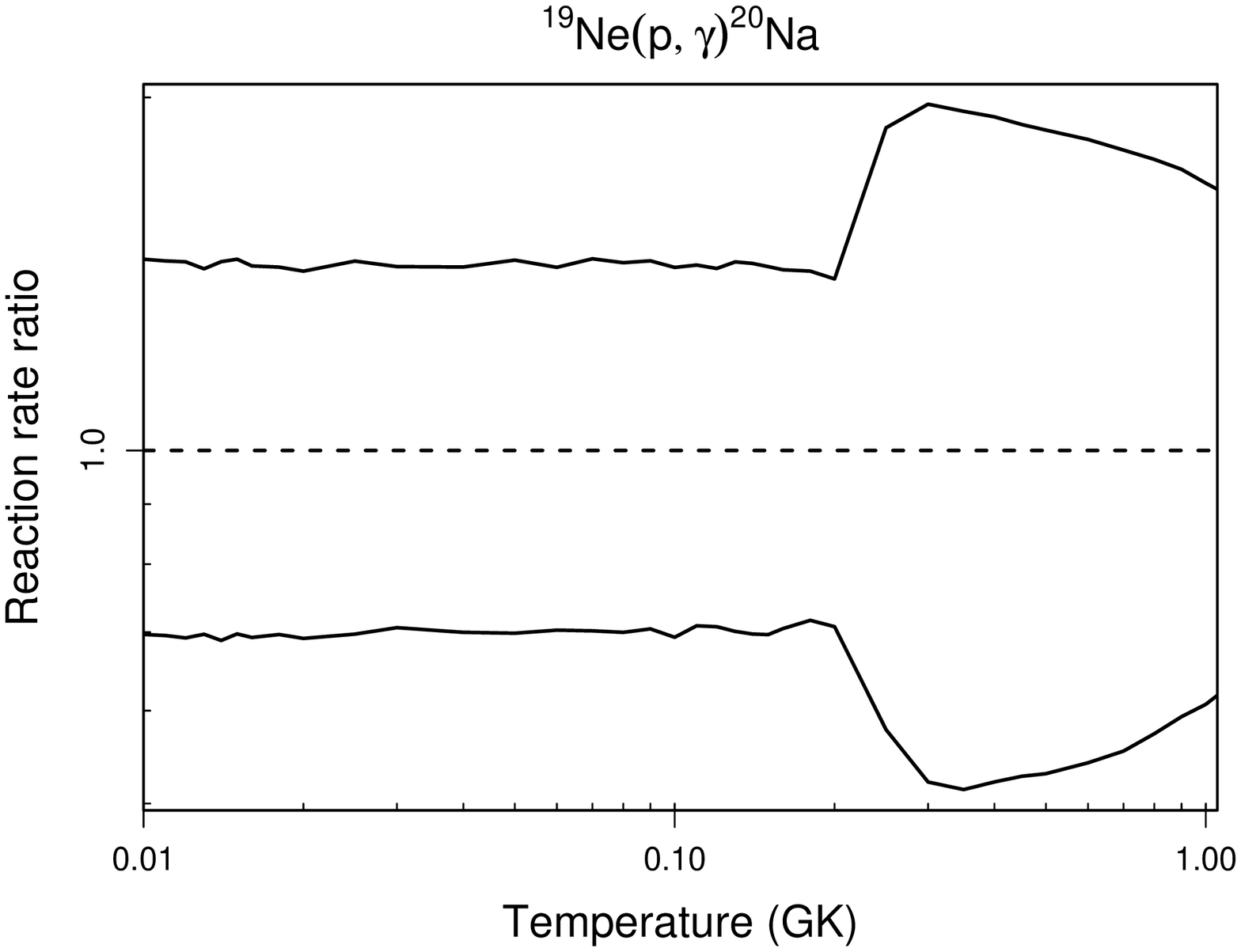}
\end{figure}
\clearpage
\begin{figure}[]
\includegraphics[height=18.5cm]{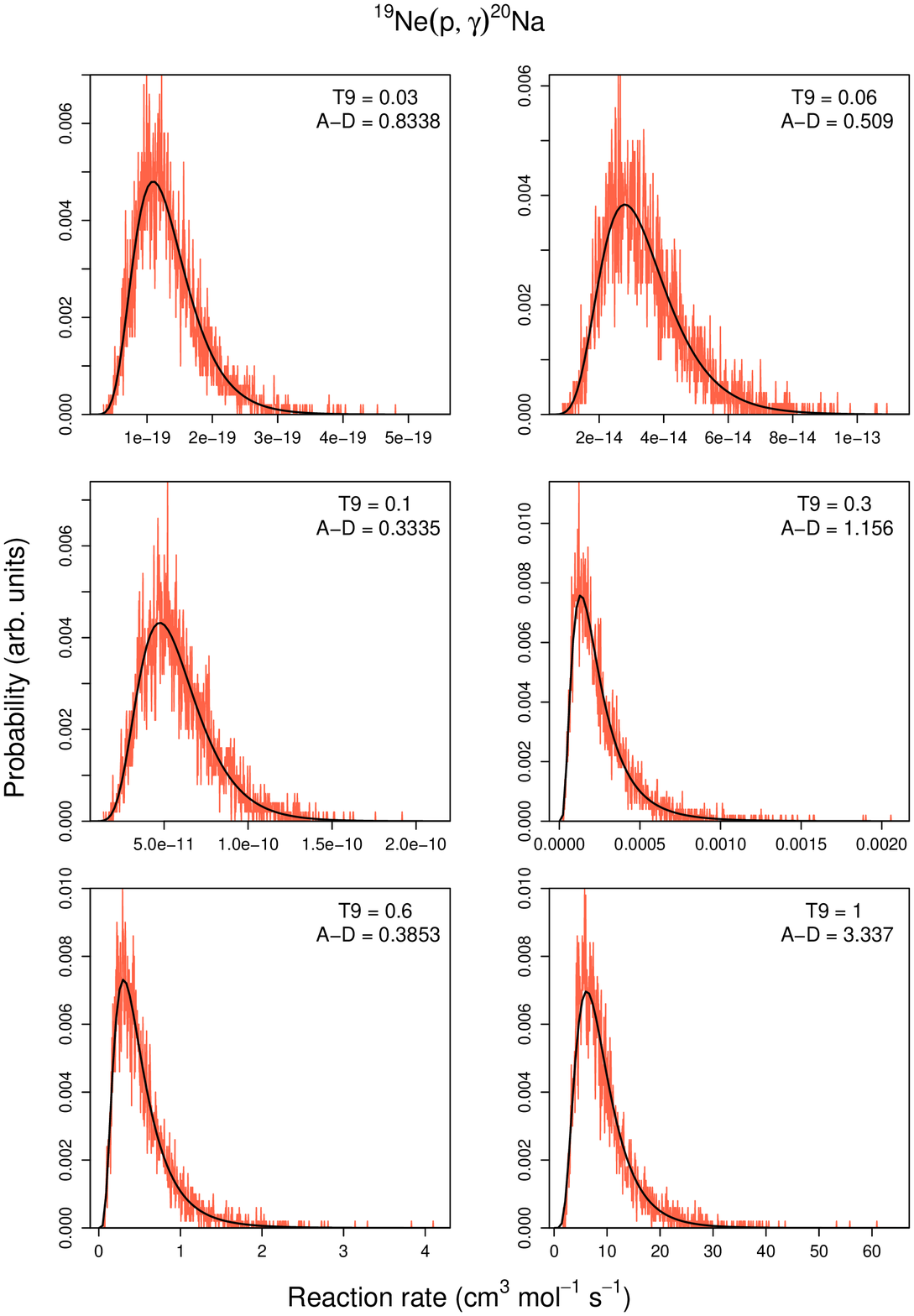}
\end{figure}
\clearpage
\setlongtables
\begin{longtable}{@{\extracolsep{\fill}}cccc | ccc}
\caption{Total thermonuclear reaction rates for $^{20}$Ne(p,$\gamma$)$^{21}$Na.}  \label{} \\
\hline \hline 
	\multicolumn{1}{c}{T (GK)} & \multicolumn{1}{c}{Low rate} & \multicolumn{1}{c}{Median rate} & \multicolumn{1}{c}{High rate}  & \multicolumn{1}{c}{lognormal $\mu$} & \multicolumn{1}{c}{lognormal $\sigma$} & \multicolumn{1}{c}{A-D} \\ \hline 
\endfirsthead
\multicolumn{6}{c}{{\tablename} \thetable{} -- continued} \\
\hline \hline 
	\multicolumn{1}{c}{T (GK)} & \multicolumn{1}{c}{Low rate} & \multicolumn{1}{c}{Median rate} & \multicolumn{1}{c}{High rate} & \multicolumn{1}{c}{lognormal $\mu$} & \multicolumn{1}{c}{lognormal $\sigma$} & \multicolumn{1}{c}{A-D} \\ \hline 
\endhead
	 \hline \hline
\endfoot
	\hline \hline
\endlastfoot
0.010 & 5.55$\times$10$^{-29}$ & 7.61$\times$10$^{-29}$ &
      1.05$\times$10$^{-28}$ &  -6.474$\times$10$^{+01}$ &
       3.22$\times$10$^{-01}$  & 6.75$\times$10$^{-01}$ \\ 
0.011 & 8.02$\times$10$^{-28}$ & 1.09$\times$10$^{-27}$ &
      1.51$\times$10$^{-27}$ &  -6.207$\times$10$^{+01}$ &
       3.19$\times$10$^{-01}$  & 7.06$\times$10$^{-01}$ \\ 
0.012 & 8.54$\times$10$^{-27}$ & 1.17$\times$10$^{-26}$ &
      1.60$\times$10$^{-26}$ &  -5.971$\times$10$^{+01}$ &
       3.16$\times$10$^{-01}$  & 7.04$\times$10$^{-01}$ \\ 
0.013 & 7.04$\times$10$^{-26}$ & 9.57$\times$10$^{-26}$ &
      1.31$\times$10$^{-25}$ &  -5.761$\times$10$^{+01}$ &
       3.12$\times$10$^{-01}$  & 1.05$\times$10$^{+00}$ \\ 
0.014 & 4.72$\times$10$^{-25}$ & 6.41$\times$10$^{-25}$ &
      8.70$\times$10$^{-25}$ &  -5.571$\times$10$^{+01}$ &
       3.09$\times$10$^{-01}$  & 8.38$\times$10$^{-01}$ \\ 
0.015 & 2.65$\times$10$^{-24}$ & 3.59$\times$10$^{-24}$ &
      4.86$\times$10$^{-24}$ &  -5.398$\times$10$^{+01}$ &
       3.07$\times$10$^{-01}$  & 1.04$\times$10$^{+00}$ \\ 
0.016 & 1.28$\times$10$^{-23}$ & 1.73$\times$10$^{-23}$ &
      2.34$\times$10$^{-23}$ &  -5.241$\times$10$^{+01}$ &
       3.05$\times$10$^{-01}$  & 8.12$\times$10$^{-01}$ \\ 
0.018 & 2.08$\times$10$^{-22}$ & 2.80$\times$10$^{-22}$ &
      3.78$\times$10$^{-22}$ &  -4.962$\times$10$^{+01}$ &
       2.99$\times$10$^{-01}$  & 9.38$\times$10$^{-01}$ \\
\end{longtable}
Comments: This rate has a substantial contribution from a subthreshold resonance, which depends critically on its energy with respect to the $^{20}$Ne+p threshold. We have revised the Q-value for this reaction by including a recent measurement of the $^{21}$Na mass excess \cite{Muk04,muk2} in a weighted average with that listed in Ref. \cite{wapstra}. Our recommended value, $Q=2431.69(14)$ keV, is about 0.5 keV higher than the result of Ref. \cite{wapstra}. Our recommended excitation energy is 2424.9(4) keV, which is a weighted average of  E$_{x}=2423.3(9)$ keV \cite{dubois} and E$_{x}=2425.2(4)$ keV \cite{Rol75}. Consequently, our revised resonance energy is E$_r^{cm}=-6.79(42)$ keV. Rolfs et al. \cite{Rol75} have measured the individual S-factors for resonance and direct capture (DC) into this state. Their absolute cross section scale was derived relative to the known strength of the E$_r^{cm}=1113$ keV resonance as well as from the cross section for the $^{16}$O(p,$\gamma$)$^{17}$F reaction \cite{Rol73}, which itself is normalized to the cross section of Tanner et al. \cite{tann}. The accepted value for the strength of the 1113 keV resonance is $\omega\gamma=1.125(75)$ eV \cite{bloch}. However, there have been two recent measurements of $\omega\gamma=1.17(6)$ eV \cite{nd1} and $\omega\gamma=1.12(2)$ eV \cite{nd2}. We have used a weighted average of these three results: $\omega\gamma=1.125(18)$ eV. For the normalization based on the $^{16}$O(p,$\gamma$)$^{17}$F reaction, we instead use the evaluated cross sections of Iliadis et al. \cite{Ili08}. Taking these results together, we find that the S-factors of Ref. \cite{Rol75} should be increased by 2.9\%, which is within their quoted uncertainties. 

We have re-fit these S-factors using a more realistic bound-state potential than was used in the original fits. This results in a slightly smaller value for the spectroscopic factor of the subthreshold state: $C^{2}S=0.82(20)$ compared to the original value of 0.9(1) \cite{Rol75}. Our uncertainty includes an estimate of the uncertainty associated with the choice of parameters for the bound-state potential. We have also re-fit the S-factor for resonance capture into the tail of this state. Taking $\Gamma_{\gamma}=0.17(5)$ eV (derived from the measured lifetime \cite{antilla}), this fit provides an independent value of $C^{2}S=0.65(19)$. Thus, we adopt an average of $C^{2}S=0.73(14)$. To calculate the contribution of the tail of this state to the total reaction rate, it is also necessary to know the dimensionless reduced width. Here we have used an ``observed" value of $\theta_{sp}^{2}=0.609$ \cite{ili2} and adopt an uncertainty of 15\%. Our value of $C^{2}S=0.73(14)$ is somewhat lower than the result of Terakawa et al. \cite{tera} ($C^{2}S=0.97$), but is consistent with the ``best-fit" value of $C^{2}S=0.65(10)$, derived from a measurement of the asymptotic normalization coefficient (ANC) by Ref. \cite{muk3}. However, this latter value assumes a resonance energy of E$_r^{cm}= -7$ keV and they show that the ANC varies greatly with the assumed resonance energy. Thus, it is not possible to scale this result to our energy of E$_r^{cm}=-6.79$ keV with a reliable estimate for the uncertainty. As a result, we have used our value derived from the data of Rolfs et al. \cite{Rol75} of $C^{2}S=0.73(14)$. It should be noted that we also obtain $C^{2}S=1.3(3)$ for the E$_{x}=332$ keV state, which is in poor agreement with $C^{2}S=0.63$ from Terakawa et al. \cite{tera}. Although the reason for this disagreement is not clear, the fit to the DC data \cite{Rol75} is extremely good and it is S$_{DC}$ that we use for the calculation of the reaction rate. Note that in this case S$_{DC}$ is not well described by a second-order polynomial. However, we are able to fit the data with two second-order polynomials, each of which having a different cutoff energy. The uncertainties assigned to each term are chosen to yield an overall uncertainty of $\approx$22\% over the entire temperature range, which accounts for uncertainties in the fits to the DC data as well as our estimate of the systematic uncertainty associated with the choice of potential parameters for the DC calculations.

There are 8 known resonances in the $^{20}$Ne(p,$\gamma$)$^{21}$Na reaction, extending up to an energy of E$_r^{cm}=2036$ keV. Resonance energies are derived from the excitation energies appearing in ENDSF \cite{endsf}. With the exception of the aforementioned 1113 keV resonance, resonance strengths are obtained from the compilation of Endt and van der Leun \cite{End78}. Since no uncertainty is listed for the strength of the 2036 keV resonance, we have arbitrarily chosen a value of 25\%. In addition to the subthreshold resonance, the contributions of the broad resonances at 1738 keV and 2036 keV were integrated numerically in our calculation of the reaction rate. 
\begin{figure}[]
\includegraphics[height=8.5cm]{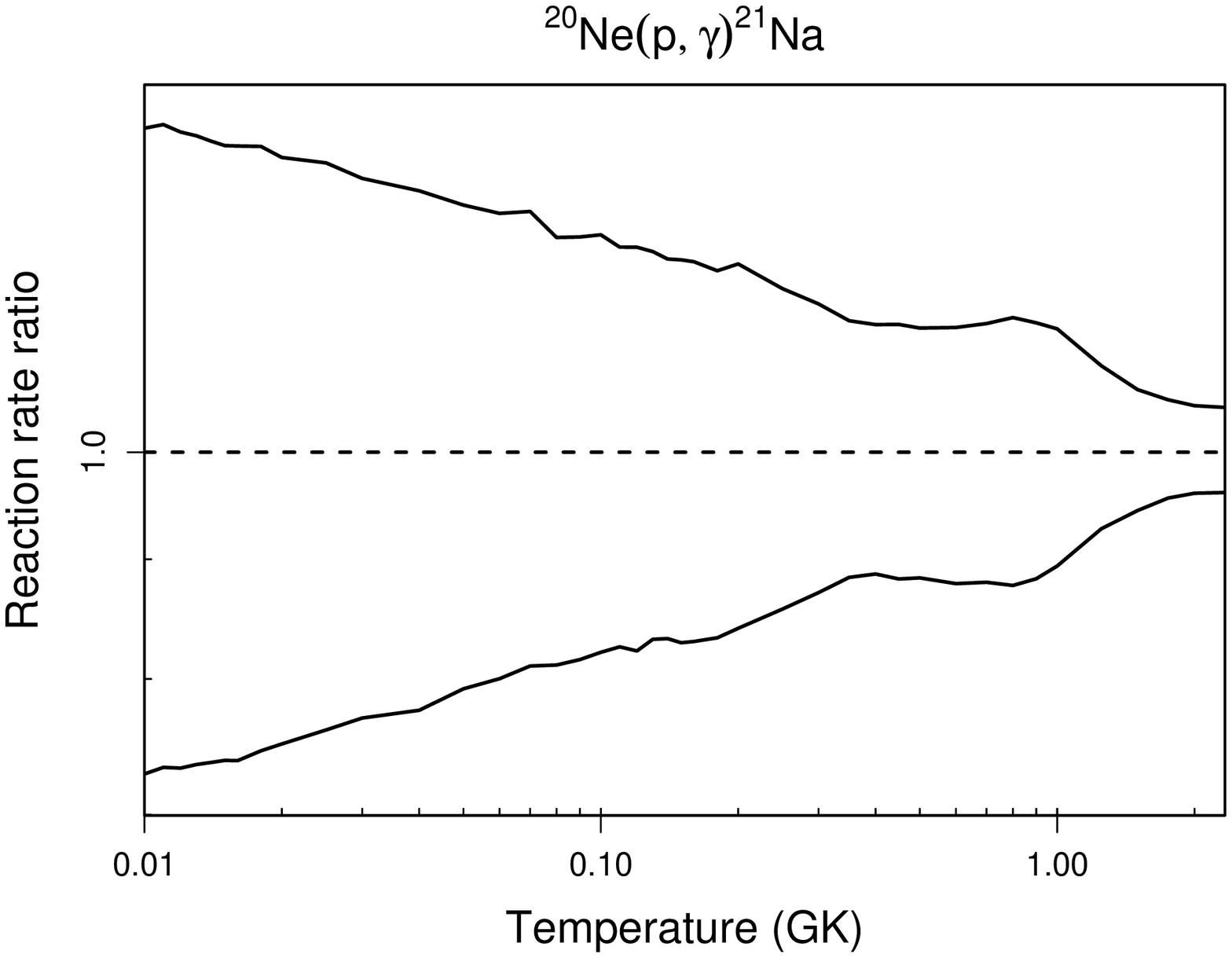}
\end{figure}
\clearpage
\begin{figure}[]
\includegraphics[height=18.5cm]{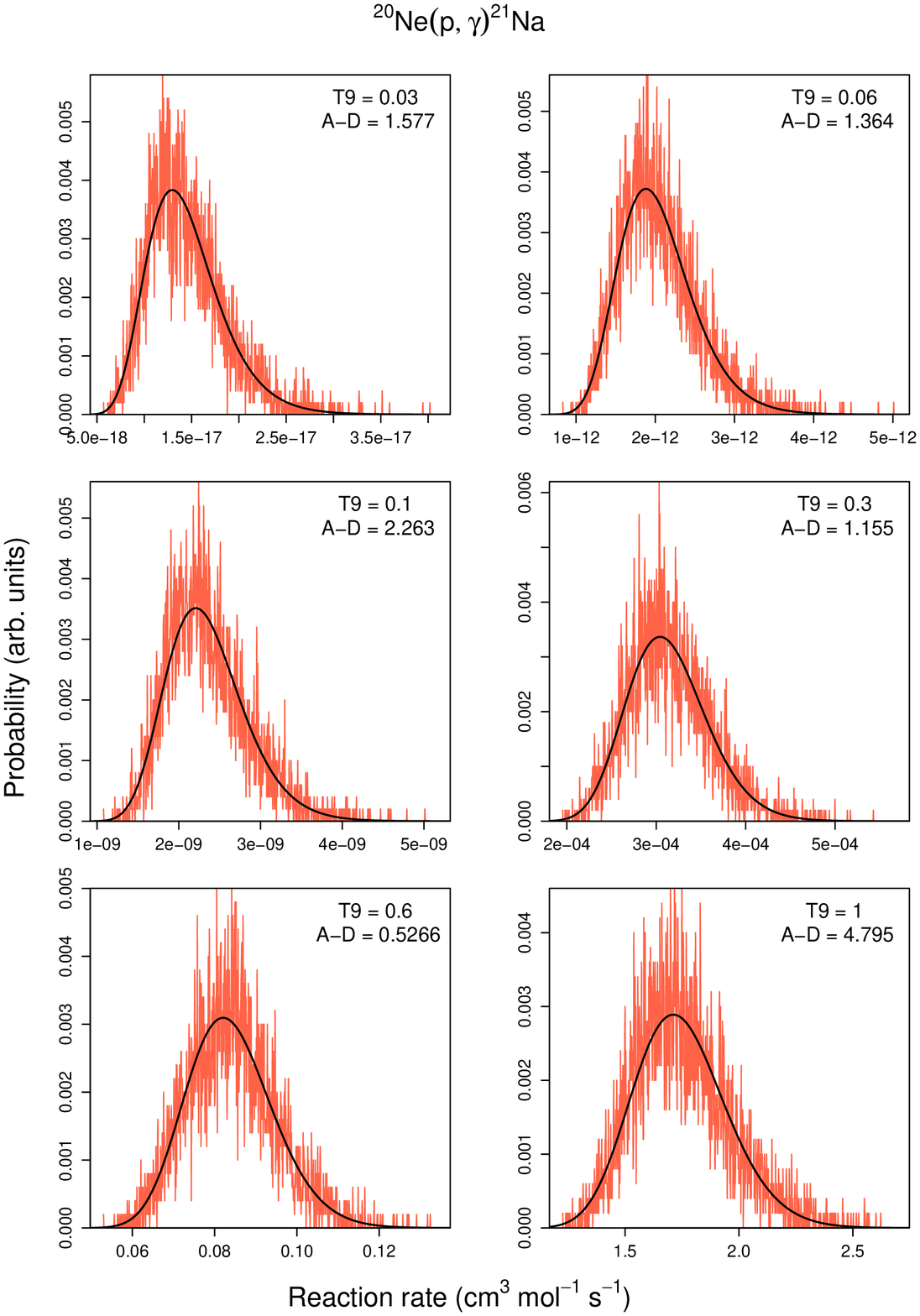}
\end{figure}
\clearpage
\setlongtables
\begin{longtable}{@{\extracolsep{\fill}}cccc | ccc}
\caption{Total thermonuclear reaction rates for $^{20}$Ne($\alpha$,$\gamma$)$^{24}$Mg.}  \label{} \\
\hline \hline 
	\multicolumn{1}{c}{T (GK)} & \multicolumn{1}{c}{Low rate} & \multicolumn{1}{c}{Median rate} & \multicolumn{1}{c}{High rate}  & \multicolumn{1}{c}{lognormal $\mu$} & \multicolumn{1}{c}{lognormal $\sigma$} & \multicolumn{1}{c}{A-D} \\ \hline 
\endfirsthead
\multicolumn{6}{c}{{\tablename} \thetable{} -- continued} \\
\hline \hline 
	\multicolumn{1}{c}{T (GK)} & \multicolumn{1}{c}{Low rate} & \multicolumn{1}{c}{Median rate} & \multicolumn{1}{c}{High rate} & \multicolumn{1}{c}{lognormal $\mu$} & \multicolumn{1}{c}{lognormal $\sigma$} & \multicolumn{1}{c}{A-D} \\ \hline 
\endhead
	 \hline \hline
\endfoot
	\hline \hline
\endlastfoot
0.010 &  1.10$\times$10$^{-81}$  &  1.96$\times$10$^{-81}$  &
       3.53$\times$10$^{-81}$  &  -1.858$\times$10$^{+02}$  &
       5.72$\times$10$^{-01}$  &  1.31$\times$10$^{+00}$  \\
0.011 &  9.14$\times$10$^{-79}$  &  1.59$\times$10$^{-78}$  &
       2.81$\times$10$^{-78}$  &  -1.791$\times$10$^{+02}$  &
       5.67$\times$10$^{-01}$  &  1.15$\times$10$^{+00}$  \\
0.012 &  3.45$\times$10$^{-76}$  &  5.99$\times$10$^{-76}$  &
       1.07$\times$10$^{-75}$  &  -1.732$\times$10$^{+02}$  &
       5.61$\times$10$^{-01}$  &  6.82$\times$10$^{-01}$  \\
0.013 &  6.94$\times$10$^{-74}$  &  1.20$\times$10$^{-73}$  &
       2.15$\times$10$^{-73}$  &  -1.679$\times$10$^{+02}$  &
       5.63$\times$10$^{-01}$  &  1.68$\times$10$^{+00}$  \\
0.014 &  8.54$\times$10$^{-72}$  &  1.46$\times$10$^{-71}$  &
       2.58$\times$10$^{-71}$  &  -1.631$\times$10$^{+02}$  &
       5.55$\times$10$^{-01}$  &  8.93$\times$10$^{-01}$  \\
0.015 &  6.59$\times$10$^{-70}$  &  1.15$\times$10$^{-69}$  &
       1.99$\times$10$^{-69}$  &  -1.587$\times$10$^{+02}$  &
       5.57$\times$10$^{-01}$  &  3.45$\times$10$^{-01}$  \\
0.016 &  3.60$\times$10$^{-68}$  &  6.24$\times$10$^{-68}$  &
       1.08$\times$10$^{-67}$  &  -1.547$\times$10$^{+02}$  &
       5.57$\times$10$^{-01}$  &  9.24$\times$10$^{-01}$  \\
0.018 &  4.18$\times$10$^{-65}$  &  7.16$\times$10$^{-65}$  &
       1.25$\times$10$^{-64}$  &  -1.477$\times$10$^{+02}$  &
       5.51$\times$10$^{-01}$  &  1.04$\times$10$^{+00}$  \\
\end{longtable}
Comments: For temperatures below T$\approx0.15$ GK, the rate of the $^{20}$Ne($\alpha$,$\gamma$)$^{24}$Mg reaction is dominated by direct capture (DC) and by possible, unobserved low-energy resonances at E$_{r}^{cm}= -15.40(8)$, $-11.16(24)$, and $215.93(10)$ keV, which correspond to states in $^{24}$Mg at E$_{x}=9301.15$, 9305.39, and 9532.48 keV, respectively \cite{endsf2}. The direct capture component is estimated by scaling a calculated DC rate using relative spectroscopic factors for $^{20}$Ne($^{6}$Li,d)$^{24}$Mg \cite{raman}. These were converted to absolute spectroscopic factors using the ratio $S_{gs}(^{24}Mg)/S_{gs}(^{20}Ne)=0.7$ \cite{raman} together with the absolute spectroscopic factor for $^{16}$O($^{6}$Li,d)$^{20}$Ne$_{gs}$ \cite{Mao96}. This procedure yields $S_{gs}(^{24}Mg)=0.67$. It should be noted that the spectroscopic factors in Ref. \cite{Mao96} agree well with those obtained from the known $\alpha$-particle widths for the 892 and 1058 keV resonances. In addition, this ground-state spectroscopic factor is consistent with $S_{gs}(^{24}Mg)=0.54$ inferred from (p,p$\alpha$) measurements \cite{cary}. 
The theoretical DC rate was calculated using the same potential parameters as used for the DWBA fits to the stripping data. This procedure assumes that the ($^{6}$Li,d) reaction proceeds solely via transfer of an $\alpha$-particle cluster. While the angular distributions are suggestive of this, to be conservative we have assigned a factor of 2 uncertainty to the DC rate. The three possible resonances were selected based on their T=0 assignments and favorable J$^{\pi}$ values, which allow for $\ell=0$ transfer in the case of the $-11.16$ keV resonance and $\ell=2$ transfer for the other two states. 

Resonances have been measured by Smulders \cite{smu}, Highland and Thwaites \cite{high}, Fifield et al. \cite{fif} and Schmalbrock et al. \cite{sch}. Resonance energies and total widths have been updated using the excitation energies and widths appearing in ENDSF \cite{endsf2}. The resonance strengths and partial widths that we have adopted are obtained from a weighted average of the published resonance strengths with the following corrections: the strengths of Smulders \cite{smu} must be multiplied by a factor of 0.83 to convert laboratory stopping powers to center-of-mass values, as pointed out by Schmalbrock et al. \cite{sch}. This also affects the results of Highland and Thwaites \cite{high}, who report strengths measured relative to that of Smulders \cite{smu} for the 2548 keV resonance. The original value of Ref. \cite{smu}, $\omega\gamma=2.4(7)$ eV, becomes $\omega\gamma=2.0(6)$ eV after the center-of-mass conversion. The weighted average of this value with the more precise result from Schmalbrock et al. \cite{sch} is $\omega\gamma=1.17(19)$ eV. Thus, we have scaled the strengths of Highland and Thwaites \cite{high} to this value. Finally, the strengths of Fifield et al. \cite{fif} were recalculated using (i) updated stopping powers from SRIM-2008 \cite{zie}, and (ii) our adopted strength for the standard resonance at E$_{r}^{cm}=5543$ keV in $^{16}$O($\alpha$,$\gamma$)$^{20}$Ne (see Paper III). To calculate partial widths, we have also made use of resonance strengths for the $^{23}$Na(p,$\alpha$)$^{20}$Ne and $^{23}$Na(p,$\gamma$)$^{24}$Mg reactions listed in Paper III as well as strengths for states populated by the $^{23}$Na(p,p$_{1}$) and $^{20}$Ne($\alpha$,$\alpha_{1}$) reactions \cite{End78}. 

Our recommended rates differ markedly from those in the NACRE compilation \cite{Ang99} for T$<$0.16 GK and T$>$ 2 GK (see Paper IV). We suspect that the difference at low temperatures stems from the fact that we have included the tails of low-lying resonances, which was apparently not done for the NACRE rate. In fact, the latter rate is similar to our classical rate. NACRE considered resonances up to E$_r^{cm}=1414$ keV and matched to a Hauser-Feshbach rate at T=1 GK. However, there are numerous resonances known above this and the level density here may not be high enough to warrant a statistical-model approach. In our case, we have included resonances up to E$_r^{cm}=5008$ keV and have matched to Hauser-Feshbach results near T=8 GK. Thus, it is likely that the difference between the rates at high temperatures is a result of the choice of matching temperatures. In our view, the matching temperature should be higher than that employed by NACRE. 
\begin{figure}[]
\includegraphics[height=8.5cm]{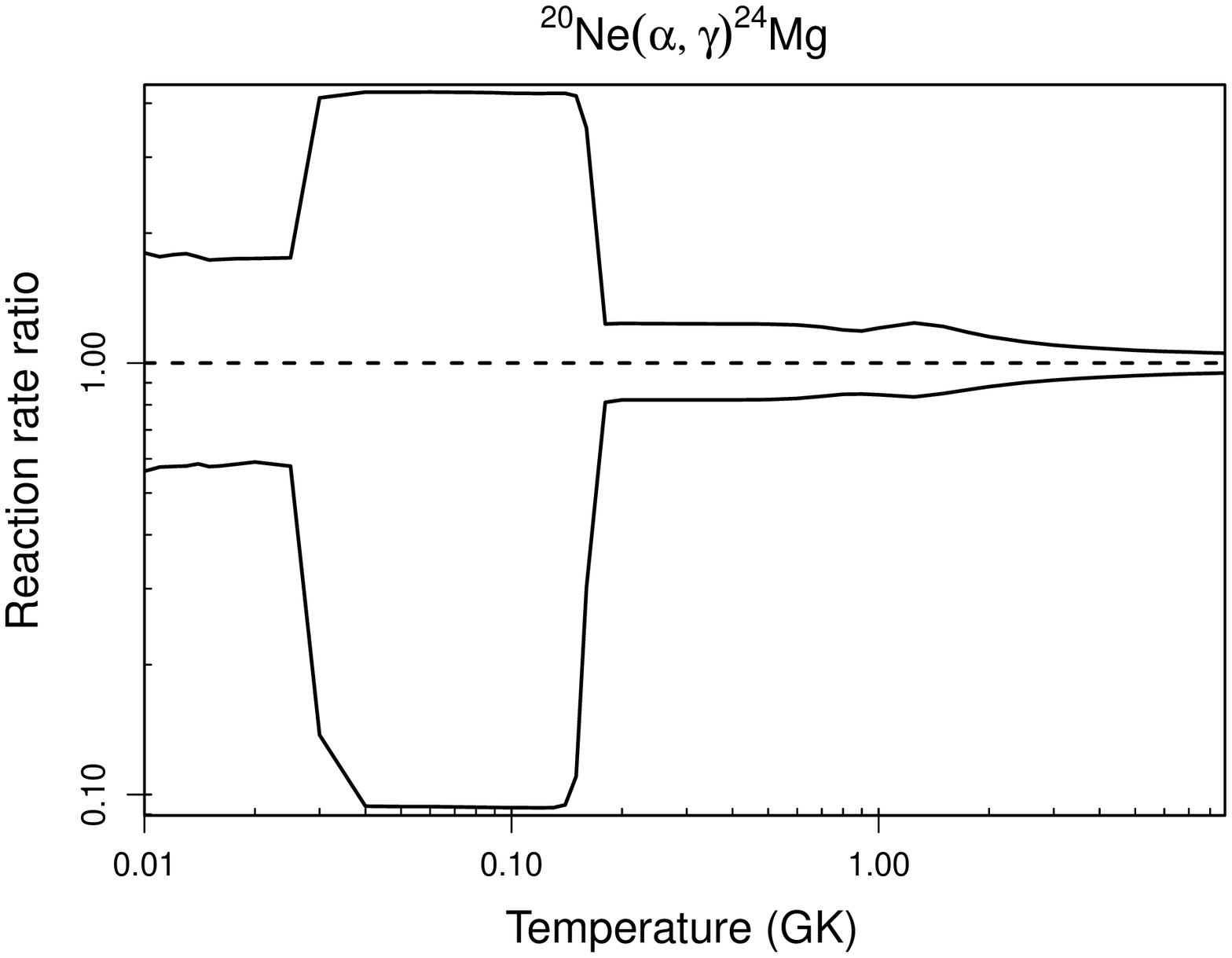}
\end{figure}
\clearpage
\begin{figure}[]
\includegraphics[height=18.5cm]{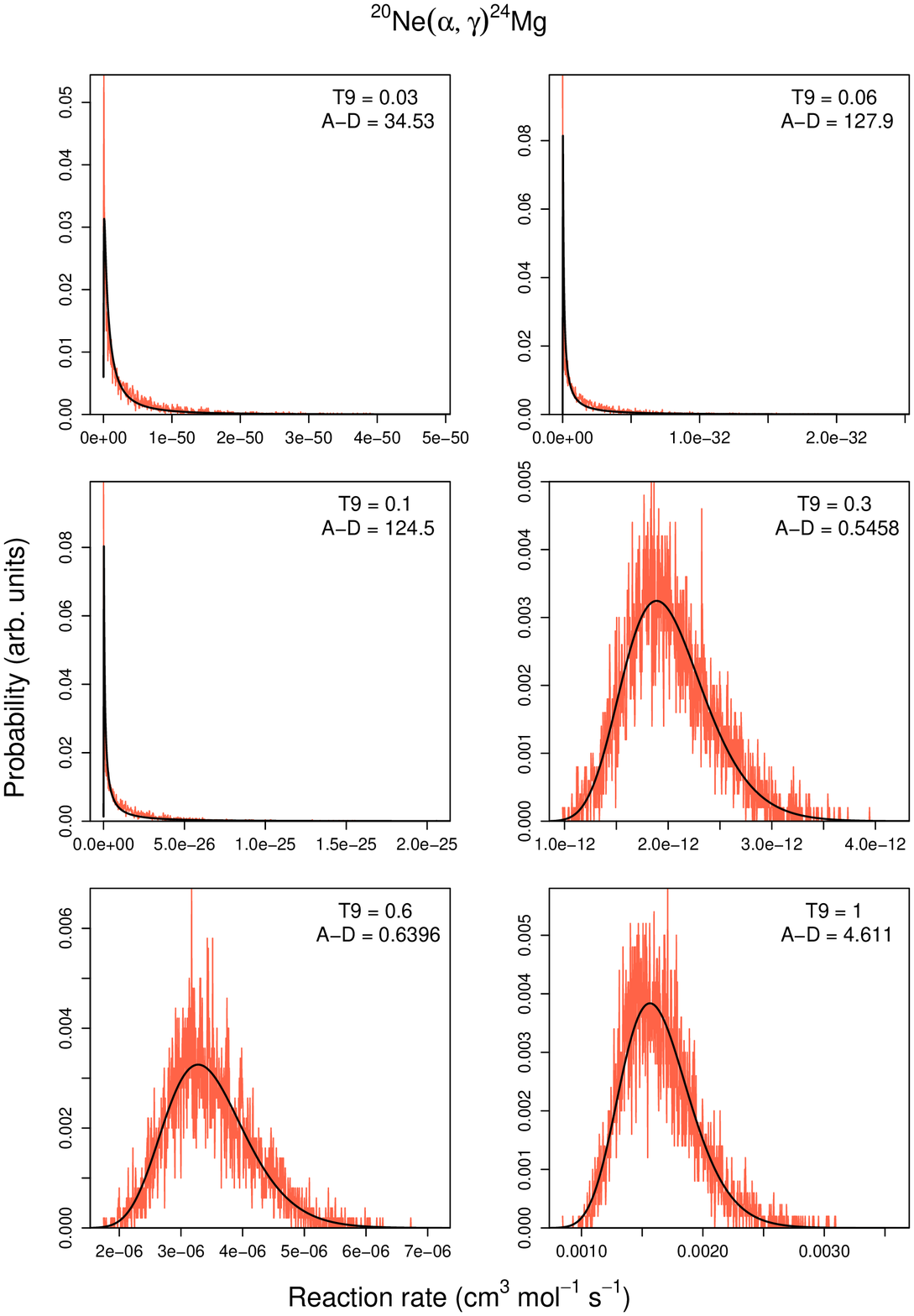}
\end{figure}
\clearpage
\setlongtables
\begin{longtable}{@{\extracolsep{\fill}}cccc | ccc}
\caption{Total thermonuclear reaction rates for $^{21}$Ne(p,$\gamma$)$^{22}$Na.}  \label{} \\
\hline \hline 
	\multicolumn{1}{c}{T (GK)} & \multicolumn{1}{c}{Low rate} & \multicolumn{1}{c}{Median rate} & \multicolumn{1}{c}{High rate}  & \multicolumn{1}{c}{lognormal $\mu$} & \multicolumn{1}{c}{lognormal $\sigma$} & \multicolumn{1}{c}{A-D} \\ \hline 
\endfirsthead
\multicolumn{6}{c}{{\tablename} \thetable{} -- continued} \\
\hline \hline 
	\multicolumn{1}{c}{T (GK)} & \multicolumn{1}{c}{Low rate} & \multicolumn{1}{c}{Median rate} & \multicolumn{1}{c}{High rate} & \multicolumn{1}{c}{lognormal $\mu$} & \multicolumn{1}{c}{lognormal $\sigma$} & \multicolumn{1}{c}{A-D} \\ \hline 
\endhead
	 \hline \hline
\endfoot
	\hline \hline
\endlastfoot
0.010 &  5.38$\times$10$^{-30}$  &  5.85$\times$10$^{-27}$  &
       4.87$\times$10$^{-25}$  &  -6.121$\times$10$^{+01}$  &
       4.75$\times$10$^{+00}$  &  2.75$\times$10$^{+02}$  \\
0.011 &  7.57$\times$10$^{-29}$  &  3.23$\times$10$^{-26}$  &
       4.47$\times$10$^{-24}$  &  -5.909$\times$10$^{+01}$  &
       4.66$\times$10$^{+00}$  &  2.79$\times$10$^{+02}$  \\
0.012 &  8.00$\times$10$^{-28}$  &  1.26$\times$10$^{-25}$  &
       2.79$\times$10$^{-23}$  &  -5.726$\times$10$^{+01}$  &
       4.53$\times$10$^{+00}$  &  3.05$\times$10$^{+02}$  \\
0.013 &  6.54$\times$10$^{-27}$  &  3.99$\times$10$^{-25}$  &
       1.31$\times$10$^{-22}$  &  -5.563$\times$10$^{+01}$  &
       4.38$\times$10$^{+00}$  &  3.40$\times$10$^{+02}$  \\
0.014 &  4.42$\times$10$^{-26}$  &  1.09$\times$10$^{-24}$  &
       5.03$\times$10$^{-22}$  &  -5.417$\times$10$^{+01}$  &
       4.20$\times$10$^{+00}$  &  3.92$\times$10$^{+02}$  \\
0.015 &  2.56$\times$10$^{-25}$  &  2.76$\times$10$^{-24}$  &
       1.53$\times$10$^{-21}$  &  -5.284$\times$10$^{+01}$  &
       4.00$\times$10$^{+00}$  &  4.47$\times$10$^{+02}$  \\
0.016 &  1.25$\times$10$^{-24}$  &  6.59$\times$10$^{-24}$  &
       4.21$\times$10$^{-21}$  &  -5.161$\times$10$^{+01}$  &
       3.80$\times$10$^{+00}$  &  5.09$\times$10$^{+02}$  \\
0.018 &  2.16$\times$10$^{-23}$  &  5.77$\times$10$^{-23}$  &
       2.20$\times$10$^{-20}$  &  -4.941$\times$10$^{+01}$  &
       3.40$\times$10$^{+00}$  &  6.42$\times$10$^{+02}$  \\
\end{longtable}
Comments: The same input information as in Iliadis et al. \cite{Ili01} is used for the calculation of the reaction rates. In total, 46 narrow resonances in the range of E$_r^{cm}$=17-1937 keV are taken into account. The direct capture S-factor is adopted from Rolfs et al. \cite{Rol75}.
\begin{figure}[]
\includegraphics[height=8.5cm]{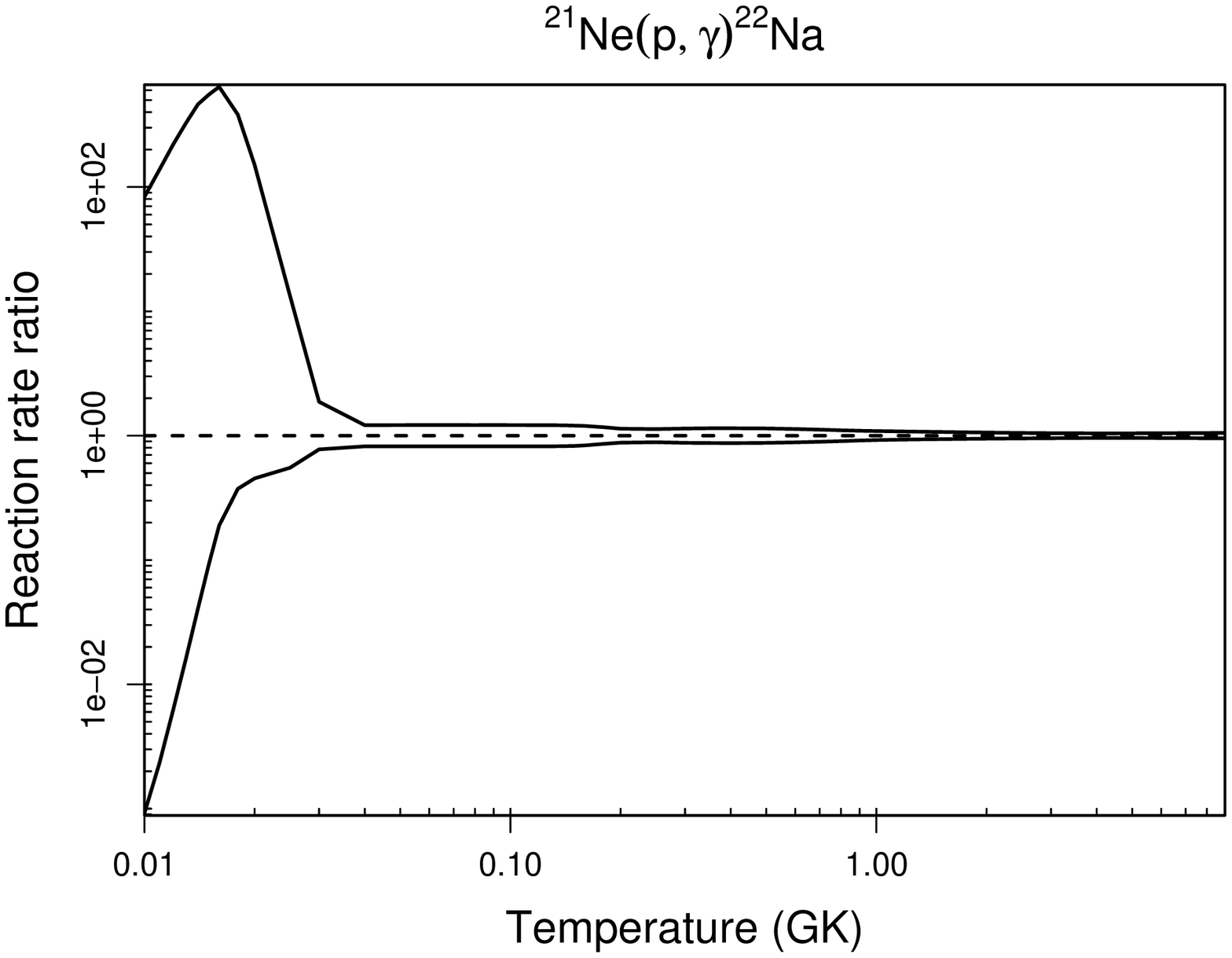}
\end{figure}
\clearpage
\begin{figure}[]
\includegraphics[height=18.5cm]{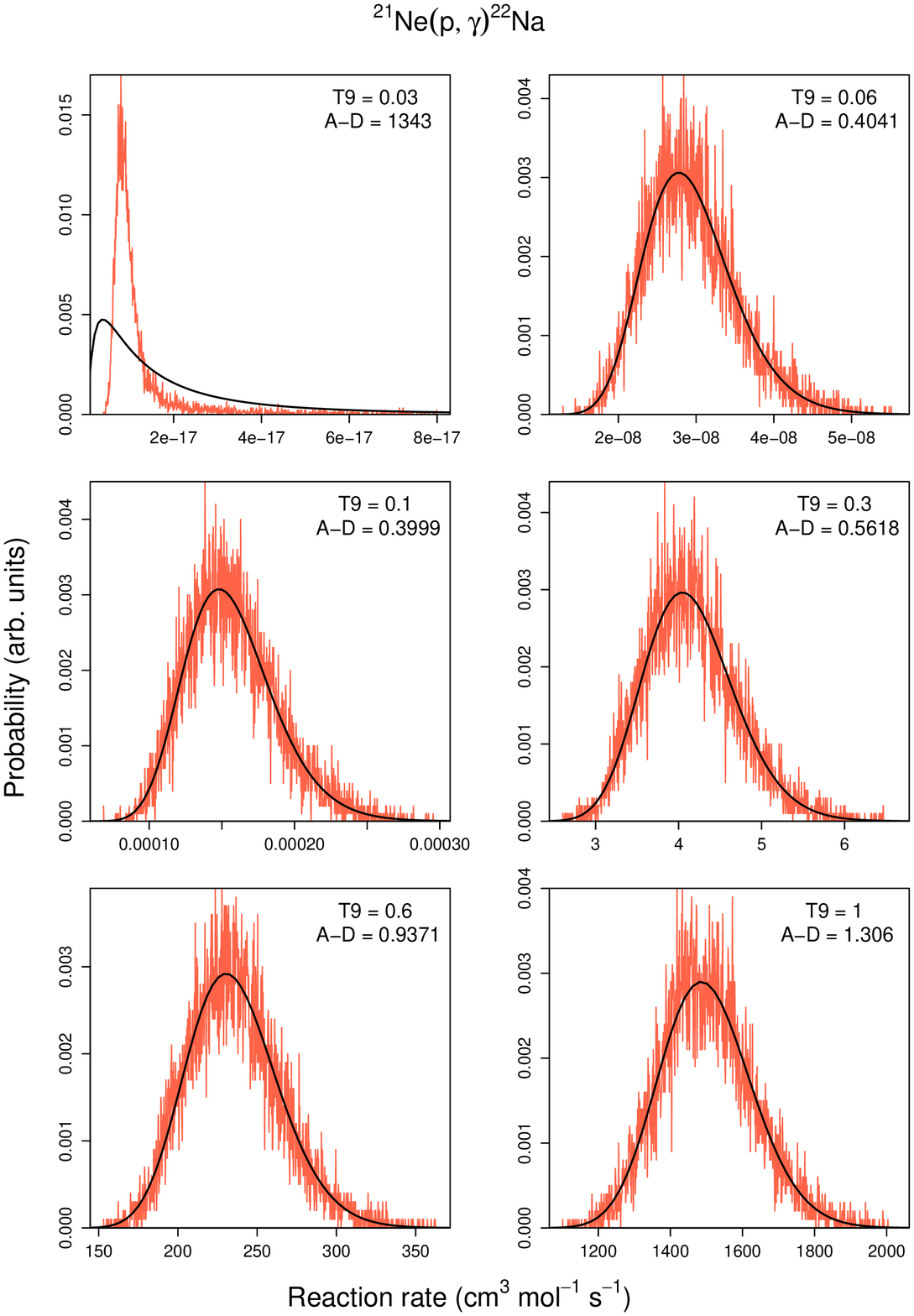}
\end{figure}
\clearpage
\setlongtables
\begin{longtable}{@{\extracolsep{\fill}}cccc | ccc}
\caption{Total thermonuclear reaction rates for $^{22}$Ne(p,$\gamma$)$^{23}$Na.}  \label{} \\
\hline \hline 
	\multicolumn{1}{c}{T (GK)} & \multicolumn{1}{c}{Low rate} & \multicolumn{1}{c}{Median rate} & \multicolumn{1}{c}{High rate}  & \multicolumn{1}{c}{lognormal $\mu$} & \multicolumn{1}{c}{lognormal $\sigma$} & \multicolumn{1}{c}{A-D} \\ \hline 
\endfirsthead
\multicolumn{6}{c}{{\tablename} \thetable{} -- continued} \\
\hline \hline 
	\multicolumn{1}{c}{T (GK)} & \multicolumn{1}{c}{Low rate} & \multicolumn{1}{c}{Median rate} & \multicolumn{1}{c}{High rate} & \multicolumn{1}{c}{lognormal $\mu$} & \multicolumn{1}{c}{lognormal $\sigma$} & \multicolumn{1}{c}{A-D} \\ \hline 
\endhead
	 \hline \hline
\endfoot
	\hline \hline
\endlastfoot
0.010 &  4.05$\times$10$^{-25}$  &  6.67$\times$10$^{-25}$  &
       1.08$\times$10$^{-24}$  &  -5.567$\times$10$^{+01}$  &
       4.80$\times$10$^{-01}$  &  5.22$\times$10$^{-01}$  \\
0.011 &  1.54$\times$10$^{-23}$  &  2.42$\times$10$^{-23}$  &
       3.77$\times$10$^{-23}$  &  -5.208$\times$10$^{+01}$  &
       4.39$\times$10$^{-01}$  &  6.15$\times$10$^{-01}$  \\
0.012 &  3.15$\times$10$^{-22}$  &  4.76$\times$10$^{-22}$  &
       7.28$\times$10$^{-22}$  &  -4.910$\times$10$^{+01}$  &
       4.12$\times$10$^{-01}$  &  7.26$\times$10$^{-01}$  \\
0.013 &  3.97$\times$10$^{-21}$  &  5.85$\times$10$^{-21}$  &
       8.79$\times$10$^{-21}$  &  -4.658$\times$10$^{+01}$  &
       3.95$\times$10$^{-01}$  &  8.32$\times$10$^{-01}$  \\
0.014 &  3.44$\times$10$^{-20}$  &  5.02$\times$10$^{-20}$  &
       7.47$\times$10$^{-20}$  &  -4.444$\times$10$^{+01}$  &
       3.85$\times$10$^{-01}$  &  8.50$\times$10$^{-01}$  \\
0.015 &  2.20$\times$10$^{-19}$  &  3.19$\times$10$^{-19}$  &
       4.76$\times$10$^{-19}$  &  -4.258$\times$10$^{+01}$  &
       3.80$\times$10$^{-01}$  &  8.60$\times$10$^{-01}$  \\
0.016 &  1.11$\times$10$^{-18}$  &  1.60$\times$10$^{-18}$  &
       2.39$\times$10$^{-18}$  &  -4.097$\times$10$^{+01}$  &
       3.79$\times$10$^{-01}$  &  7.64$\times$10$^{-01}$  \\
0.018 &  1.60$\times$10$^{-17}$  &  2.33$\times$10$^{-17}$  &
       3.49$\times$10$^{-17}$  &  -3.829$\times$10$^{+01}$  &
       3.85$\times$10$^{-01}$  &  5.69$\times$10$^{-01}$  \\
\end{longtable}
Comments: In total, 55 resonances with energies of E$_r^{cm}$=28-1822 keV are taken into account. Above E$_r^{cm}=400$ keV, the resonance strengths are adopted from Ref. \cite{End90}, which have been normalized relative to the strength of the E$_r^{cm}=1222$ keV resonance ($\omega\gamma=10.5\pm1.0$ eV \cite{Kei77}). The direct capture S-factor is adopted from Ref. \cite{Gor83}, with an estimated uncertainty of $\pm$40\% \cite{Hal01}. Our treatment of the threshold states differs in two main respects from the analysis of Hale et al. \cite{Hal01}. First, for the E$_r^{cm}=151$ keV resonance, we consider the spectroscopic factor of $C^2S_{\ell=3}=0.0011$ as a mean value rather than an upper limit, in agreement with the original interpretation in Hale's Ph.D. thesis. Second, we entirely disregard the levels at $E_x=8862, 8894$ and 9000 keV that were reported by Powers et al. \cite{Pow71}, who concluded that the existence of these states should be considered as tentative. These levels should have been populated in the much more sensitive study of Ref. \cite{Hal01}, but no evidence for their existence was seen. 
\begin{figure}[]
\includegraphics[height=8.5cm]{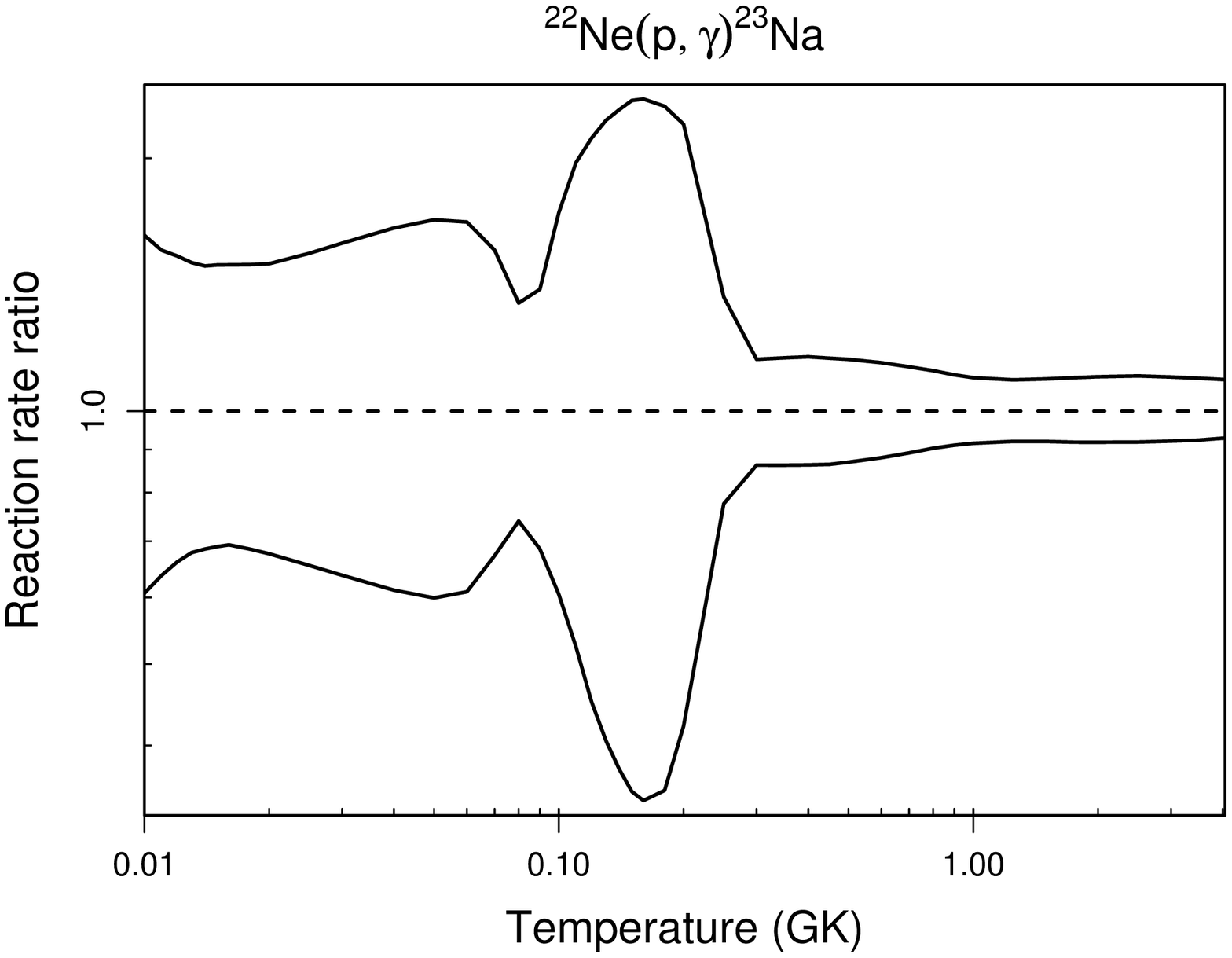}
\end{figure}
\clearpage
\begin{figure}[]
\includegraphics[height=18.5cm]{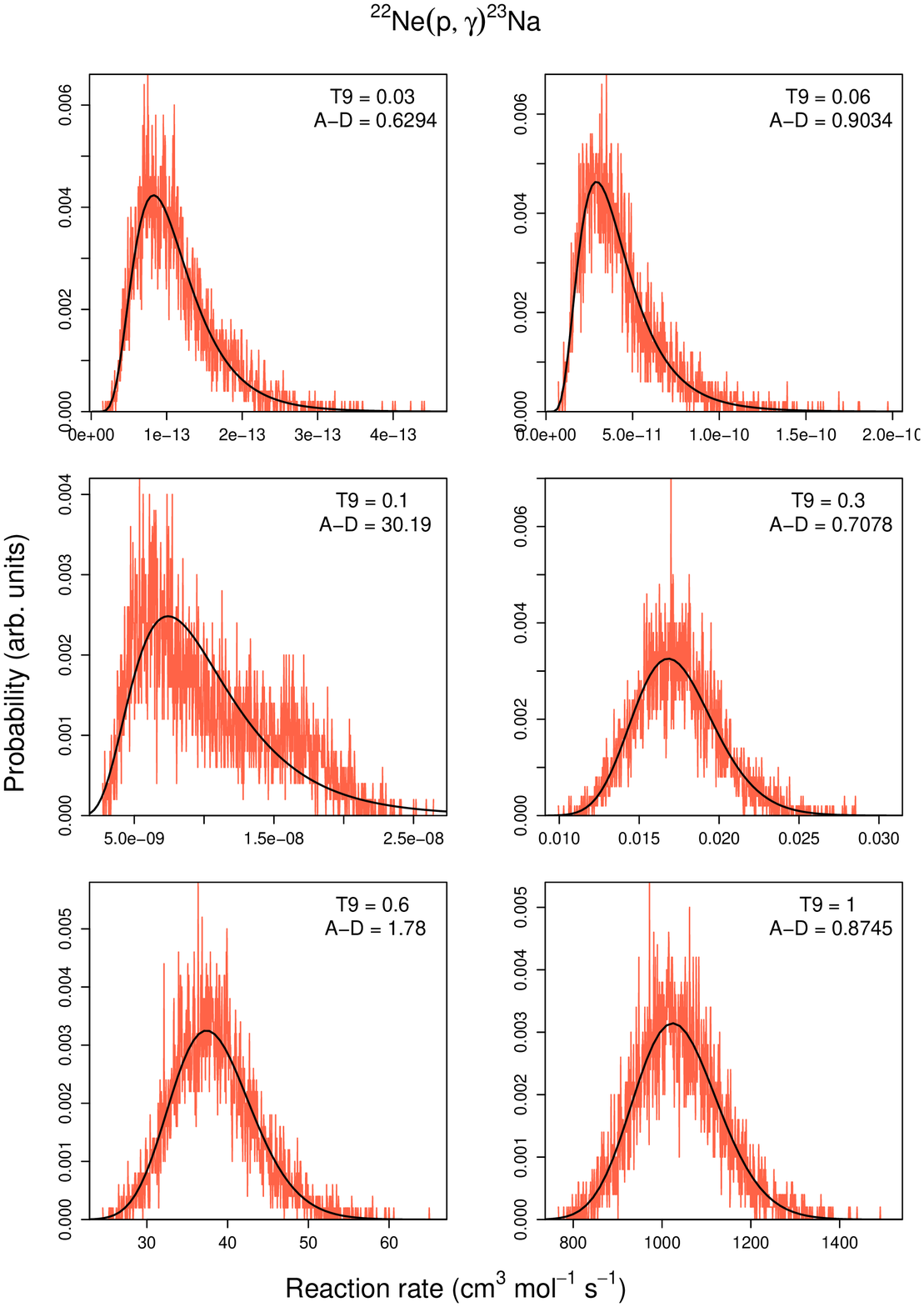}
\end{figure}
\clearpage
\setlongtables
\begin{longtable}{@{\extracolsep{\fill}}cccc | ccc}
\caption{Total thermonuclear reaction rates for $^{22}$Ne($\alpha$,$\gamma$)$^{26}$Mg.}  \label{} \\
\hline \hline 
	\multicolumn{1}{c}{T (GK)} & \multicolumn{1}{c}{Low rate} & \multicolumn{1}{c}{Median rate} & \multicolumn{1}{c}{High rate}  & \multicolumn{1}{c}{lognormal $\mu$} & \multicolumn{1}{c}{lognormal $\sigma$} & \multicolumn{1}{c}{A-D} \\ \hline 
\endfirsthead
\multicolumn{6}{c}{{\tablename} \thetable{} -- continued} \\
\hline \hline 
	\multicolumn{1}{c}{T (GK)} & \multicolumn{1}{c}{Low rate} & \multicolumn{1}{c}{Median rate} & \multicolumn{1}{c}{High rate} & \multicolumn{1}{c}{lognormal $\mu$} & \multicolumn{1}{c}{lognormal $\sigma$} & \multicolumn{1}{c}{A-D} \\ \hline 
\endhead
	 \hline \hline
\endfoot
	\hline \hline
\endlastfoot
0.010 &  3.42$\times$10$^{-78}$  &  8.27$\times$10$^{-78}$  &
       1.98$\times$10$^{-77}$  &  -1.775$\times$10$^{+02}$  &
       8.89$\times$10$^{-01}$  &  5.21$\times$10$^{-01}$  \\
0.011 &  1.26$\times$10$^{-74}$  &  2.79$\times$10$^{-74}$  &
       6.00$\times$10$^{-74}$  &  -1.694$\times$10$^{+02}$  &
       7.93$\times$10$^{-01}$  &  3.23$\times$10$^{-01}$  \\
0.012 &  1.14$\times$10$^{-71}$  &  2.39$\times$10$^{-71}$  &
       4.92$\times$10$^{-71}$  &  -1.626$\times$10$^{+02}$  &
       7.35$\times$10$^{-01}$  &  1.71$\times$10$^{-01}$  \\
0.013 &  3.54$\times$10$^{-69}$  &  7.10$\times$10$^{-69}$  &
       1.43$\times$10$^{-68}$  &  -1.569$\times$10$^{+02}$  &
       7.08$\times$10$^{-01}$  &  1.07$\times$10$^{-01}$  \\
0.014 &  4.70$\times$10$^{-67}$  &  9.45$\times$10$^{-67}$  &
       1.89$\times$10$^{-66}$  &  -1.520$\times$10$^{+02}$  &
       7.02$\times$10$^{-01}$  &  1.62$\times$10$^{-01}$  \\
0.015 &  3.16$\times$10$^{-65}$  &  6.45$\times$10$^{-65}$  &
       1.30$\times$10$^{-64}$  &  -1.478$\times$10$^{+02}$  &
       7.11$\times$10$^{-01}$  &  2.23$\times$10$^{-01}$  \\
0.016 &  1.25$\times$10$^{-63}$  &  2.60$\times$10$^{-63}$  &
       5.30$\times$10$^{-63}$  &  -1.441$\times$10$^{+02}$  &
       7.28$\times$10$^{-01}$  &  1.93$\times$10$^{-01}$  \\
0.018 &  5.49$\times$10$^{-61}$  &  1.21$\times$10$^{-60}$  &
       2.60$\times$10$^{-60}$  &  -1.380$\times$10$^{+02}$  &
       7.75$\times$10$^{-01}$  &  2.37$\times$10$^{-01}$  \\
\end{longtable}
Comments: In total, 40 resonances below E$_r^{cm}=1728$ keV are taken into account. Of these, 10 have been directly measured by Wolke et al. \cite{Wol89}. Another 7 resonances have measured strengths in the competing ($\alpha$,n) reaction \cite{Jae01}. Most of these rate contributions are integrated numerically, with $\Gamma_{\alpha}$ and $\Gamma_n$ computed from results presented in Refs. \cite{Jae01,Dro93} assuming $\Gamma_n\gg\Gamma_{\alpha}$ and an average $\gamma$-ray partial width of $\Gamma_{\gamma}=3$ eV. One more level has been observed below the neutron threshold by Giesen et al. \cite{Gie93} at E$_r^{cm}=78$ keV ( $J^{\pi}=4^+$). In addition, there are upper limit contributions from 22 states, 18 of which have been estimated indirectly. For these levels, upper limits of $\Gamma_{\alpha}$ are obtained from the excitation function shown in Ref. \cite{Wol89}, with $\Gamma_n$ and $\Gamma_{\gamma}$ adopted from Koehler \cite{Koe02}. The remaining 4 upper limit contributions are obtained from Ugalde et al. \cite{Uga07}. Results from a ($\gamma$,$\gamma$) experiment \cite{Lon09} unambiguously determine the $J^{\pi}$ values of the E$_r^{cm}=191$ and 334 keV resonances, and also show that the level corresponding to a previously assumed E$_r^{lab}=630$ keV resonance has in fact unnatural parity. Furthermore, according to Ref. \cite{Lon09}, a doublet exists near 330 keV and, consequently, we apply the spectroscopic factor upper limit from Ref. \cite{Uga07} to both states. 
\begin{figure}[]
\includegraphics[height=8.5cm]{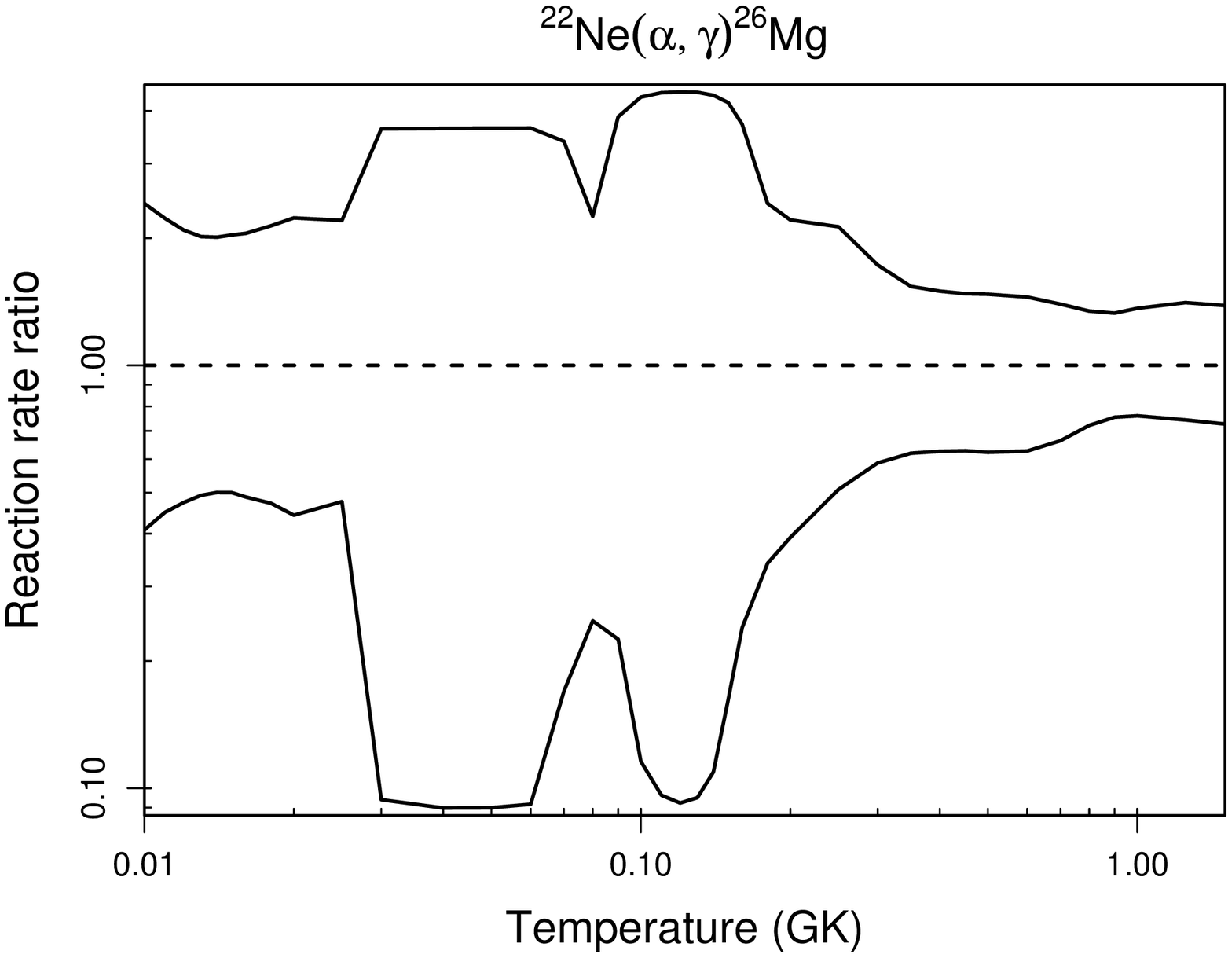}
\end{figure}
\clearpage
\begin{figure}[]
\includegraphics[height=18.5cm]{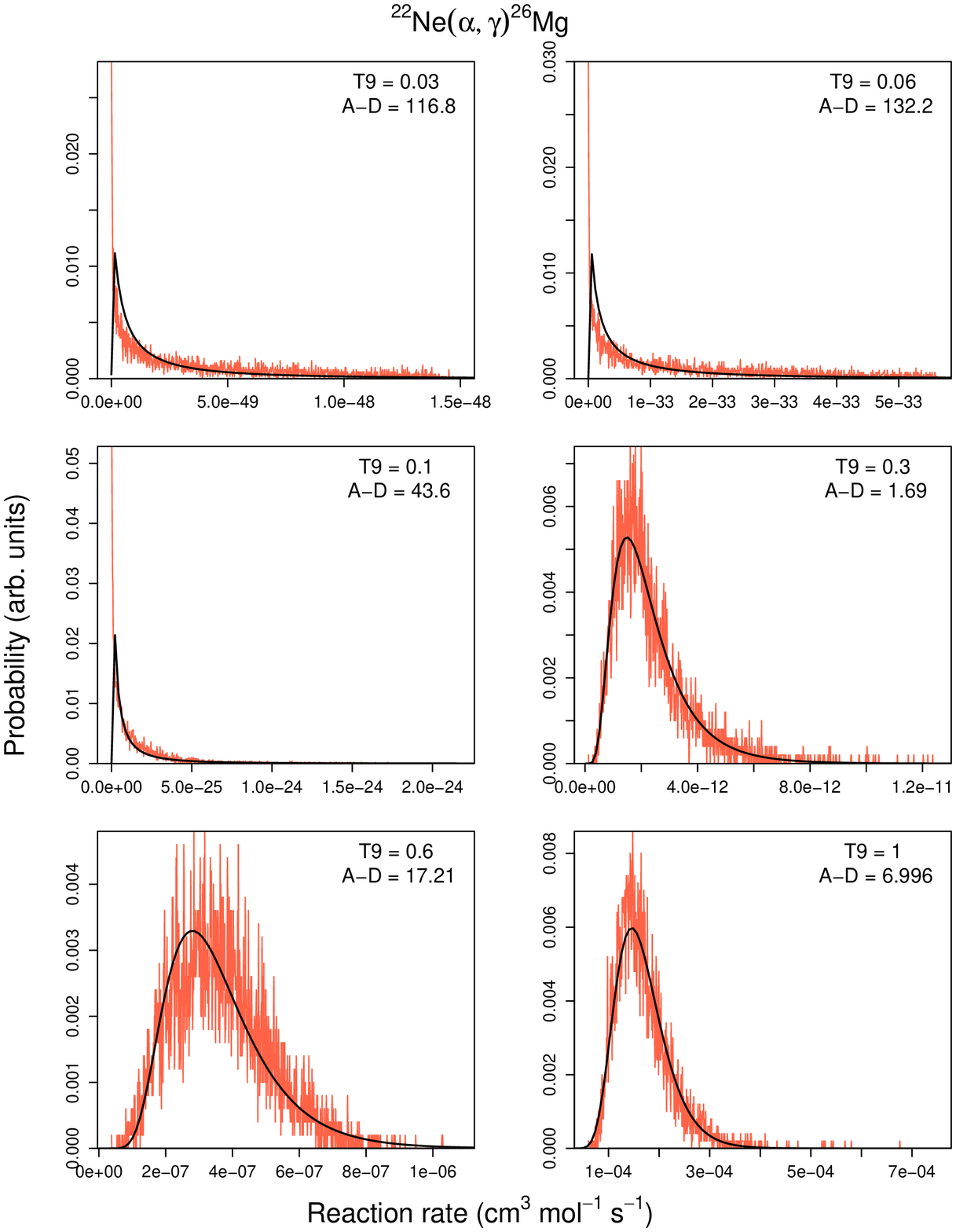}
\end{figure}
\clearpage
\setlongtables

Comments: In total, 41 resonances between the neutron threshold at 479 keV and E$_r^{cm}=1937$ keV are taken into account. Of these, 22 have been directly measured by Refs. \cite{Wol89,Har91,Dro93,Jae01}. Those studies were performed in the same laboratory with continuing improvements on the target and detection systems. Thus we assume that the most recent work \cite{Jae01} supersedes the other three studies. For resonances beyond the energy range covered by Jaeger et al., the results of Ref. \cite{Dro93} are adopted. Neutron and $\gamma$-ray partial widths are obtained from Koehler \cite{Koe02} when available. For all other wide resonances an average value of $\Gamma_{\gamma}=3$ eV is used for the $\gamma$-ray partial width. Of the remaining resonances, only upper limits for $\Gamma_{\alpha}$ are available. The values are adopted from Ref. \cite{Jae01}, except for the E$_r^{cm}=497$ keV resonance, for which the results of the $^{22}$Ne($^{6}$Li,d)$^{26}$Mg measurement by Ugalde et al. \cite{Uga07} are used. Since this study was performed at one angle only, we interpret the reported spectroscopic factor as an upper limit rather than a mean value. Results from a ($\gamma$,$\gamma$) experiment \cite{Lon09} demonstrated that the level corresponding to a previously assumed E$_r^{lab}=630$ keV resonance has in fact unnatural parity.
\begin{figure}[]
\includegraphics[height=8.5cm]{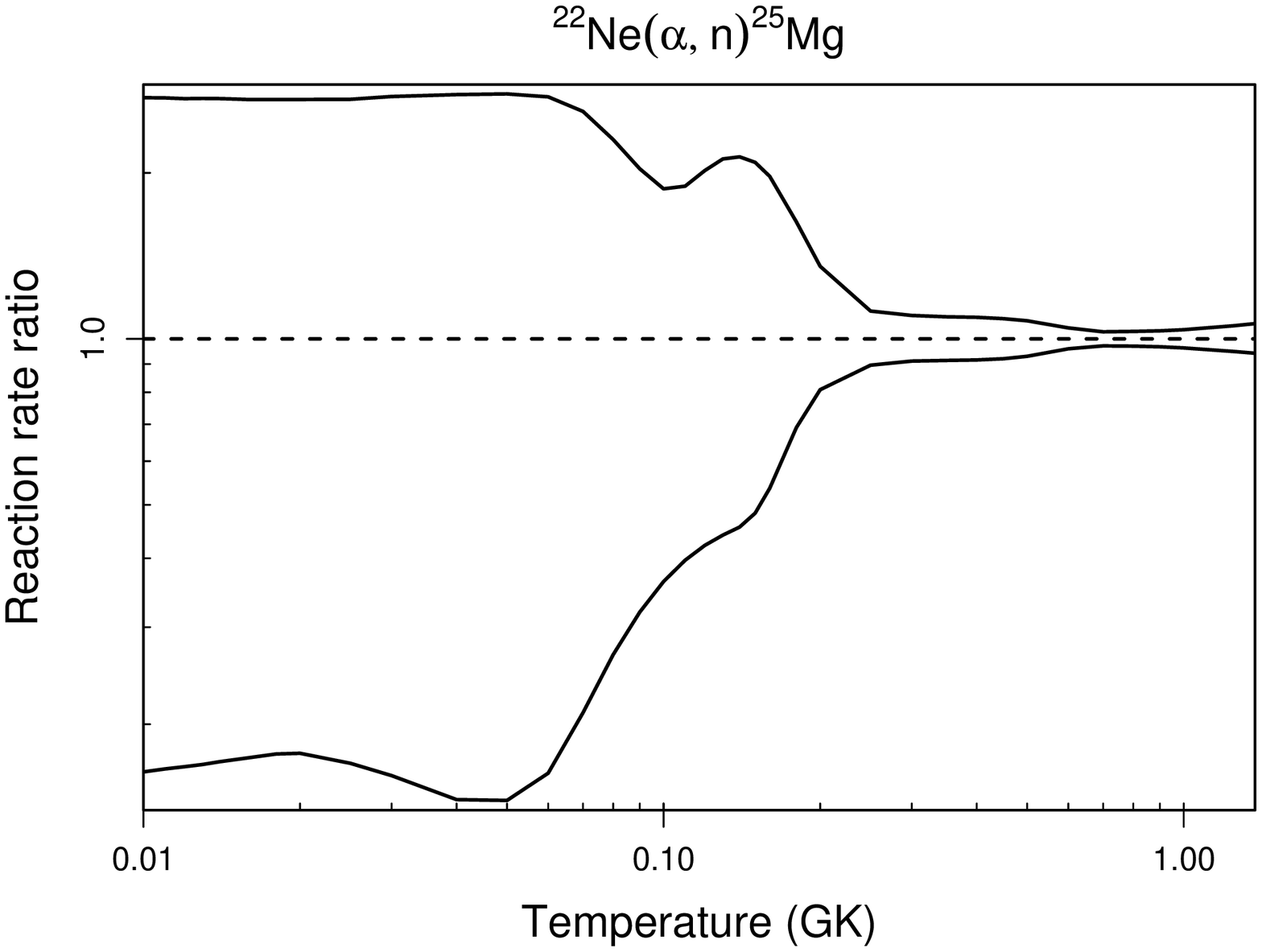}
\end{figure}
\clearpage
\begin{figure}[]
\includegraphics[height=18.5cm]{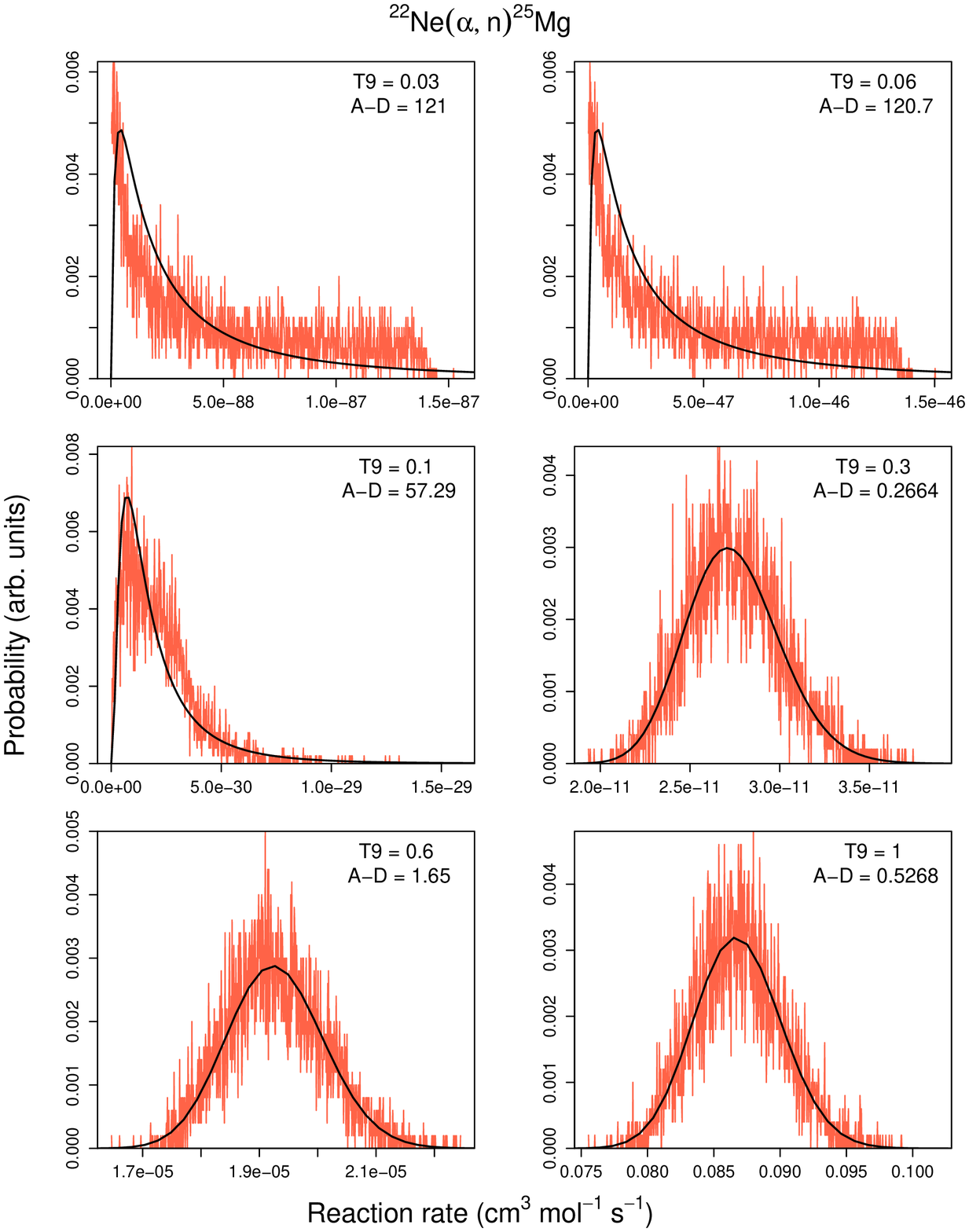}
\end{figure}
\clearpage
\setlongtables

Comments: The value of Q$_{p\gamma}$=5504.18$\pm$0.34 keV is obtained from the mass excesses reported by Mukherjee et al. \cite{Muk04}. The resonance energies are either calculated from excitation energies (Ruiz et al. \cite{Rui05}) or are directly adopted from experiment (D'Auria et al. \cite{DAu04}), depending on which procedure yielded a smaller uncertainty. In total, 6 narrow resonances in the range of E$_r^{cm}$=206-1101 keV are taken into account. The resonance at E$_r^{cm}$=329 keV, corresponding to a $^{22}$Mg level at E$_x$=5837 keV, has been disregarded according to the suggestion of Seweryniak et al. \cite{Sew05}. The direct capture S-factor is adopted from Bateman et al. \cite{Bat01}.
\begin{figure}[]
\includegraphics[height=8.5cm]{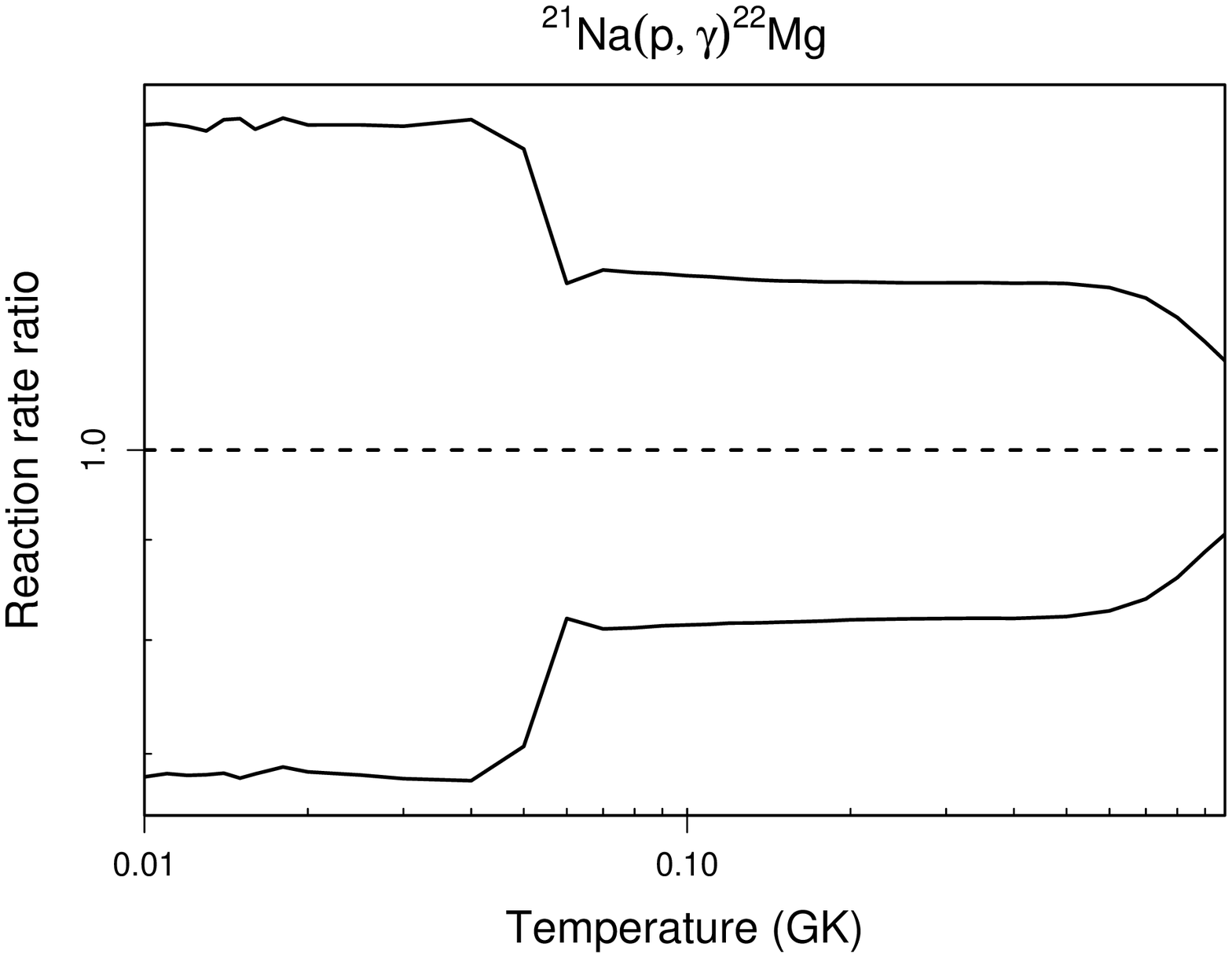}
\end{figure}
\clearpage
\begin{figure}[]
\includegraphics[height=18.5cm]{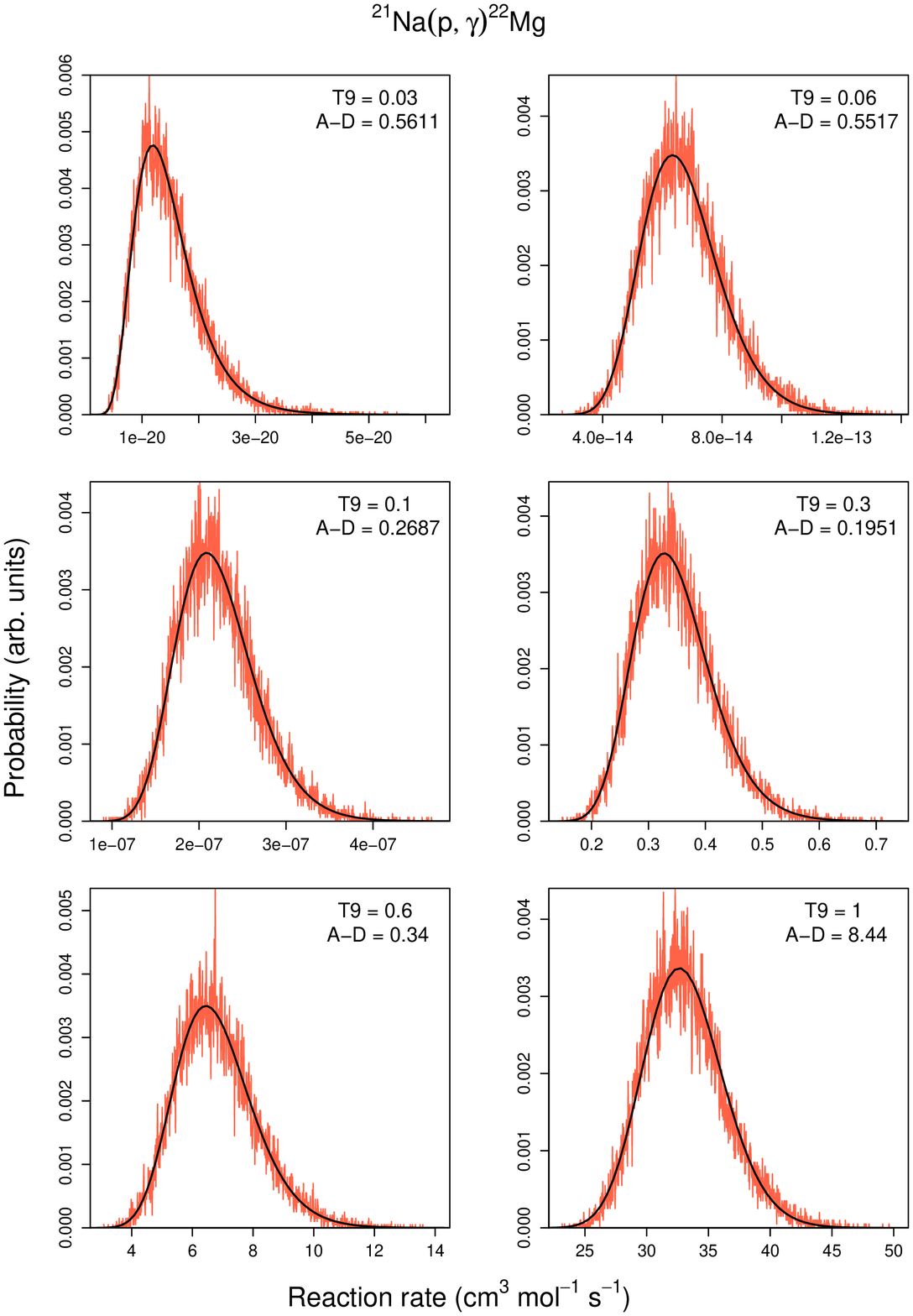}
\end{figure}
\clearpage
\setlongtables

Comments: Measured resonance energies and strengths are adopted from Seuthe et al. \cite{Seu90} and Stegm\"uller et al. \cite{Ste96}. For some of the observed resonances we calculated resonance energies from the excitation energies reported in Jenkins et al. \cite{Jen04}. For the observed E$_r^{cm}$=204 keV resonance, the experimental yield \cite{Ste96} together with the measured branching ratio (100\% \cite{Jen04}) results in a resonance strength of $\omega\gamma$=(1.4$\pm$0.3)$\times$10$^{-3}$ eV (i.e., any unobserved $\gamma$-ray transitions can be excluded). The unobserved E$_r^{cm}$=200 keV resonance can most likely be disregarded, but the unobserved E$_r^{cm}$=189 keV resonance strongly influences the reaction rates. For the latter resonance, we estimate an upper limit of $\omega\gamma$$\leq$3$\times$10$^{-3}$ eV from the results of the ($^3$He,d) work of Schmidt et al. \cite{Sch95}. The measurement of Jenkins et al. \cite{Jen04} unambiguously identified the E$_x$=7623 and 7647 keV levels in $^{23}$Mg as (unobserved) d-wave resonances (i.e., the s-wave contributions assumed in Ref. \cite{Sch95} can be disregarded; see also the arguments based on the shell model in Comisel et al. \cite{Com07}). In total, 13 narrow resonances with energies of E$_r^{cm}$=43--761 keV are taken into account. The unobserved resonance at E$_r^{cm}$=4 keV and the direct capture process (Seuthe et al.  \cite{Seu90}) do not contribute significantly to the total rates.
\begin{figure}[]
\includegraphics[height=8.5cm]{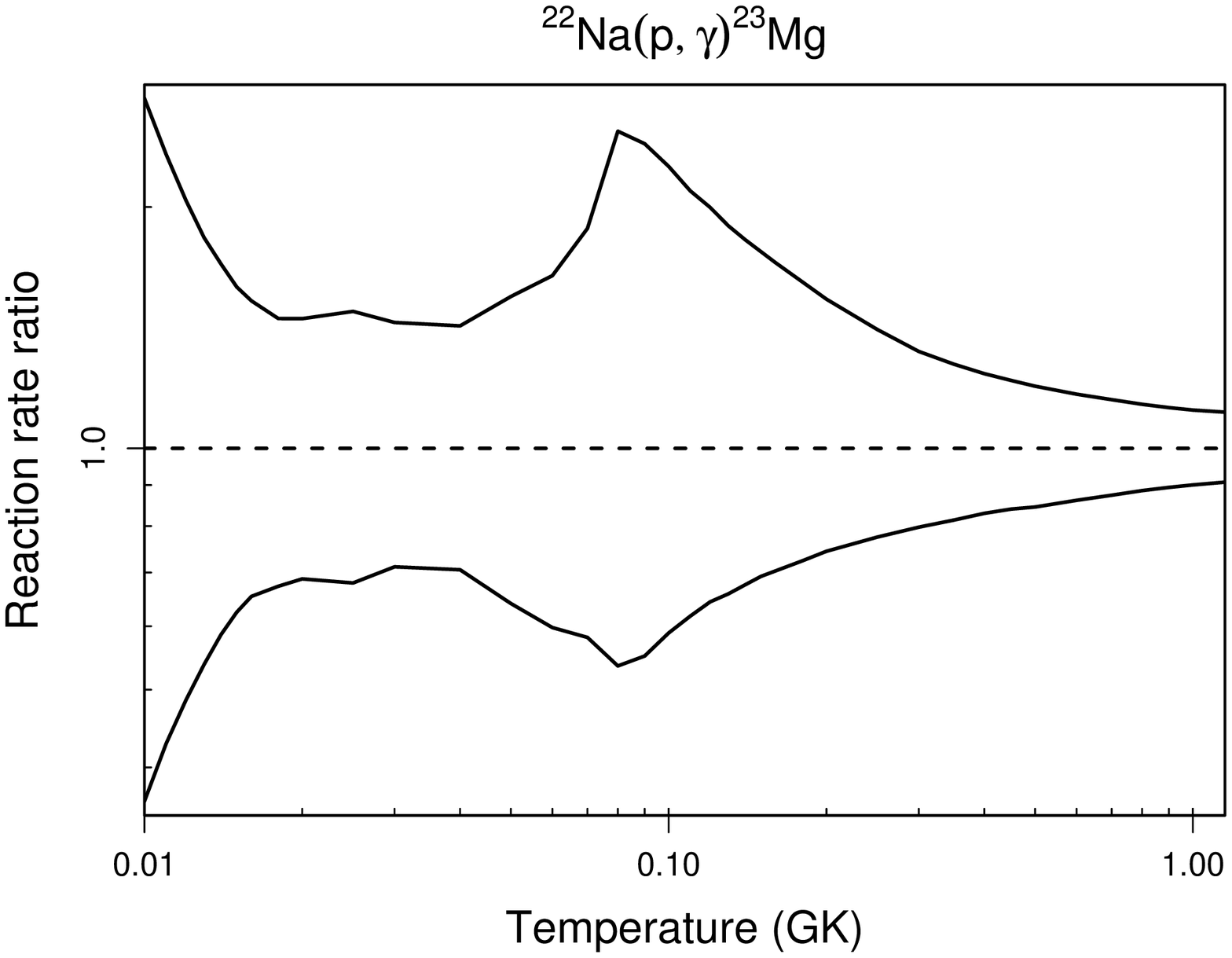}
\end{figure}
\clearpage
\begin{figure}[]
\includegraphics[height=18.5cm]{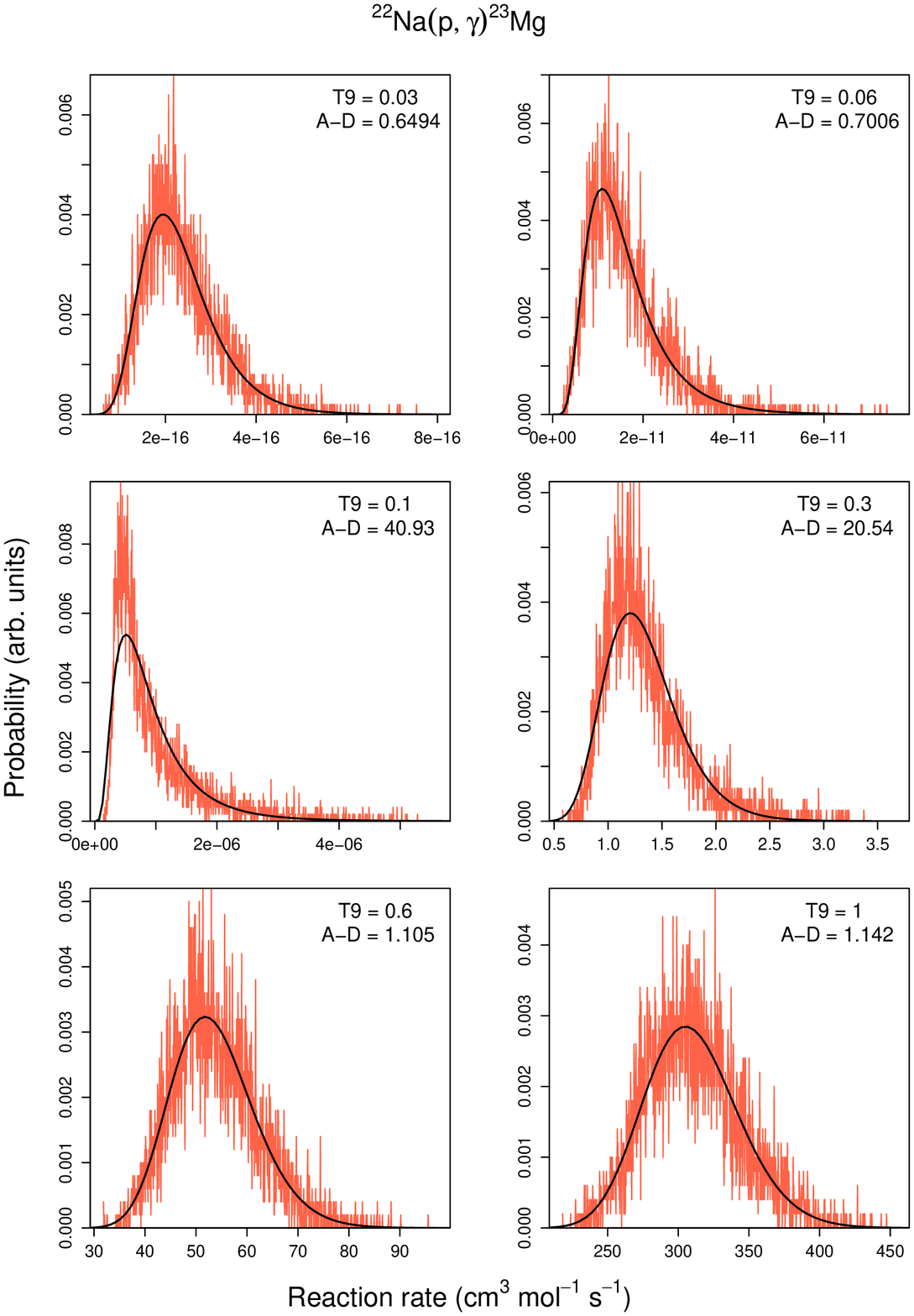}
\end{figure}
\clearpage
\setlongtables

Comments: Excitation energies and spectroscopic factors of threshold states are presented in Hale et al. \cite{Hal04}. For the E$_r^{cm}$=138 keV resonance, the directly measured upper limit of the resonance strength (Rowland et al. \cite{Row04}) is taken into account. The measured resonance strengths for E$_r^{cm}\geq$240 keV are adopted from Endt \cite{End90}, but are renormalized relative to the E$_r^{lab}$=512 keV resonance (see Tab.1 in Iliadis et al. \cite{Ili01}). In total, 54 resonances with energies in the range of E$_r^{cm}$=6-2256 keV are taken into account. The direct capture cross section is adopted from Hale et al. \cite{Hal04}.
\begin{figure}[]
\includegraphics[height=8.5cm]{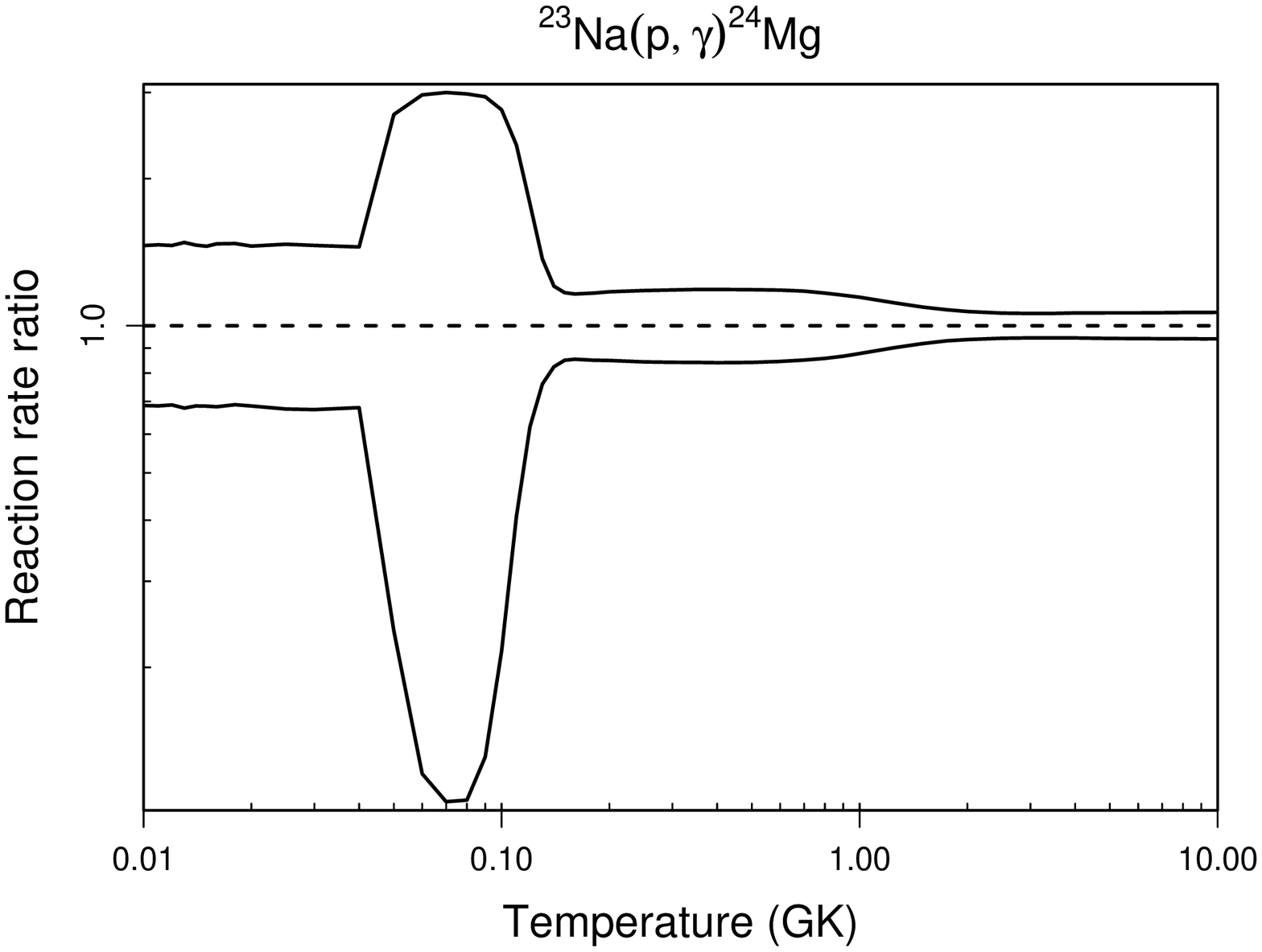}
\end{figure}
\clearpage
\begin{figure}[]
\includegraphics[height=18.5cm]{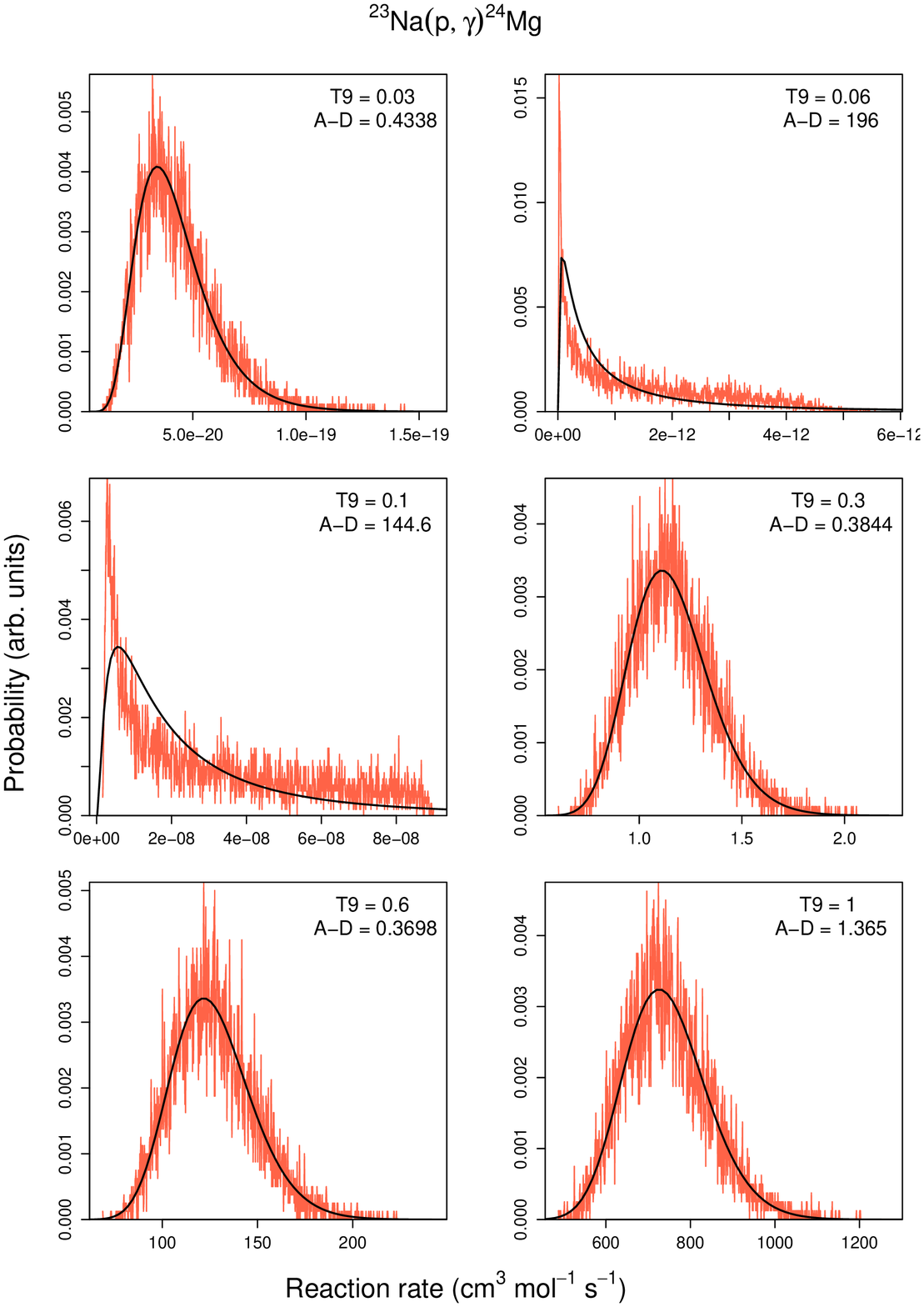}
\end{figure}
\clearpage
\setlongtables

Comments: Excitation energies and spectroscopic factors of threshold states are presented in Hale et al. \cite{Hal04}. For the E$_r^{cm}$=138 keV resonance, the directly measured upper limit for the (p,$\gamma$) resonance strength (Rowland et al. \cite{Row04}) also influences the (p,$\alpha$) reaction rates. For E$_r^{cm}\leq$968 keV, the resonance strengths are adopted from Tab. VI of Hale et al. \cite{Hal04}. The resonance strengths for E$_r^{cm}>$968 keV are renormalized relative to the E$_r^{lab}$=338 keV resonance (Rowland et al.  \cite{Row02}). In total, 52 resonances with energies in the range of E$_r^{cm}$=6-2328 keV are taken into account.
\begin{figure}[]
\includegraphics[height=8.5cm]{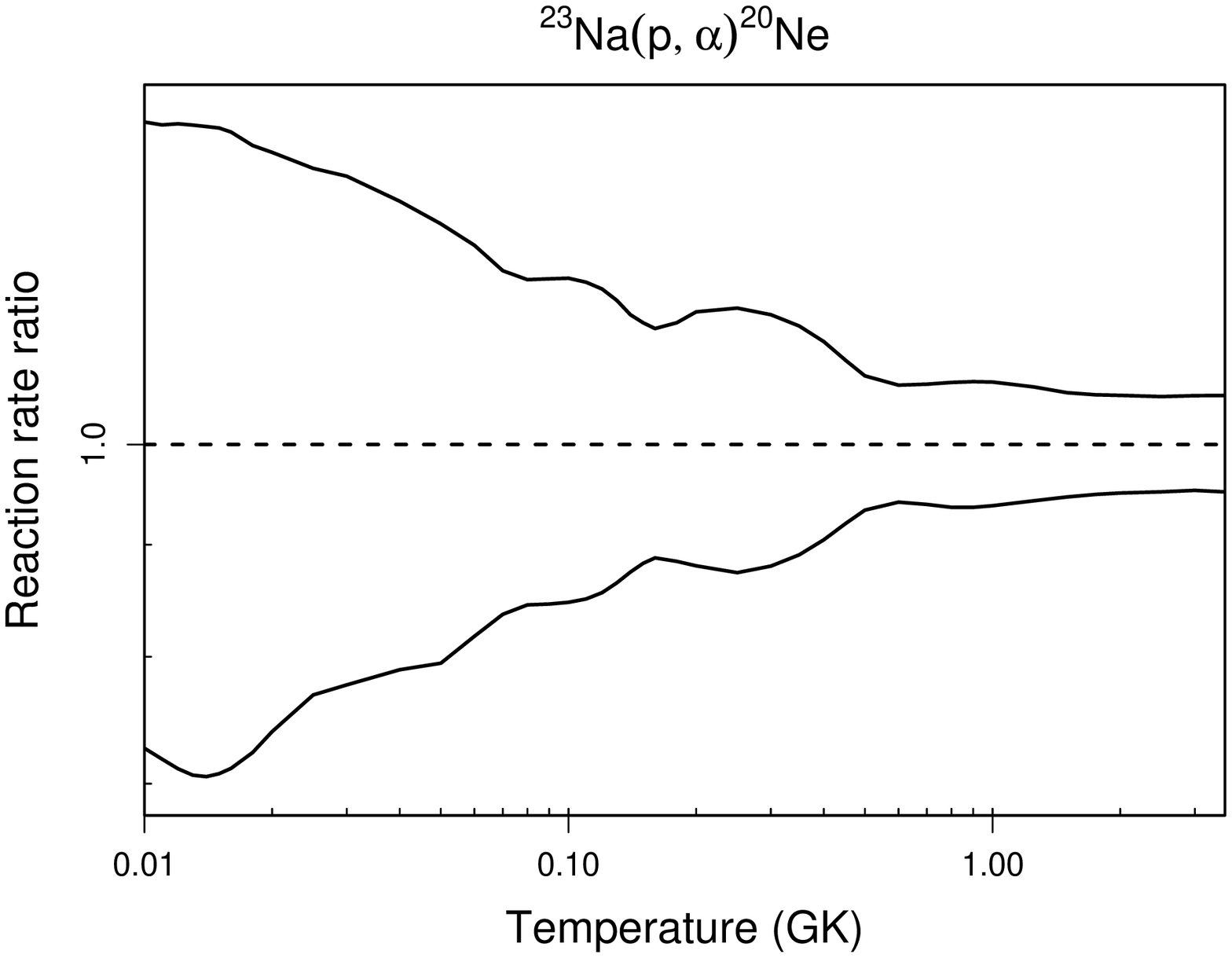}
\end{figure}
\clearpage
\begin{figure}[]
\includegraphics[height=18.5cm]{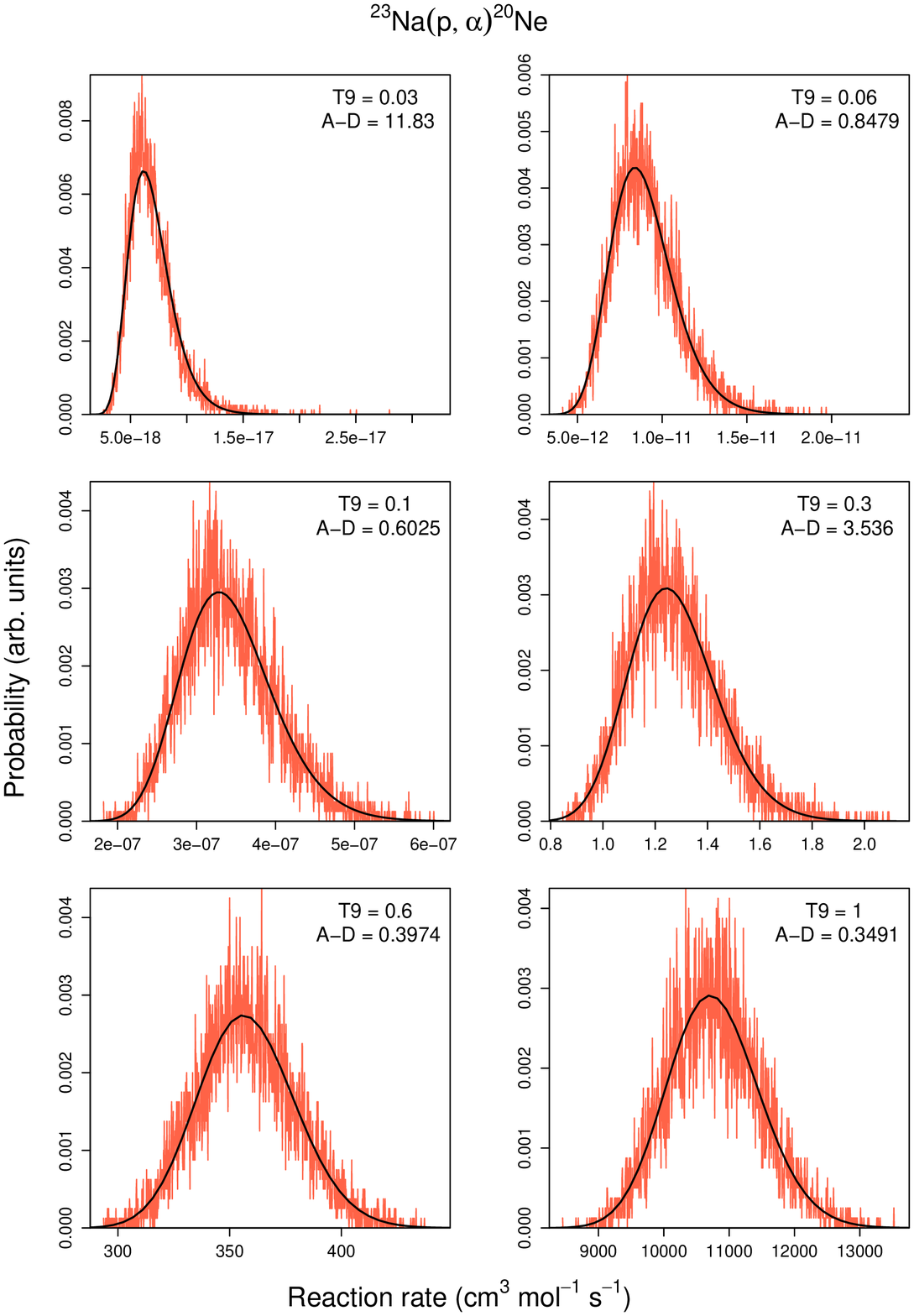}
\end{figure}
\clearpage
\setlongtables

Comments: The contributions of two resonances and the direct capture process are taken into account for the calculation of the reaction rates. The S-factor for the direct capture to the $^{23}$Al ground state (J$^{\pi}$=5/2$^+$) is calculated here by using the measured spectroscopic factor of $C^2S_{\ell=2}=0.22$ \cite{How70} for the mirror state in $^{23}$Ne. (The shell model value of $C^2S_{\ell=2}=0.34$ was used previously). The two resonances are located at E$_r^{cm}=405\pm27$ \cite{Cag01} and 1651$\pm$40 keV. The first resonance corresponds to an excitation energy of E$_x=528$ keV and has a spin-parity of 1/2$^+$. The proton width is adopted from Ref. \cite{Cag01}, while the $\gamma$-ray partial width has been measured in a Coulomb dissociation experiment \cite{Gom05}. The second resonance corresponds to E$_x=1773$ keV and represents most likely the 3/2$^+_1$ shell model state. We obtain its proton and $\gamma$-ray partial widths using the shell model results of Ref. \cite{He07}. A g-wave ($\ell=4$) resonance, corresponding to the $^{23}$Ne mirror state at E$_x=1702$ keV, is expected to occur near E$_r^{cm}=1.5-2$ MeV. Neither has the level been observed in $^{23}$Al, nor has a value been reported for the shell model $\gamma$-ray partial width and, consequently, we disregard this state. Note that it is highly unlikely that any of the resonances observed in the scattering study of He et al. \cite{He07} have a significant effect on the total reaction rates since they are located too high in energy (E$_r^{cm}=2.9-3.8$ MeV).
\begin{figure}[]
\includegraphics[height=8.5cm]{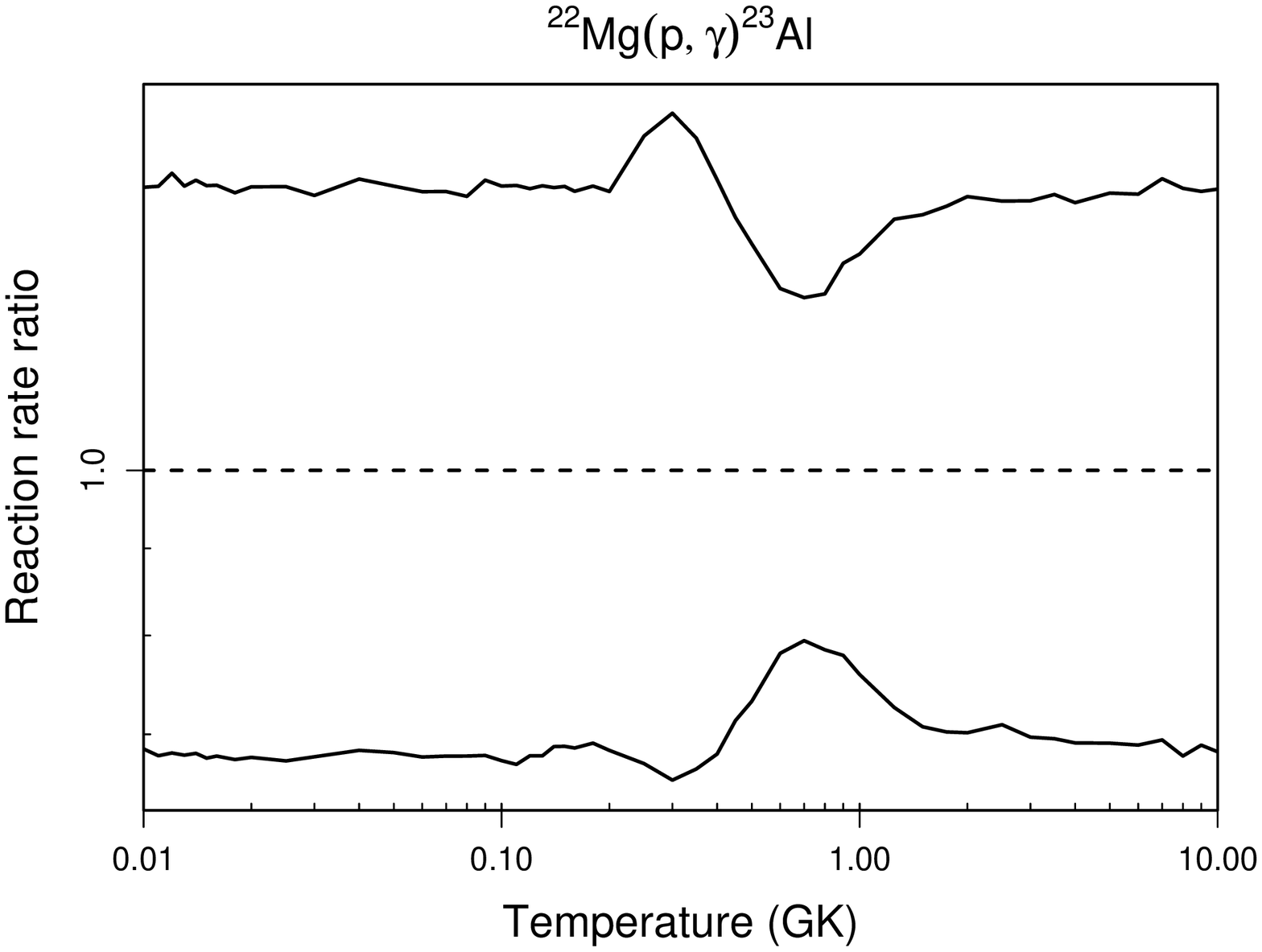}
\end{figure}
\clearpage
\begin{figure}[]
\includegraphics[height=18.5cm]{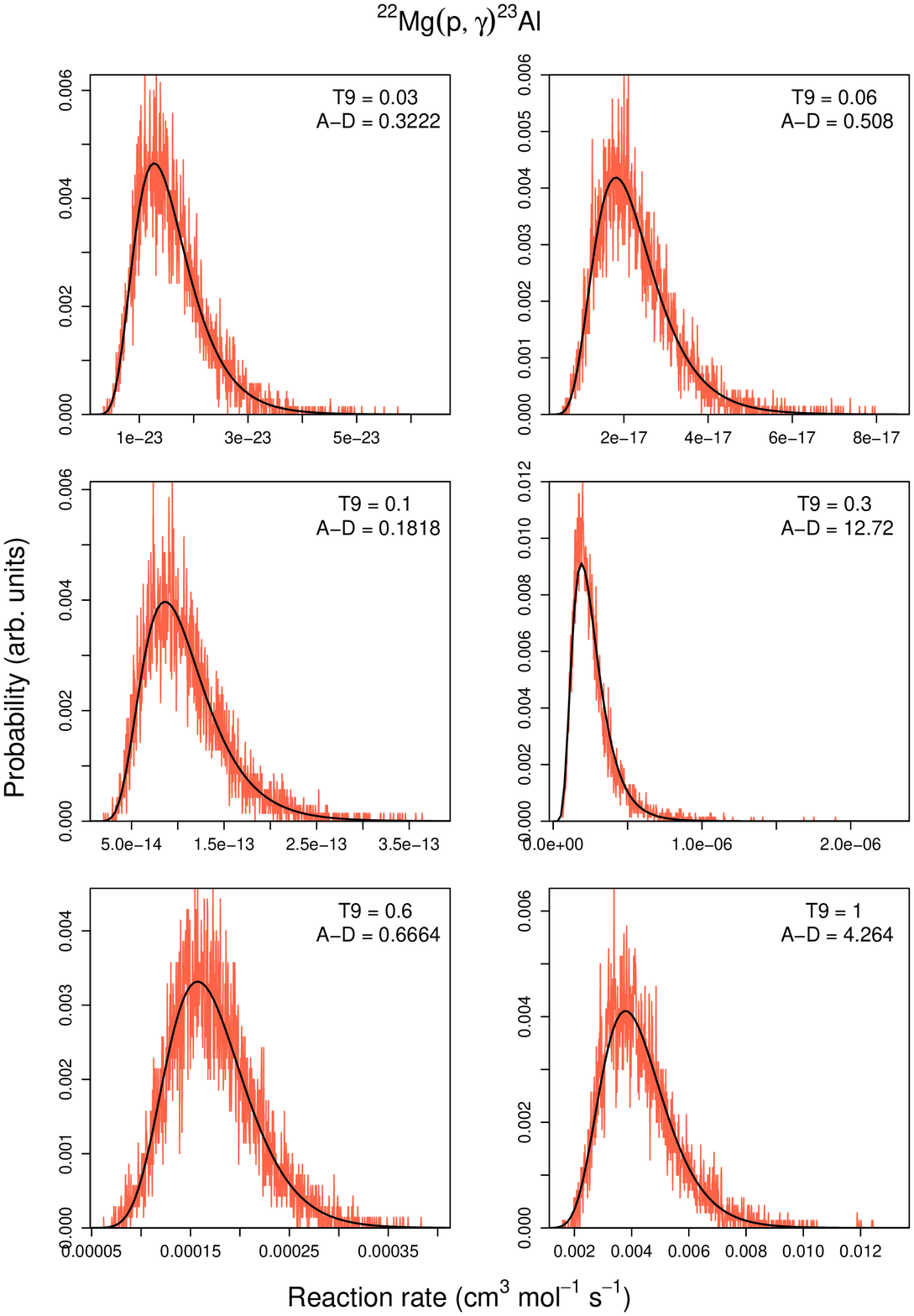}
\end{figure}
\clearpage
\setlongtables

Comments: In total, 6 resonances with energies of E$_r^{cm}\leq1.0$ MeV are taken into account. The resonance energies are deduced from the $^{24}$Al excitation energies measured by Lotay et al. \cite{Lot08} and Visser et al. \cite{Vis07,Vis08}. Spectroscopic factors are adopted from the mirror states in $^{24}$Na, measured using the (d,p) reaction (Tomandl et al. \cite{Tom04}). Gamma-ray partial widths are obtained from the shell model (Herndl et al. \cite{Her98}) for unbound states, or from Endt \cite{End90} for bound states. Information on $\gamma$-ray branching ratios is adopted from Refs. \cite{Tom04,Lot08}. The rate uncertainties result from the following uncertainties of input parameters: (i) resonance energies ($\pm$3--4 keV), (ii) spectroscopic factors ($\pm$40\%), (iii) $\gamma$-ray transition strengths ($\pm$50\%), and (iv) direct capture S--factor ($\pm$40\%). Note that the rate uncertainties presented here do not take into account the fact that the experimental spin and parity assignments for some of the low-energy resonances are not unambiguous.
\begin{figure}[]
\includegraphics[height=8.5cm]{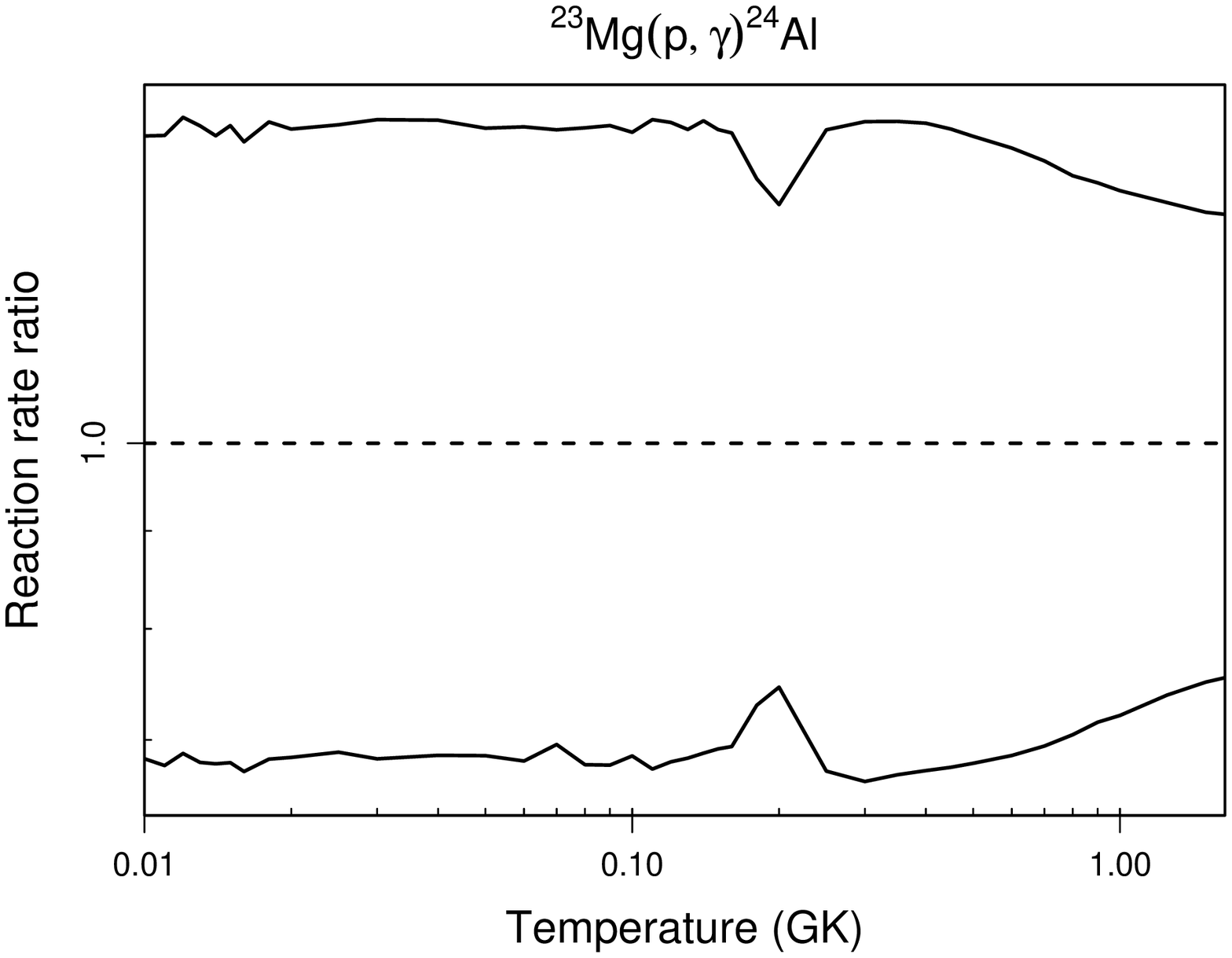}
\end{figure}
\clearpage
\begin{figure}[]
\includegraphics[height=18.5cm]{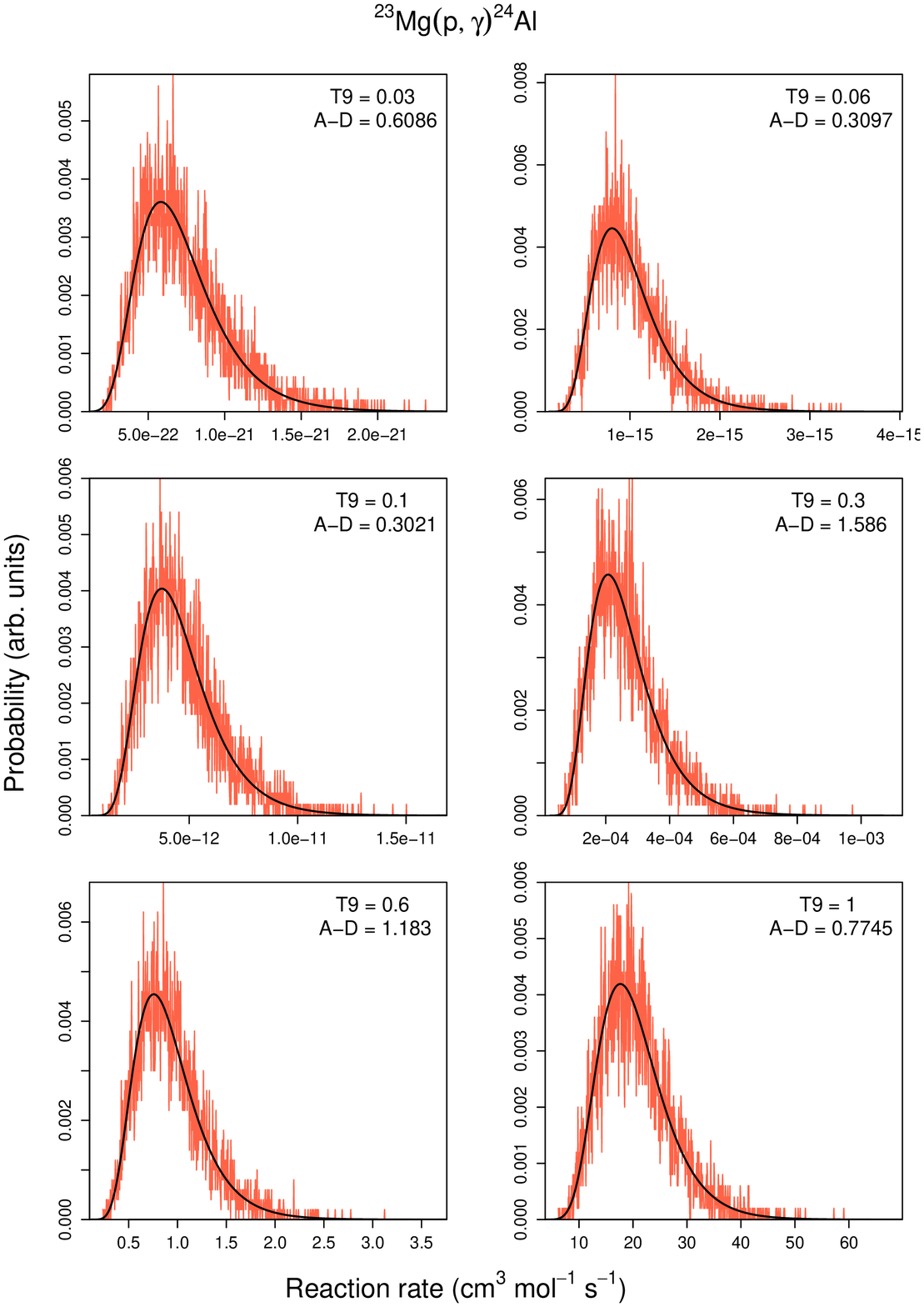}
\end{figure}
\clearpage
\setlongtables

Comments: The reaction rate is calculated from the same input information as in Powell et al. \cite{Pow99}, except that (i) the strengths of the higher-lying resonances (E$_r^{cm}\geq$1 MeV) have been normalized to the weighted average strength of E$_r^{cm}$=790 keV from Trautvetter \cite{Tra75} and Engel et al. \cite{Eng05}, and (ii) the new rate incorporates the updated Q-value (see Tab. \ref{tab:master}). In total, 9 resonances with energies in the range of E$_r^{cm}$=214-2311 keV are taken into account. The partial rates for the E$_r^{cm}$=214 and 402 keV resonances have been found by numerical integration in order to account for the low-energy resonance tails (see Powell et al. \cite{Pow99}).
\begin{figure}[]
\includegraphics[height=8.5cm]{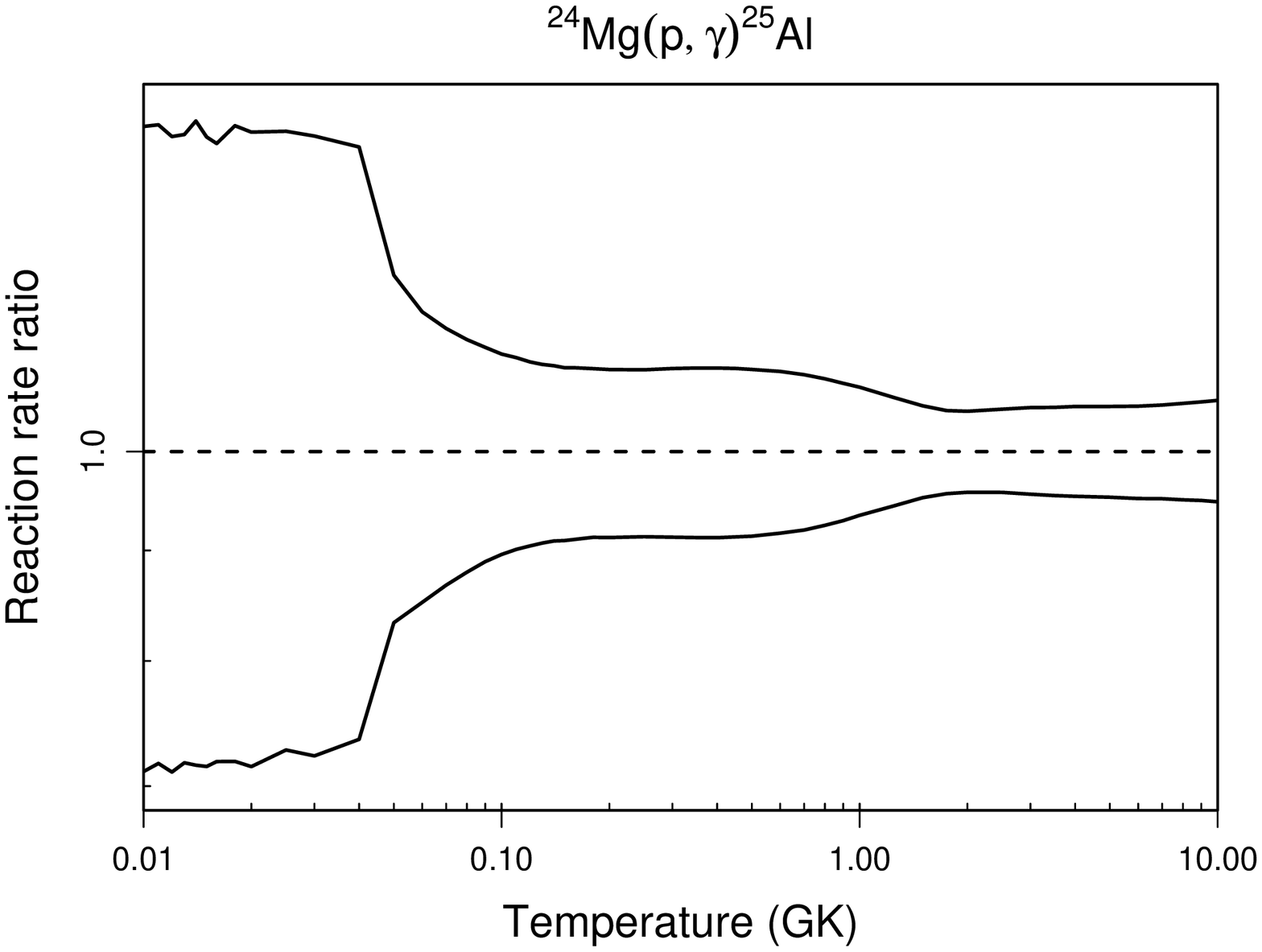}
\end{figure}
\clearpage
\begin{figure}[]
\includegraphics[height=18.5cm]{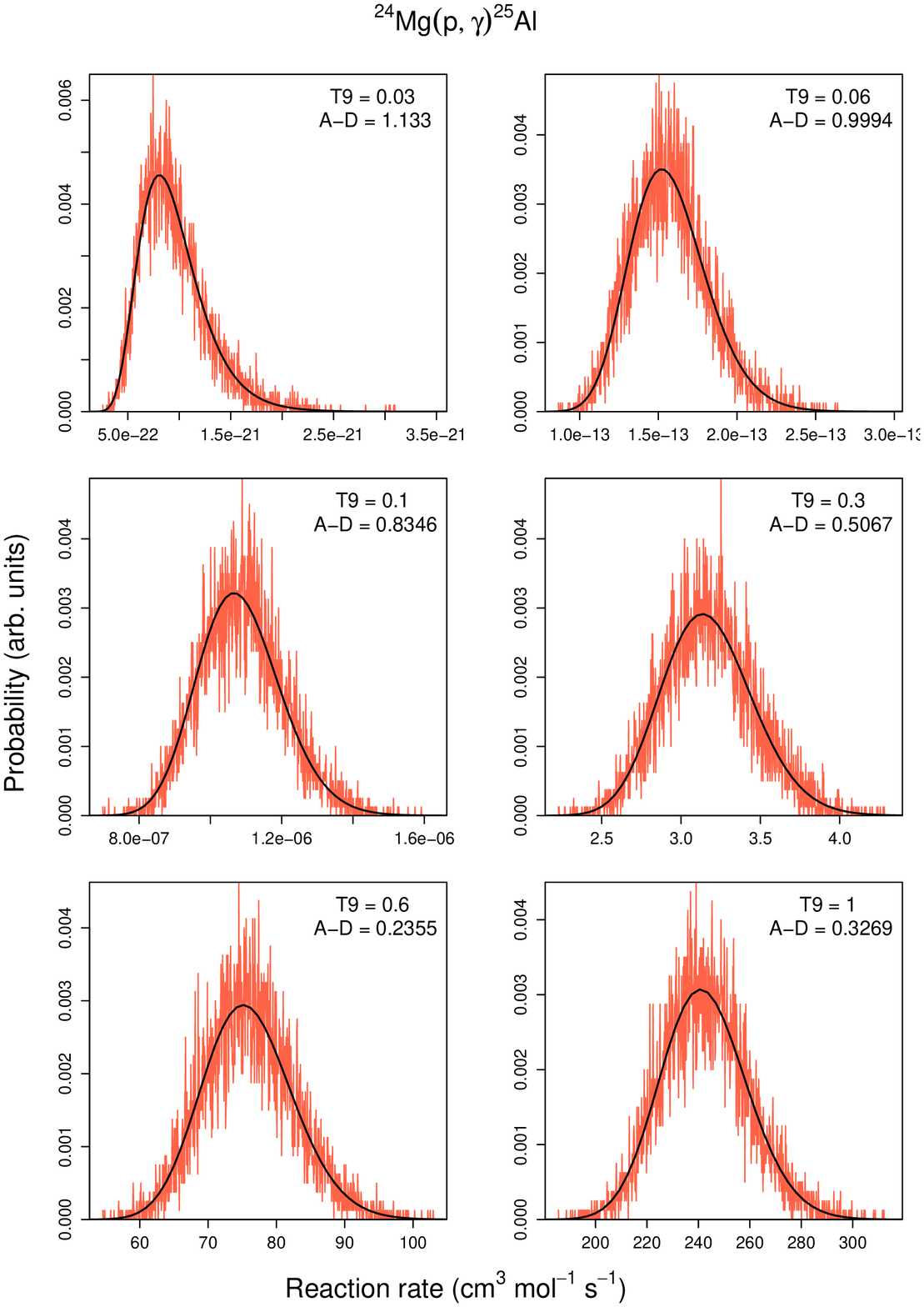}
\end{figure}
\clearpage
\setlongtables

Comments: For temperatures below T$_{9}\approx0.15$, the rate of the $^{24}$Mg($\alpha$,$\gamma$)$^{28}$Si reaction is dominated by direct capture (DC) and by possible, unobserved low-energy resonances at E$_r^{cm}=197$, 530, 821, 932 and 969 keV, which correspond to known states in $^{28}$Si at E$_{x}=10.182$, 10.514, 10.806, 10.916, and 10.953 MeV, respectively \cite{endsf}. The DC component was estimated by scaling a calculated DC rate using relative spectroscopic factors for $^{20}$Ne($^{6}$Li,d)$^{24}$Mg \cite{dr74,ta83}. These were converted to absolute spectroscopic factors using the same procedure described for the $^{20}$Ne($\alpha$,$\gamma$)$^{24}$Mg reaction. Again, to be conservative we have assigned a factor of 2 uncertainty to the DC rate. The 5 possible resonances were selected based on their T = 0 assignments and favorable J$^{\pi}$ values, which allow for $\ell\leq3$ transfer. Upper limits on the $\alpha$-particle widths were calculated using a potential model for the lower three states. For the upper two states, the results of direct ($\alpha,\gamma$) measurements \cite{str08} were used to determine upper limits on the $\alpha$-particle widths.

Resonances have been measured by Smulders and Endt \cite{smE}, Weinman et al. \cite{we64}, Lyons et al. \cite{ly69}, Maas et al. \cite{maas}, Cseh et al. \cite{cseh}, and Strandberg et al. \cite{str08}. Resonance energies and total widths have been updated using the excitation energies and widths appearing in ENDSF \cite{endsf} and Endt \cite{End90,End98}. The resonance strengths and partial widths that we adopted are obtained from a weighted average of the published resonance strengths, excluding those of Weinman et al. \cite{we64}, which were reported without uncertainties. The earler studies used stopping powers and standard resonance strengths that are now considered to be archaic. Nonetheless, there is excellent agreement amongst the various data sets and thus no corrections were made. To calculate partial widths, we also made use of resonance strengths for the $^{27}$Al(p, $\alpha$)$^{24}$Mg and $^{27}$Al(p,$\gamma$)$^{28}$Si reactions (see Paper III for input values).

Overall, our classical reaction rate is similar to that reported by Strandberg et al. \cite{str08}. However, at their lowest temperature, T$_{9}=0.15$, our rate is significantly larger because we have also included the direct capture and the possible low-energy resonances listed above, which have a negligible effect at higher temperatures. 
\begin{figure}[]
\includegraphics[height=8.5cm]{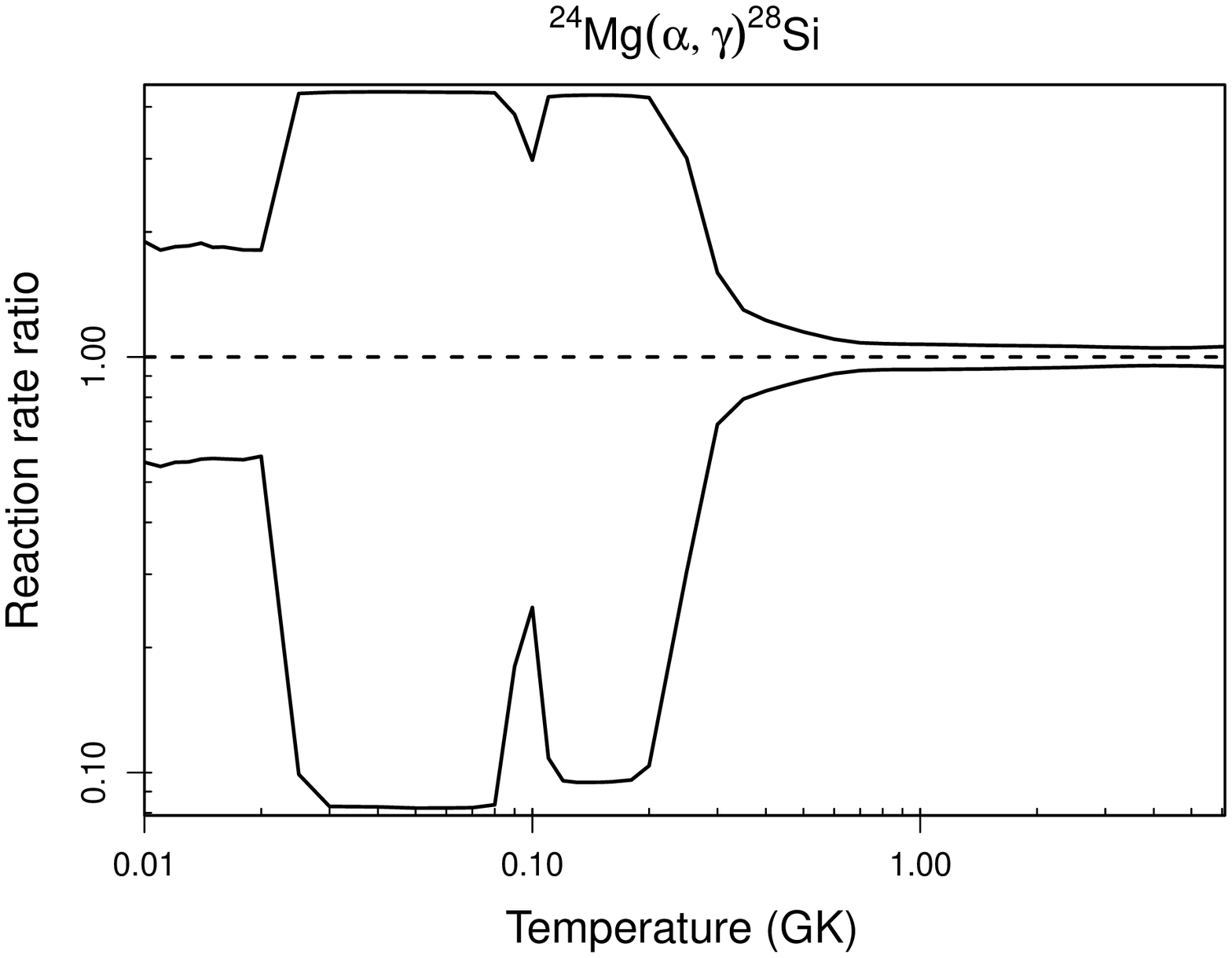}
\end{figure}
\clearpage
\begin{figure}[]
\includegraphics[height=18.5cm]{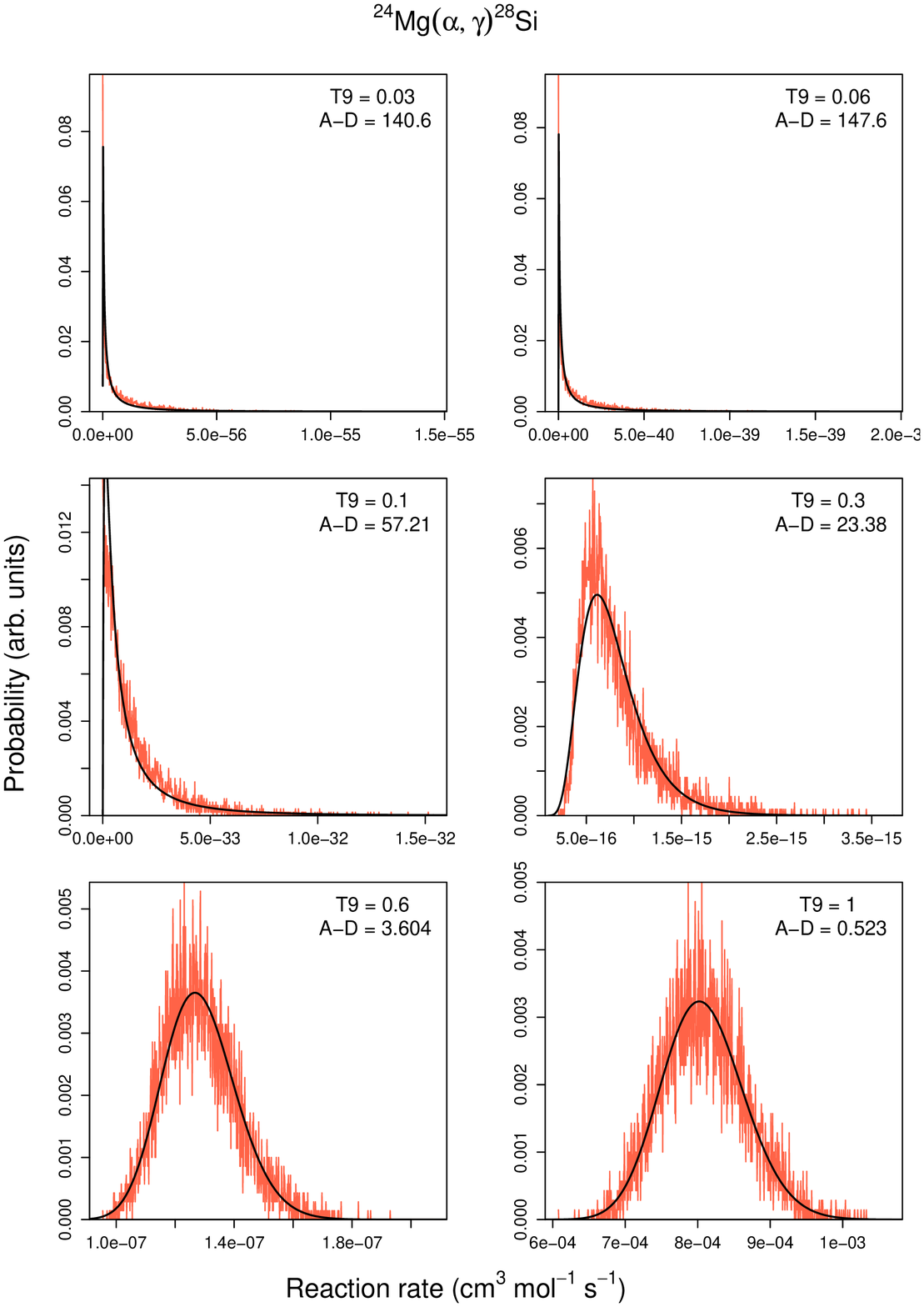}
\end{figure}
\clearpage
\setlongtables

Comments: Altogether, 82 resonances at energies of E$_r^{cm}=37-1761$ keV are taken into account in calculating the total reaction rates for the formation of either the ground or isomeric $^{26}$Al state. For measured resonances (E$_r^{cm}\ge189$ keV) the energies and strengths are adopted from Endt \cite{End90}, but the latter values are renormalized using the standard values presented in Tab. 1 of Iliadis et al. \cite{Ili01}. For the threshold states the spin-parity assignments and the proton partial widths are adopted from Iliadis et al.  \cite{Ili96}. For the E$_r^{cm}=37$ keV resonance the parity is not known experimentally; we adopt J$^{\pi}$=4$^{+}$ as predicted by the shell model \cite{Ili96}, implying $\ell=2$ transfer. (Note that this assumption differs from the one adopted in Ref. \cite{Ili01} where an upper limit on the proton partial width was derived for a p-wave resonance). We disregarded the (ground state) resonance strengths measured by Arazi et al. \cite{Ara06} using accelerator mass spectrometry since their strength of the 189 keV resonance in particular seems far too small (see also Formicola et al. \cite{For08}). The direct capture S-factor is adopted from Endt and Rolfs \cite{EnR87}.
\begin{figure}[]
\includegraphics[height=8.5cm]{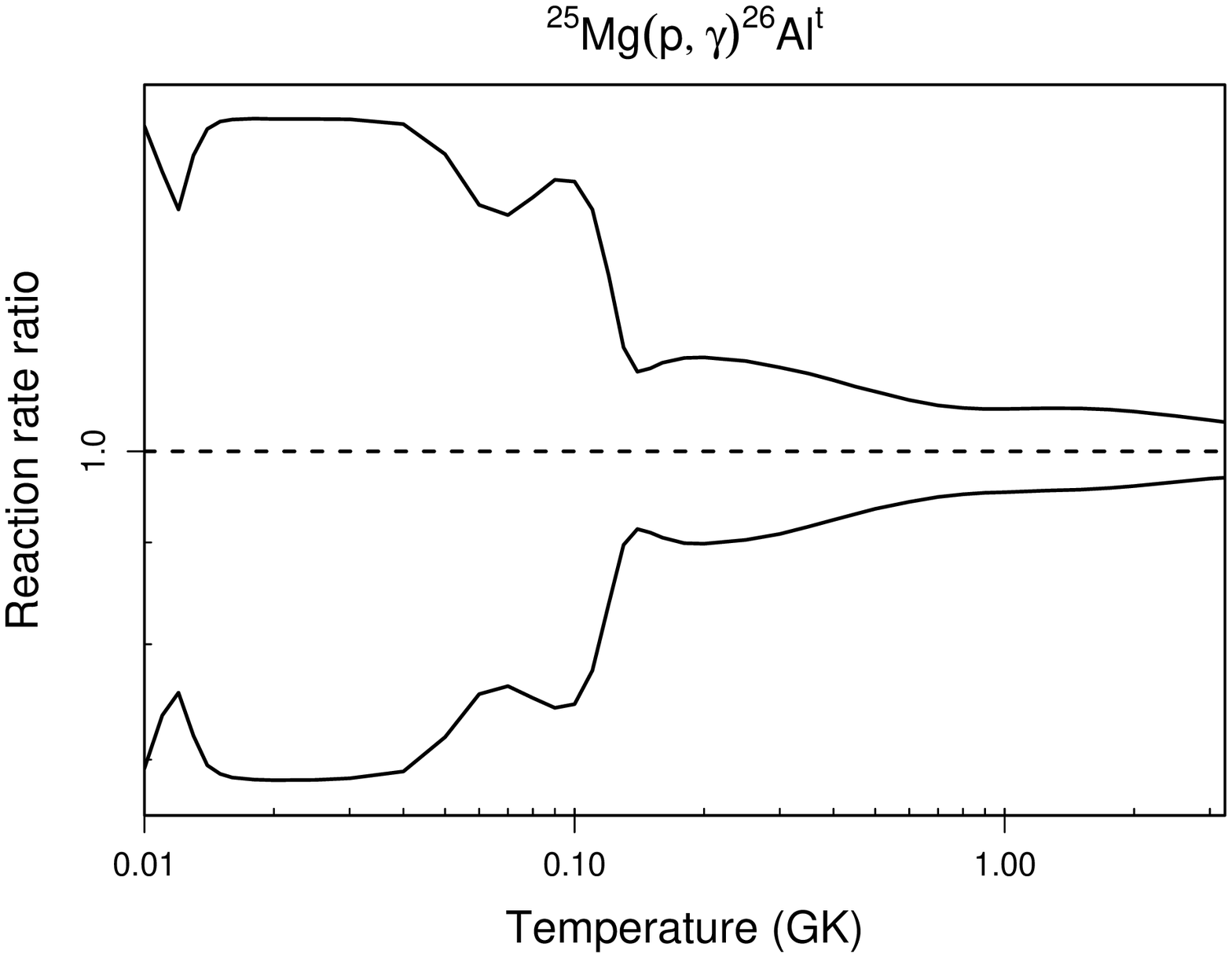}
\end{figure}
\clearpage
\begin{figure}[]
\includegraphics[height=18.5cm]{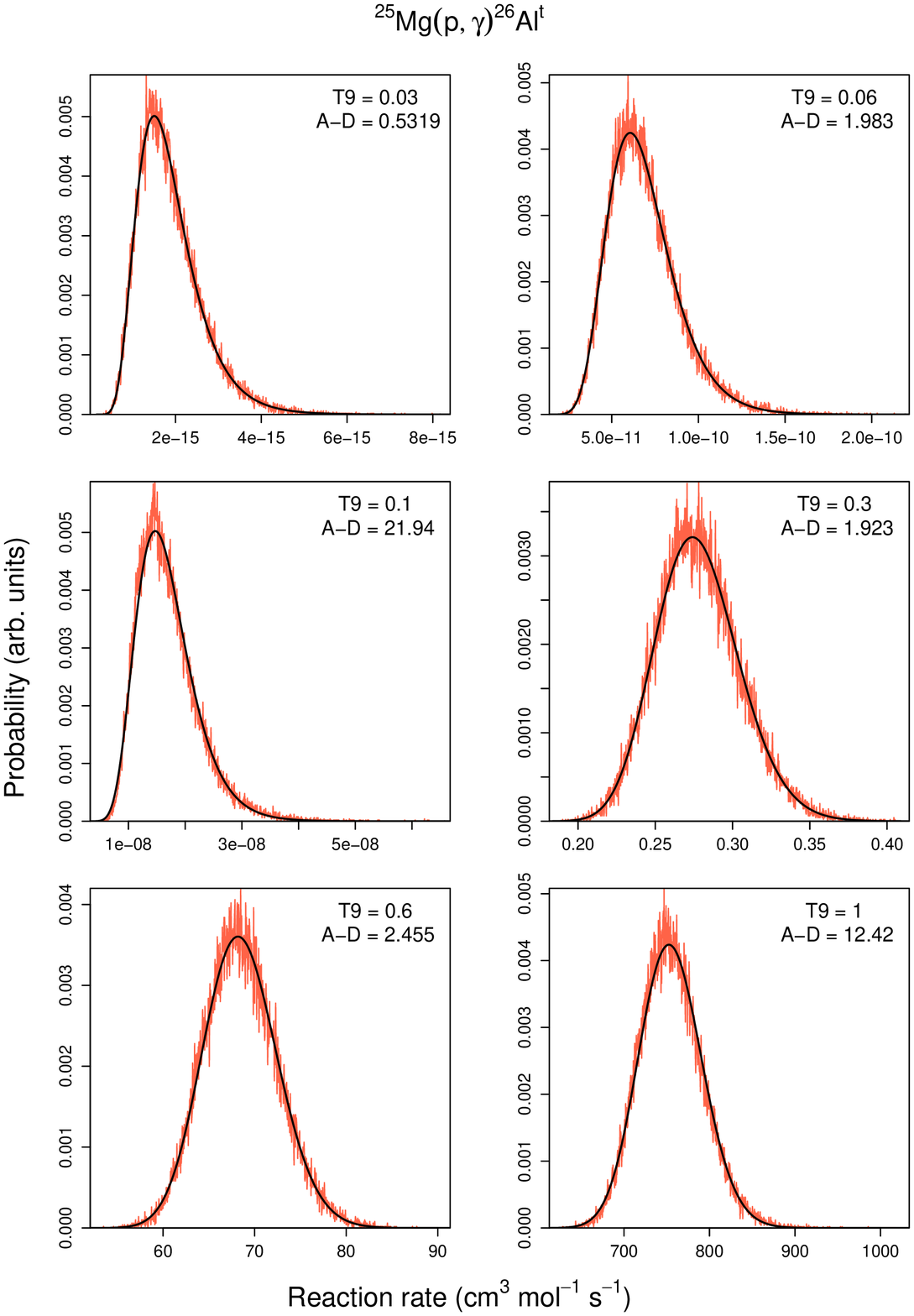}
\end{figure}
\clearpage
\setlongtables

Comments: The reaction rates for the formation of the $^{26}$Al ground state are calculated from the same information used to compute the total reaction rates (see Tab. \ref{tab:mgpgt}), but in addition the ground state $\gamma$-ray branching ratios, $f_0$, have to be taken into account; these are adopted from Endt and Rolfs \cite{EnR87}, except for the E$_r^{cm}=189, 244$ and 292 keV resonances, for which the more accurate results of Iliadis \cite{Ili89} are used ($f_0=0.66, 0.76$ and 0.79, respectively). 
\begin{figure}[]
\includegraphics[height=8.5cm]{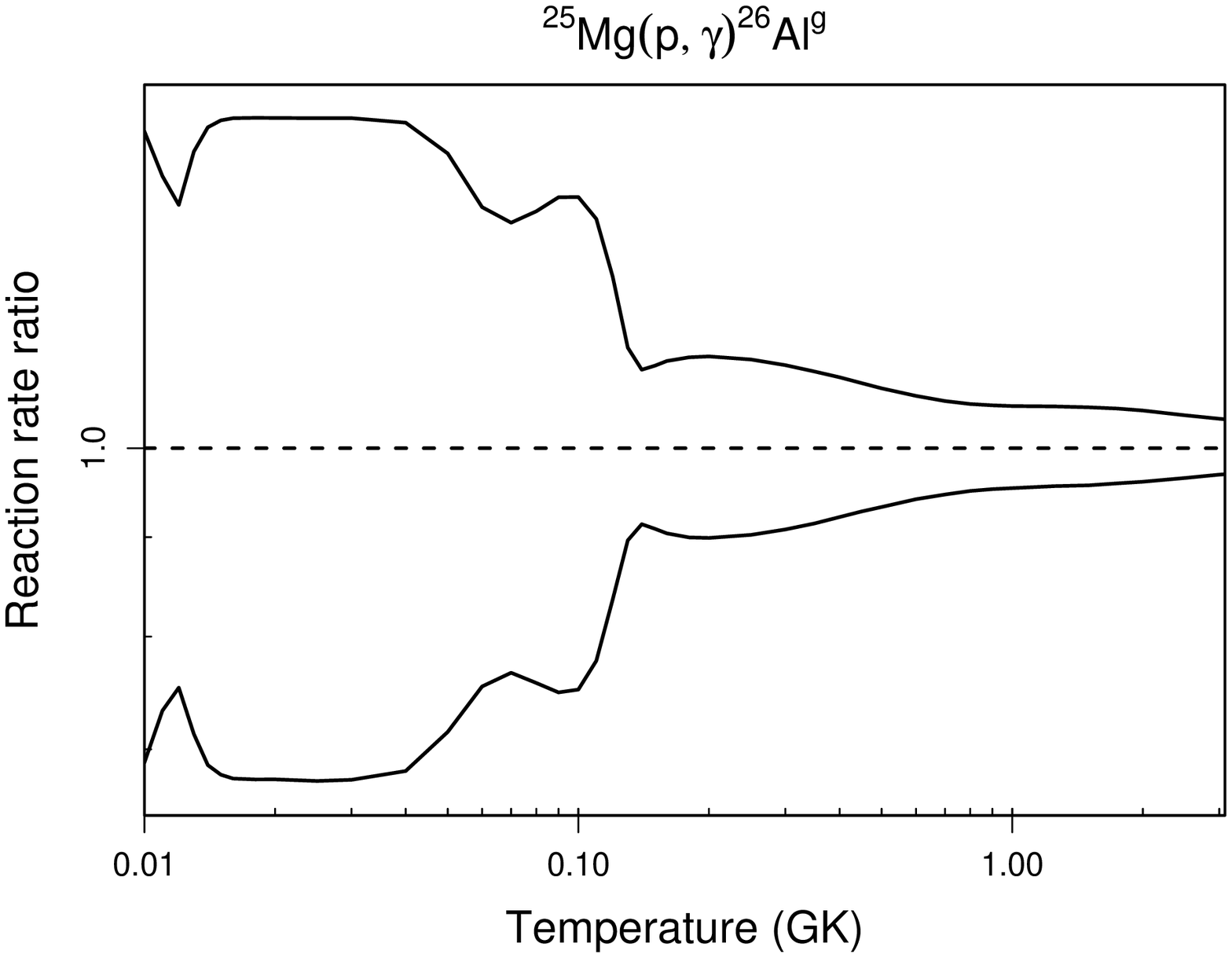}
\end{figure}
\clearpage
\begin{figure}[]
\includegraphics[height=18.5cm]{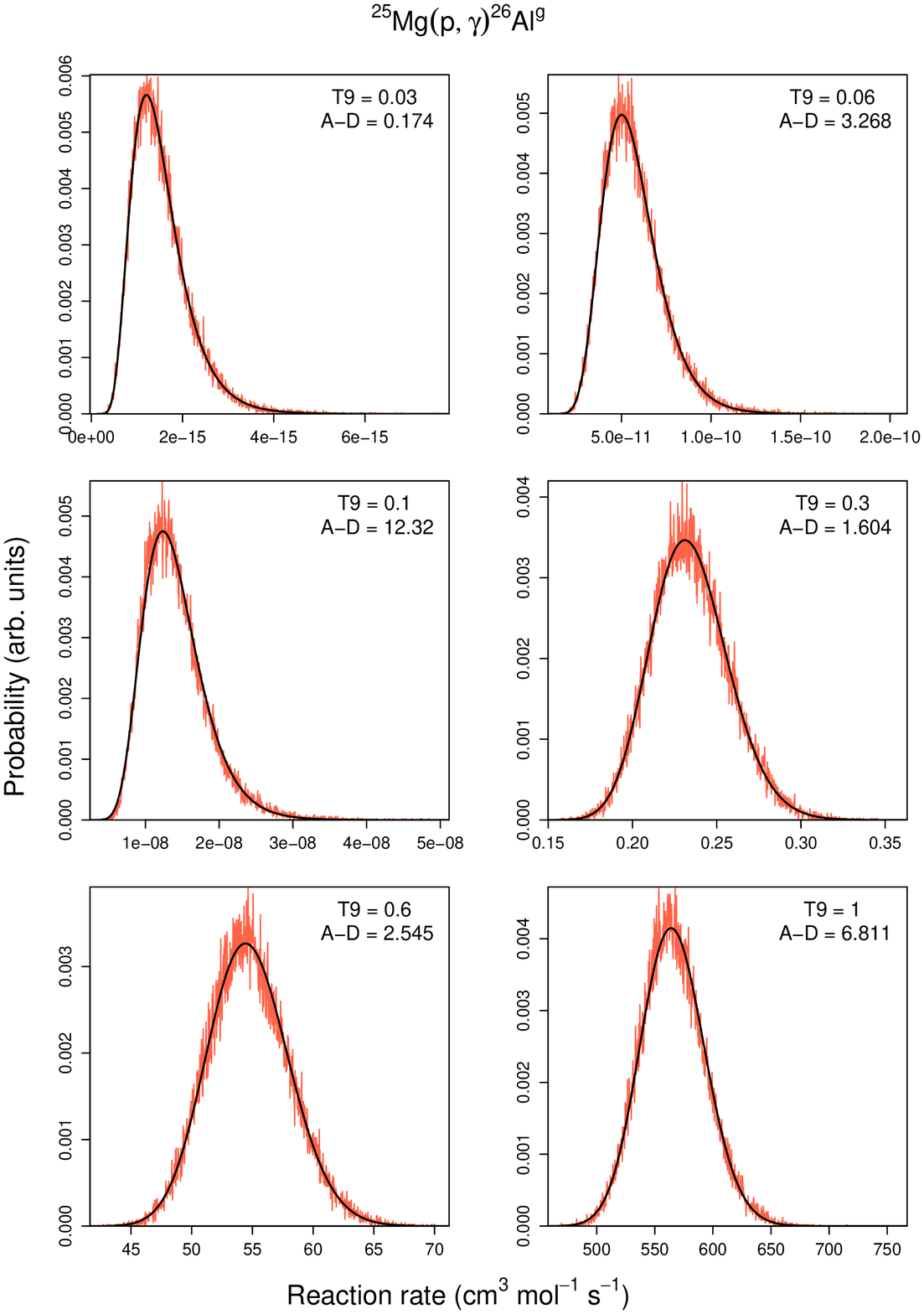}
\end{figure}
\clearpage
\setlongtables

Comments: The reaction rates for the formation of the $^{26}$Al isomeric state at E$_x=228$ keV are calculated from the same information used to compute the total reaction rates (see Tab. \ref{tab:mgpgt}), but in addition the isomeric state $\gamma$-ray branching ratios, $1-f_0$, have to be taken into account; these are adopted from Endt and Rolfs  \cite{EnR87}, except for the E$_r^{cm}=189, 244$ and 292 keV resonances, for which the more accurate results of Iliadis \cite{Ili89} are used ($f_0=0.66, 0.76$ and 0.79, respectively). 
\begin{figure}[]
\includegraphics[height=8.5cm]{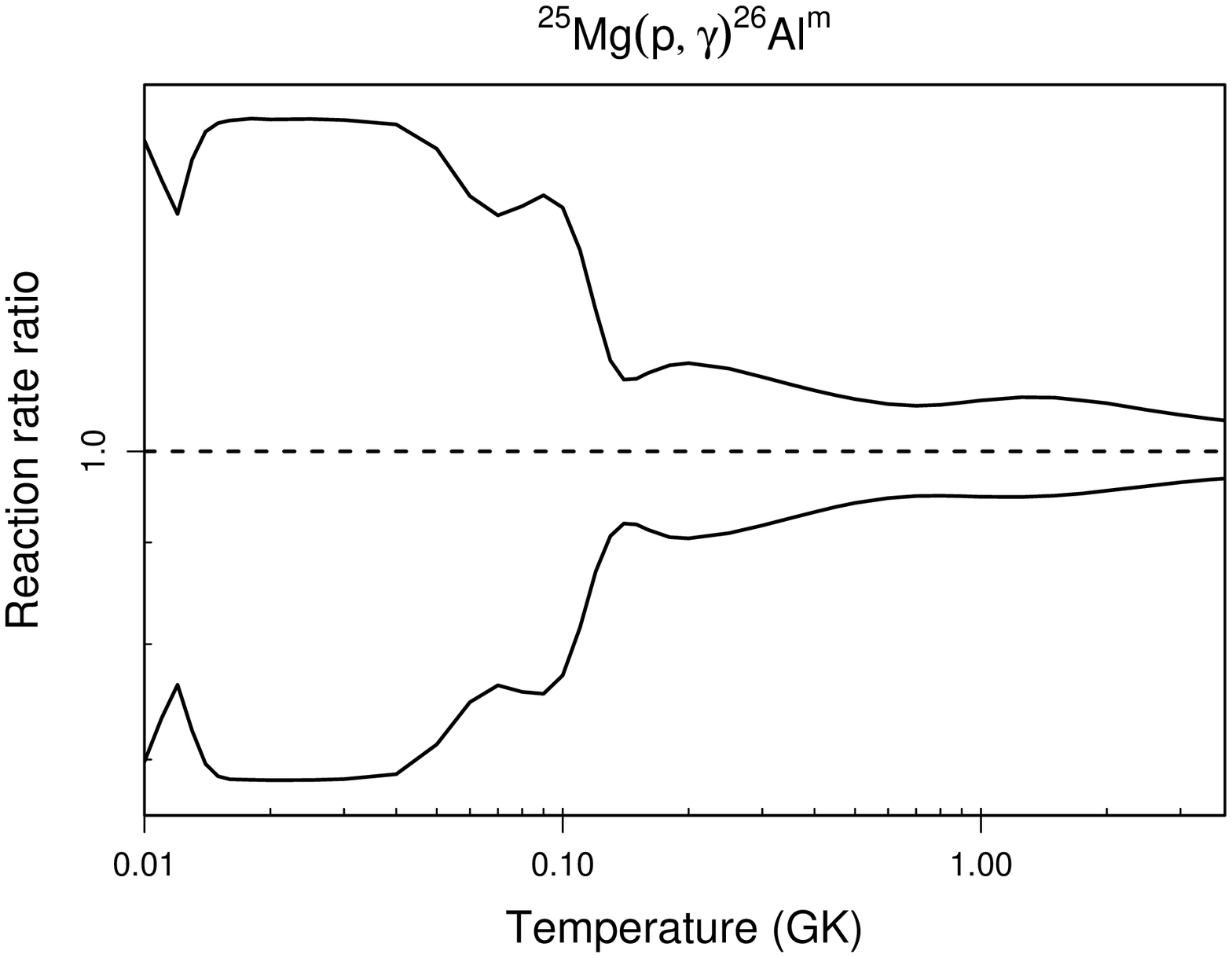}
\end{figure}
\clearpage
\begin{figure}[]
\includegraphics[height=18.5cm]{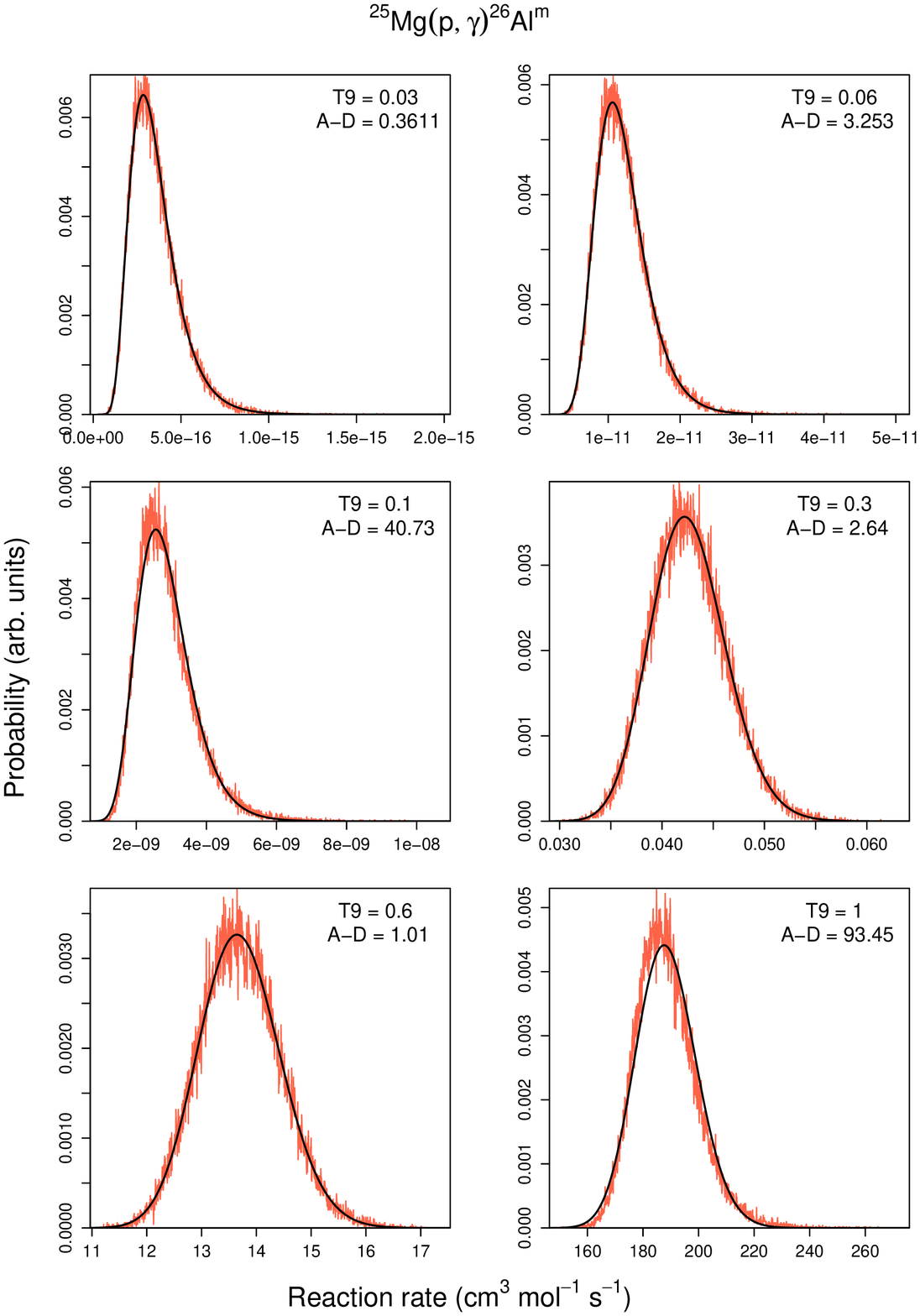}
\end{figure}
\clearpage
\setlongtables

Comments: In total, 133 resonances in the range of E$_r^{cm}$=16-2867 keV are taken into account for the calculation of the total rates. The direct capture component is adopted from Iliadis et al. \cite{Ili90}. The measured resonance strengths listed in Endt \cite{End90} have been normalized to the standard strengths given in Tab. 1 of Iliadis et al. \cite{Ili01}. The rate contribution of threshold states is estimated by using information from the $^{26}$Mg($^3$He,d)$^{27}$Al study of Champagne et al. \cite{Cha90}. These stripping data have been reanalyzed in 2000 (C. Rowland, priv. comm.) with a modern version of the DWBA code DWUCK4. For E$_x$=8324 keV (E$_r^{cm}$=53 keV) we adopt the $\ell=2$ assignment from Ref. \cite{Cha90} (on which the value of $J^{\pi}=5/2^+$ listed in Ref. \cite{End90} was based); however, this assignment must be regarded as tentative since a $\ell=3$ angular distribution (implying $J^{\pi}=5/2^-$) fits the stripping data almost equally well. Similar arguments apply to E$_x$=8376 keV (E$_r^{cm}$=105 keV). Both $\ell=1$ and $\ell=2$ angular distributions fit the stripping data. We adopt here $\ell=1$ which seems to fit slightly better, although $\ell=2$ was reported in the original analysis of Ref. \cite{Cha90} (note also that $J^{\pi}=(3/2, 5/2)^+$ listed in Ref. \cite{End90} was based on the originally reported $\ell=2$ value). The quantum numbers of these two levels, and those for E$_x$=8361 keV (E$_r^{cm}$=90 keV), need to be determined unambiguously in future work.
\begin{figure}[]
\includegraphics[height=8.5cm]{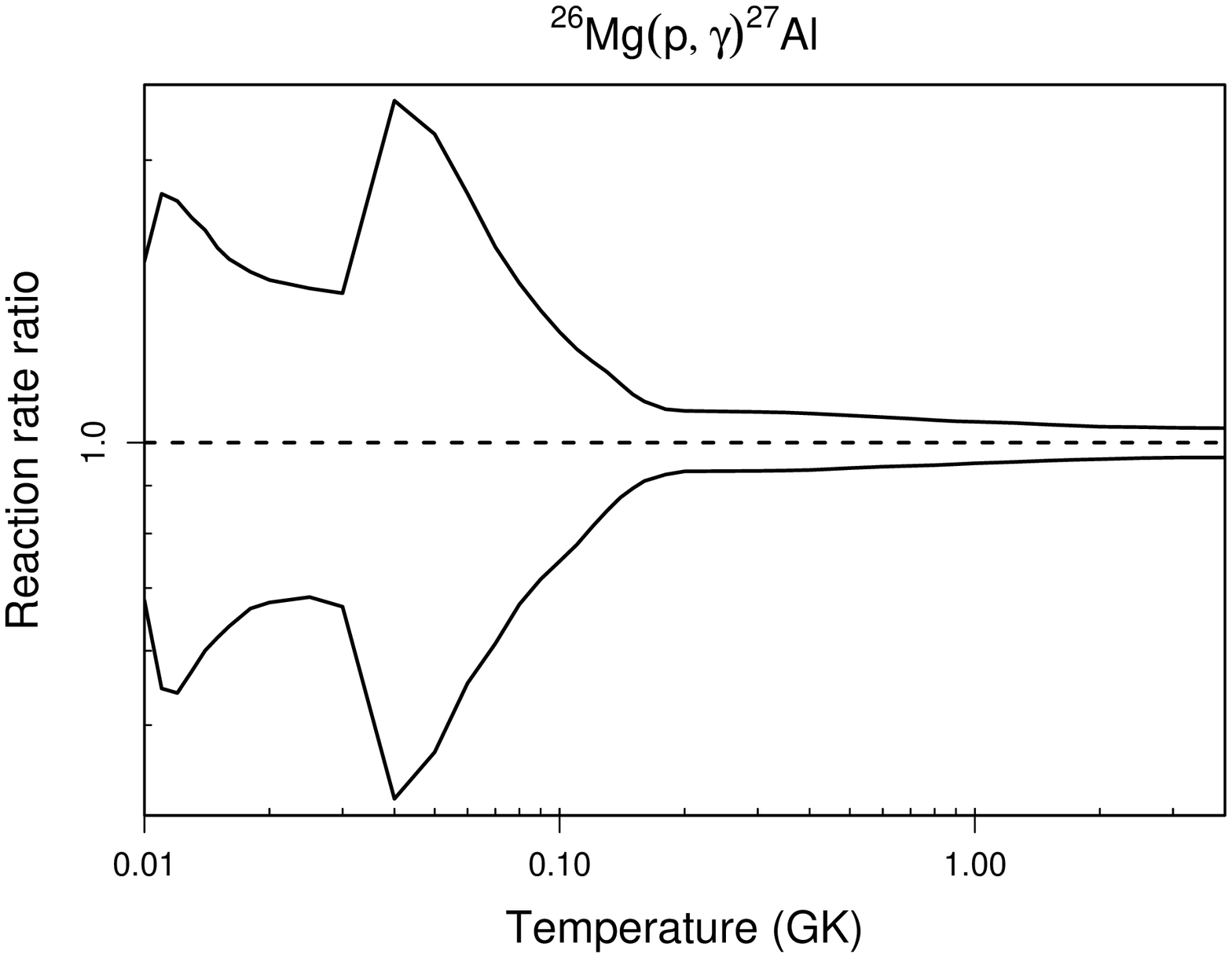}
\end{figure}
\clearpage
\begin{figure}[]
\includegraphics[height=18.5cm]{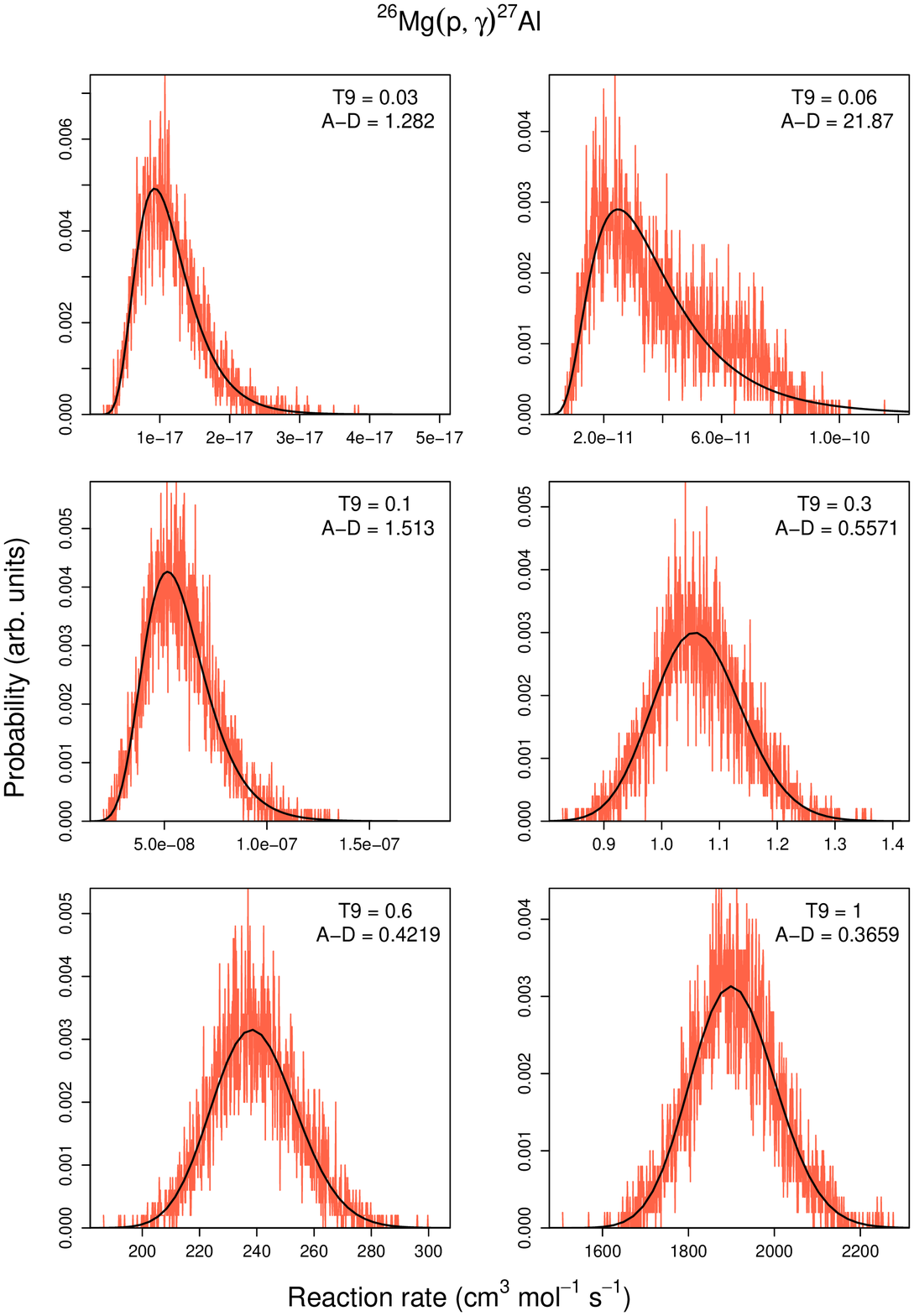}
\end{figure}
\clearpage
\setlongtables

Comments: Four resonances, none of them observed directly, are taken into account for calculating the total reaction rates. The energy of the lowest-lying resonance (E$_r^{cm}=137\pm29$ keV) is calculated from the measured excitation energy of E$_x=3441\pm10$ keV \cite{Sch97}. Based on a comparison to the shell model and to the $^{24}$Ne mirror nucleus structure, this level corresponds most likely to $J^{\pi}=2_2^+$. The energies of the remaining three resonances are computed by using the Coulomb shift calculations of Ref. \cite{Her95}. For the energy uncertainties we assume a value of $\pm150$ keV, which should be regarded as a rough estimate only. (Note that the Coulomb shift calculations of Ref. \cite{Her95} overpredict the energy of the $2_2^+$ level by 180 keV.) All proton and $\gamma$-ray partial widths used here are based on the shell model results of Ref. \cite{Her95}. The direct capture S-factor is calculated using shell model spectroscopic factors and is in reasonable agreement with Ref. \cite{Her95}.
\begin{figure}[]
\includegraphics[height=8.5cm]{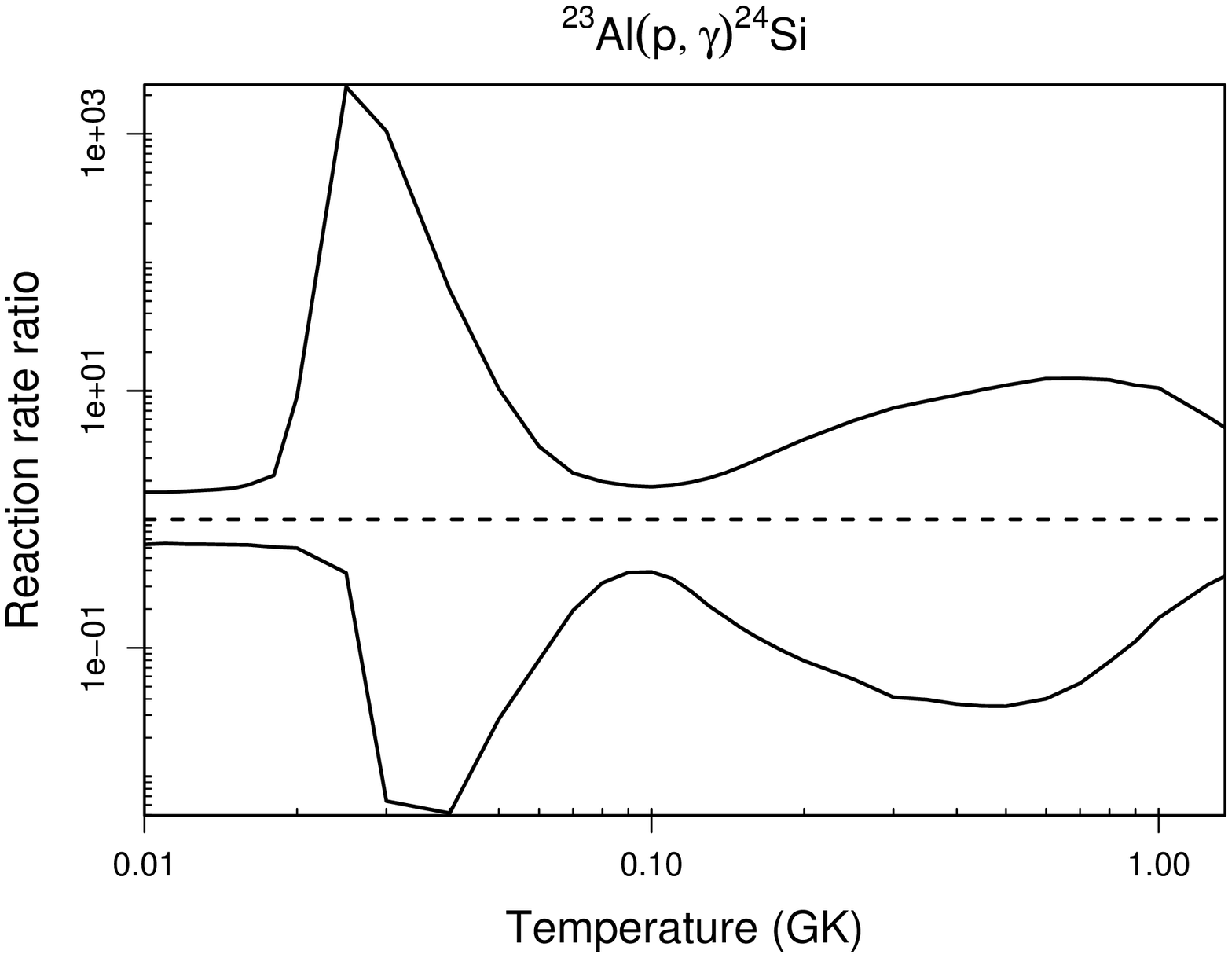}
\end{figure}
\clearpage
\begin{figure}[]
\includegraphics[height=18.5cm]{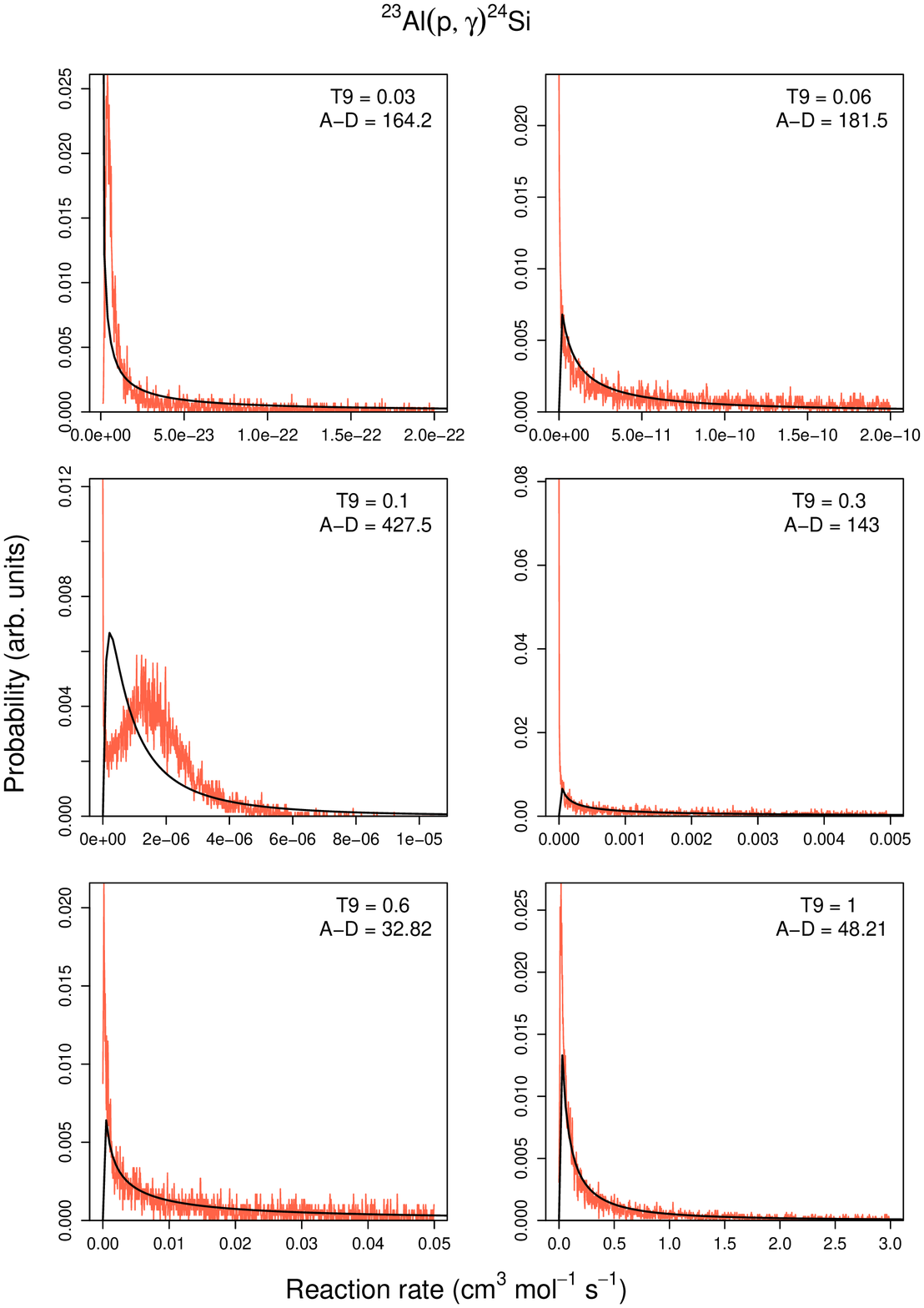}
\end{figure}
\clearpage
\setlongtables

Comments: The total rate has contributions from the direct capture process and from 5 resonances located at E$_r^{cm}=190-730$ keV. The direct capture S-factor as well as the proton and $\gamma$-ray partial widths of the resonances are based on the shell-model \cite{Her95}. Our rate does not take into account several potentially important systematic effects. First, only one level has been observed in $^{25}$Si within 1 MeV of the proton threshold, at E$_x=3820\pm20$ keV, but its spin-parity is unknown. Second, the energies of the resonances at E$_r^{cm}=190$, 500, 510 and 730 keV are not based on experimental excitation energies, but are derived from Coulomb shift calculations \cite{Her95}; the adopted value of 100 keV for the resonance energy uncertainty must be regarded as a rough value only. Third, the Coulomb displacement energy calculations of Ref. \cite{Her95} use as a starting point the experimental excitation energies of the $^{25}$Na mirror nucleus; however, for several of these states the spin-parities are also unknown and thus have been based on a comparison to the shell model. Fourth, a number of levels observed in the mirror nucleus $^{25}$Na remain unaccounted for in the derivation of the reaction rates, although their contributions are expected to be small \cite{Her95}.
\begin{figure}[]
\includegraphics[height=8.5cm]{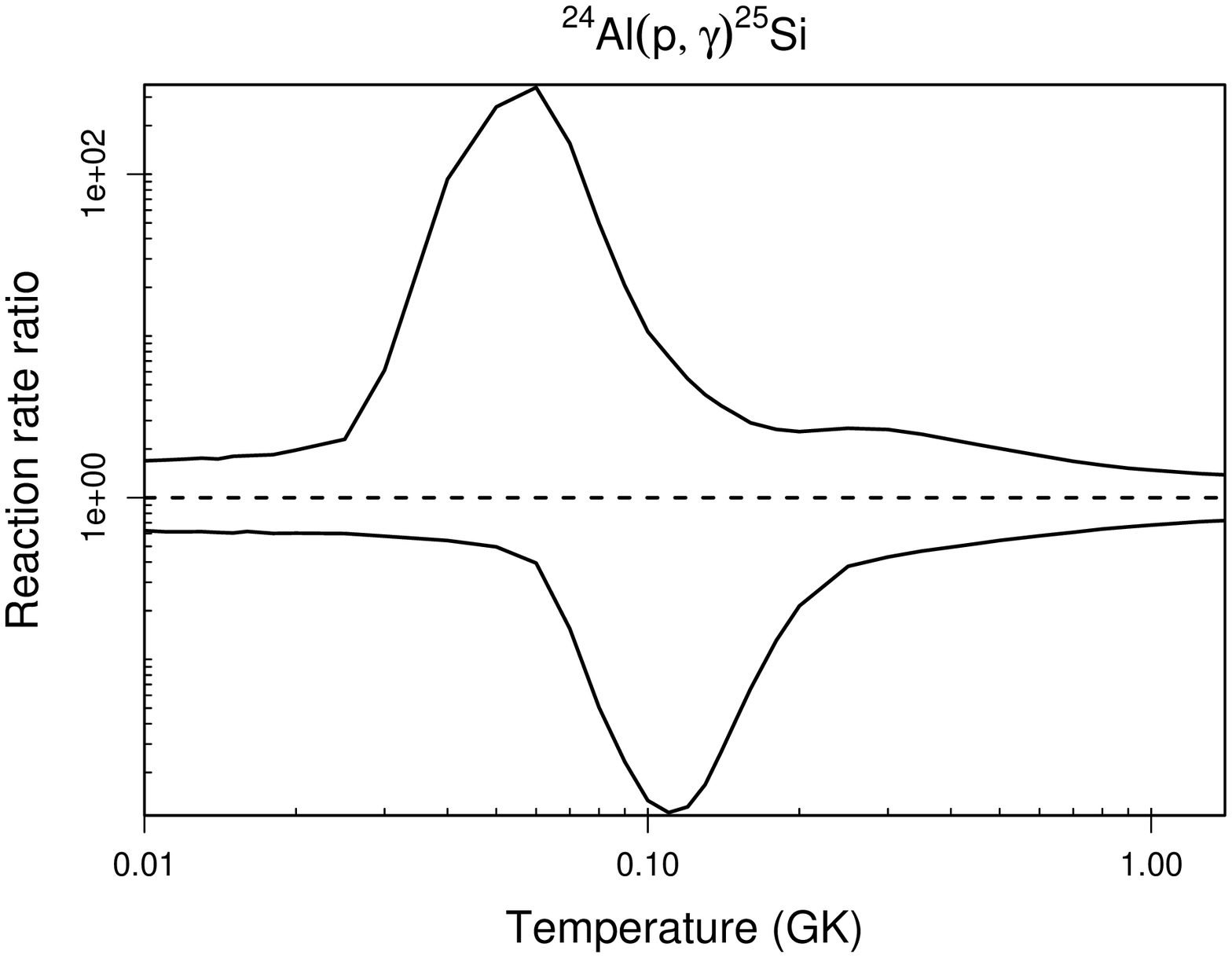}
\end{figure}
\clearpage
\begin{figure}[]
\includegraphics[height=18.5cm]{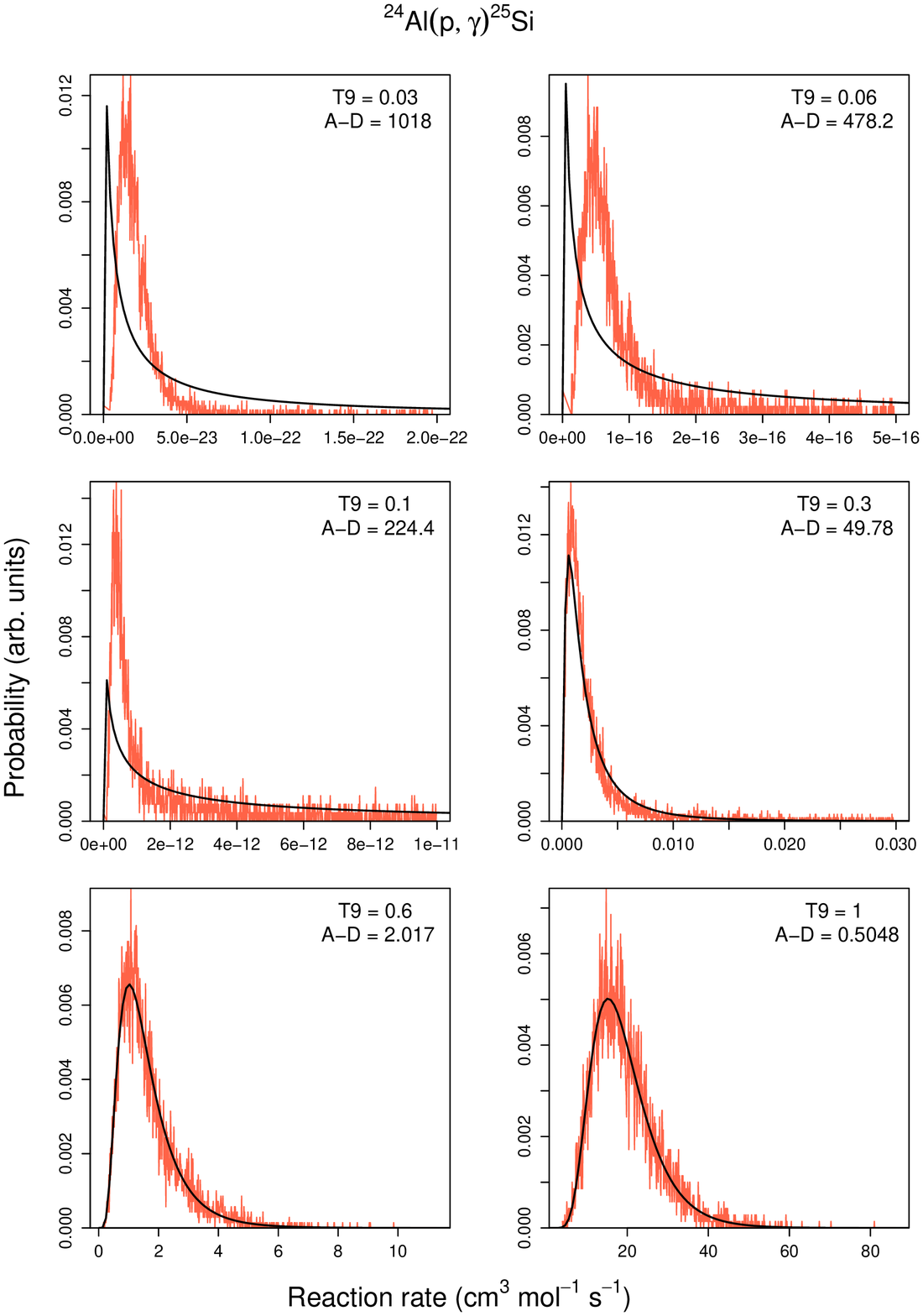}
\end{figure}
\clearpage
\setlongtables

Comments: The value of Q$_{p\gamma}$=5513.7$\pm$0.5 keV is obtained from Eronen et al. \cite{Ero09}. In total, 6 resonances in the range of E$_r^{cm}$=163-965 keV are taken into account. For E$_r^{cm}=163$, 407 and 438 keV the energies are obtained from the adjusted excitation energies of Wrede \cite{Wre09b}. Note that, unlike Ref. \cite{Wre09b}, we prefer to calculate all resonance energies from excitation energies. In particular, we do not use the proton energy from the $\beta$-delayed proton decay of $^{26}$P since it is less precise than the value derived from the excitation energy. For E$_r^{cm}=806$, 882 and 965 keV the energies are found from the excitation energies listed in column 1 of Tab. I in Parpottas et al. \cite{Par04}; however, we added an average value of 8 keV, by which the excitation energies measured in Ref. \cite{Par04} seem to be too low (see comments in Ref. \cite{Wre09b}). For the spin and parity assignments of the first three resonances, we follow the suggestions of Parpottas et al. \cite{Par04} and Bardayan et al.  \cite{Bar06}. In particular, the assignments advocated in Caggiano et al. \cite{Cag02} and Bardayan et al. \cite{Bar02} disagree with the measured $^{28}$Si(p,t)$^{26}$Si angular distribution of the E$_{x}$=5921 keV level (Bardayan et al. \cite{Bar06}). This level has most likely a spin and parity of J$^{\pi}$=3$^{+}$. For the last three resonances the J$^{\pi}$ values are uncertain, but note that the reported values of 2$^+$, 2$^+$, 0$^+$ \cite{Par04} are inconsistent both with the known level scheme of the $^{26}$Mg mirror and the shell model. Spectroscopic factors, $\gamma$-ray partial widths and the direct capture S-factor are adopted from Iliadis et al. \cite{Ili96}. In particular, we use for the former two quantities the shell-model values listed in their Tab. XI. The resonance at E$_r^{cm}$=5 keV, corresponding to a $^{26}$Si level at E$_x$=5518 keV, makes a negligible contribution to the total rates. 
\begin{figure}[]
\includegraphics[height=8.5cm]{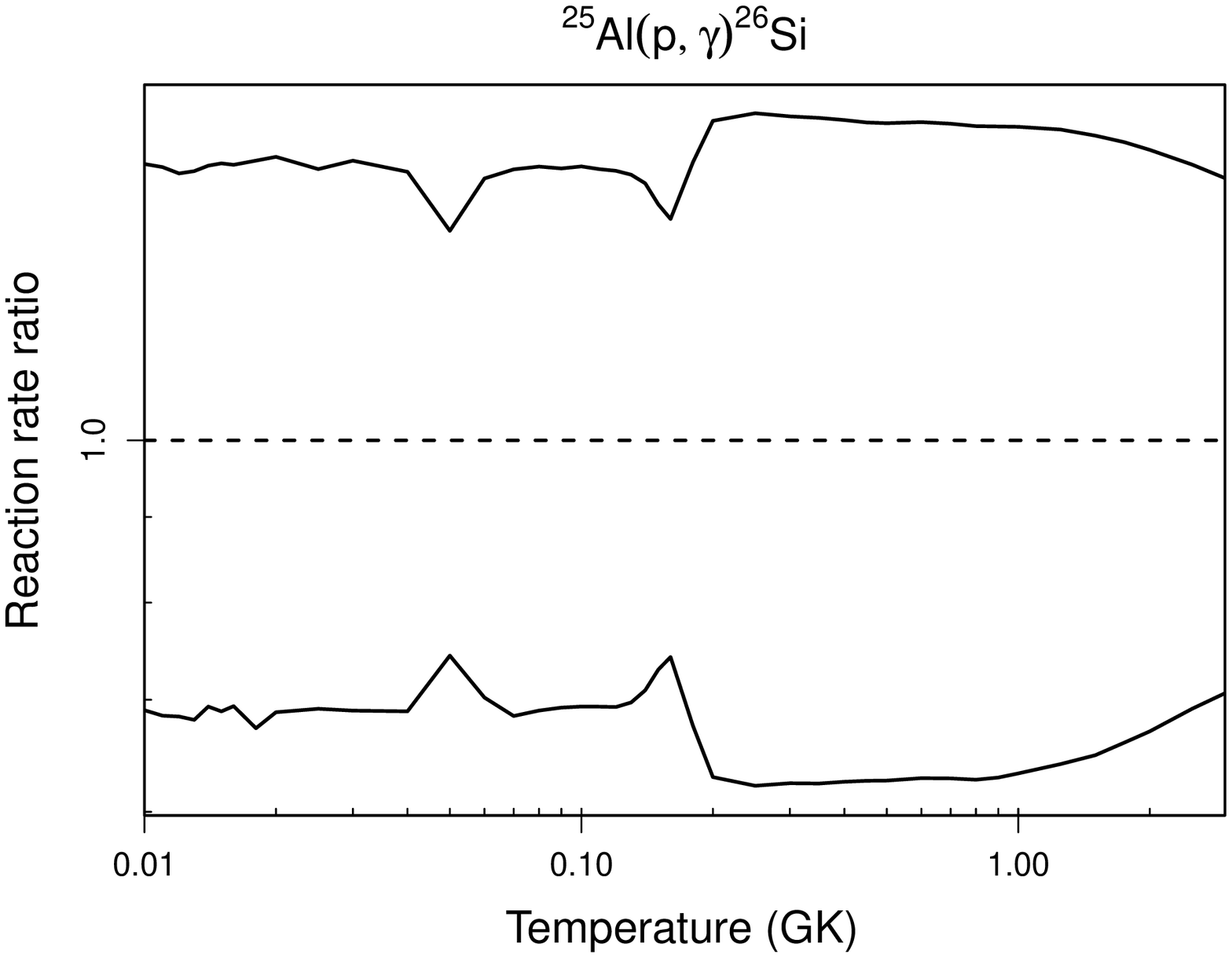}
\end{figure}
\clearpage
\begin{figure}[]
\includegraphics[height=18.5cm]{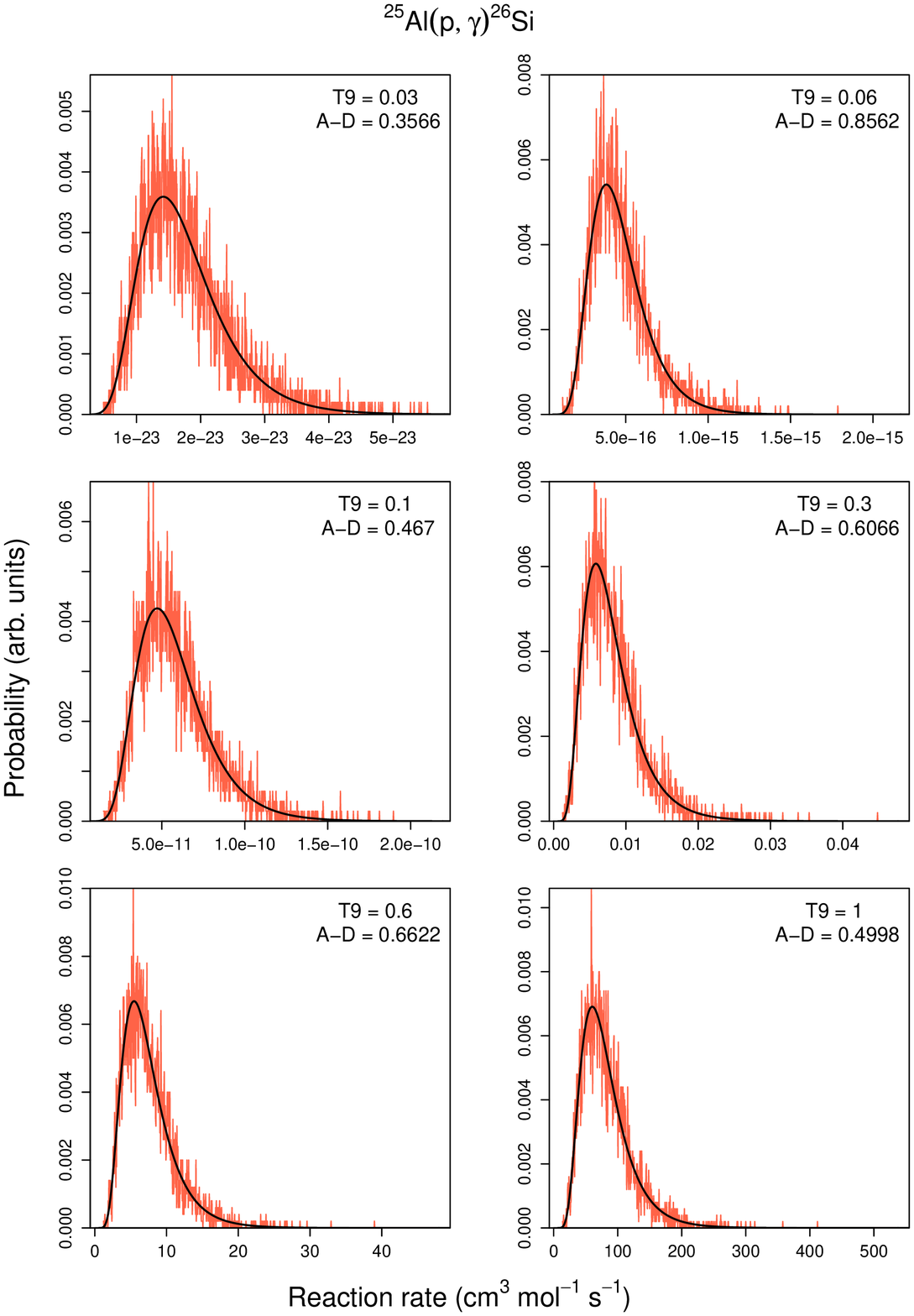}
\end{figure}
\clearpage
\setlongtables

Comments: Measured resonance energies and strengths are reported in Vogelaar \cite{Vog89}. For the strength of the E$_r^{cm}$=189 keV resonance, the weighted average of  the values presented in Vogelaar \cite{Vog89} and Ruiz et al. \cite{Rui06} has been adopted. For almost all resonances below E$_r^{cm}=500$ keV, including the E$_r^{cm}$=189 keV resonance and the threshold states, resonance energies are computed from the excitation energies determined by Lotay et al. \cite{Lot09}. For unobserved low-energy resonances, we compute proton partial widths using C$^2$S$\leq$1 and $\ell=0$, except: (i) for E$_r^{cm}=68$ keV, for which the lowest possible orbital angular momentum is $\ell=2$ \cite{Lot09}; (ii) for E$_r^{cm}=127$ keV, for which an s-wave spectroscopic factor of C$^2$S$\leq$0.002 has been measured by Ref. \cite{Vog96} (assuming $J^{\pi}=9/2^+$ \cite{Lot09}); and (iii) for E$_r^{cm}=231$ keV, where a proton width upper limit has been computed from the experimental upper limit of the resonance strength \cite{Vog89} (assuming $J^{\pi}=5/2^+$ \cite{Lot09}). In total, 19 resonances with E$_r^{cm}$=6-895 keV are taken into account. The direct capture component is adopted from the calculation of Champagne et al. \cite{Cha93}. 
\begin{figure}[]
\includegraphics[height=8.5cm]{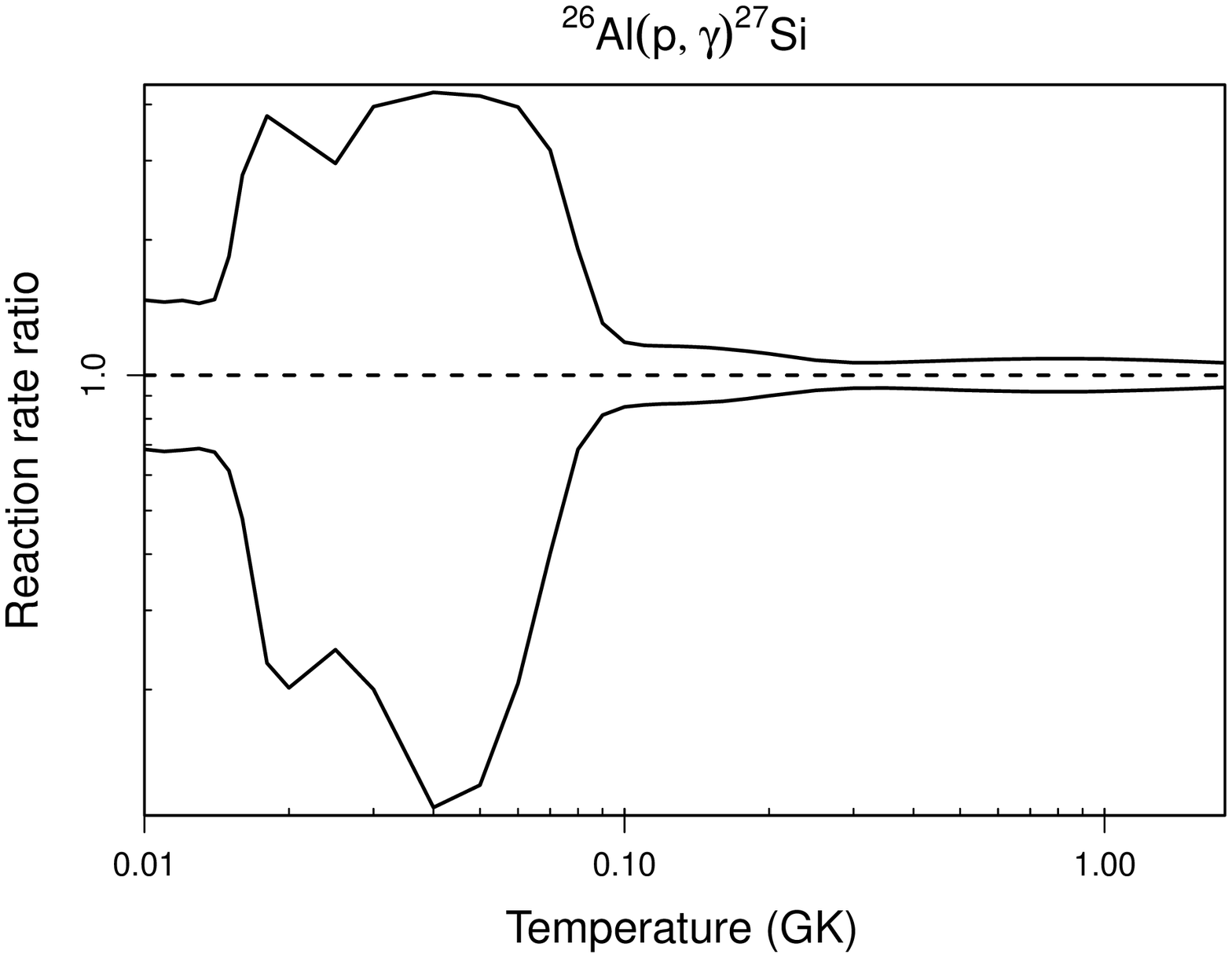}
\end{figure}
\clearpage
\begin{figure}[]
\includegraphics[height=18.5cm]{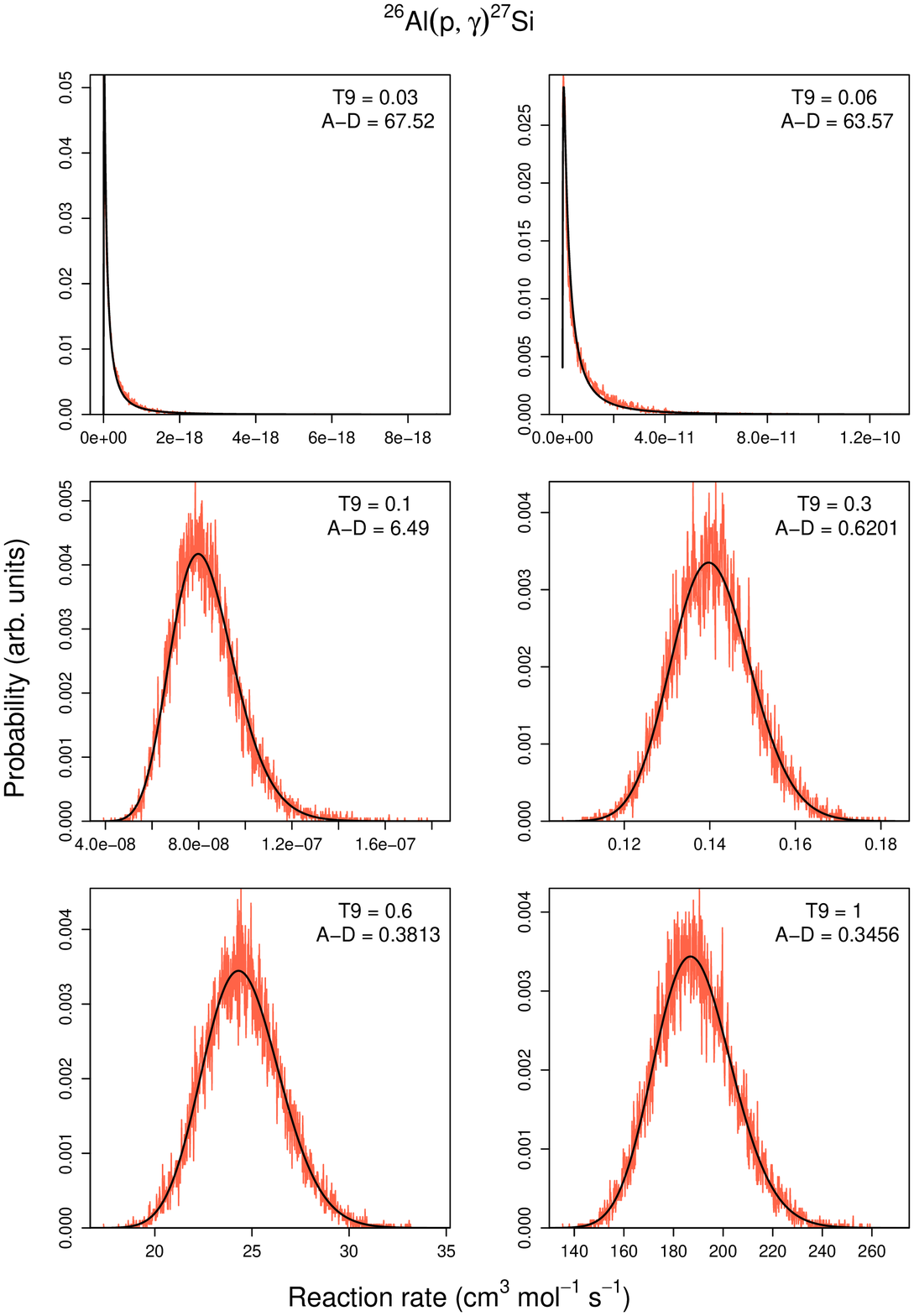}
\end{figure}
\clearpage
\setlongtables

Comments: A total of 109 resonances at energies of E$_r^{cm}=72-3818$ keV are taken into account for calculating the reaction rates. The strengths of measured resonances (E$_r^{cm}\ge196$ keV \cite{End98, Chr99, Har00}) are renormalized to the standard strengths listed in Tab. 1 of Iliadis et al. \cite{Ili01}. For the contribution of threshold states and direct capture, see the comments in Ref. \cite{Ili01}.   
\begin{figure}[]
\includegraphics[height=8.5cm]{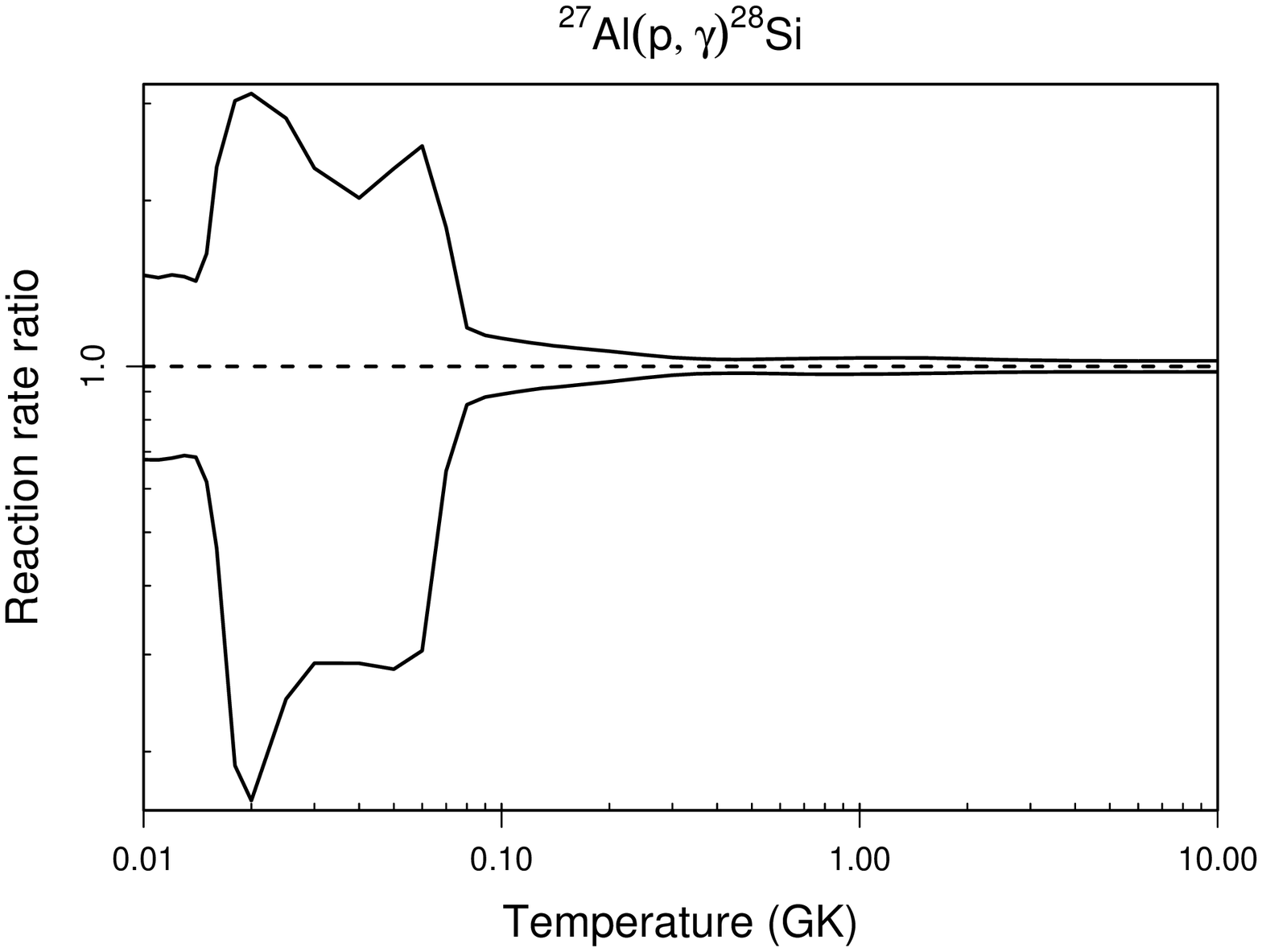}
\end{figure}
\clearpage
\begin{figure}[]
\includegraphics[height=18.5cm]{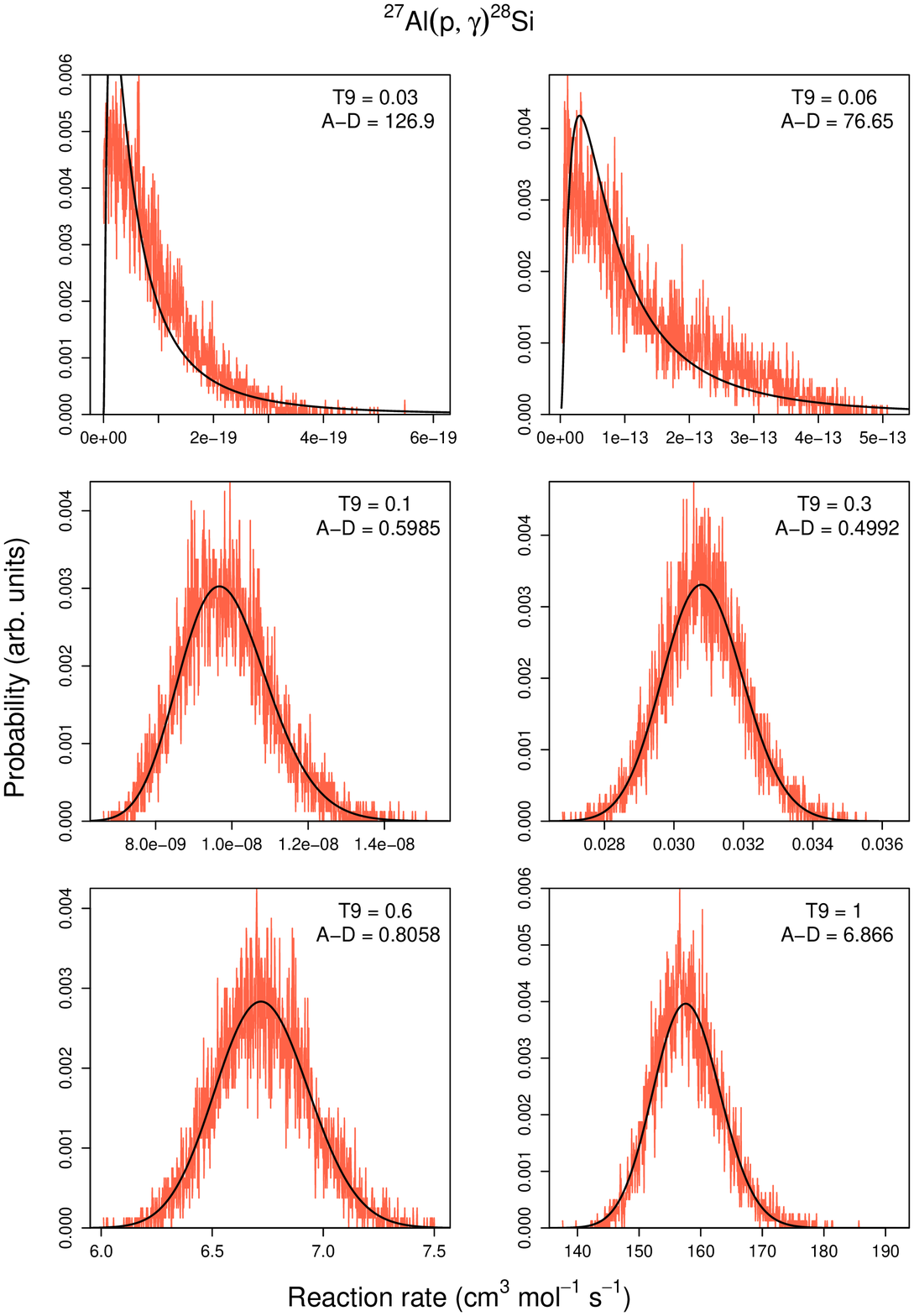}
\end{figure}
\clearpage
\setlongtables

Comments: A total of 91 resonances at energies of E$_r^{cm}=72-2966$ keV are taken into account for calculating the reaction rates. The strengths of measured resonances (E$_r^{cm}\ge487$ keV) are adopted from Endt \cite{End98}. Note that no carefully measured (that is, standard) strength exists for this reaction. For the contribution of threshold states, see the comments in Iliadis et al. \cite{Ili01}. The levels at E$_x=11933.5$, 11985 and 12015.2 keV (E$_r^{cm}=348$, 400, 431 keV) have ambiguous $J^{\pi}$ assignments. Here we disregard these states based on the unnatural parity assignments ($5^+_3$, $1^+_5$, $3^+_6$) from the shell model; see Endt and Booten \cite{EnB93}. 
\begin{figure}[]
\includegraphics[height=8.5cm]{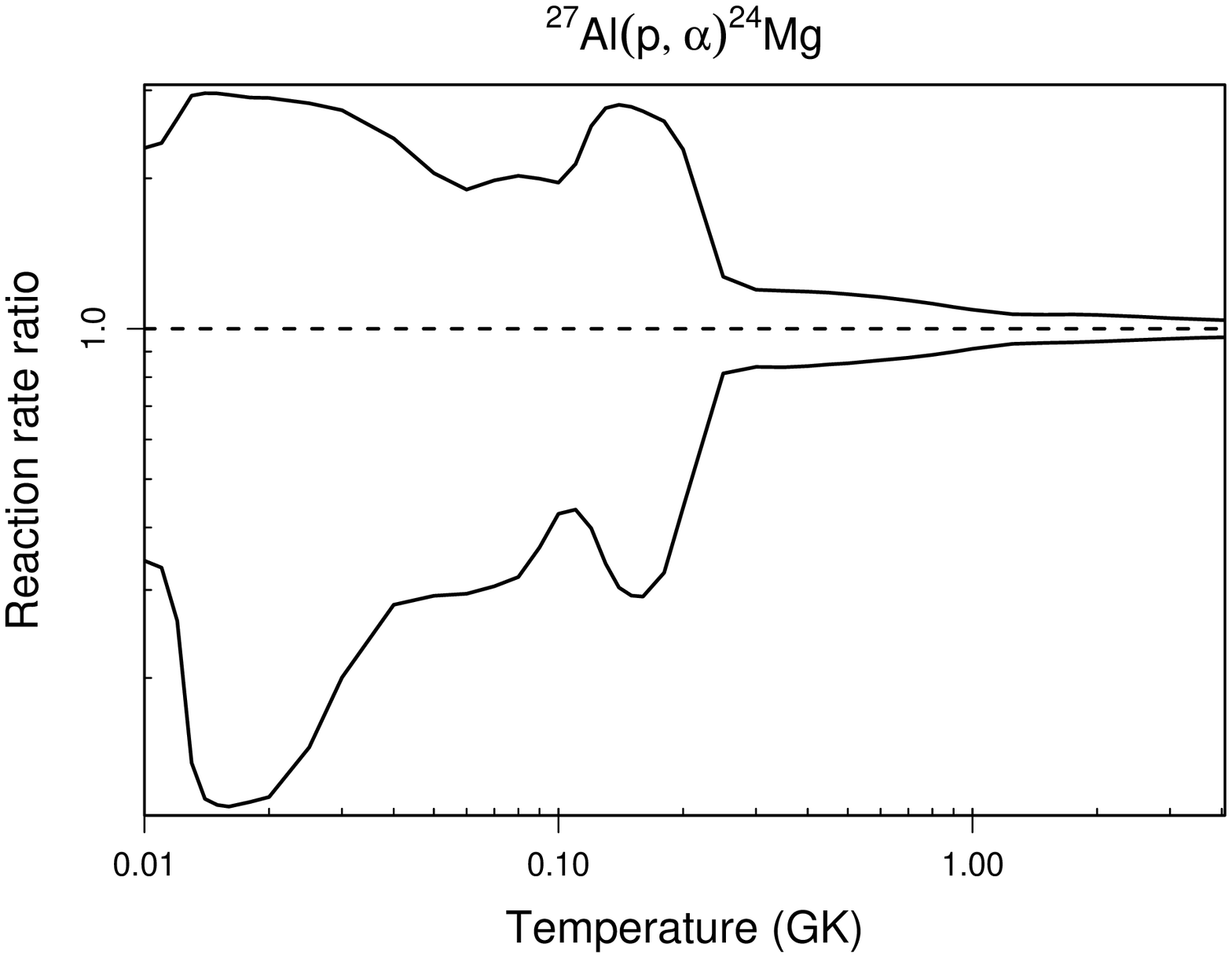}
\end{figure}
\clearpage
\begin{figure}[]
\includegraphics[height=18.5cm]{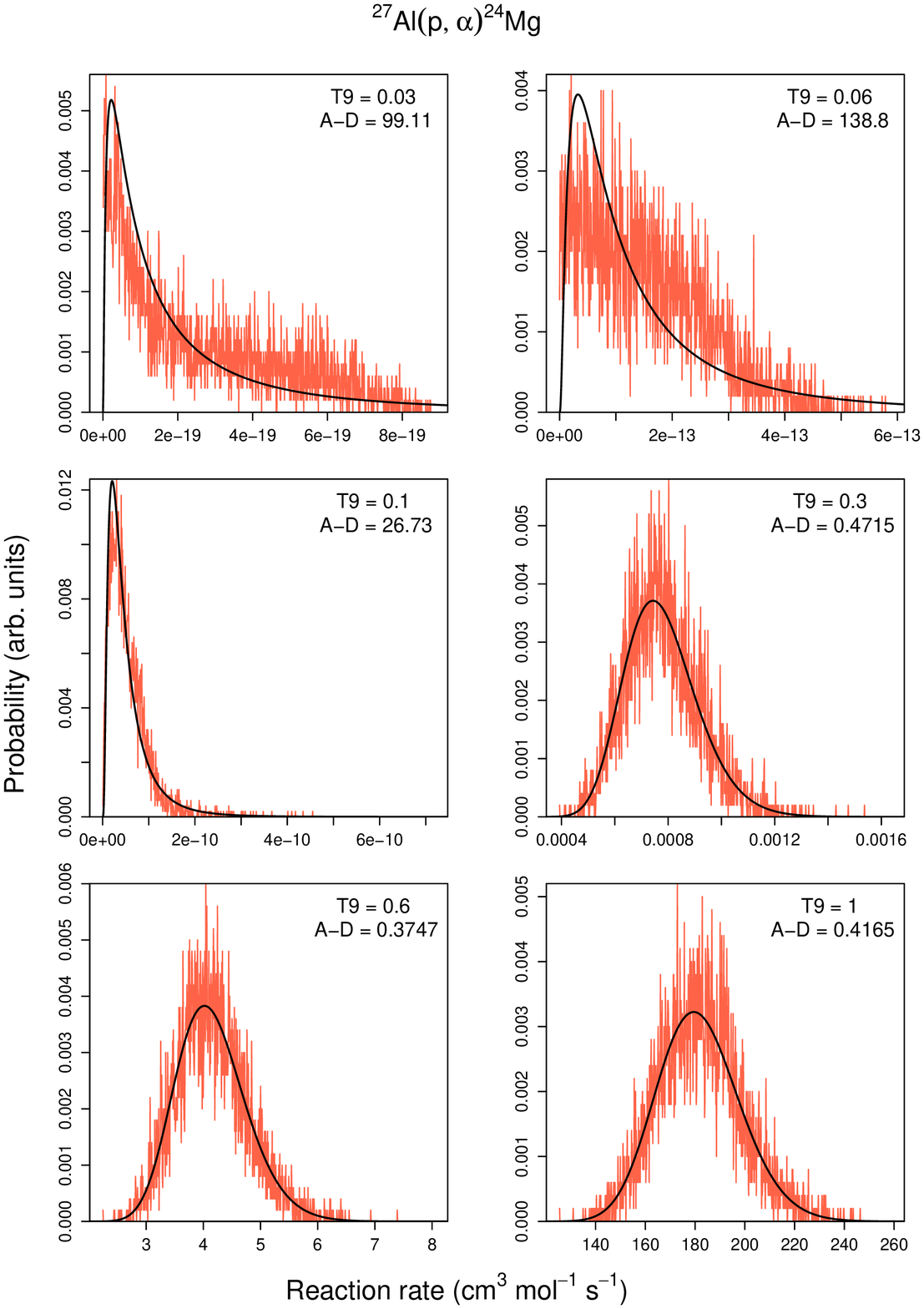}
\end{figure}
\clearpage
\setlongtables

Comments: The contributions from the direct capture to the ground state and from 3 resonances are taken into account for the calculation of the total rates. The direct capture S-factor is obtained here using the experimental spectroscopic factor of $C^2S=0.58$ \cite{End77} from the $^{27}$Mg mirror state. Our S-factor is slightly lower than what has been reported by Guo et al. \cite {Guo06} (using the ANC method) and slightly higher than the shell model based value of Herndl et al. \cite{Her95}. The resonances are located at E$_r^{cm}=259\pm28$, 772$\pm$33 and 1090$\pm$100 keV. The energy of the first resonance (3/2$^+_1$) is calculated using a measured excitation energy of E$_x=1120\pm8$ keV \cite{Gad08} (from $\gamma$-ray spectroscopy) which disagrees with the previously reported value by Caggiano et al. \cite{Cag01}. The energy of the second resonance ($5/2^+_1$) is adopted from Ref. \cite{Cag01}. The energy of the third resonance ($5/2^+_2$) is a rough estimate that is based on the excitation energy of E$_x=1951$ keV listed in Moon et al.  \cite{Moo05} (which, in turn, was extracted from Fig. 4 of Ref. \cite{Cag01}). All proton partial widths are computed using experimental spectroscopic factors from the $^{27}$Mg mirror levels \cite{End77}. The $\gamma$-ray partial widths are either adopted from the shell model \cite{Cag01} or are calculated using the measured lifetime of the mirror state. Note that no other resonances are expected to occur below an energy of E$_r^{cm}=2234$ keV, according to the study of Moon et al. \cite{Moo05}.
\begin{figure}[]
\includegraphics[height=8.5cm]{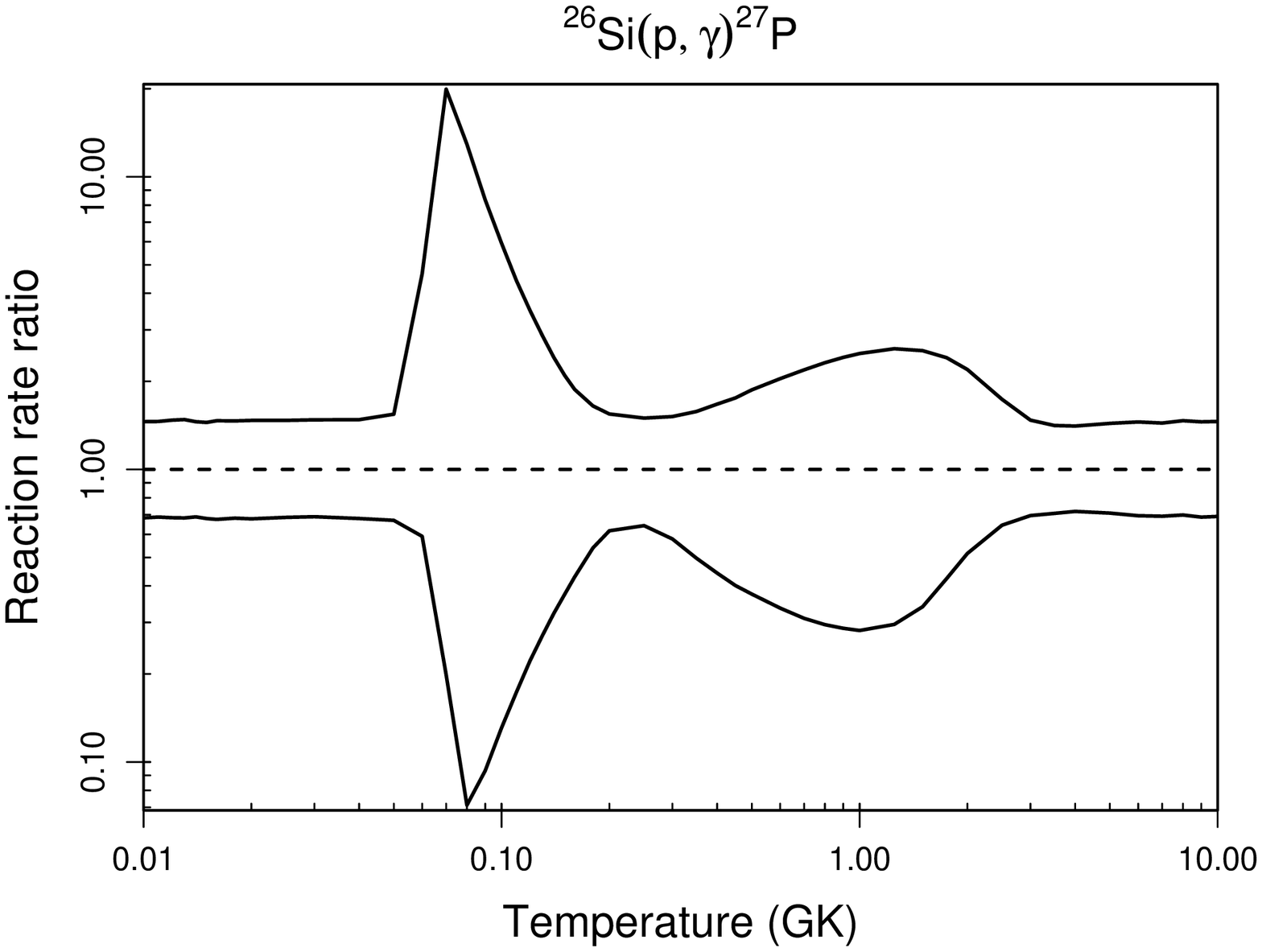}
\end{figure}
\clearpage
\begin{figure}[]
\includegraphics[height=18.5cm]{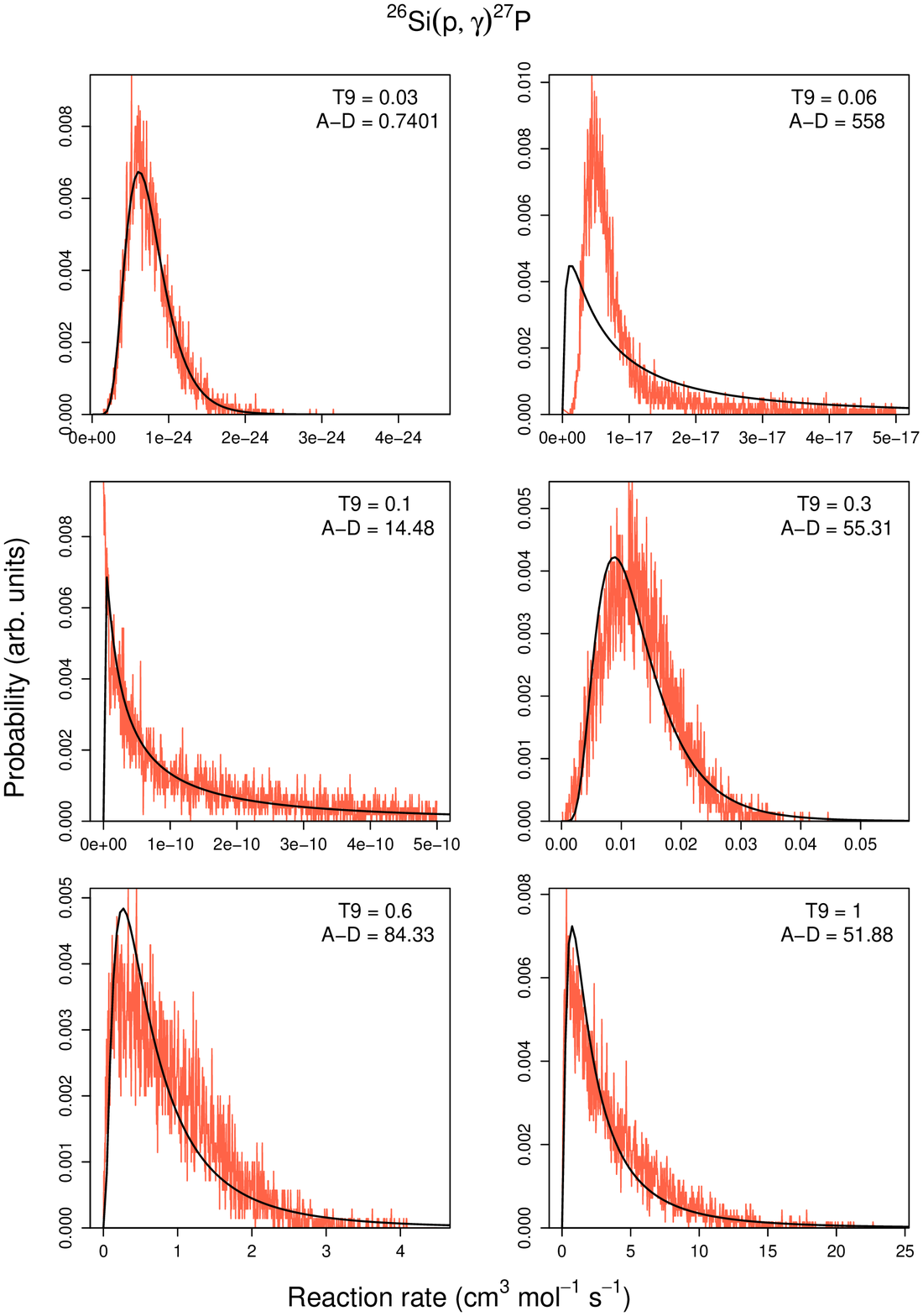}
\end{figure}
\clearpage
\setlongtables

Comments: The reaction rate, including uncertainties, is calculated from the same input information as in
Iliadis et al. \cite{Ili99}.
\begin{figure}[]
\includegraphics[height=8.5cm]{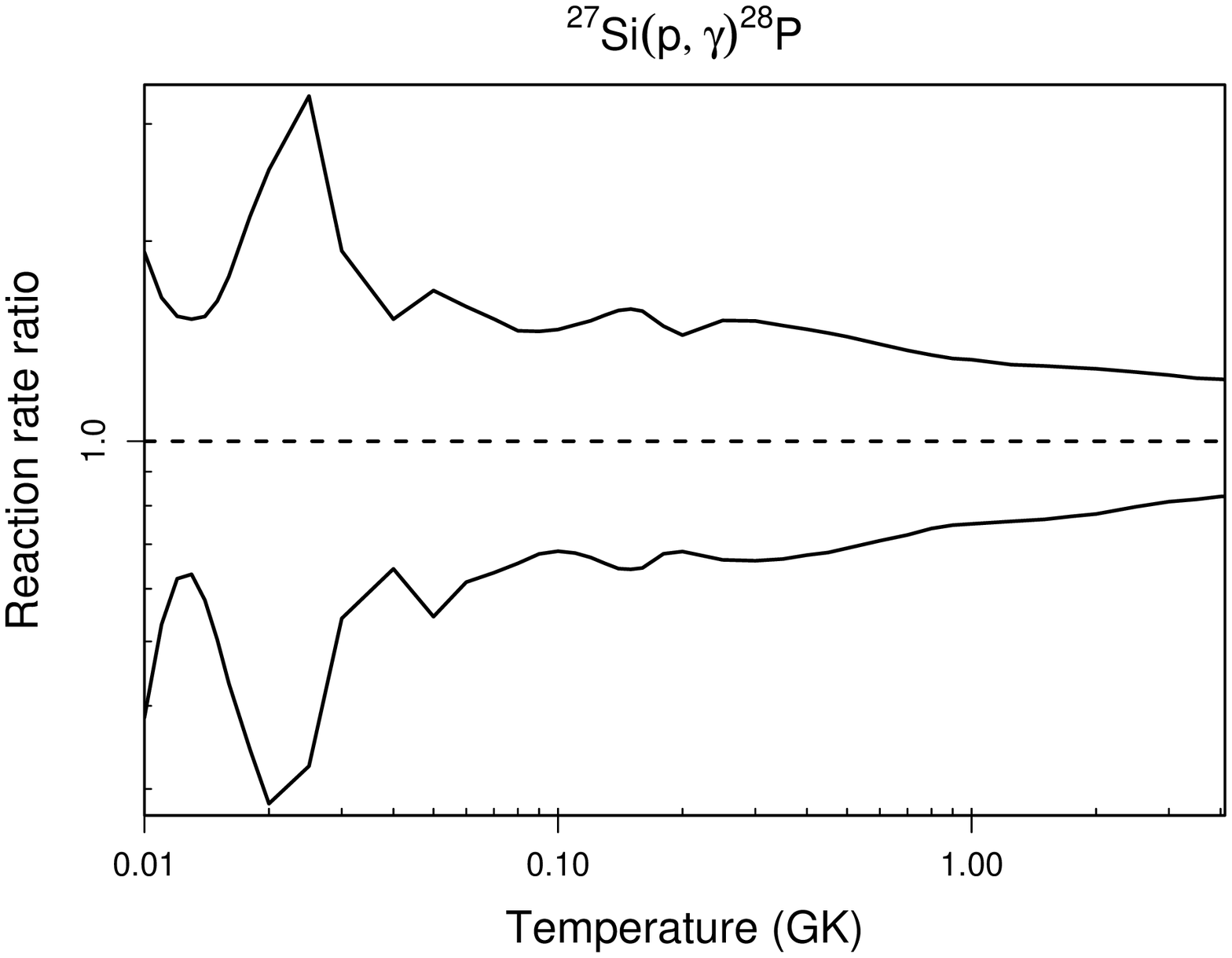}
\end{figure}
\clearpage
\begin{figure}[]
\includegraphics[height=18.5cm]{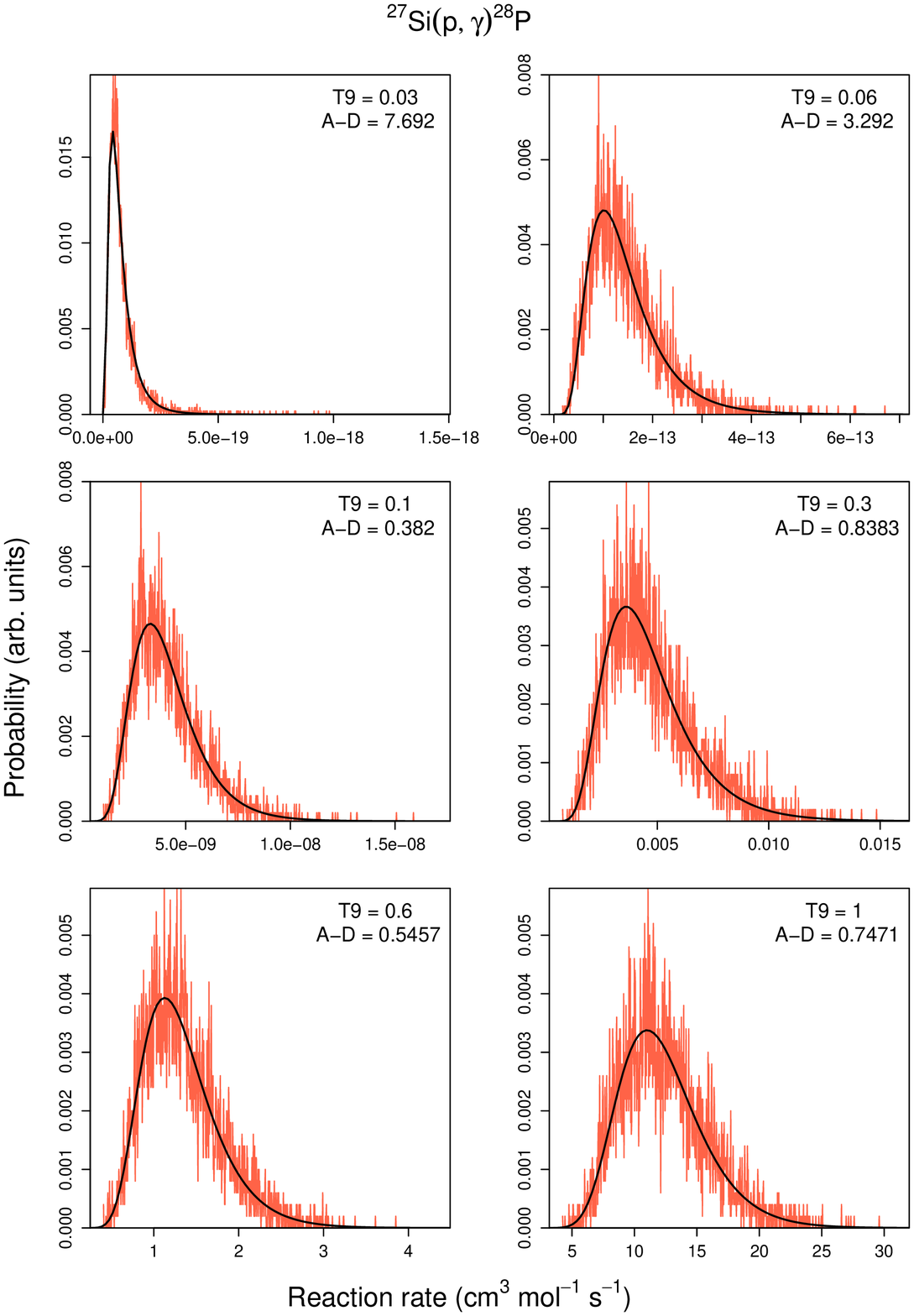}
\end{figure}
\clearpage
\setlongtables

Comments: In total, 10 resonances with E$_r^{cm}$=357-2991 keV are taken into account for the estimation of the rates. The resonance energies are calculated from the excitation energies listed in Endt \cite{End90} and the Q-value (see Tab. \ref{tab:master}), except for two broad resonances (E$_r^{cm}$=1594 and 2009 keV) whose energies are adopted from Graff et al. \cite{Gra90}. For the resonance strengths we use the average values presented in Angulo et al. \cite{Ang99}. Below E$^{cm}$=1.1 MeV the nonresonant rate contribution is dominated by direct capture and the tail of the broad E$_r^{cm}$=1594 keV resonance. The corresponding S-factor is adopted from Graff et al. \cite{Gra90}. The tail of the E$_r^{cm}$=357 keV resonance is found to be negligible compared to other contributions. The present recommended rates are in agreement with those of Angulo et al. \cite{Ang99}, but our rate uncertainties are smaller.
\begin{figure}[]
\includegraphics[height=8.5cm]{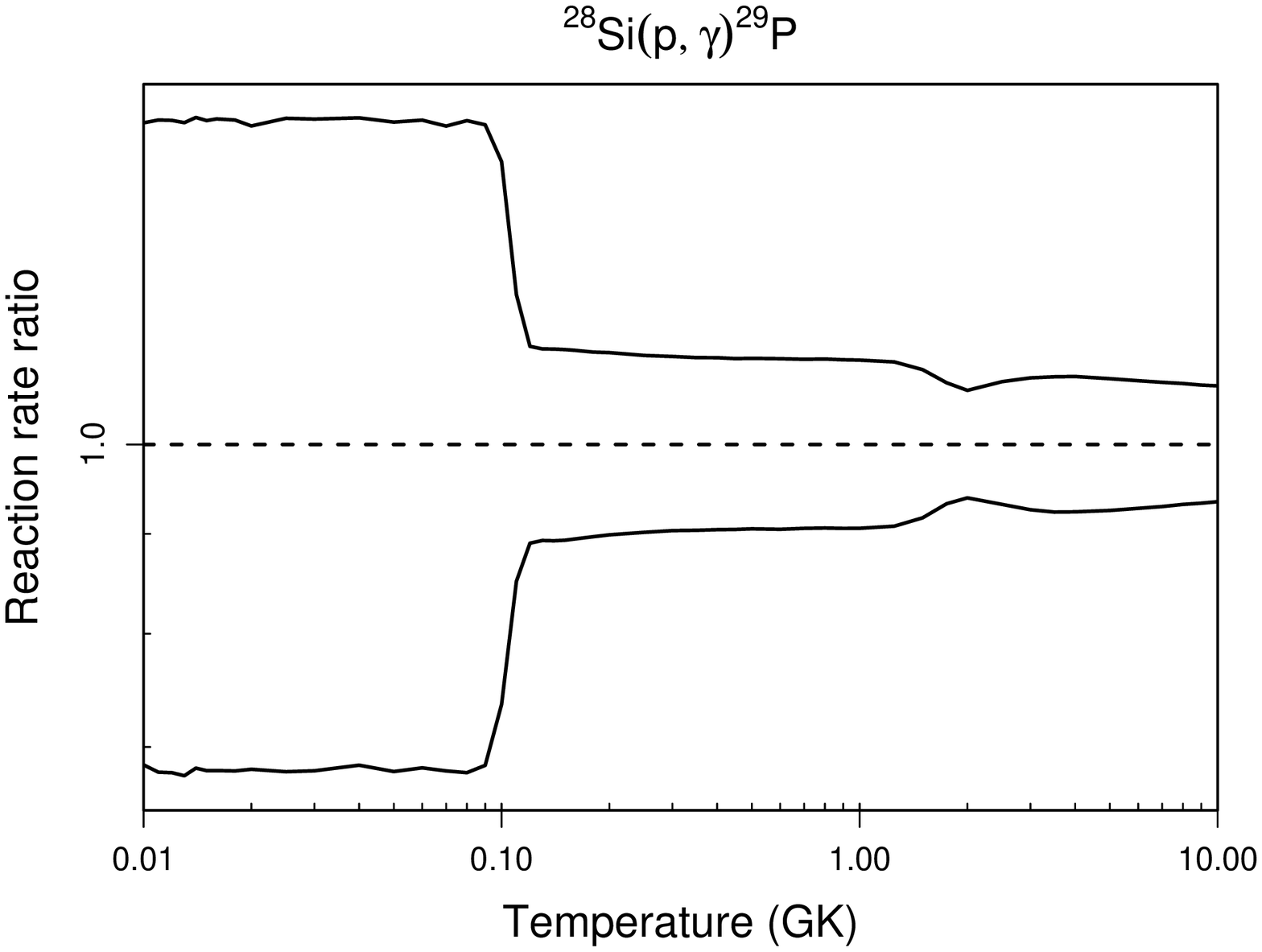}
\end{figure}
\clearpage
\begin{figure}[]
\includegraphics[height=18.5cm]{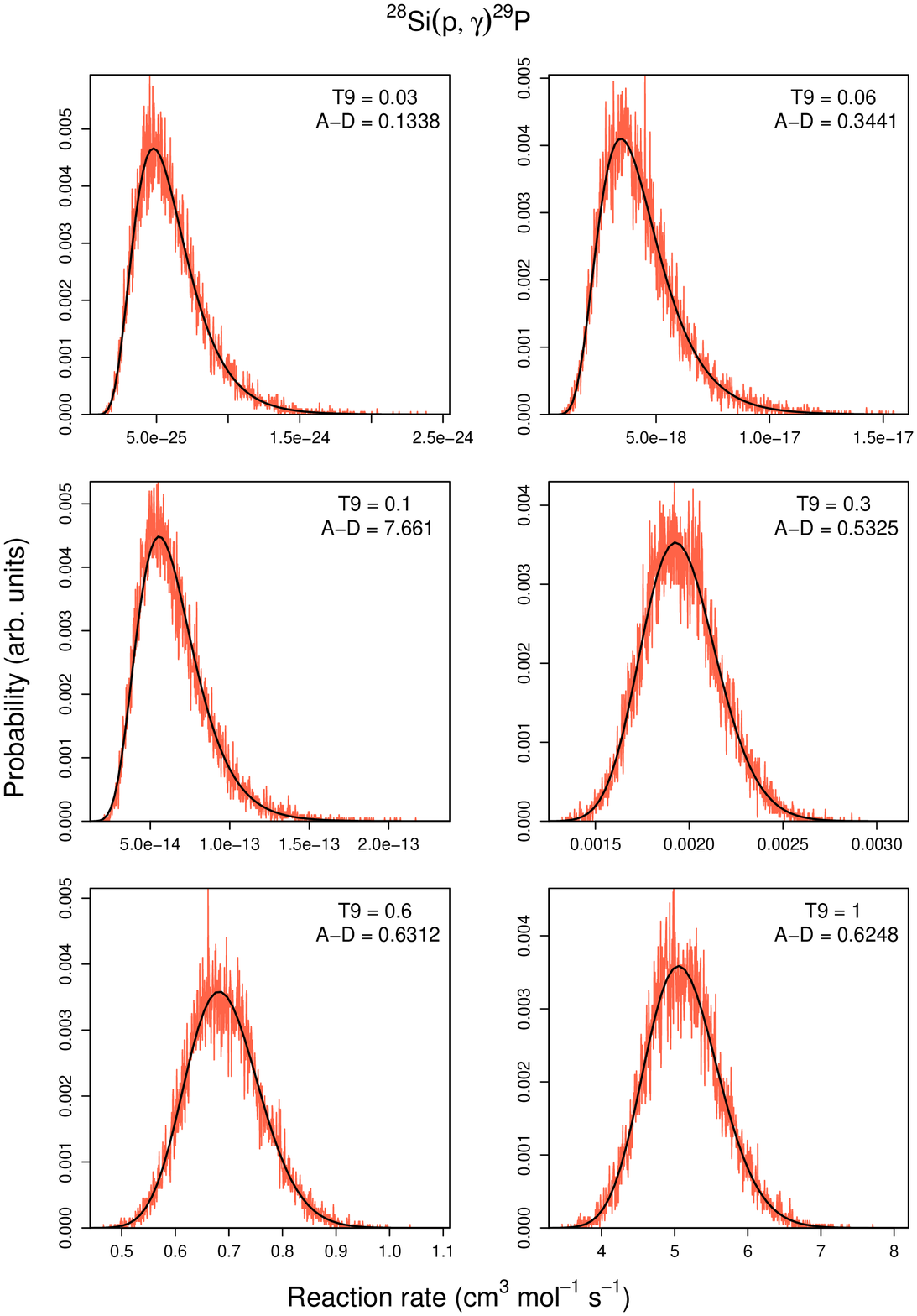}
\end{figure}
\clearpage
\setlongtables

Comments: The same input information as in Iliadis et al. \cite{Ili01} is used for the calculation of the rates, with the exception of the unobserved E$_r^{cm}$=296 keV resonance, which has previously been disregarded in the reaction rate calculation. According to Endt \cite{End98}, the corresponding E$_x$=5890 keV state in $^{30}$P is the analog of the E$_x$=5232 keV state (J$^\pi$;T=3$_2^+$;1) in $^{30}$Si. With this assumption, a spectroscopic factor of S$_{\ell=2}\approx$0.025 can be deduced from the results of a $^{29}$Si(d,p)$^{30}$Si study (Mackh et al. \cite{Mac73}), in reasonable agreement with the shell model value reported in Baxter and Hinds \cite{Bax73}. In total, 79 resonances with energies in the range of E$_r^{cm}$=107-3075 keV are taken into account. 
\begin{figure}[]
\includegraphics[height=8.5cm]{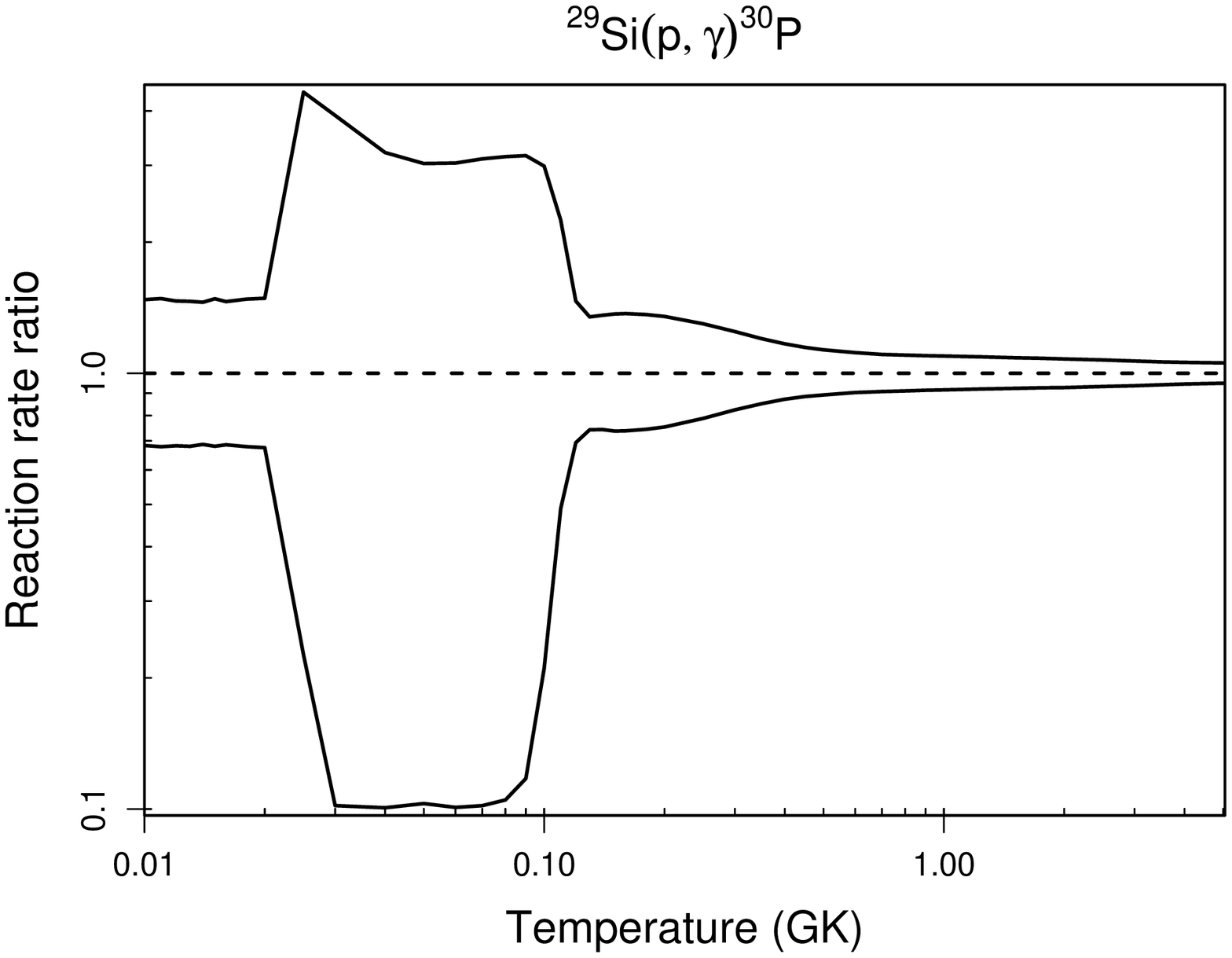}
\end{figure}
\clearpage
\begin{figure}[]
\includegraphics[height=18.5cm]{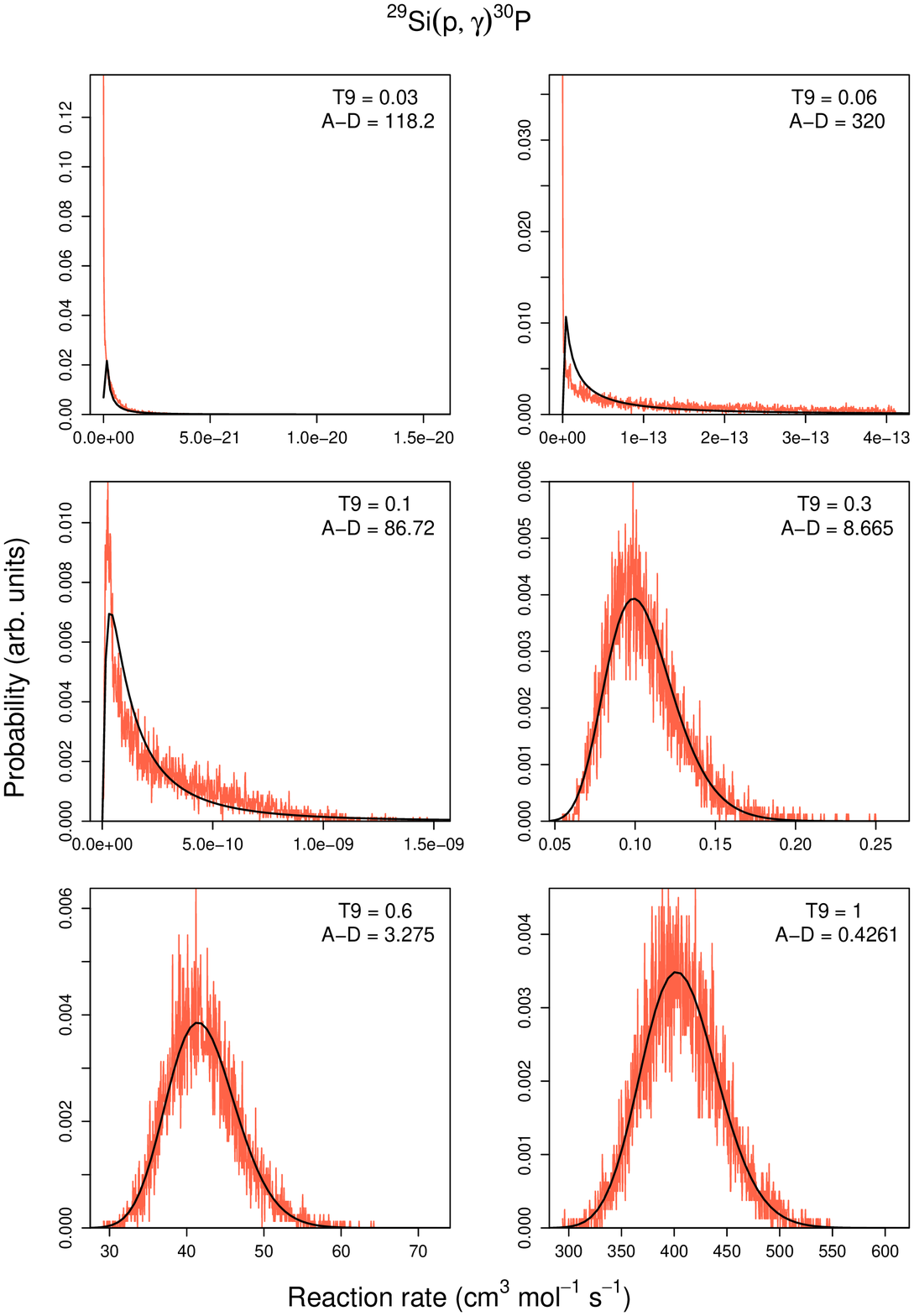}
\end{figure}
\clearpage
\setlongtables

Comments: The same input information as in Iliadis et al. \cite{Ili01} is used for the calculation of the rates, except that the energies and resonance strengths of the threshold states have been adjusted according to a slight change in the reaction Q-value (Audi, Wapstra and Thibault \cite{Aud03}). In total, 97 resonances with energies in the range of E$_r^{cm}$=17-2929 keV are taken into account. 
\begin{figure}[]
\includegraphics[height=8.5cm]{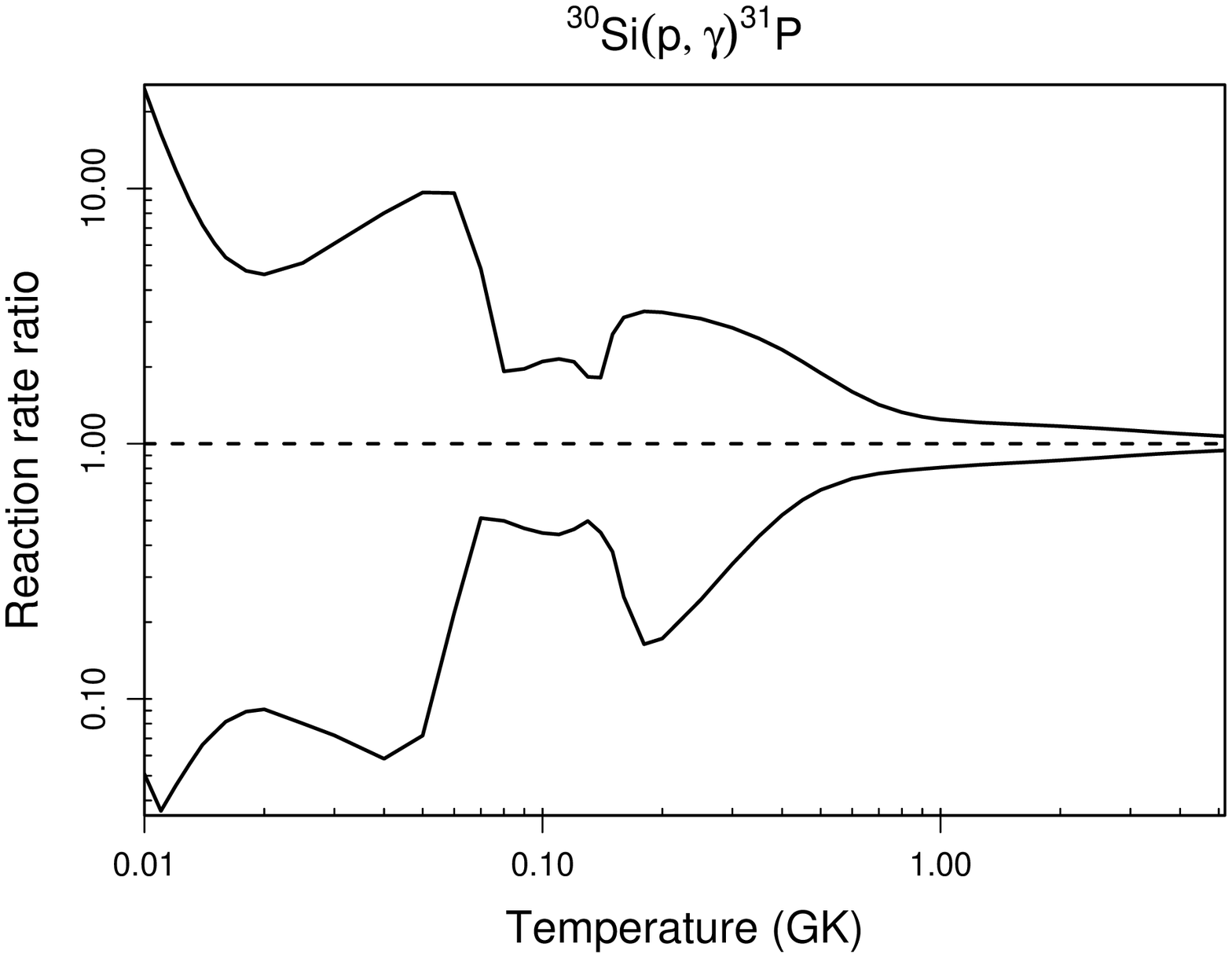}
\end{figure}
\clearpage
\begin{figure}[]
\includegraphics[height=18.5cm]{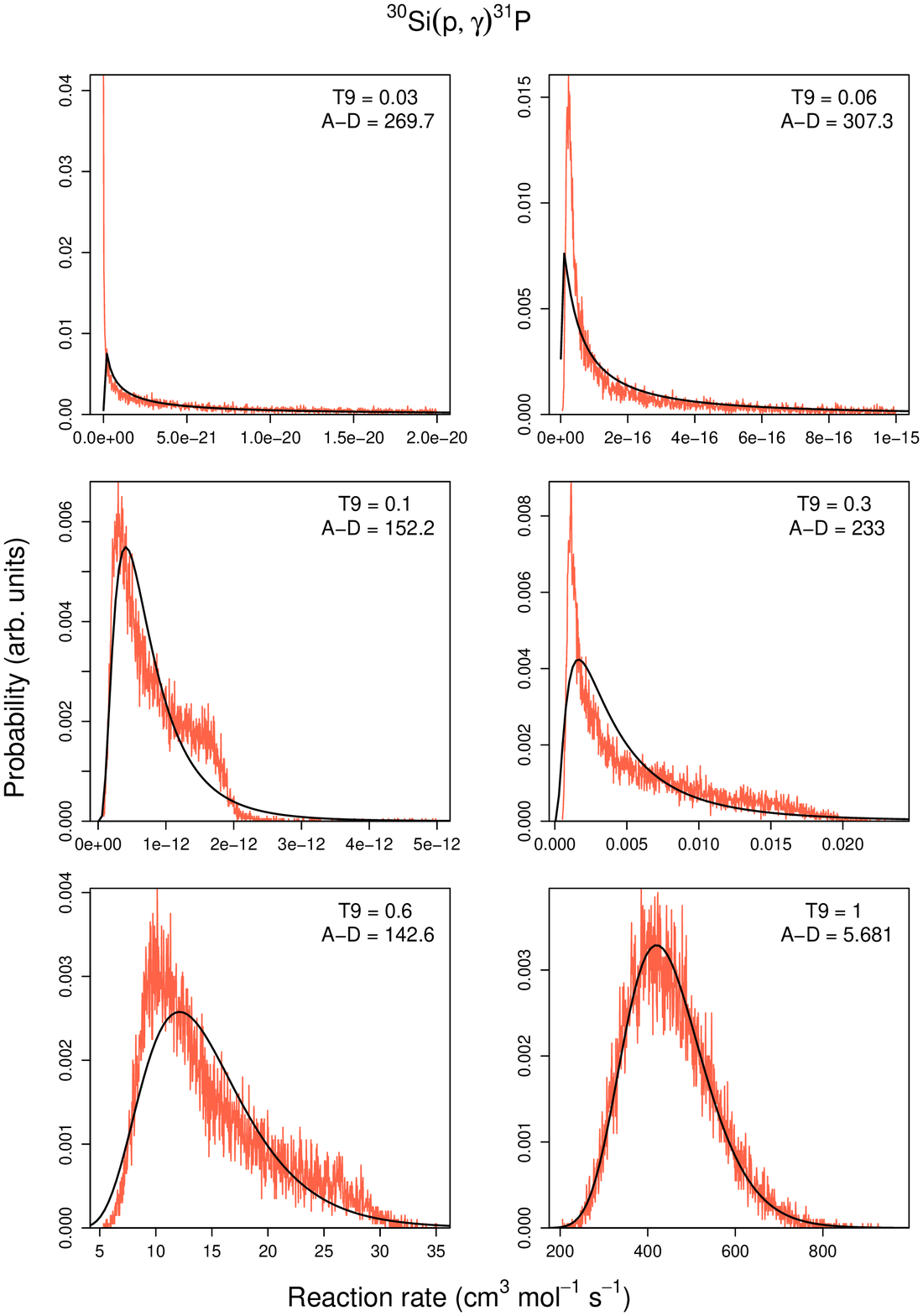}
\end{figure}
\clearpage
\setlongtables

Comments: No unbound states have been observed in the compound nucleus $^{28}$S. Based on the comparison to the structure of the $^{28}$Mg mirror nucleus it can be concluded that the lowest-lying resonance in $^{27}$P+p is expected near a relatively high energy of 1 MeV. In fact, a Coulomb displacement energy calculation \cite{Her95} finds E$_r^{cm}=1100$ keV, which is adopted in the present work; we estimate an uncertainty of 100 keV, although this value should be regarded as a rough guess only. The proton and $\gamma$-ray partial widths are calculated by using the shell model \cite{Her95}. The total reaction rates are dominated by the direct capture process into the ground and first excited states of $^{28}$S at all temperatures of interest. We adopt the direct capture S-factor from Ref. \cite{Her95}, which is based on shell model spectroscopic factors. Higher lying resonances are expected to make a minor contribution to the total rate.
\begin{figure}[]
\includegraphics[height=8.5cm]{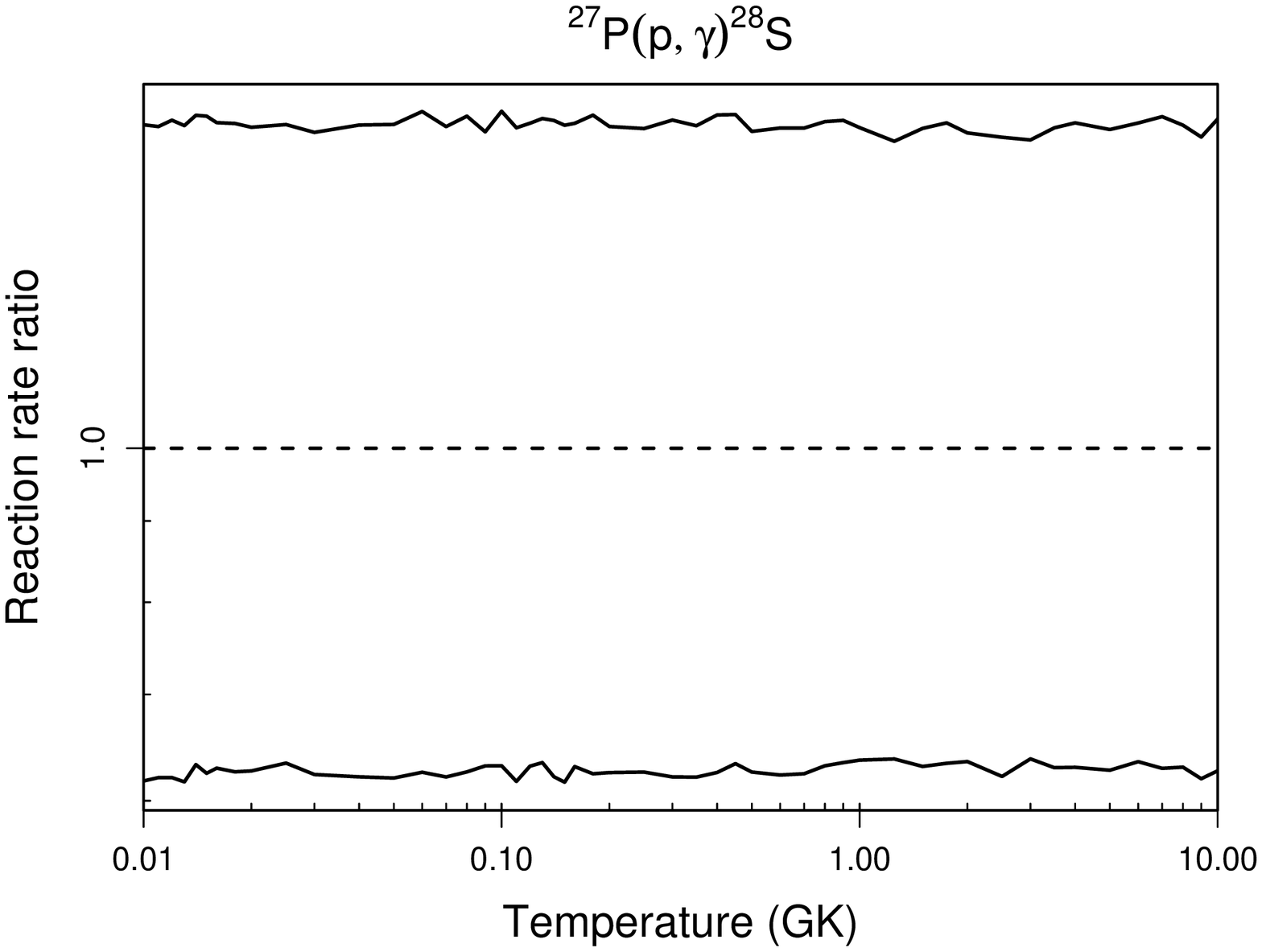}
\end{figure}
\clearpage
\begin{figure}[]
\includegraphics[height=18.5cm]{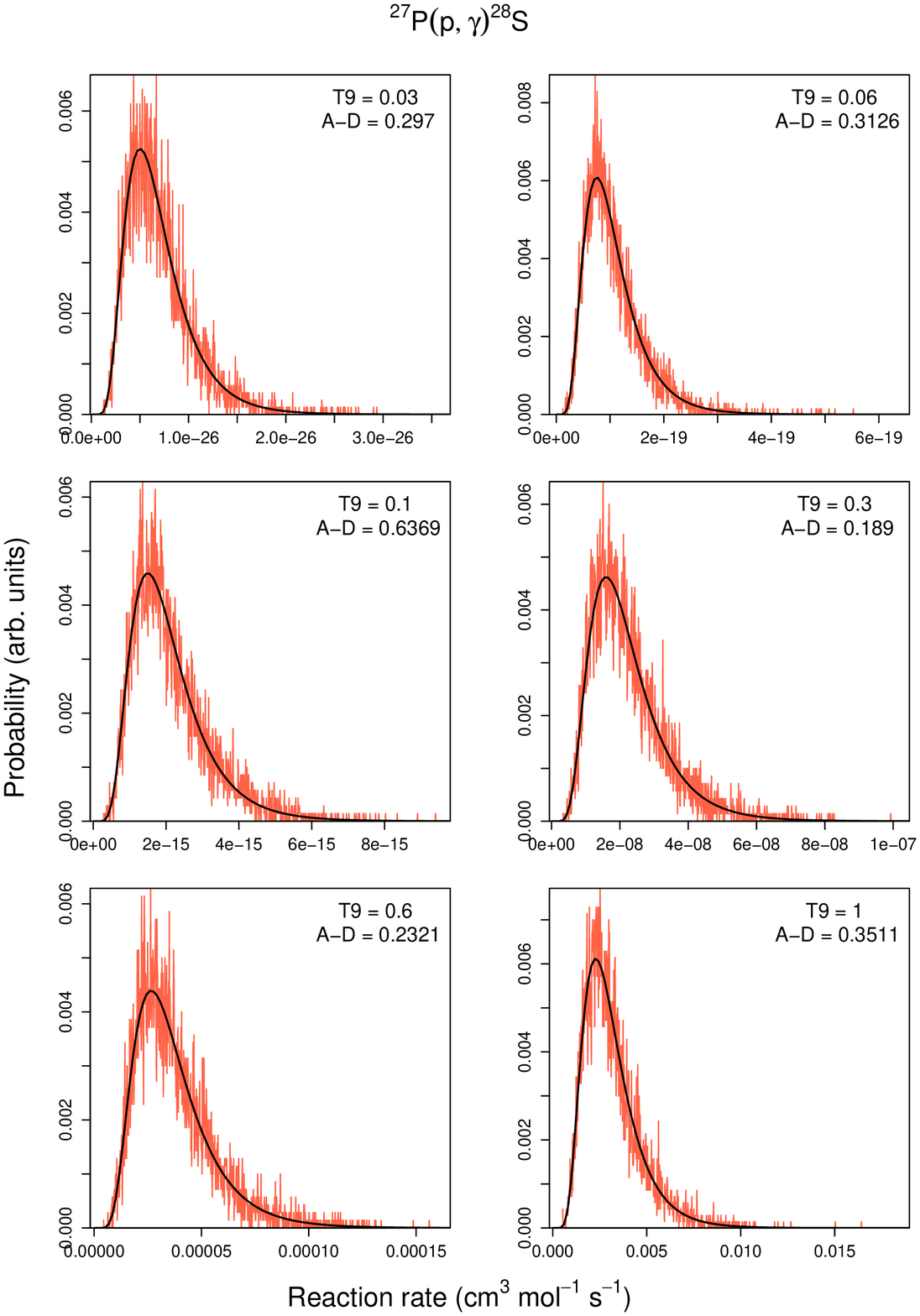}
\end{figure}
\clearpage
\setlongtables

Comments: Altogether 8 resonances at energies of E$_r^{cm}=305-1443$ keV are considered for calculating the reaction rates. Their energies are obtained from measured excitation energies \cite{End90,Bar07} and the reaction Q-value. The exception is the E$_r^{cm}=489$ keV resonance for which the corresponding level has not been observed. Its excitation energy is obtained by applying the Isobaric Multiplet Mass Equation (IMME), with a corresponding uncertainty of about 40 keV (see Iliadis et al. \cite{Ili01} for details). The spins and parities of the resonances are not known unambiguously. They are based here on experimental $J^{\pi}$ restrictions, comparison of mirror reaction cross sections, and the application of the IMME. Note that the present resonance energies and some $J^{\pi}$ assignments differ in general from those of Iliadis et al. \cite{Ili01} since new information became recently available \cite{Fyn00,Bar07}. In particular we assume that the E$_x=5217$ keV level from Fynbo et al. \cite{Fyn00} is the same as the E$_x=5288$ keV (3$^-$) level of Yokota et al.  \cite{Yok82} (comparison of the excitation energies in Refs. \cite{Yok82,Fyn00} shows that the former values are too high by about $30-50$ keV). Obviously, an experimental verification of the present $J^{\pi}$ assignments is desirable. Proton partial widths are calculated by using $C^2S$ values of $^{30}$Si mirror states (we estimate a value of $C^2S=0.02$ for E$_r^{cm}=737$ keV from the published (d,p) angular distribution of Mackh et al. \cite{Mac73}). Gamma-ray partial widths are estimated using measured lifetimes of $^{30}$Si mirror states, except for E$_r^{cm}=990$ keV for which no measured lifetime is available; here we roughly estimate the transition strengths using RUL's and the known $\gamma$-ray branching and mixing ratios of the E$_x=5614$ keV mirror in $^{30}$Si. The direct capture S-factor is also computed using experimental $C^2S$ values of $^{30}$Si mirror states.
\begin{figure}[]
\includegraphics[height=8.5cm]{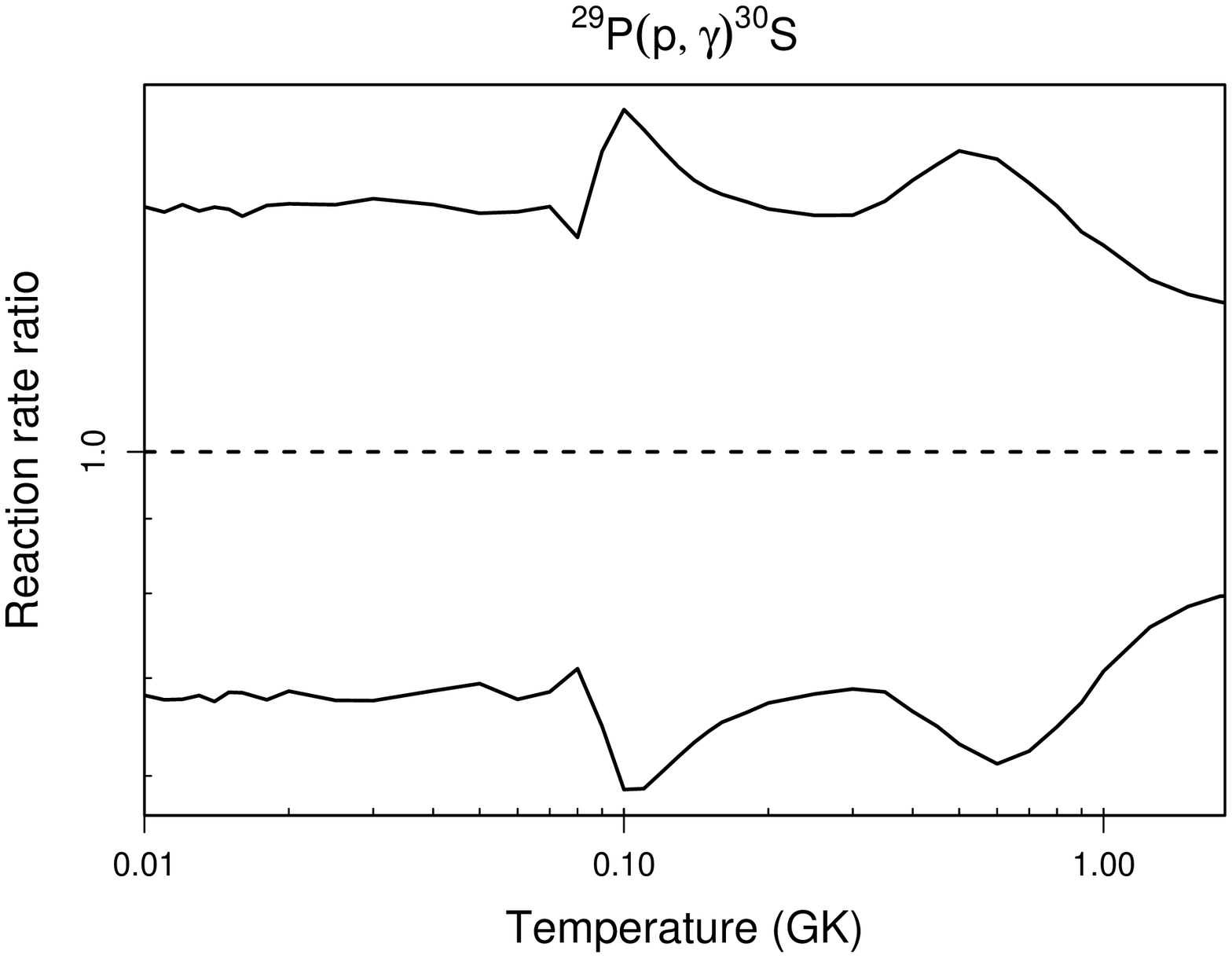}
\end{figure}
\clearpage
\begin{figure}[]
\includegraphics[height=18.5cm]{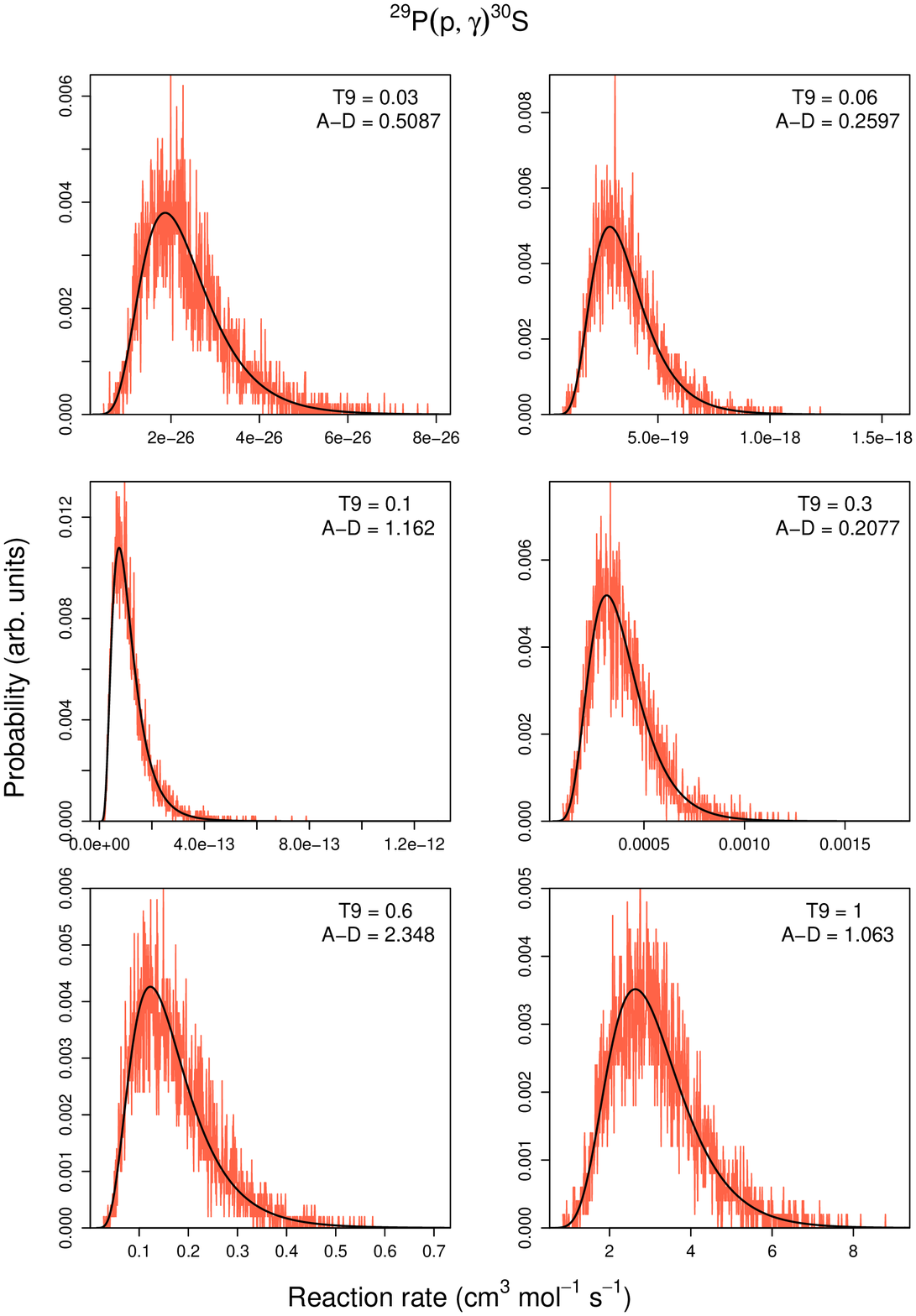}
\end{figure}
\clearpage
\setlongtables

Comments: In total, 41 resonances at E$_r^{cm}=159-1963$ keV are taken into account for the calculation of the total reaction rate. The direct capture S-factor is adopted from Iliadis et al. \cite{Ili93}. Resonance energies and strengths are adopted from Endt \cite{End90,End98}, where the latter values are renormalized using the standard strengths given in Tab. 1 of Iliadis et al. \cite{Ili01}. The subthreshold level at E$_x=8861$ keV (2$^+$) has not been taken into account since its contribution is negligible \cite{Ili93}. For information on unobserved low-energy resonances, see the comments in Ref. \cite{Ili93}. Two levels are omitted in the present work: (i) E$_x=9138$ keV, since its observation has only been reported as a private communication \cite{Kal78} (it has not been observed in proton transfer), and (ii) E$_x=9196$ keV, since Ref. \cite{End98} considers it to be identical to the E$_x=9208$ keV level. The interference of the two 1$^-$ resonances at E$_r^{cm}=372$ and 1468 keV is explicitly taken into account (the interference has an unknown sign and is thus sampled by using a binary probability density; see Sec 4.4 in Paper I).
\begin{figure}[]
\includegraphics[height=8.5cm]{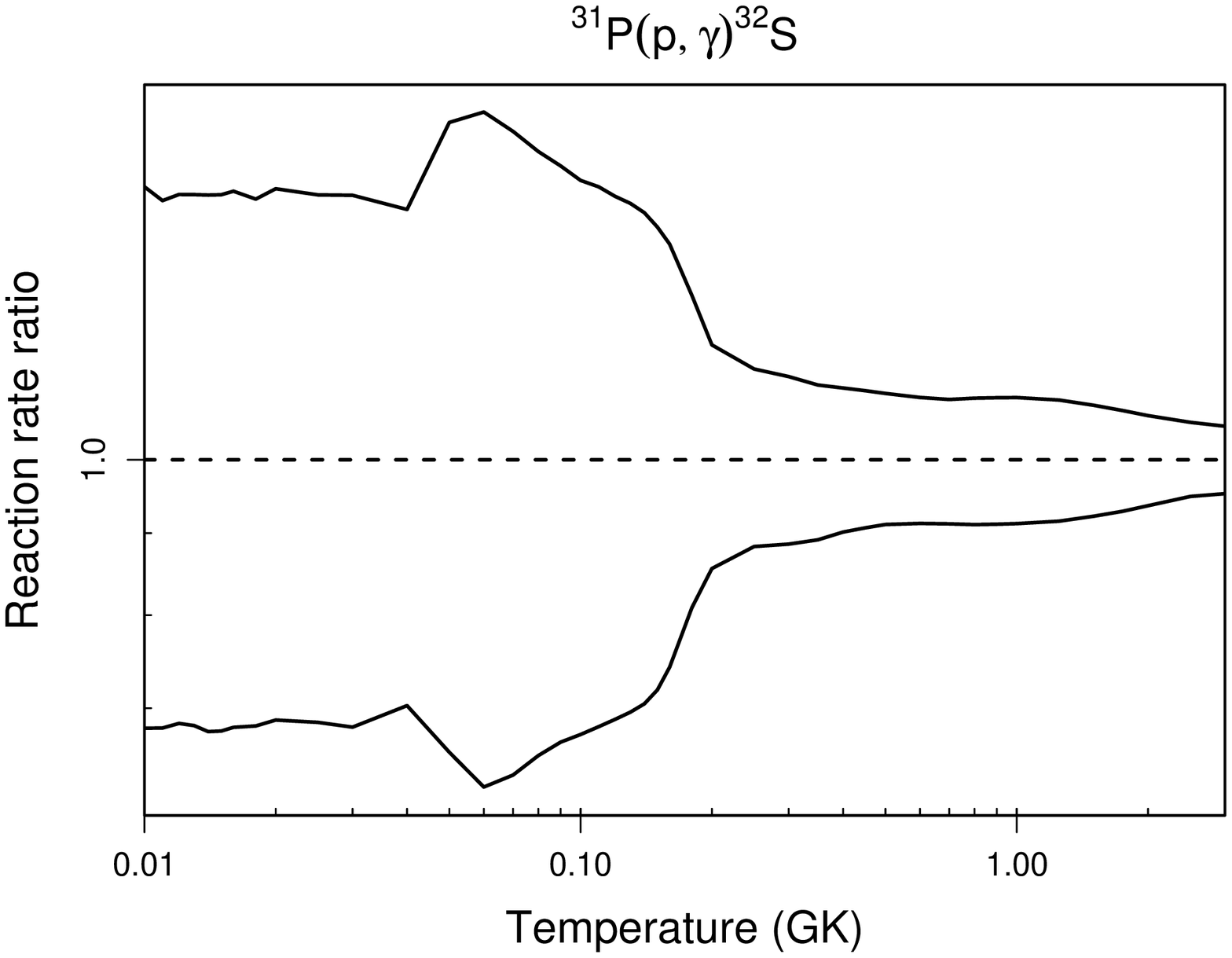}
\end{figure}
\clearpage
\begin{figure}[]
\includegraphics[height=18.5cm]{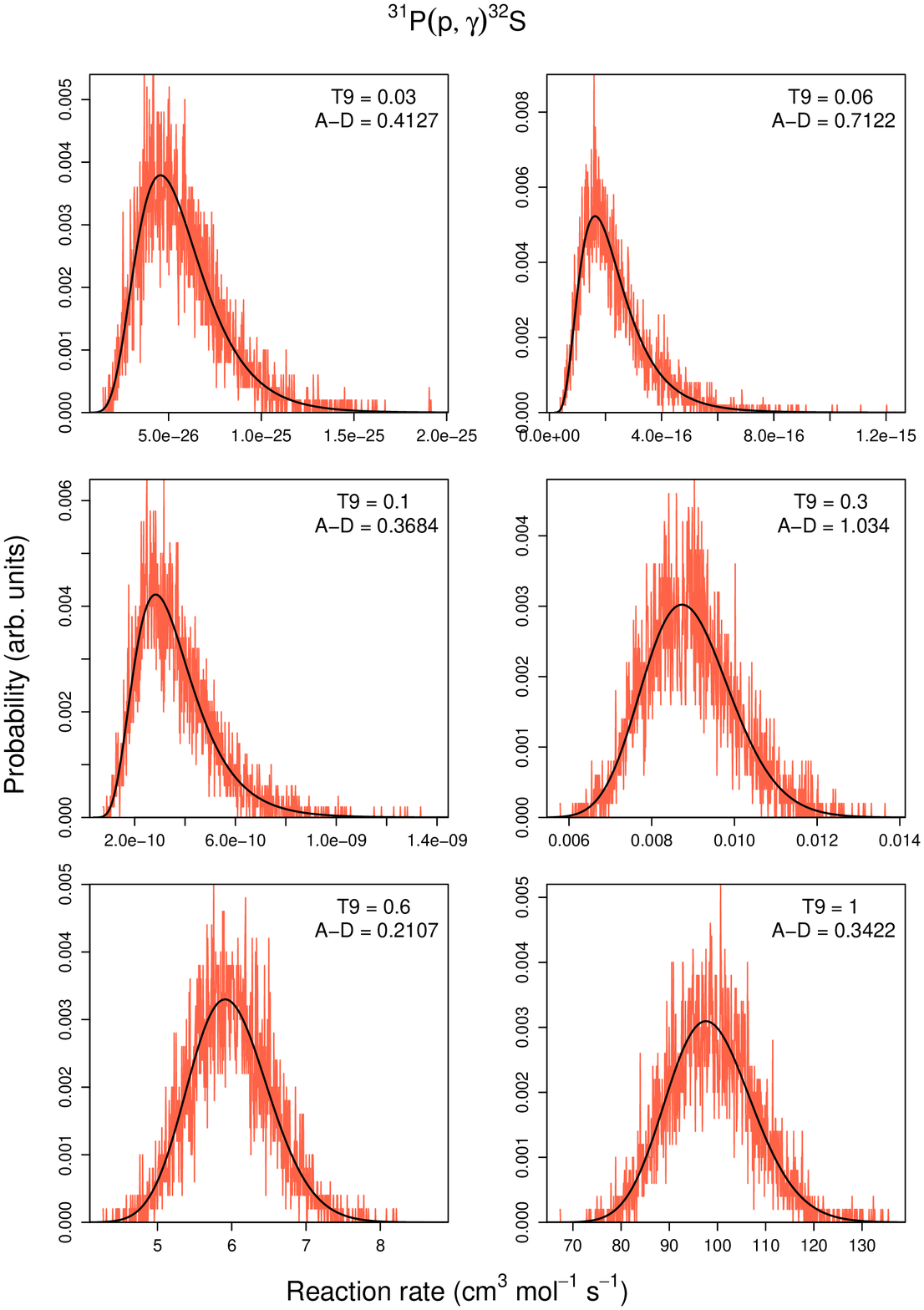}
\end{figure}
\clearpage
\setlongtables

Comments: In total, 41 resonances at E$_r^{cm}=159-2942$ keV are taken into account for the calculation of the total reaction rate. Resonance energies and strengths are adopted from Endt \cite{End90,End98}. Note that no reliable resonance strength standard is available for this reaction. The subthreshold level at E$_x=8861$ keV (2$^+$) has not been taken into account since its contribution is expected to be small \cite{Ili93}. For information on unobserved low-energy resonances, see the comments in Ref. \cite{Ili93}. Two levels are omitted in the present work: (i) E$_x=9138$ keV, since its observation has only been reported as a private communication \cite{Kal78} (it has not been observed in proton transfer), and (ii) E$_x=9196$ keV, since Ref. \cite{End98} considers it to be identical to the E$_x=9208$ keV level. The existence of some $\alpha$-particle strength near the proton threshold is apparent from the measured $^{28}$Si($^{6}$Li,d)$^{32}$S spectrum, shown in Fig. 4 of Ref. \cite{Tan81}. The interference of the two 1$^-$ resonances at E$_r^{cm}=372$ and 1468 keV is explicitly taken into account, giving rise to the bimodal reaction rate probability density function at temperatures below 40 MK, as can be seen in the panel below. (The interference has an unknown sign and is thus sampled by using a binary probability density; see Sec 4.4 in Paper I).
\begin{figure}[]
\includegraphics[height=8.5cm]{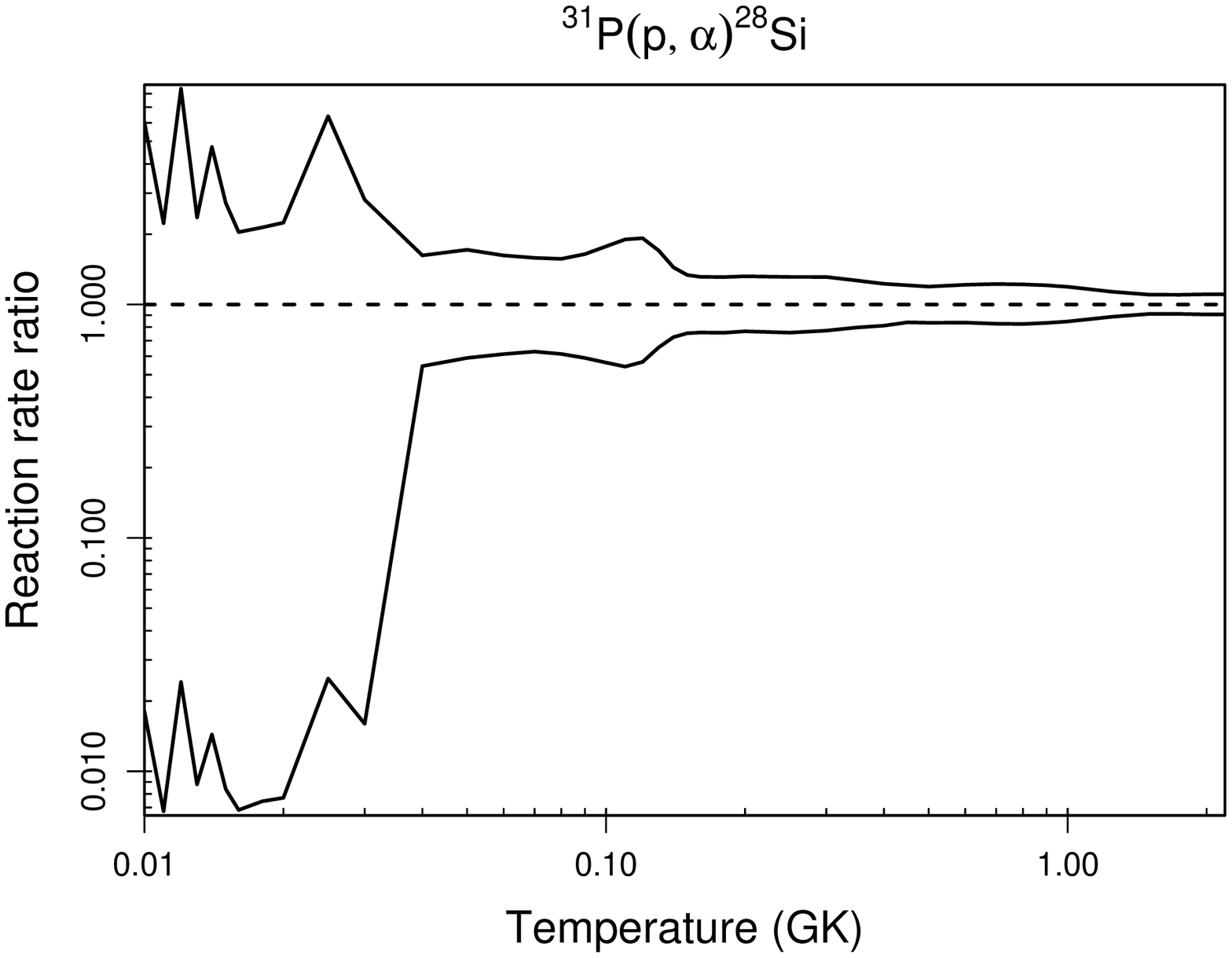}
\end{figure}
\clearpage
\begin{figure}[]
\includegraphics[height=18.5cm]{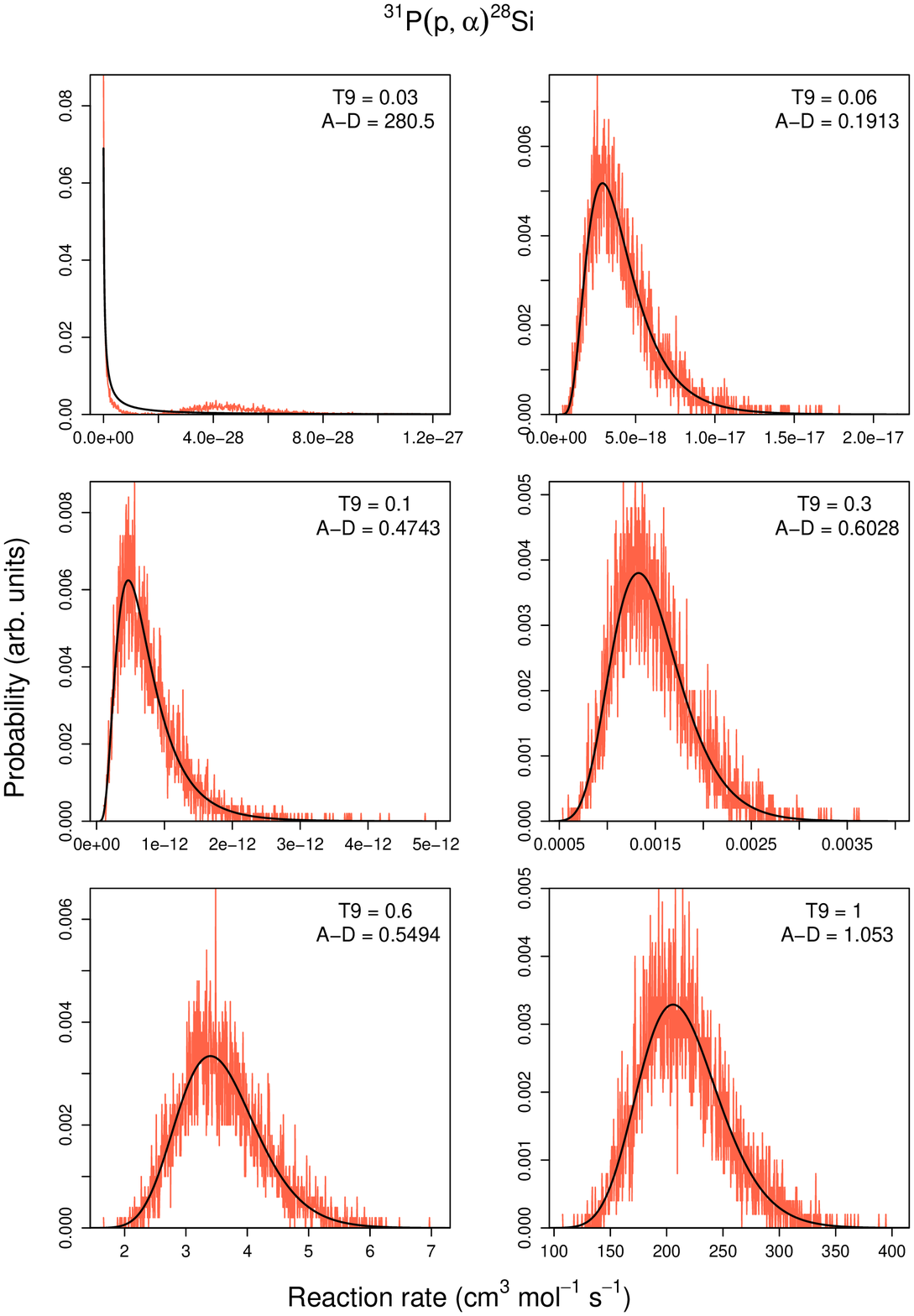}
\end{figure}
\clearpage
\setlongtables

Comments: Because of the small Q-value (Tab. \ref{tab:master}) only the first and second excited states in $^{31}$Cl are expected to contribute to the resonant reaction rate. The second excited state has been observed at an excitation energy of E$_x=1754\pm3$ keV by Fynbo et al. \cite{Fyn00}; from their proton energy observed in the $^{31}$Ar $\beta$-delayed decay spectrum (E$_p=1416\pm2$ keV) we find a resonance energy of E$_r^{cm}=1463\pm2$ keV. The resonance energy for the astrophysically more important first excited state was recently estimated using the isobaric multiplet mass equation (IMME), resulting in a value of E$_r^{cm}=453\pm8$ keV \cite{Wre09}. This value is consistent with the energy of the proton peak (E$_p=446\pm15$ keV) observed in the $^{31}$Ar $\beta$-delayed decay work of Axelsson et al. \cite{Axe98};
the observed energy yields a resonance energy of E$_r^{cm}=461\pm15$ keV. Although we adopt this value for the resonance energy, an independent measurement would be highly desirable in view of the possibility that the observed proton peak in Ref. \cite{Axe98} may arise from the $\beta$-delayed 2p-decay (instead of single proton emission) of $^{31}$Ar. The proton and $\gamma$-ray partial widths of these two resonances, as well as the direct capture S-factor, are derived from shell model results \cite{Her95}.
\begin{figure}[]
\includegraphics[height=8.5cm]{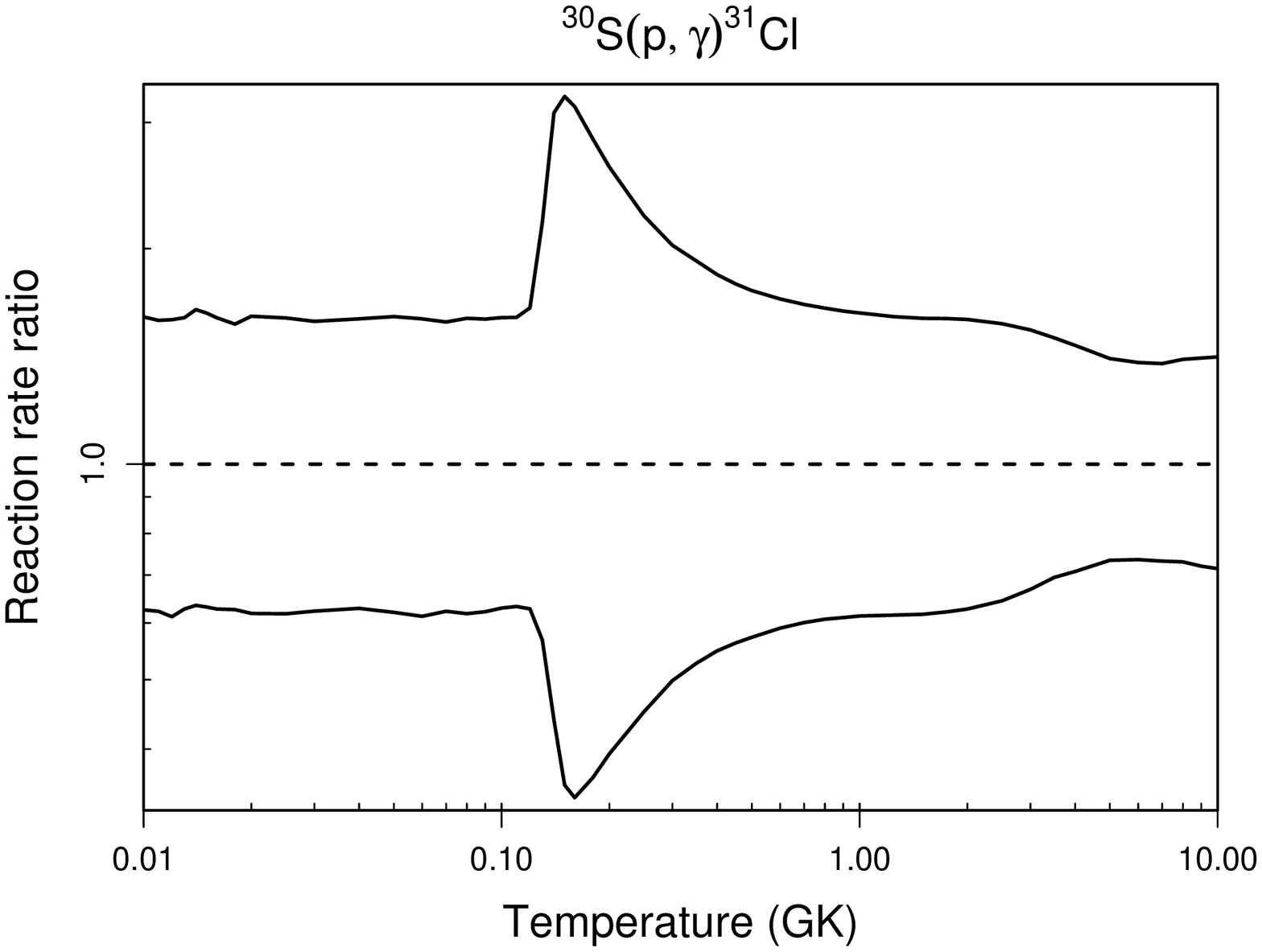}
\end{figure}
\clearpage
\begin{figure}[]
\includegraphics[height=18.5cm]{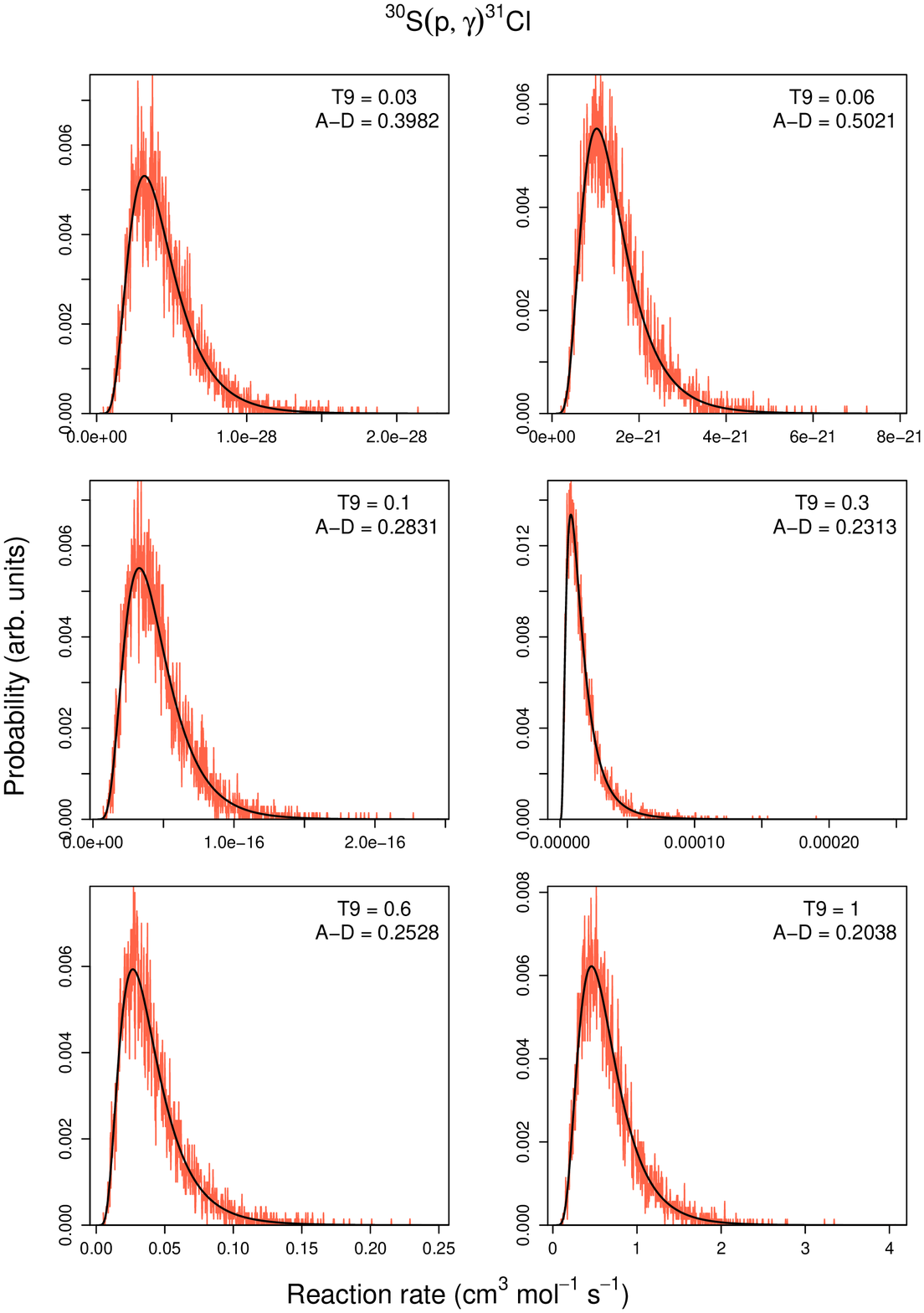}
\end{figure}
\clearpage
\setlongtables

Comments: The reaction rate, including uncertainties, is calculated from the same input information as in
Iliadis et al. \cite{Ili99}.
\begin{figure}[]
\includegraphics[height=8.5cm]{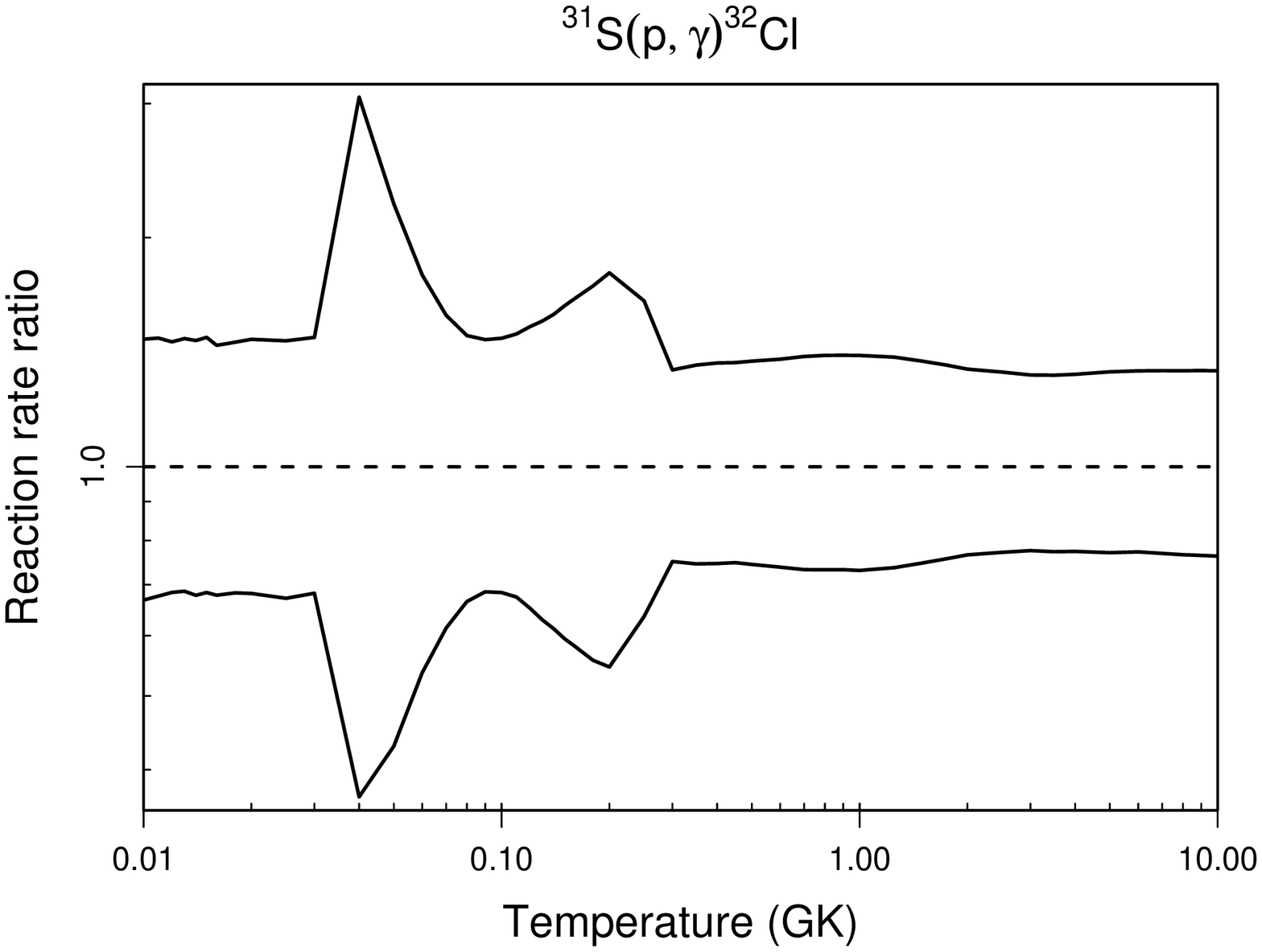}
\end{figure}
\clearpage
\begin{figure}[]
\includegraphics[height=18.5cm]{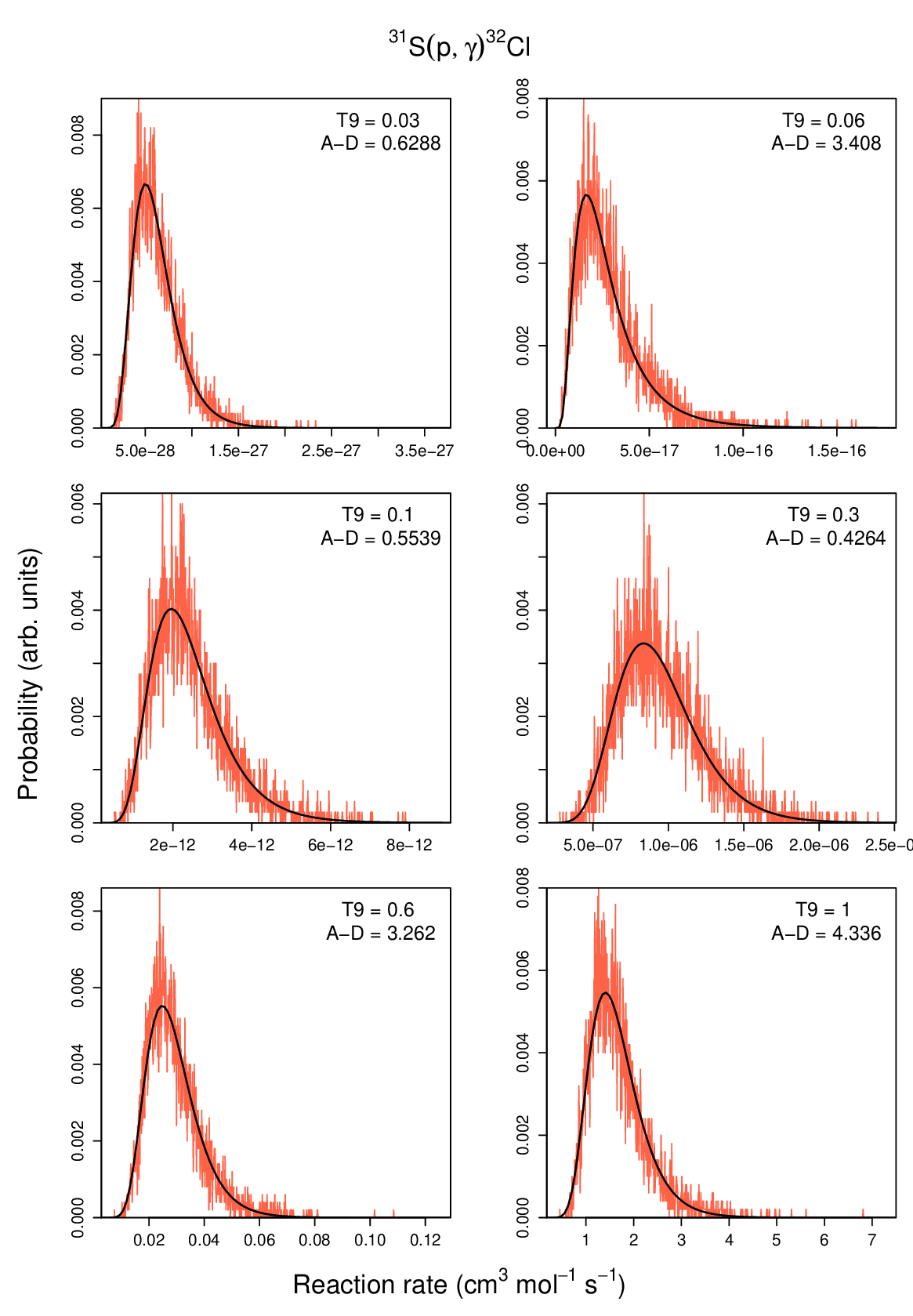}
\end{figure}
\clearpage
\setlongtables
%
Comments: The reaction rate, including uncertainties, is calculated from the same input information as in
Iliadis et al. \cite{Ili01}.
\begin{figure}[]
\includegraphics[height=8.5cm]{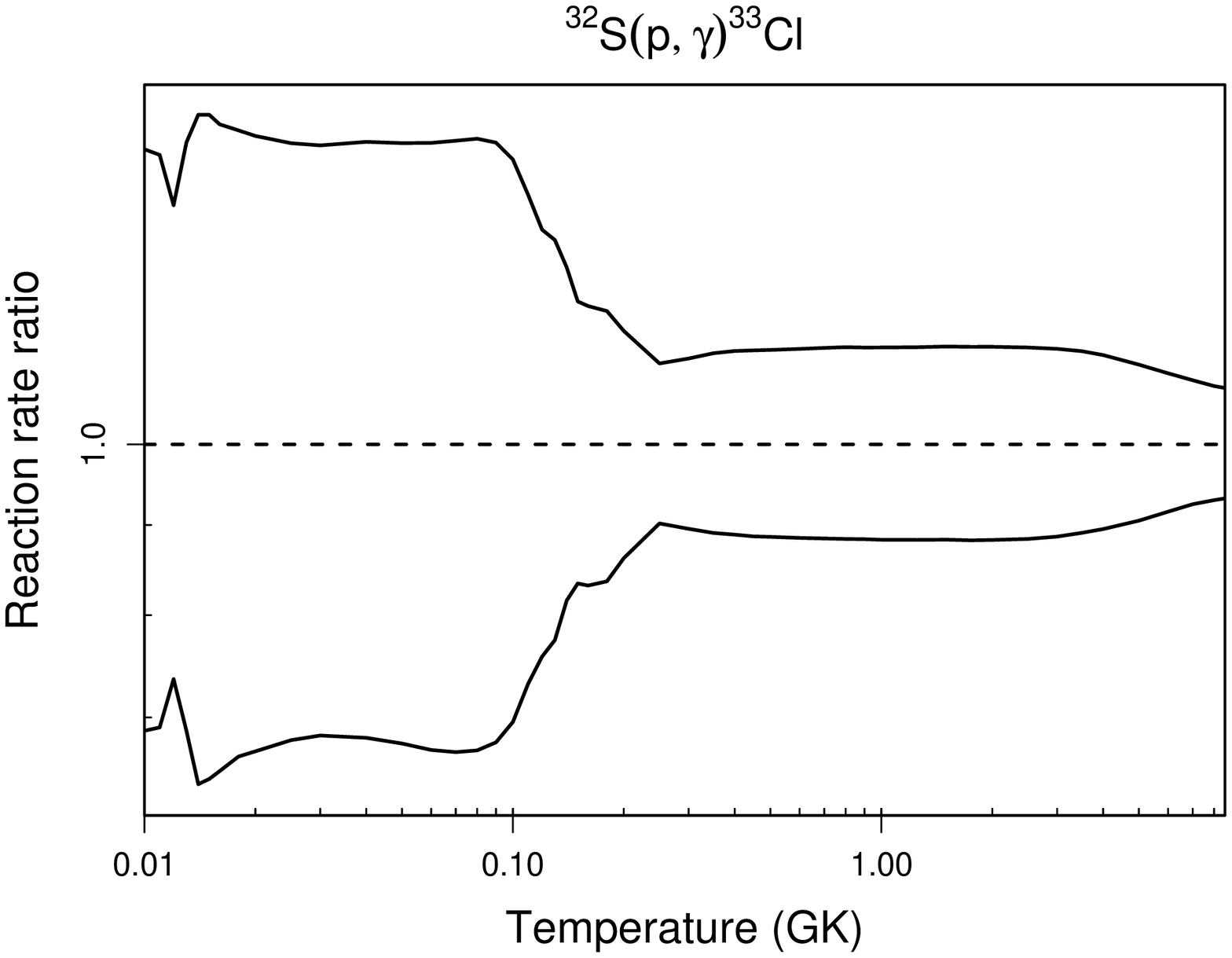}
\end{figure}
\clearpage
\begin{figure}[]
\includegraphics[height=18.5cm]{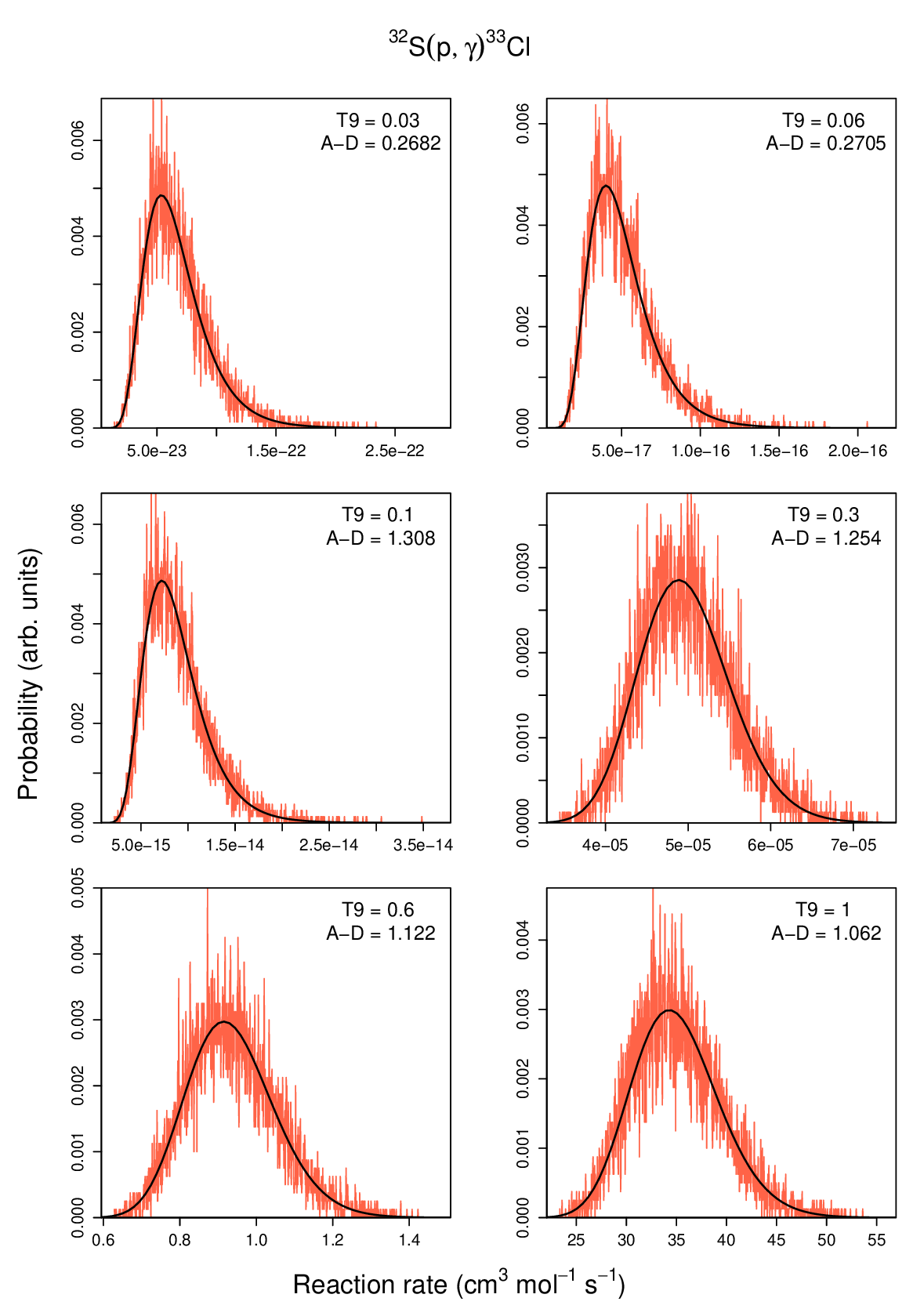}
\end{figure}
\clearpage
\setlongtables
%
Comments: No unbound states have been observed in the compound nucleus $^{32}$Ar. Based on the comparison to the structure of the $^{32}$Si mirror nucleus it can be concluded that the lowest-lying resonance in $^{31}$Cl+p is expected to occur above an energy of 1 MeV. In fact, a Coulomb displacement energy calculation \cite{Her95} yields E$_r^{cm}=1600$ keV, which is adopted in the present work; we estimate an uncertainty of 100 keV, although this value should be regarded as a rough guess only. The proton and $\gamma$-ray partial widths are calculated by using the shell model \cite{Her95}. The total reaction rates are dominated by the direct capture process into the ground and first excited states of $^{32}$Ar at all temperatures of interest. We adopt the direct capture S-factor from Ref. \cite{Her95}, which is based on shell model spectroscopic factors. Higher lying resonances are expected to make a minor contribution to the total rate.
\begin{figure}[]
\includegraphics[height=8.5cm]{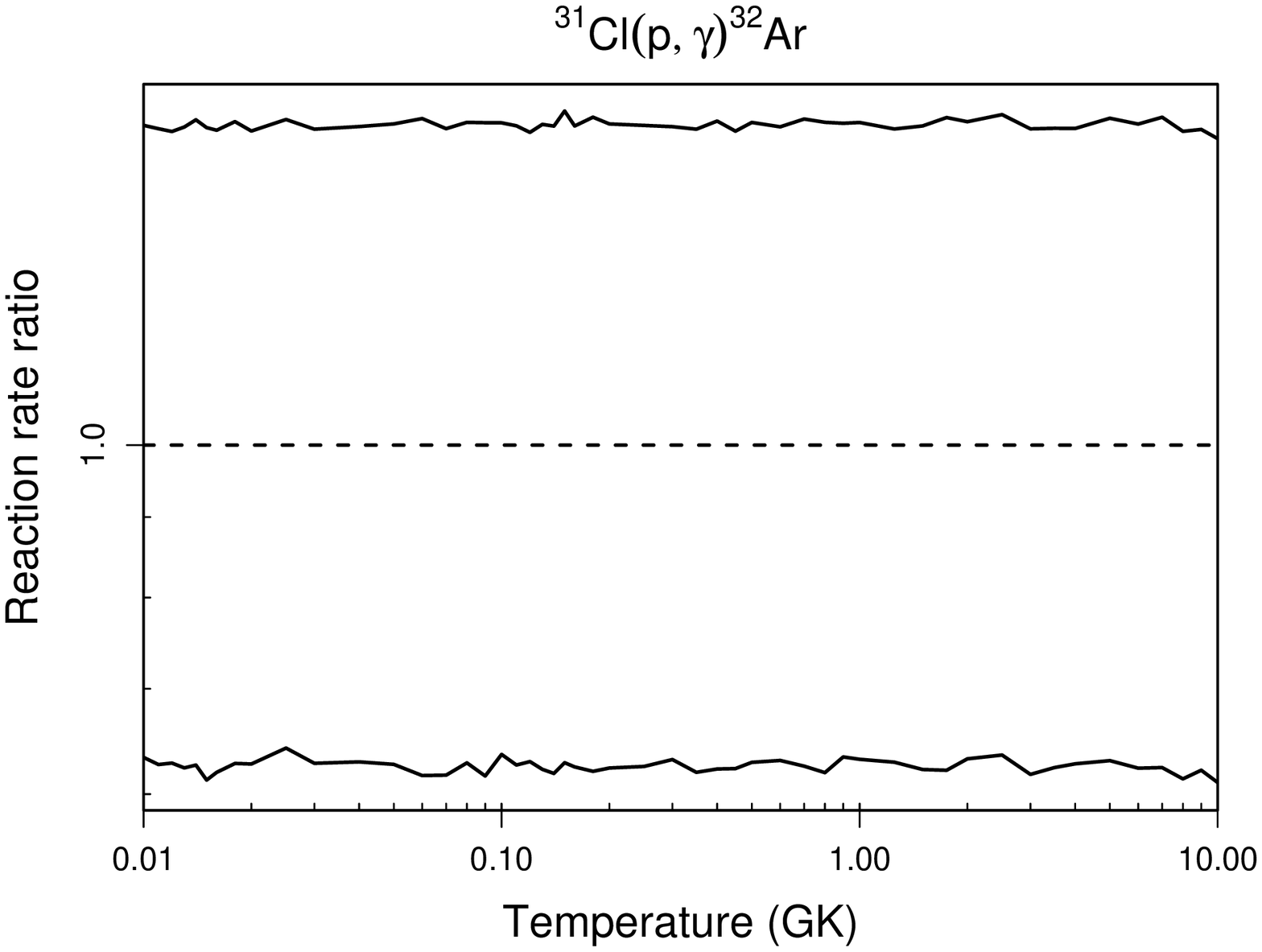}
\end{figure}
\clearpage
\begin{figure}[]
\includegraphics[height=18.5cm]{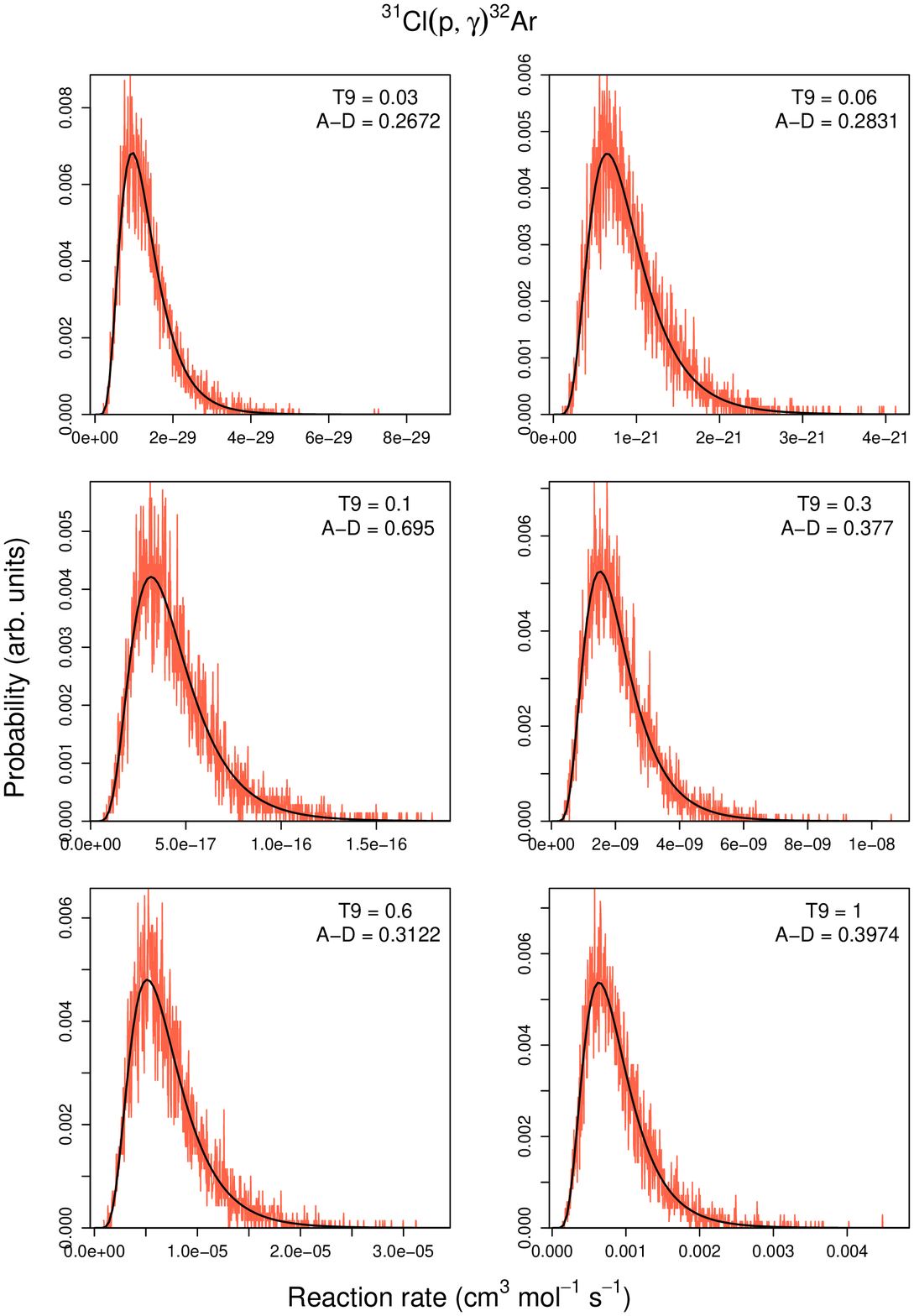}
\end{figure}
\clearpage
\setlongtables

Comments: The total rate has contributions from the direct capture process and from 5 resonances located at E$_r^{cm}=21-1387$ keV. The direct capture S-factor as well as the proton and $\gamma$-ray partial widths of the resonances are based on the shell-model \cite{Her95,Sch05}. Our rate does not take into account three potentially important systematic effects. First, the spin-parity assignments of the measured levels at E$_x=3364, 3456$ and 3819 keV \cite{Cle04}, corresponding to the lowest-lying resonances, are not unambiguously known; the assignments $J^{\pi}=5/2^+_2$, $7/2^+_1$ and $5/2^+_3$ for these levels are based on the ``analogy with the $^{33}$P analog nucleus"  \cite{Cle04}. Second, the energies of the resonances at E$_r^{cm}=847$ and 1387 keV are not based on experimental excitation energies, but are derived from Coulomb shift calculations \cite{Her95}; the adopted value of 100 keV \cite{Sch05} for the resonance energy uncertainty must be regarded as a rough value only. Third, comparison to the structure of the $^{33}$P mirror nucleus reveals that two more unobserved levels ($J^{\pi}=5/2^+_4$ and $7/2^-_1$) are expected as resonances near E$_r^{cm}=1$ MeV; these remain at present unaccounted for in the total rate. 
\begin{figure}[]
\includegraphics[height=8.5cm]{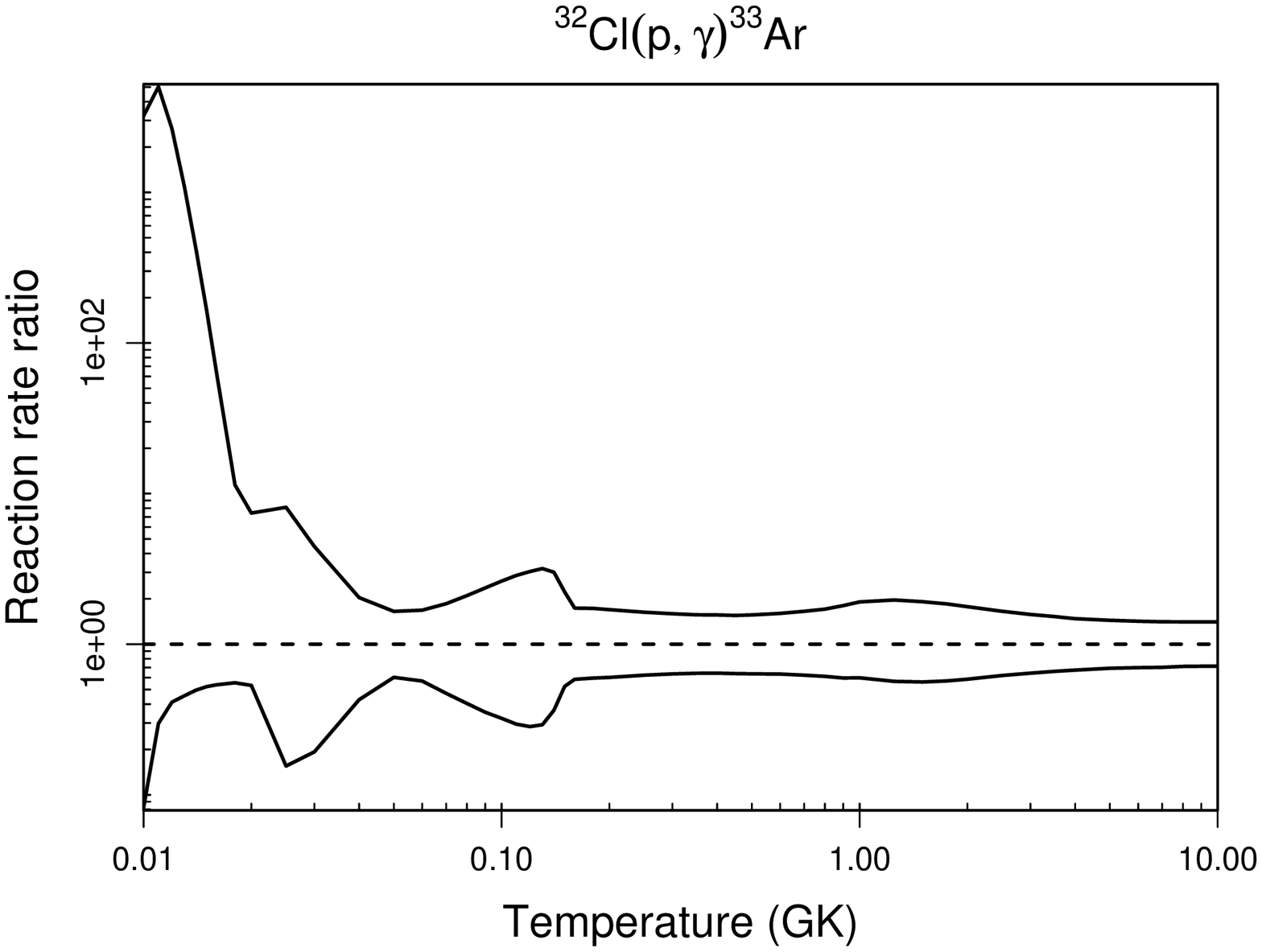}
\end{figure}
\clearpage
\begin{figure}[]
\includegraphics[height=18.5cm]{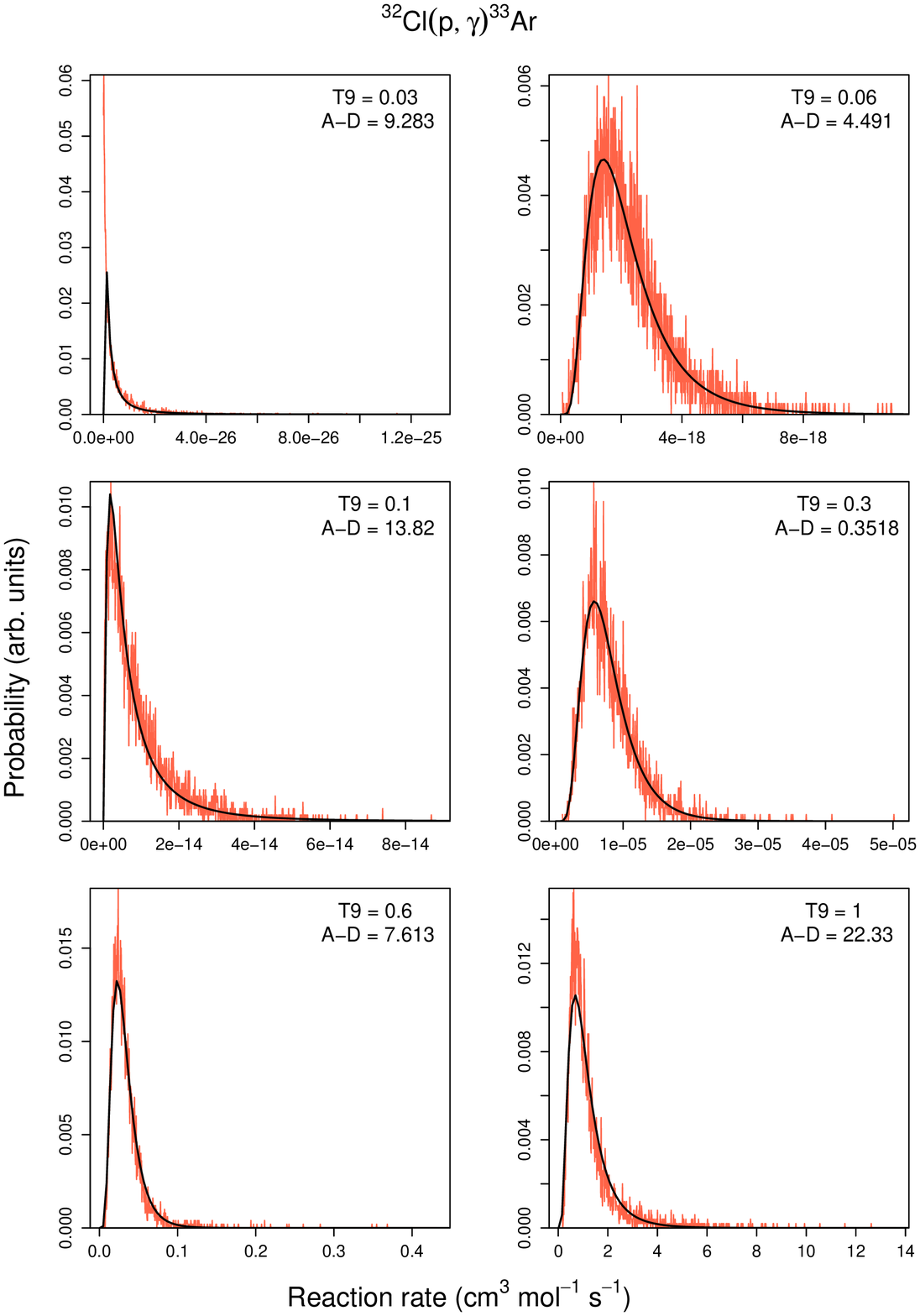}
\end{figure}
\clearpage
\setlongtables

Comments: In total, 92 resonances at energies of E$_r^{cm}=49-2675$ keV are taken into account for calculating the total reaction rates. The direct capture S-factor is adopted from Iliadis et al. \cite{Ili94}. Measured resonance energies and strengths are from Johnson, Meyer and Reitmann \cite{Joh74}, Endt and van der Leun \cite{End78} and Iliadis et al. \cite{Ili94}, where all the strengths are normalized by using the standard value listed in Tab. 1 of Iliadis et al. \cite{Ili01}. When no uncertainties of resonance strengths are reported, we adopt a value of 20\%. Three $^{36}$Ar threshold levels are reported in R\"opke, Brenneisen and Lickert \cite{Roe02} at E$_x=8593$, 8739 and 8919 keV: we assume that the latter state is identical to the E$_x=8923$ keV level \cite{Ili94}, while the former two states are new, corresponding to resonance energies of E$_r^{cm}=86$ and 232 keV.
\begin{figure}[]
\includegraphics[height=8.5cm]{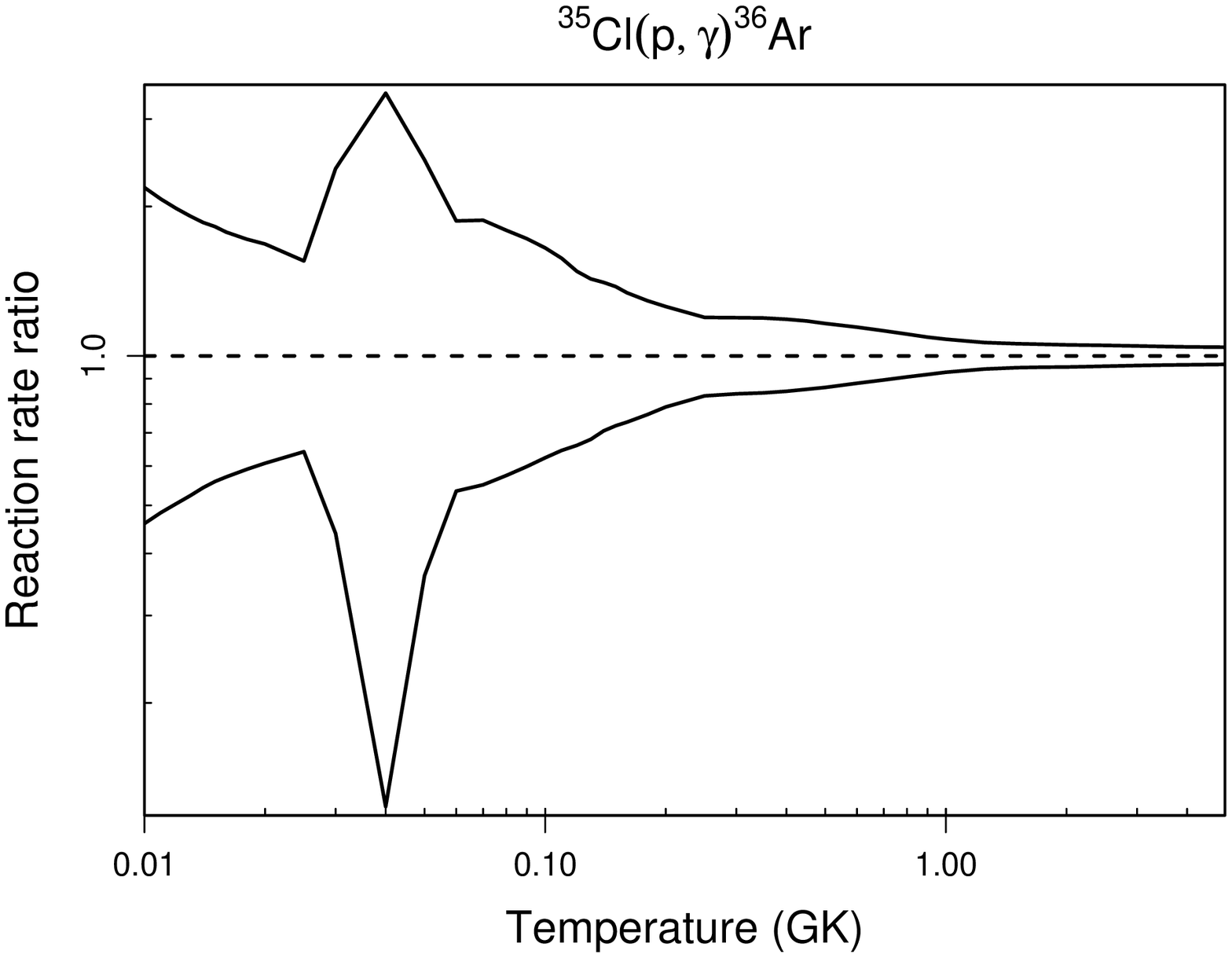}
\end{figure}
\clearpage
\begin{figure}[]
\includegraphics[height=18.5cm]{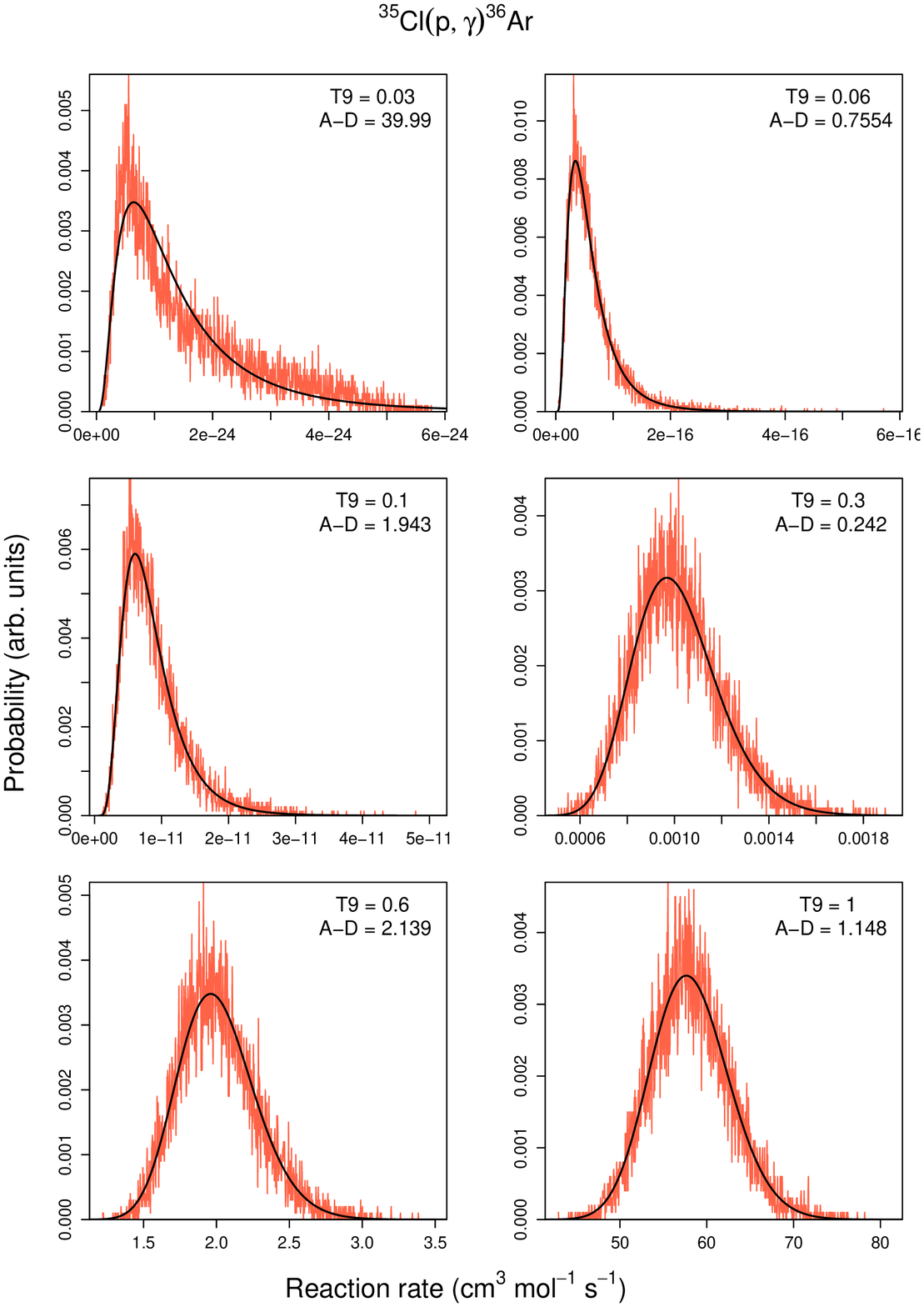}
\end{figure}
\clearpage
\setlongtables

Comments: In total, 95 resonances at energies of E$_r^{cm}=49-2837$ keV are taken into account for calculating the total reaction rates. Measured resonance energies and strengths are from Bosnjakovic et al. \cite{Bos68}, Endt and van der Leun \cite{End78} and Iliadis et al. \cite{Ili94}. Note that no carefully measured standard resonance strength exists for this reaction. When no uncertainties of resonance strengths are reported, we adopt a value of 25\%. The S-factor describing the low-energy tails of broad, higher-lying resonances is also taken into account (see comments in Ref. \cite{Ili01}). It must be emphasized that no direct measurement exists for energies below E$_r^{cm}=840$ keV and that the excitation energy region corresponding to E$_r^{cm}=610-840$ keV has neither been studied in proton transfer \cite{Ili94} nor in coincidence measurements \cite{Ros95}. For these levels we adopt single-particle estimates for the upper limits of proton and $\alpha$-particle widths. Three $^{36}$Ar threshold levels are reported in R\"opke, Brenneisen and Lickert \cite{Roe02} at E$_x=8593$, 8739 and 8919 keV: we assume that the latter state is identical to the E$_x=8923$ keV level \cite{Ili94}, while the former two states are new, corresponding to resonance energies of E$_r^{cm}=86$ and 232 keV. 
\begin{figure}[]
\includegraphics[height=8.5cm]{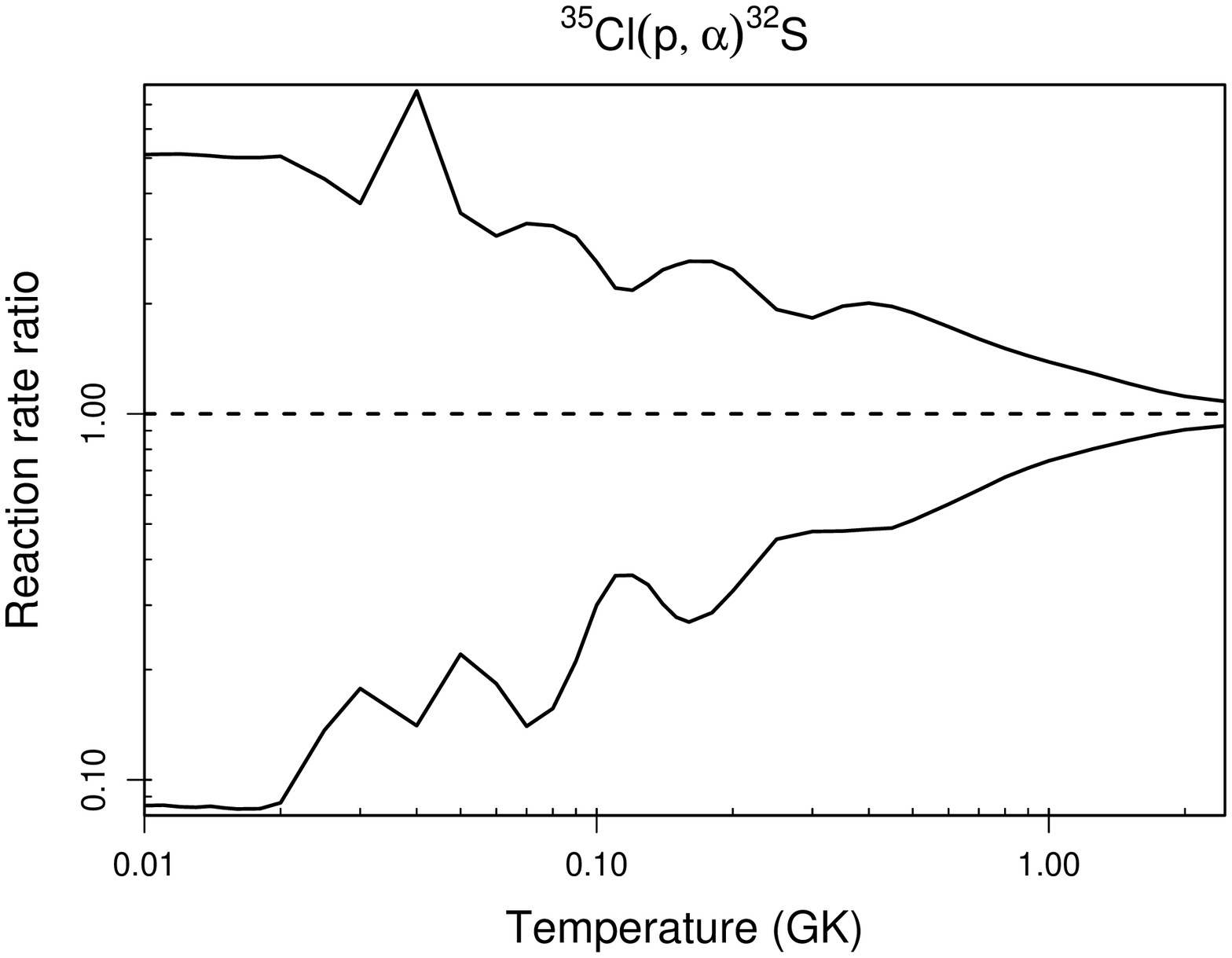}
\end{figure}
\clearpage
\begin{figure}[]
\includegraphics[height=18.5cm]{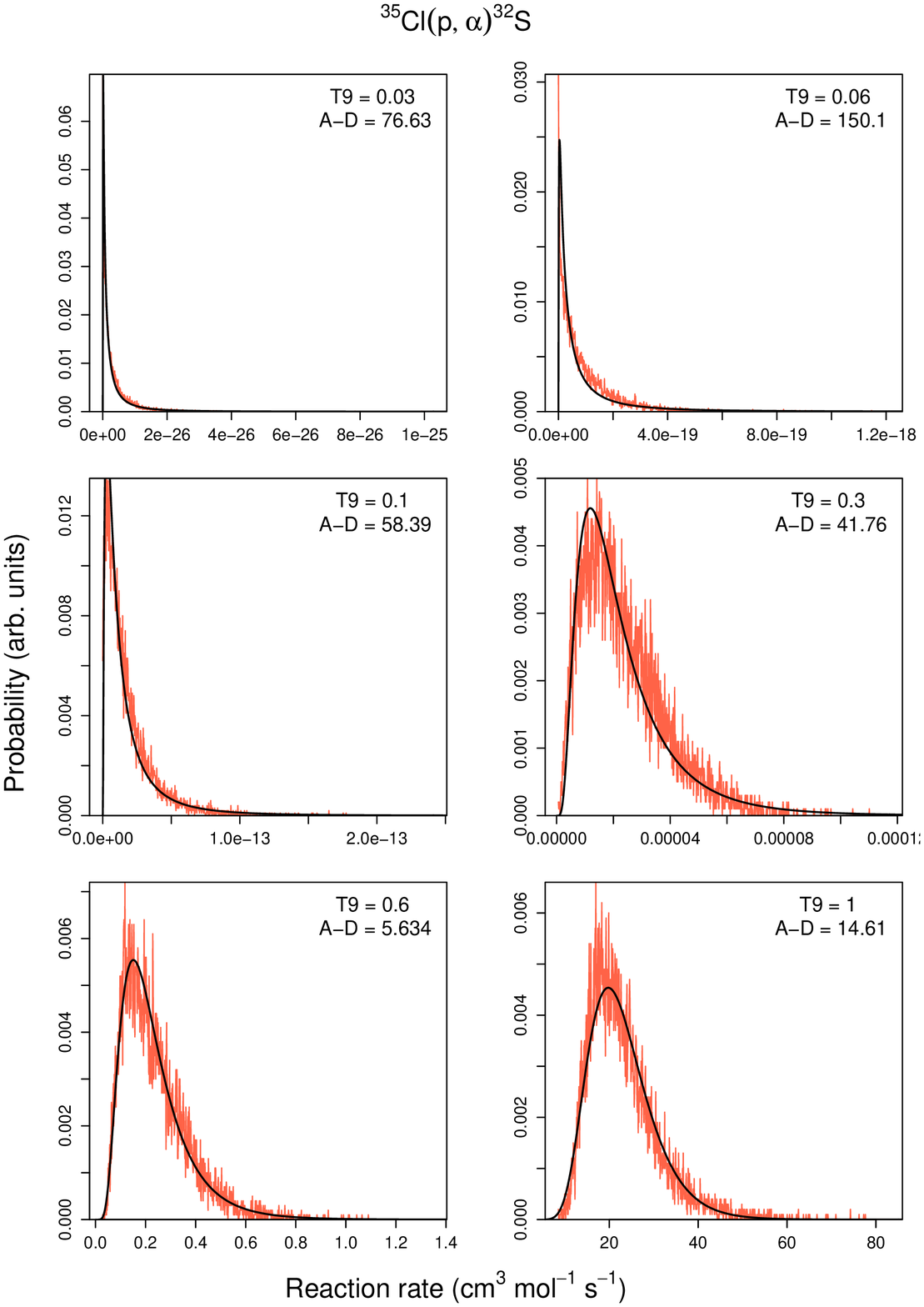}
\end{figure}
\clearpage
\setlongtables

Comments: A significantly improved Q-value (Q=$84.5\pm0.7$ keV; Tab. \ref{tab:master}) is obtained by using the measured $^{35}$K mass excess of Yazidjian et al. \cite{Yaz07}. Because of the small Q-value only the first excited state (1/2$^+$) in $^{35}$K is expected to contribute to the resonant reaction rate. We calculate a corresponding resonance energy of E$_r^{cm}=1469.0\pm5.0$ keV directly from the observed energy of $\beta$-delayed protons from the decay of $^{35}$Ca, as reported in Trinder et al. \cite{Tri99}. (This energy also agrees with the measured excitation energy reported by Benenson et al. \cite{Ben76}). The proton and $\gamma$-ray partial width, as well as the direct capture S-factor, are adopted from the shell model calculation of Ref. \cite{Her95}. The total reaction rates are dominated by the direct capture process. Higher lying resonances are expected to make a minor contribution to the total rate.
\begin{figure}[]
\includegraphics[height=8.5cm]{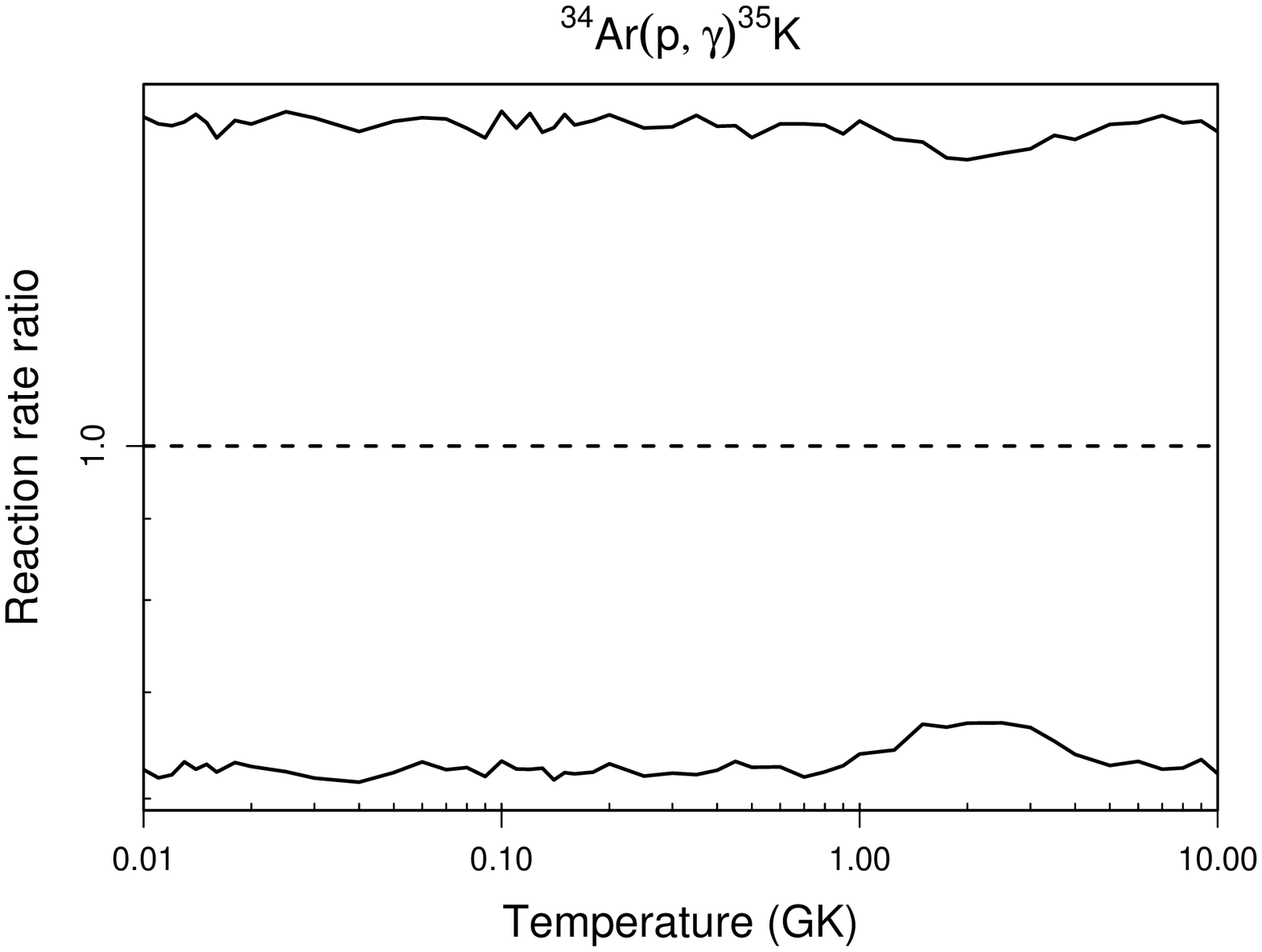}
\end{figure}
\clearpage
\begin{figure}[]
\includegraphics[height=18.5cm]{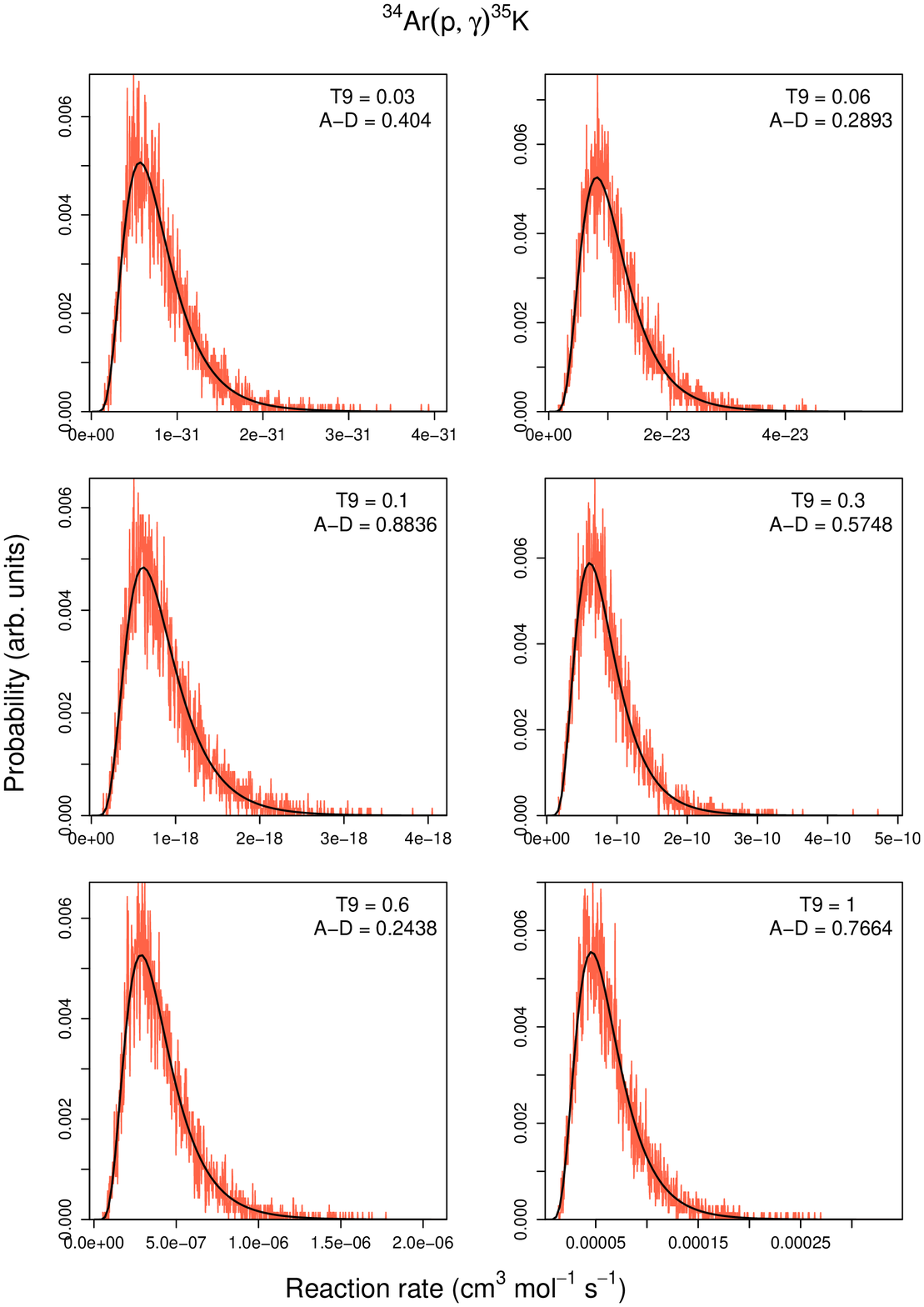}
\end{figure}
\clearpage
\setlongtables

Comments: The reaction rate, including uncertainties, is calculated from the same input information as in
Iliadis et al. \cite{Ili99}. Note that the highest-lying resonance for which reliable input information is available is located at a relatively low energy of E$_r^{cm}$=744 keV. Consequently, the Hauser-Feshbach model has to be used for calculating the rates beyond a relatively low temperature. The rate uncertainties presented here do not include the systematic error introduced by 5 additionally expected resonances with E$_r^{cm}\leq$1 MeV. According to Iliadis et al. \cite{Ili99}, these resonances may increase the high rate at T$\geq$1.5 GK by a factor of 2.
\begin{figure}[]
\includegraphics[height=8.5cm]{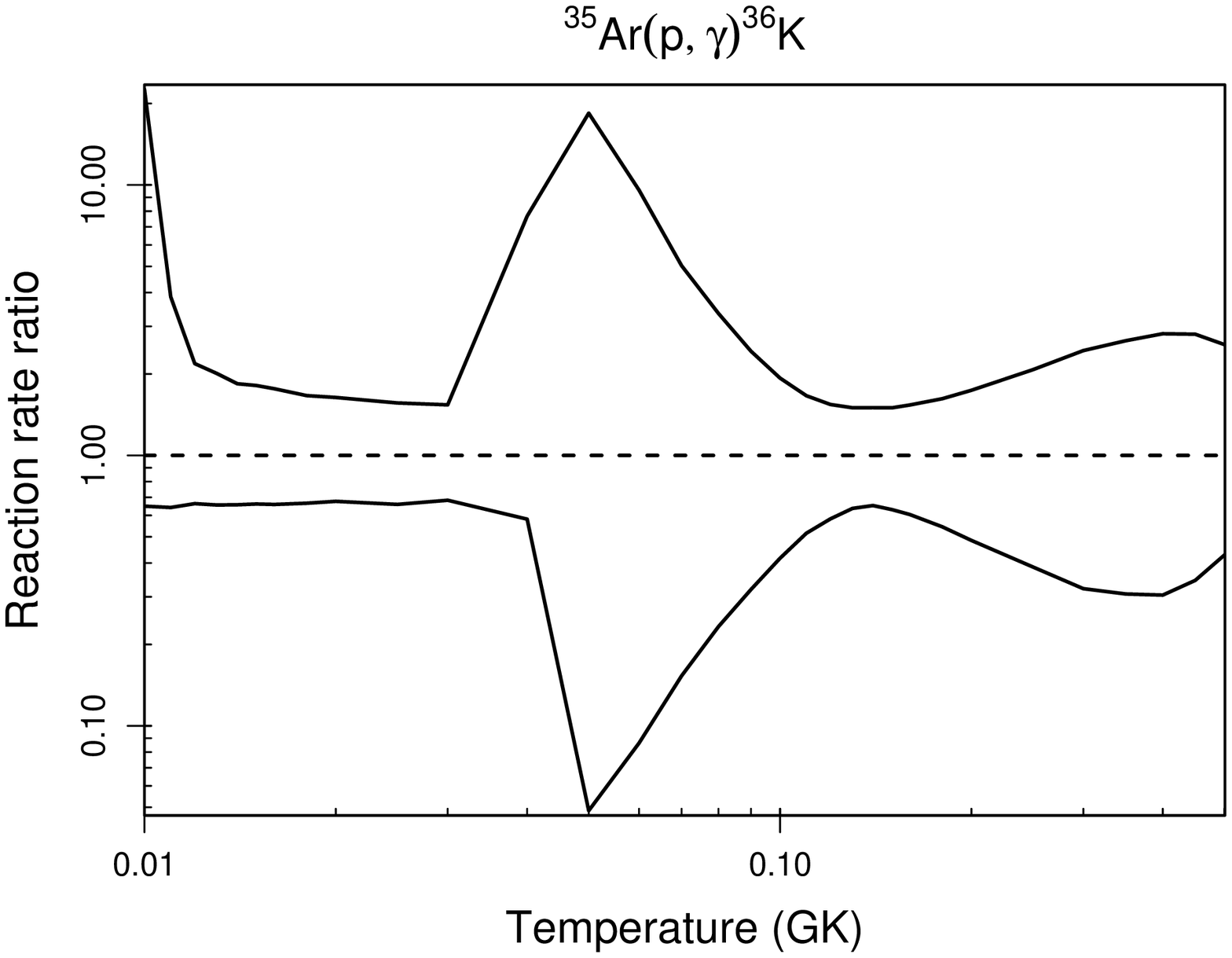}
\end{figure}
\clearpage
\begin{figure}[]
\includegraphics[height=18.5cm]{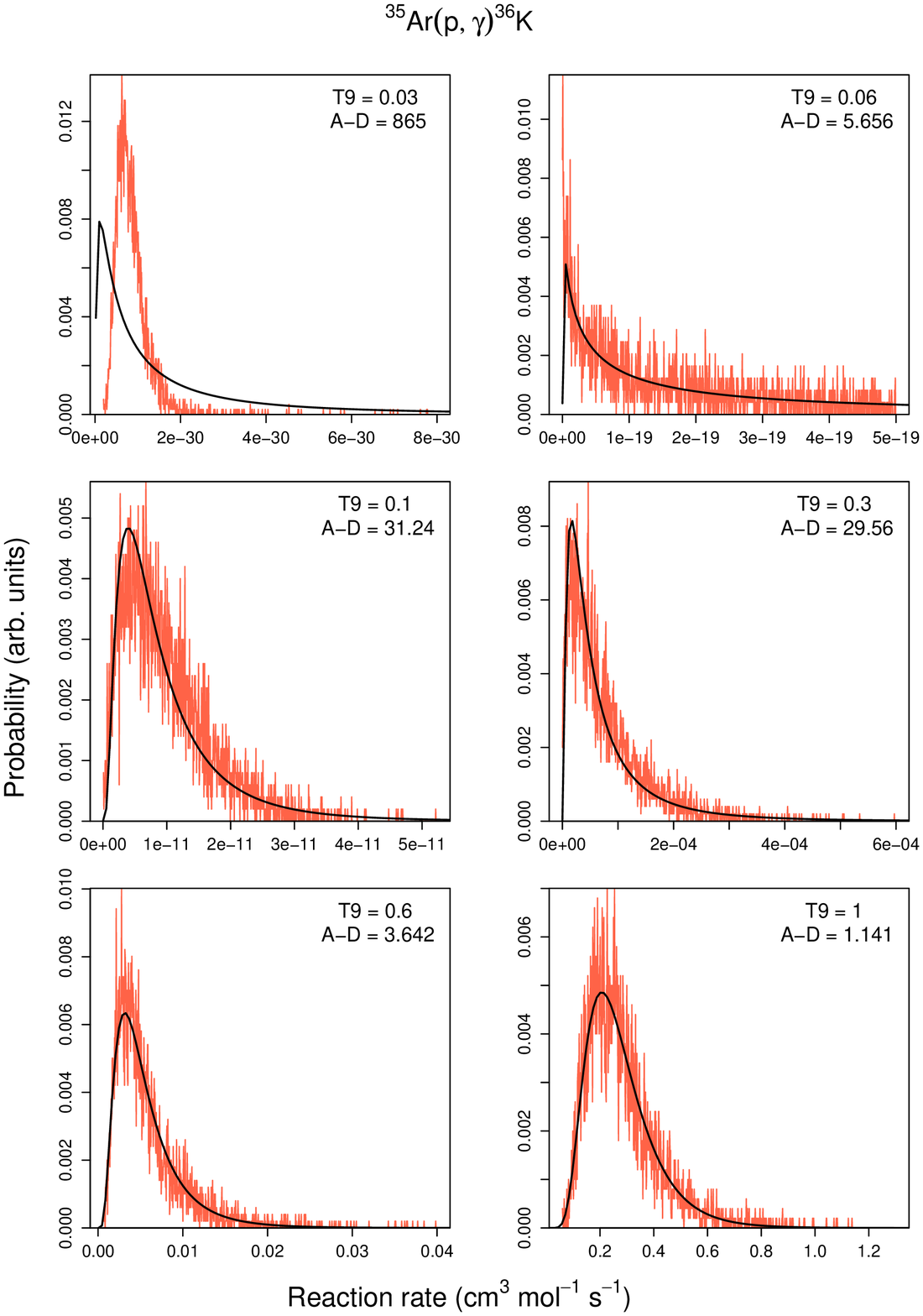}
\end{figure}
\clearpage
\setlongtables

Comments: The reaction rate is calculated from the same input information as in
Iliadis et al. \cite{Ili01}.
\begin{figure}[]
\includegraphics[height=8.5cm]{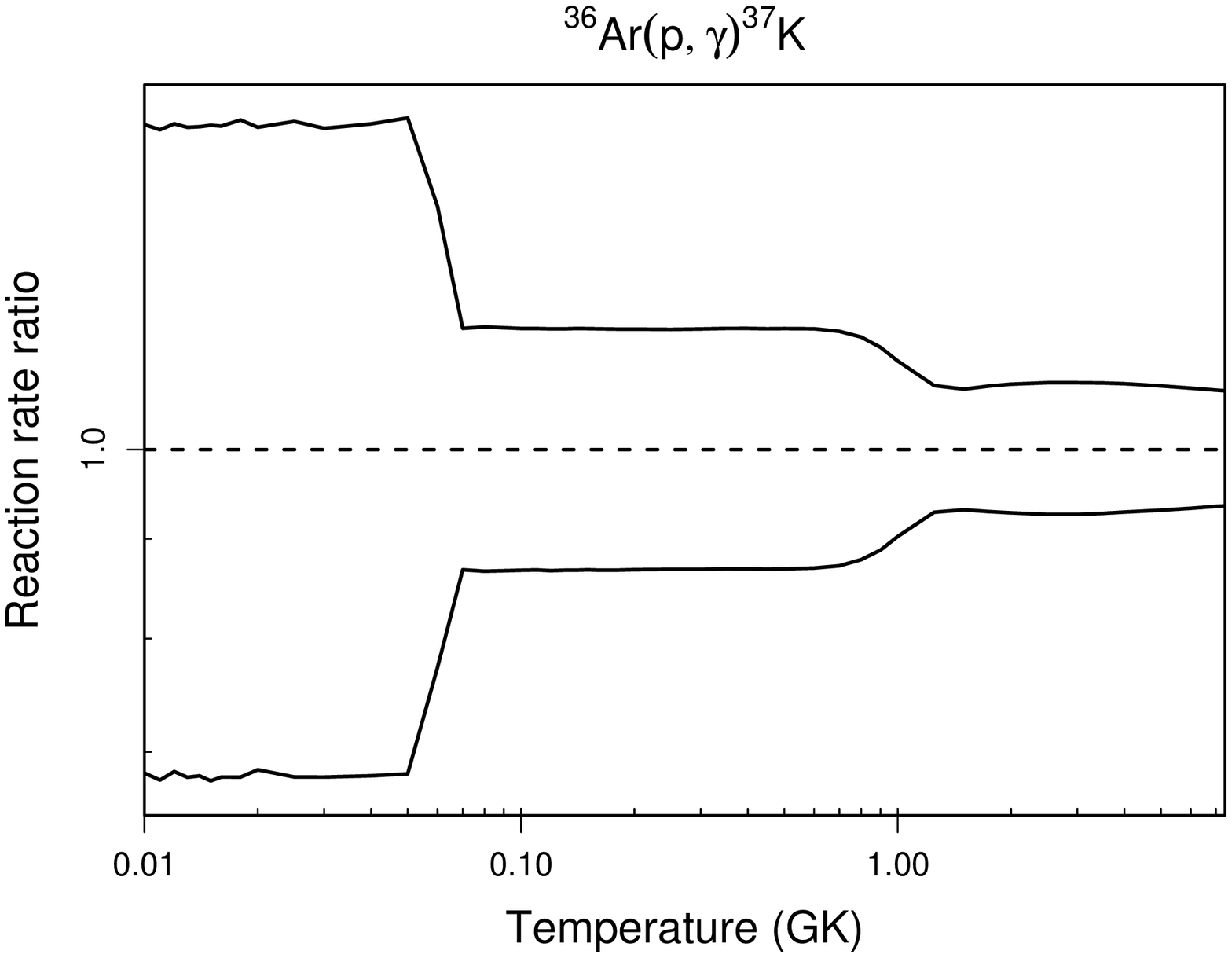}
\end{figure}
\clearpage
\begin{figure}[]
\includegraphics[height=18.5cm]{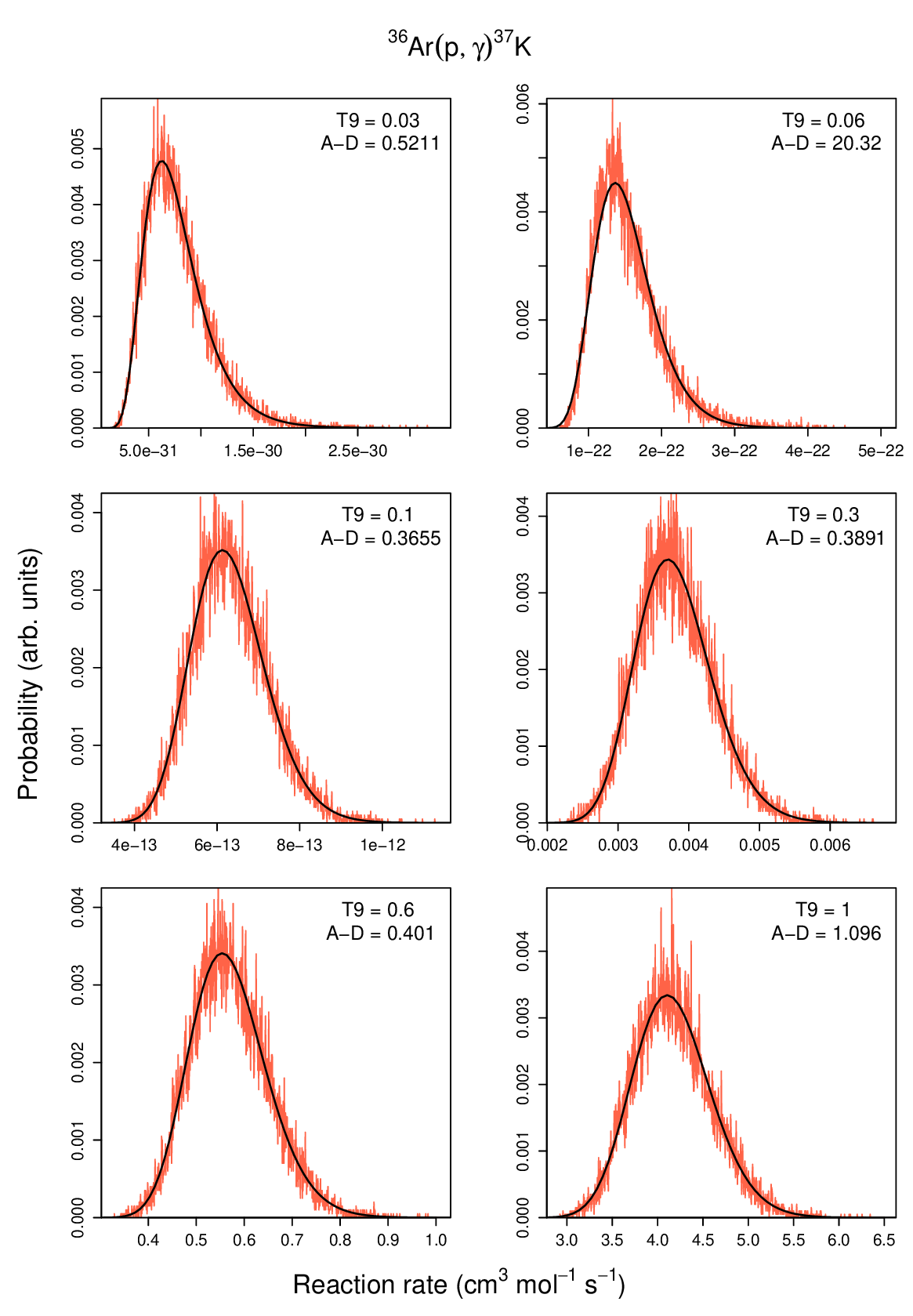}
\end{figure}
\clearpage
\setlongtables
%
Comments: A single resonance, located at E$_r^{cm}=459\pm43$ keV, is taken into account in the evaluation of the total rate. Its energy is obtained from the measured excitation energy of the first-excited (2$^+$) state in $^{36}$Ca (E$_x=3015\pm16$ keV \cite{Doo07}) and the reaction Q-value (see Tab. \ref{tab:master}) that is obtained from the mass differences presented in Ref. \cite{Yaz07}. Note that the new resonance energy is significantly lower than previous estimates \cite{Her95} and this strongly affects both the proton and the $\gamma$-ray partial width. For these quantities we adopt the shell model result of Herndl et al. \cite{Her95}, but correct the published values for the new excitation energy. The direct capture S-factor is also adopted from Ref. \cite{Her95}. We assume uncertainties of 50\% for the partial widths and the direct capture S-factor. Our Monte Carlo rates do not take higher-lying resonances into account. Their influence is presumably small since, based on the known level scheme of the $^{36}$S mirror nucleus, they are expected to occur at much higher energies (see discussion in Ref. \cite{Her95}). 
\begin{figure}[]
\includegraphics[height=8.5cm]{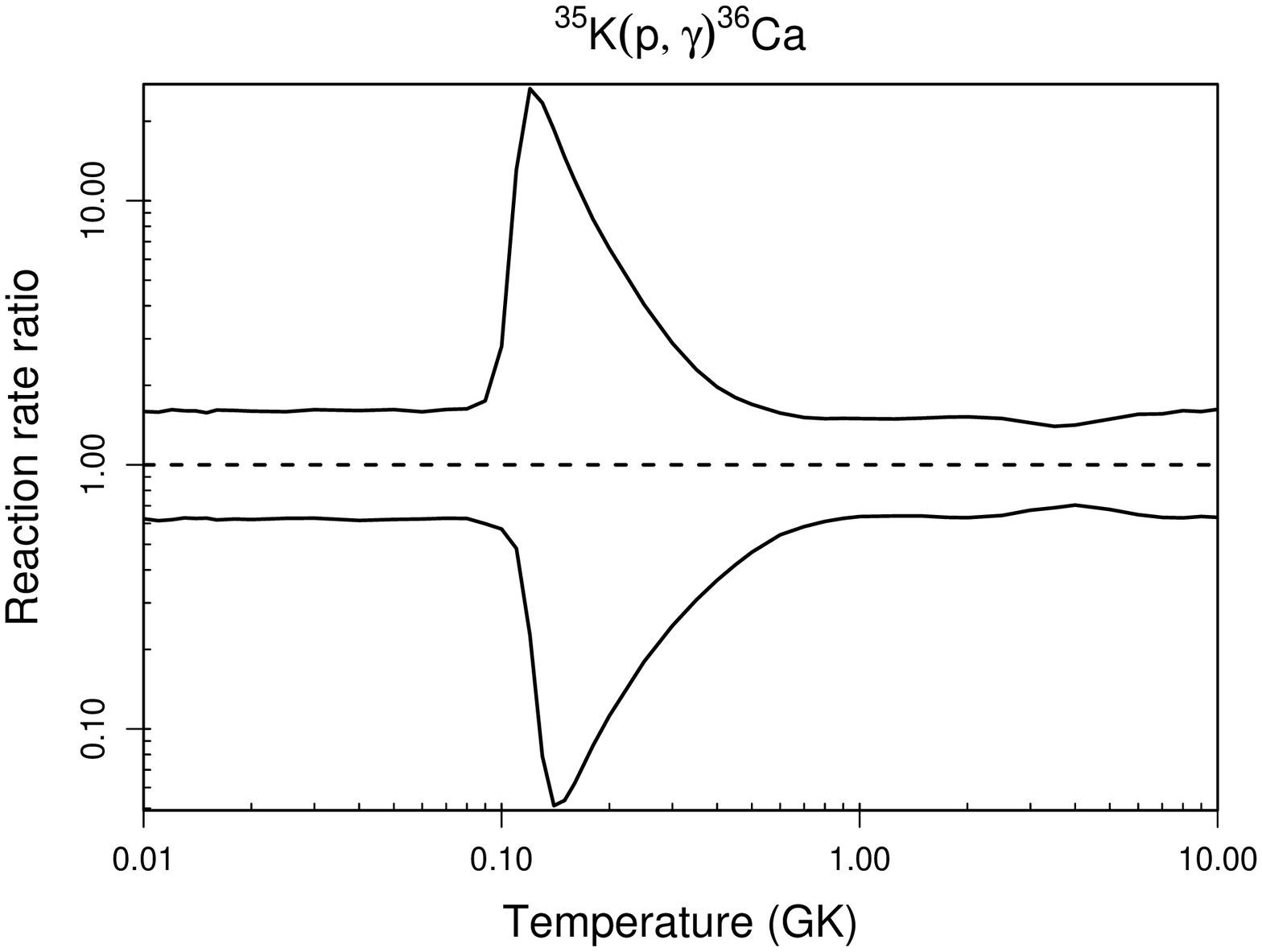}
\end{figure}
\clearpage
\begin{figure}[]
\includegraphics[height=18.5cm]{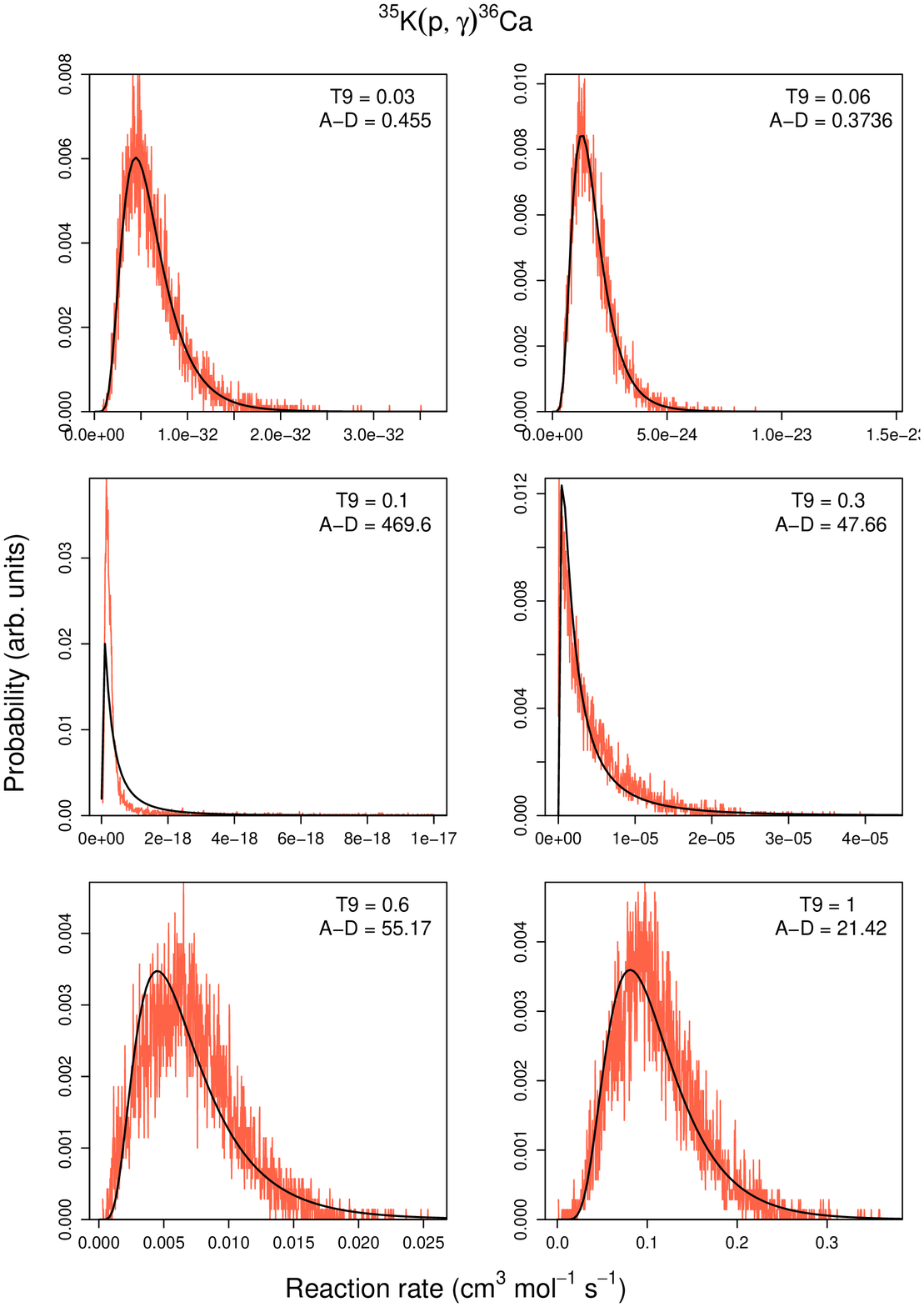}
\end{figure}
\clearpage
\setlongtables

Comments: The reaction rate is calculated by using the $^{40}$Sc excitation energies and A=40 mirror state assignments presented in Tabs. I and II of Hansper et al. \cite{Han00}. Additional information, including the direct capture component, is adopted from Iliadis et al. \cite{Ili99}. We specifically assume for the E$_x$=1671, 1703 and 1797 keV levels in $^{40}$Sc assignments of J$^{\pi}$=(2$^-_2$), (1$^-_1$) and (3$^-_3$), respectively. In total, 5 resonances with energies in the range of E$_r^{cm}$=234-1259 keV are taken into account. The rate uncertainties presented here disregard the fact that these assignments are not unambiguous. Also not included in the rate uncertainties is the unobserved 3$^-_2$ level which the isobaric multiplet mass equation predicts at an excitation energy of E$_x$$\approx$1.45 MeV (Hansper et al. \cite{Han00}). This level may increase the total rates above T=0.8 GK by up to a factor of $\approx$4.

Interestingly, recent calculations by Descouvemont \cite{Des00} using a microscopic cluster model predict total reaction rates that are larger by 1--2 orders of magnitude below T=0.5 GK. This strong disagreement can be traced back to the different values of spectroscopic factors used in the two approaches. For the dominant 2$^-_1$ level, Descouvemont's model predicts values of C$^2$S$_{\ell=1}$=1.05 and C$^2$S$_{\ell=3}$=0.03 (see his Tab. 3). In contrast, we use the {\it experimental} values of C$^2$S$_{\ell=1}$=0.014 and C$^2$S$_{\ell=3}$=0.92 which were measured in the $^{39}$K(d,p)$^{40}$K neutron-transfer to the mirror state (Fink and Schiffer \cite{Fin74}). The measured (d,p) angular distribution (see their Fig. 3) clearly reveals a dominant $\ell$=3 transfer with a relatively small $\ell$=1 component, whereas Descouvemont finds that ``...the $\ell$=1 component is strongly dominant in the wave function...". We prefer using experimental rather than calculated spectroscopic factors when estimating reliable reaction rates. But we agree with Descouvemont that a new measurement of spectroscopic factors in $^{40}$K (and perhaps $^{40}$Sc) would be helpful to clarify the situation.
\begin{figure}[]
\includegraphics[height=8.5cm]{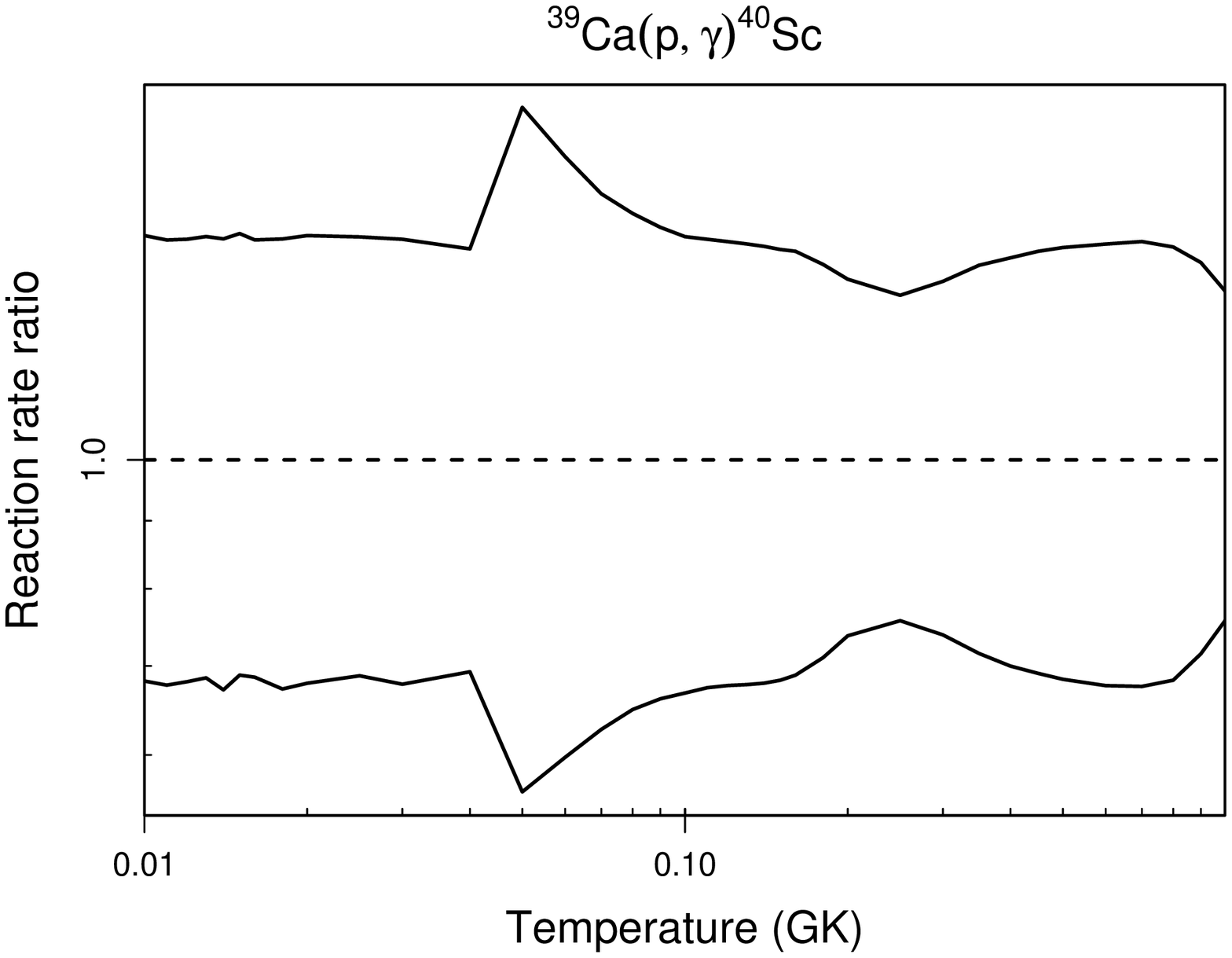}
\end{figure}
\clearpage
\begin{figure}[]
\includegraphics[height=18.5cm]{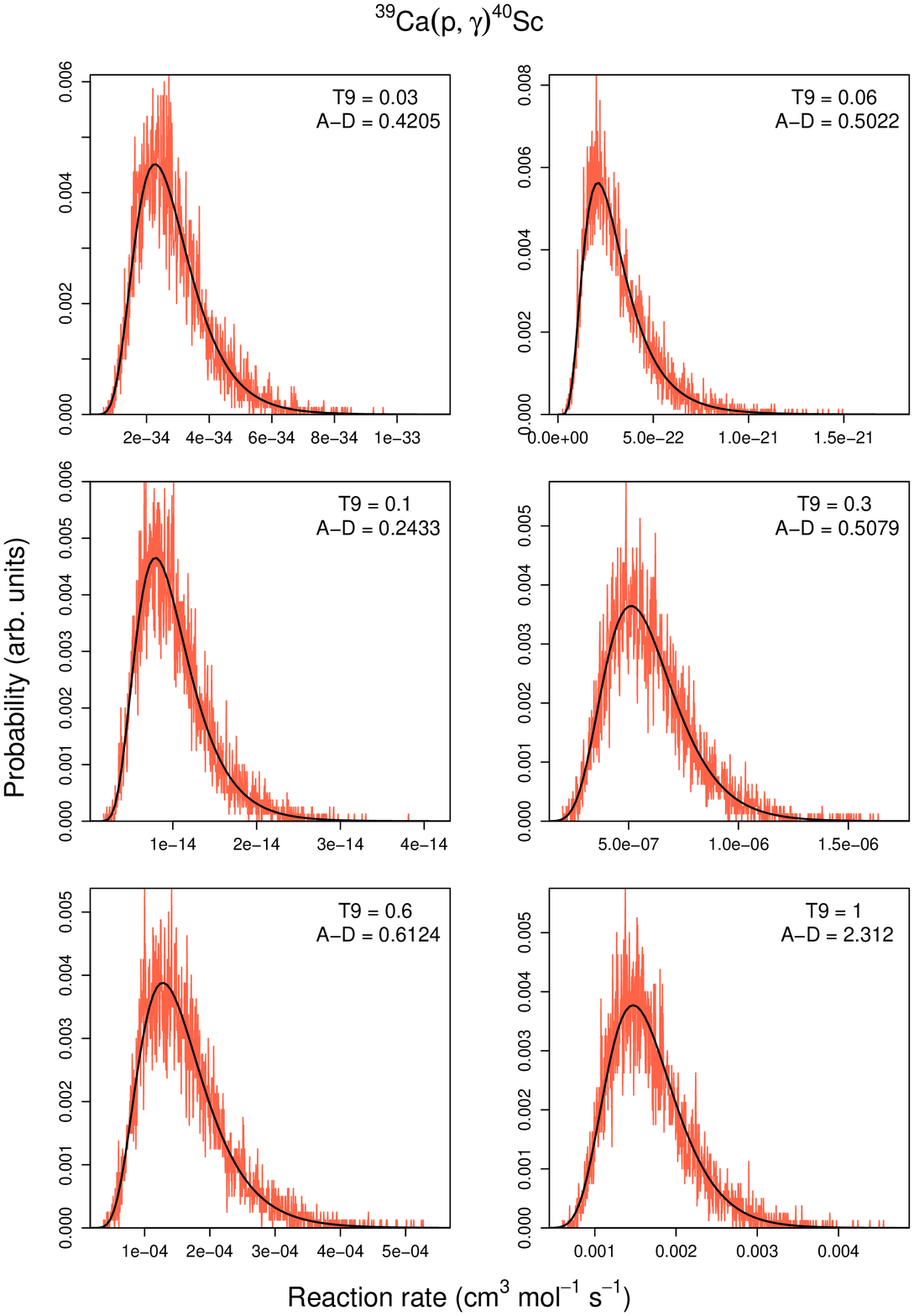}
\end{figure}
\clearpage
\setlongtables

Comments: The reaction rate, including uncertainties, is calculated from the same input information as in
Iliadis et al. \cite{Ili01}.
\begin{figure}[]
\includegraphics[height=8.5cm]{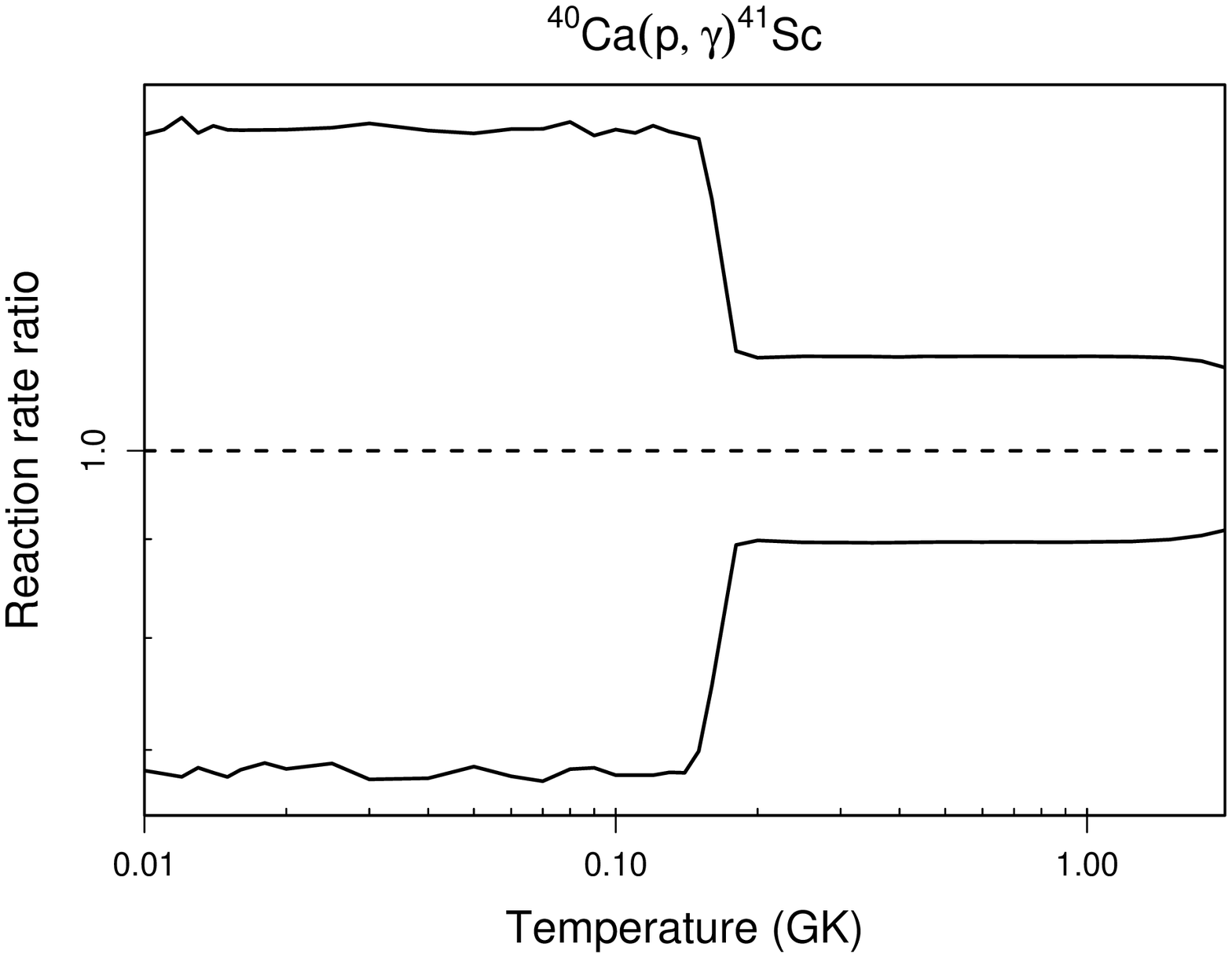}
\end{figure}
\clearpage
\begin{figure}[]
\includegraphics[height=18.5cm]{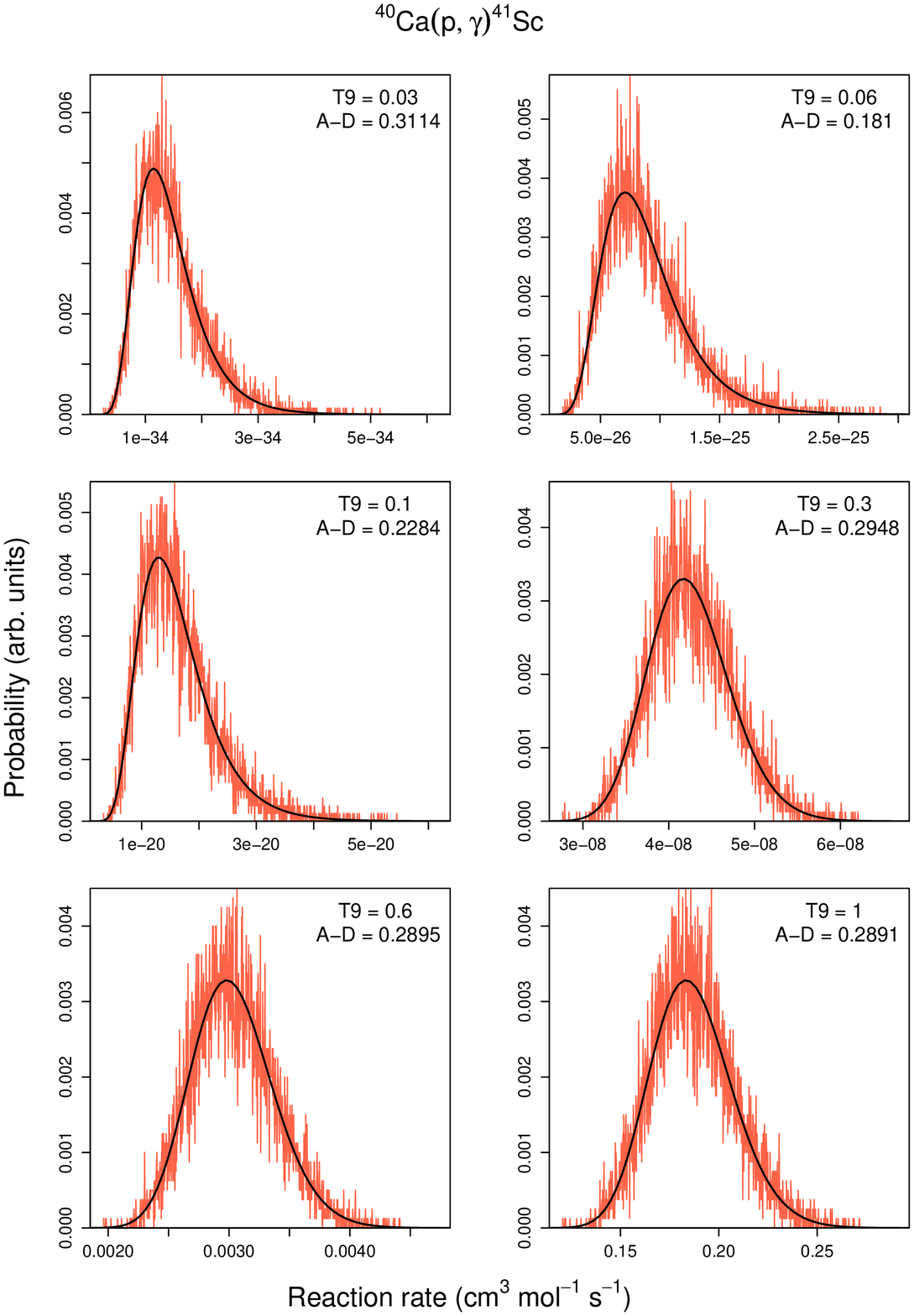}
\end{figure}
\clearpage


\end{document}